# Whole genome sequencing identifies putative associations between genomic polymorphisms and clinical response to the antiepileptic drug *levetiracetam*

Vavoulis DV[1*], Pagnamenta AT[1], Knight SJL[1], Pentony MM[1], Armstrong M[2], Galizia EC[3], Balestrini S[3,4], Sisodiya SM[3,4] & Taylor JC[1*]

1. NIHR Oxford Biomedical Research Centre, Wellcome Centre for Human Genetics, University of Oxford, Oxford, UK
2. New Medicines, UCB Pharma, Slough, UK
3. Department of Clinical and Experimental Epilepsy, UCL Queen Square Institute of Neurology, London, UK
4. Chalfont Centre for Epilepsy, London, UK.

*corresponding authors

## ABSTRACT
In the context of pharmacogenomics, whole genome sequencing provides a powerful approach for identifying correlations between response variability to specific drugs and genomic polymorphisms in a population, in an unbiased manner. In this study, we employed whole genome sequencing of DNA samples from patients showing extreme response (n=72) and non-response (n=27) to the antiepileptic drug levetiracetam, in order to identify genomic variants that underlie response to the drug. Although no common SNP (MAF>5%) crossed the conventional genome-wide significance threshold of $5 \times 10^{-8}$, we found common polymorphisms in genes *SPNS3*, *HDC*, *MDGA2*, *NSG1* and *RASGEF1C*, which collectively predict clinical response to levetiracetam in our cohort with ~91% predictive accuracy (~94% positive predictive value, ~85% negative predictive value). Among these genes, *HDC*, *NSG1*, *MDGA2* and *RASGEF1C* are potentially implicated in synaptic neurotransmission, while *SPNS3* is an atypical solute carrier transporter homologous to *SV2A*, the known molecular target of levetiracetam. Furthermore, we performed gene- and pathway-based statistical analysis on sets of rare and low-frequency variants (MAF<5%) and we identified associations between genes or pathways and response to levetiracetam. Our findings include a) the genes *PRKCB* and *DLG2*, which are involved in glutamatergic neurotransmission, a known target of anticonvulsants, including levetiracetam; b) the genes *FILIP1* and *SEMA6D*, which are involved in axon guidance and modelling of neural connections; and c) pathways with a role in synaptic neurotransmission, such as *WNT5A-dependent internalization of FZD4* and *disinhibition of SNARE formation*. Targeted analysis of genes involved in neurotransmitter release and transport further supports the possibility of association between drug response and genes *NSG1* and *DLG2*. In summary, our approach to utilise whole genome sequencing on subjects with extreme response phenotypes is a feasible route to generate plausible hypotheses for investigating the genetic factors underlying drug response variability in cases of pharmaco-resistant epilepsy.

## FUNDING
The research was supported by the National Institute for Health Research (NIHR) Oxford Biomedical Research Centre based at Oxford University Hospitals NHS Trust and University of Oxford. This work was supported by Epilepsy Society, UK. UCB Pharma (Brussels, Belgium) funded the sequencing and bioinformatics work and provided input into the analytical approaches used. Part of this work was undertaken at University College London Hospitals, which received a proportion of funding from the NIHR Biomedical Research Centres funding scheme. S Balestrini was supported by the Muir Maxwell Trust. The views expressed are those of the author(s) and not necessarily those of the NHS, the NIHR or the Department of Health. This publication presents independent research commissioned by the Health Innovation Challenge Fund (R6-388 / WT 100127), a parallel funding partnership between the Wellcome Trust and the Department of Health. The views expressed in this publication are those of the authors and not necessarily those of the Wellcome Trust or the Department of Health.

## ACKNOWLEDGMENTS
The authors would like to thank Dr Chris Spencer of Genomics PLC Oxford for critically reviewing an earlier version of the paper.

# INTRODUCTION

The advent of next-generation sequencing (NGS) has made possible the routine reconstruction of an individual's genetic variation profile across their whole genome[1,2], while the introduction of NGS to clinical practice brings closer the promise of personalised medicine for diagnostic sensitivity and therapeutic precision[3,4]. In the context of pharmacogenomics, whole exome and genome sequencing combined with appropriate bioinformatics and statistical analysis has the potential to identify variants that correlate with clinical response to specific drugs, in a comprehensive, high-resolution and unbiased manner[5–12], i.e. without the need for a prior hypothesis regarding the type (e.g. common or rare), location or distribution of genomic polymorphisms across the whole extent of the genome. We employed whole genome sequencing to better understand response variability to the antiepileptic drug levetiracetam (LEV), a third-generation first-line drug for the treatment of both focal and generalised epilepsies, for which high variability of clinical response is observed.

Experiments in mice show that *SV2A*, but not its paralogs *SV2B* and *SV2C*, is the molecular target of LEV[13]. *SV2A* is a synaptic glycoprotein with widespread distribution in the brain[14] and a crucial role in synaptic vesicle exocytosis[15]. Mice deficient in SV2 functionality exhibit severe seizures with a concomitant reduction in (inhibitory) GABAergic neurotransmission[16] and an abnormal presynaptic accumulation of calcium leading to increased neurotransmitter release[17]. LEV inhibits presynaptic calcium channels[18] and calcium-dependent vesicle exocytosis[19], and it reverses synaptic deficits due to overexpression of *SV2A*[20]. However, its exact mechanism of action as an antiepileptic drug is not understood.

It is natural to hypothesize that LEV may act by modifying deregulated *SV2A*-dependent neurotransmission and that variability in SV2A functionality may explain differential responsiveness to treatment with LEV. This view is supported by reports showing that partial loss of SV2A functionality is linked to decreased LEV efficacy in several mice seizure models[21], or that levels of SV2A expression in tumour and peri-tumoral tissue predicts clinical response to LEV in patients with glioma[22]. However, neither common nor rare polymorphisms in *SV2A* (including polymorphisms overlapping its binding site with LEV) are associated with clinical response to the drug, based on targeted sequencing approaches[23,24]. Any role of genetic variation (either rare or common) in other genomic loci as potential predictors of LEV efficacy remains to be elucidated.

We analysed whole genome sequencing (WGS) data from 99 people with epilepsy, classified as extreme responders (n=72) or non-responders to LEV (n=27), aiming to explore the genetic differences between the two groups and to identify rare or common polymorphisms that may be predictive of the response/non response phenotype. Using whole genome sequencing (instead of targeted sequencing or genome-wide SNP arrays) facilitates the search for genetic predictors to LEV in a complete, high-resolution and unbiased manner. At the same time, a targeted search for genomic features associated with response to LEV is still possible. Here, we identified common polymorphisms which collectively predict a substantial fraction of clinical response to the drug in our cohort of patients with epilepsy. Furthermore, analysis of groups of low-frequency variants highlights significant associations between response to LEV and genes involved in synaptic neurotransmission, axon guidance and modelling of neural connections.

# METHODS

## Sample acquisition and whole genome sequencing

The study was approved by the relevant local ethics committee. Patients provided written informed consent, or in the case of people unable to provide consent, assent was obtained from parents or guardians as permitted within the approved protocol.

Ninety-nine unrelated adults with a range of types of epilepsy were recruited from the National Hospital for Neurology and Neurosurgery. Non-responders (n=27; ~27%) were defined as patients who had failed to respond to at least two of the currently established epilepsy treatments and had not responded to maximum tolerated doses of levetiracetam used for at least 12 months. Extreme responders were defined as patients who became seizure-free for at least 12 months after initiation of levetiracetam and who had not previously responded to at least three appropriately chosen and used antiepileptic drugs (AEDs; n=72; ~73%).

Samples from the above subjects were sequenced at the Oxford Genomics Centre using the HiSeq2500 platform, v3 chemistry and the 100bp paired-end read format (Illumina, San Diego, CA). Sequencing was performed across 2.3 lanes per sample at depth 30X (Figure 1).

## Bioinformatics analysis

Reads were mapped to hs37d5 using BWA[25] and duplicate reads were removed using the MarkDuplicates option from the Picard toolkit[26] all with default options. Variants were called simultaneously across all 99 samples with Platypus[27] v0.7.9.3 resulting in a multi-sample VCF file. Read alignments were checked visually using the Integrative Genomics Viewer v2.3.5[28].

In total, ~20M variants were called across all samples (Figure 1). We excluded variants in multi-allelic loci or in sex chromosomes, variants with FILTER flag other than PASS, and variants in homopolymers with running length larger than 8 base pairs (HP>8). We excluded genotypes of low quality (PHRED score GQ<20), and with less than 10 reads covering the variant location (DP<10). We also excluded variants not in Hardy-Weinberg Equilibrium[29] (p-value less than $10^{-6}$) and with missing genotypes in more than 2 individuals (~2%). Furthermore, we excluded variants in low complexity regions[30], in poor mappability regions[31], in segmental duplications[32] and in the top 1% most variable genes according to Ingenuity IVA[33]. On the remaining ~8.4M variants, we conducted principal component analysis using the *prcomp* function in R[34] and we identified 7 outlier samples, which were excluded from further analysis (Figure 2A,B). We did not find evidence of association between clinical response and sex in the remaining 92 patients (Figure 2C). The filtered data were annotated using the Ensembl Variant Effect Predictor[35] software v90.5 with allele frequency annotations provided by gnomAD r2.0.1[36] and variant IDs provided by dbSNP[37] build 150. Overall, we reviewed ~3.9M common variants (MAF>5%), and ~4M low-frequency (1%<MAF<5%) and rare variants (MAF<1%) (Figure 2D,E).

## Statistical analysis

We conducted single-variant tests on common variants, and gene- and pathway-based tests on low-frequency and rare variants (Figure 1). In the case of common variants, we calculated SNP-specific p-values by applying a two-tailed Fisher's exact test on each common variant (Figure 3A). In a pre-specified second stage, we selected a small subset of variants by using all variants with p-value less than the conventional suggestive genome-wide significance threshold of $10^{-5}$ (n=23 variants; Table S1 and Figure 3A, white dots) as predictors (along with sex) in a penalised logistic regression model[38] (known as the *LASSO*; Figure 3B). An optimal penalisation parameter was estimated using leave-one-out cross-validation. This resulted in the selection of 10 out of 23 variants with maximal predictive power (Figure 3B, red dots). An additional selection step was applied by filtering out all variants (among those selected by the LASSO in the previous step) that had non-protein-coding gene annotation or were annotated as *intergenic*. This resulted in the final selection of 5 variants with protein-coding gene annotations (Table S1). The reason for this final selection step was to avoid overfitting during the downstream analyses described below and because the selection of variants in protein-coding genes (instead of non-protein coding or intergenic variants) facilitates the subsequent investigation of their possible biological relevance. After variant selection, we conducted an analysis of deviance by examining a series of logistic regression models using response to LEV as the dependent variable (Table S2). The BASIC model includes, besides the intercept, a single predictor, sex. The FULL model includes in addition the genotypes of the previously selected variants. A number of intermediate models are simple extensions of the BASIC model through the inclusion of just one of these variants. Finally, we calculated the predictive power of the FULL model using leave-one-out cross-validation and the accuracy (ACC), sensitivity (TPR), specificity (TNR), positive (PPV) and negative (NPV) predictive values, and Matthews correlation coefficient (MCC) as metrics of predictive power. For completeness, we also conducted auxiliary statistical analyses, which included a genome-wide Bayesian analysis and calculation of bespoke genome-wide significance thresholds (see Supplementary Material for more details).

In the case of rare and low-frequency variants, we first calculated a variant-specific p-value by applying a two-tailed Fisher's exact test, as in the case of the common variants. Subsequently, we aggregated all variant-specific p-values in a gene- or pathway-specific statistic using an appropriately corrected Fisher's product method[39] (see Supplementary Material), which takes into account the effective number of independent variants in a group of variants, thus correcting for correlations between variants in the same gene or pathway. The resulting statistic was used to calculate a gene- (Table S3) or pathway-specific (Table S4) p-value for testing the null hypothesis that none of the variants in the gene/pathway are associated with response to LEV, against the alternative hypothesis that at least one variant in the set is associated with response to LEV. P-values were corrected for multiple hypothesis testing across all genes or pathways using Sidak's method.

Finally, we conducted a targeted analysis of common and rare variants in a set of genes implicated in neurotransmitter transport and release and in a set of genes associated with epilepsy (Table S5). For the common variants, we tested each variant individually using a two-tailed Fisher's exact test of independence, as above. We used Sidak's method for multiplicity correction across all genes in each of the two sets. The effective number of independent variants was estimated by first calculating a gene-specific estimate of the number of independent variants using four alternative methods[39–43], followed by summing these estimates over all genes. All four methods returned consistent results. For the rare variants, we calculated gene-specific p-values followed by multiplicity correction using the Sidak method, as before.

More details on the statistical analysis are given in the Supplementary Material.

## RESULTS

### Common polymorphisms in genes *SPNS3*, *HDC*, *NSG1*, *MDGA2* and *RASGEF1C* predict clinical response to LEV in our cohort with overall accuracy ~91%

We constructed a statistical model that utilises common genomic variation to predict response to LEV in our cohort. Towards this aim, we first assessed the significance of association between each SNP and response to LEV (Figure 3A). The smallest SNP-specific p-value calculated at this stage was $1.6 \times 10^{-7}$, i.e. no p-value crossed the conventional genome-wide significance threshold of $5 \times 10^{-8}$ (Table S1). This was followed by a principled SNP selection process (see Methods) to identify a minimal set of highly predictive variants (n=5 variants). These are located in the protein-coding genes *SPNS3*, *HDC*, *NSG1*, *MDGA2* and *RASGEF1C*, as indicated by the non-zero coefficients in Figure 1B. Variants with non-zero coefficients in the non-coding genes *RP11-284F21.8*, *RP11-446J8.1* and *RP11-650J17.1*, as well as two intergenic variants in chromosome 15, were not included, in order to keep the model small and avoid overfitting (see Methods for rationale). All these variants are listed in Supplementary Table S1.

At the next stage, we conducted an analysis of deviance on the polymorphisms identified in the previous step (see Methods and Table S2). We found that the inclusion of these SNPs in a logistic regression model reduces the residual deviance from ~107 (BASIC model) to ~28 (FULL model), thus significantly improving the goodness of fit (p-value=$1.15 \times 10^{-15}$ based on a $\chi^2$ test) of the model to the data. The fraction of explained deviance in the data was assessed using a pseudo-$R^2$ metric, the adjusted $D^2$, as described in Guisan & Zimmermann[44]. The BASIC and FULL models have an adjusted $D^2$ equal to 1% and 73%, respectively, which implies that the identified variants in genes *SPNS3*, *HDC*, *NSG1*, *MDGA2* and *RASGEF1C* collectively explain ~72% of the total deviance (Table S2). When considering just a single gene as predictor (as in any of the intermediate models between BASIC and FULL), the improvement in model fit is significant (as indicated by the low p-values). Furthermore, the proportion of explained deviance by SNPs in each gene ranges between 10% (*HDC*) and 21% (*SPNS3*), as inferred by comparing the adjusted $D^2$ value for each of the intermediate models to the adjusted $D^2$ value of the BASIC model.

Subsequently, we assessed the predictive power of the FULL model using leave-one-out cross-validation. In brief, this involves fitting the FULL model in all but one subjects and predicting the response phenotype of the held-out subject using the fitted model. This process of model fitting and prediction is repeated until all 92 subjects have been used for prediction. We found that the FULL model correctly predicts clinical response to LEV in 62 responders and 22 non-responders, which corresponds to ~94% sensitivity (TPR) and positive predictive value (PPV), ~85% specificity (TNR) and negative predictive value (NPV), and ~91% overall predictive accuracy (ACC). The Matthews correlation coefficient (MCC), a balanced performance metric for binary classifiers even when the two classes are of very different size, was equal to ~79%.

*Local genomic structure near the identified variants and possible biological relevance*

For gene *NSG1* on chromosome 4, three highly correlated SNPs (rs7695197, rs3981 and rs12641832) are located ~5kb upstream of the gene, less than 3kb upstream or downstream of transcription factor binding sites (TFBS) and DNaseI hypersensitivity sites (DHS), and less than 5kb upstream of a small cluster of conserved elements (CE; Figure 4A). The odds ratio for a recessive model (with respect to the ALT allele) is ~23 times in favour of the non-responders, while the corresponding odds ratio for a dominant model is ~2.7 (see Table S1 for the number of homozygous/heterozygous cases in each group). In other words, non-responders to LEV are ~23 times more likely to be homozygous for the alternative allele than responders. *NSG1* (Neuronal Vesicle Trafficking Associated 1) is abundantly expressed in the brain[45,46] and it plays a role in synaptic neurotransmission and plasticity due to its involvement in recycling and trafficking of receptors, such as the glutamate receptor AMPA, the amyloid precursor protein (APP), and the L1 cell adhesion molecule (L1CAM)[47].

The intronic variant rs34570575 in gene *RASGEF1C* on chromosome 5 overlaps a DHS and it is located ~5kb upstream of a TFBS and a cluster of CE (Figure 4B). The odds ratio for a dominant model of inheritance (with respect to the ALT allele) is slightly higher than that of a recessive model (~9.5 and ~8, respectively; Table S1). *RASGEF1C* (RAS guanyl-nucleotide exchange factor domain family member 1C) is abundantly expressed in the brain[45,46]. It belongs to a family of proteins containing the RASGEF domain, which regulates the GTPase activity of RAS-like proteins. These comprise a superfamily of membrane-associated signalling molecules involved in a variety of essential cellular processes, including vesicle trafficking and synaptic function[48–50].

In gene *MDGA2* on chromosome 14, rs1952220 is an intronic variant, less than ~4kb from CE, TFBS and DHS (Figure 4C). The odds ratios for recessive and dominant models (with respect to the ALT allele) are 0.11 and 0.61 in favour of the non-responders, respectively, suggesting a recessive model where non-responders to LEV are ~9 times less likely to be homozygous for the alternative allele than responders (Table S1). The *MDGA2* (MAM Domain Containing Glycosylphosphatidylinositol Anchor 2) mRNA is expressed in the cerebral cortex[45,46]. MDGAs are Ig superfamily cell adhesion molecules that contribute to the radial migration of cortical neurons during early neural development. They play an important, neuroglin-2-dependent role in controlling the function of inhibitory synapses, and they have been associated with autism spectrum disorders and schizophrenia[51,52].

In gene *HDC* on chromosome 15, rs7182203 is an intronic variant that overlaps a TFBS and a DHS, and it is within 5kb of upstream or downstream CE (Figure 4D). From Table S1, the odds ratios for recessive and dominant models (with respect to the ALT allele) are 1.1 and 0.12 in favour of the non-responders, respectively. This implies that patients that respond to LEV are ~8 times more likely to be homozygous or heterozygous for the alternative allele in comparison to non-responders. *HDC* (histidine decarboxylase) is expressed in the brain[45,46], and it catalyses the synthesis of histamine, which is implicated, among others, in neurotransmission and smooth muscle tone. Elevated levels of histamine in the brain appear to suppress seizures and confer neuroprotection, thus antiepileptic agents that boost the levels of histamine in the brain may act by increasing *HDC* activity[53]. Furthermore, *HDC* has been linked to the pathogenesis of Tourette's syndrome[54]. Interestingly, LEV has been used for the treatment of Tourette's syndrome, although its efficacy has not been established[55–57].

Finally, in gene *SPNS3* on chromosome 17, the intronic variants rs2047231, rs2047232 and rs2047233 overlap a DHS and a cluster of CE, and they are located within 5kb of upstream or downstream TFBS (Figure 4E). From Table S1, the odds ratio for a recessive model (with respect to the ALT allele) ranges among these three SNPs between 0.07 and 0.09 in favour of non-responders. This implies that patients responding to LEV are between ~11 and ~14 times more likely to be homozygous for the alternative allele than non-responders. *SPNS3* (a putative sphingolipid transporter 3) is expressed in the cerebral cortex[45,46]. Both *SPNS3* and *SV2A,* the known target of LEV, are atypical solute carrier (SLC) transporters. They belong to the Major Facilitator Superfamily (MFS) of membrane transporters, and they share a common structure consisting of 12 transmembrane segments, which is necessary for optimal transporter activity[58,59].

**Tests on sets of low frequency variants (MAF<5%)**

Next, we studied variants with MAF<5%, i.e. low-frequency and rare variants. Among the approximately 4M variants with MAF<5%, we focused on the top 5% genotypically most variable variants across all 92 samples in our cohort. These included ~182K variants with MAF between 0.003% and 5%. A common strategy for increasing statistical power when studying low-frequency and rare variants is to analyse sets of variants, instead of individual variants. Therefore, we examined gene- and pathway-based sets of variants (see Methods).

*Gene-based tests indicate that low-frequency variants in genes PRKCB, DLG2, FILIP1, SEMA6D and LINC01090 are associated with response to LEV*

We conducted 19,824 gene-based tests, which is the number of genes harbouring at least one of the ~182K low-frequency and rare variants in our data. We found that four protein-coding genes (*PRKCB*, *DLG2*, *FILIP1* and *SEMA6D*) and a long intergenic non-protein-coding RNA (*LINC1090*) had a Family-Wise Error Rate (FWER) less than 10%, and they were kept for further study (Table S3 and Figure 5).

The top hit, *PRKCB*, encodes a protein kinase C, a family of serine- and threonine-specific protein kinases, which can be activated by calcium and second messenger diacylglycerol[47]. There are 78 variants in *PRKCB* with MAF between 2.2% and 4.9%. Forty-five of them have p-values less than 0.05 and they aggregate towards the 5' end of the gene (Figure 5A). Associated Reactome pathways are *glutamate binding*, *activation of AMPA receptors* and *synaptic plasticity*[60]. *PRKCB* is implicated in the trafficking of GluR2-containing AMPA receptors[60]. It is known that fast synaptic excitation relevant to epilepsy is mediated mainly by AMPA receptors, thus rendering the latter potential targets of antiepileptic treatment[61]. There is evidence suggesting that LEV interacts with AMPA receptors[62] and that its antiepileptic action is mediated by inhibiting glutamatergic neurotransmission through presynaptic calcium channels[63], but the precise molecular mechanism that mediates its action remains unclear.

A second hit of interest, *DLG2*, encodes a membrane-associated guanylate kinase, which is implicated in the clustering of receptors (including NMDARs), ion channels, and associated signalling proteins at postsynaptic sites of excitatory synapses[47]. We found 208 variants in this gene with MAF between 0.96% and 4.97%, 53 of which have p-values less than 0.05 (Figure 5B). A related Reactome pathway is *protein-protein interactions at synapses*[60]. There is evidence supporting the role of NMDARs in epilepsy, and as a potential therapeutic target of antiepileptic drugs, including LEV[64]. It is possible that LEV blocks epileptiform bursting induced by NMDA *in vitro* without affecting normal synaptic transmission[65] and that it inhibits NMDA-dependent excitatory postsynaptic currents[63], although its precise molecular mechanism of action remains unclear.

Among the remaining three hits (Figure 5C-E), *FILIP1* includes 12 variants (11 with p-values less than 0.01) with MAF between 1.8% and 5%, *SEMA6D* has 41 variants (17 with p-values less than 0.05) with MAF between 1.9% and 5% and *LINC01090* harbours 35 variants (18 with p-values less than 0.01) with MAF between 1.7% and 5%. *FILIP1* encodes a protein that stimulates filamin A degradation, which may regulate cortical neuron migration, dendritic spine morphology, and normal excitatory signalling[47]. *SEMA6D* encodes a transmembrane semaphorin, a class of proteins involved in axon guidance, and maintenance and remodelling of neural connections[47]. Finally, *LINC01090* is transcribed into a long intergenic non-protein-coding RNA[47], which is associated with post-traumatic stress disorder[66].

***Pathway-based tests indicate that associations between response to LEV and Reactome pathways are driven mainly by low-frequency variants in gene PRKCB***

We conducted tests using gene sets, instead of single genes, as the organisational unit for grouping individual variants together. We have used all pathways from *Reactome*, a curated, peer-reviewed database of interacting signalling and metabolic molecules, which are organised into groups of higher order structures (pathways) with well-defined biological relevance[60]. In total, we considered 2,028 pathways, of which 1,979 harboured at least one of the ~182K low-frequency highly-variable variants in our data. Among these, we identified six pathways with FWER<5% and one pathway with FWER<10% (Table S4).

The top hit is *activation of NF-kappaB (nuclear factor kappaB) in B cells*. NF-kappaB is a ubiquitous transcription factor, which is instrumental in gene regulation relevant to cell death and survival and to the immune system's response to inflammation. The next hit is *WNT5A-dependent internalization of FZD4*. WNT5A regulates multiple intracellular signalling cascades via internalisation of its receptors. These include FZD4, a member of the frizzled gene family, which encode seven-transmembrane domain proteins[47]. Importantly, the WNT5A-dependent uptake of FZD4 occurs in a clathrin-dependent manner[67]. Clathrins are adaptor proteins, which are essential in the formations of synaptic vesicles, and which are known to interact with SV2A, the molecular target of LEV[68].

Another interesting pathway is *disinhibition of SNARE formation*. SNARE is a family of proteins, which are important components of the mechanism responsible for membrane fusion, thus playing an important role in docking of synaptic vesicles with the presynaptic membrane, and neurotransmitter release. It is known that SV2A, the target of LEV, regulates the formation of SNARE complexes: kindling epileptogenesis triggers the long-term accumulation of both SNARE and SV2A in the ipsilateral hippocampus, a molecular process which is reversed by LEV[69].

Next, we asked which genes underlie these findings. In total, in these pathways, there are 108 genes harbouring low-frequency mutations (Table S4). In Figure 6, we illustrate these genes, as well as their pathway membership. *PRKCB* is mutated in all but the least significant pathway with FWER<10%, followed by its paralog, *PRKCA*, which is mutated in three pathways (central panel). The remaining genes are mutated in only 1 or 2 pathways. Furthermore, *PRKCB* and *PRKCA* harbour the largest number of low-frequency mutations, along with *RUNX1* (top panel). However, in *PRKCB*, more than half of these mutations have p-values less than 0.05 (see also Table S3) leading to a low gene- and pathway-based p-value (Tables S3 and S4), while only a very small proportion of mutations in *PRKCA* and *RUNX1* have p-values less than 0.05. Among the other highly mutated genes, *LPR6* and *IKBKB* also harbour a large proportion of mutations with low p-values, but they participate in only 1 and 2 pathways, respectively. We conclude that the significant associations in Table S4 are driven mainly by *PRKCB* in all but the least significant pathway with FWER<10%. Associations in this last pathway (*disassembly of the destruction complex and recruitment of AXIN to the membrane*) are driven mainly by *LRP6*, a transmembrane low density lipoprotein (LDL) receptor[47]. Neuronal *LRP6*-mediated Wnt signalling is critical for synaptic function and cognition[70,71].

### Targeted analysis of genes implicated in neurotransmitter transport and release highlights the previously identified genes *NSG1* and *DLG2*, but not *SV2A*, *SV2B* or *SV2C*

Whole genome sequencing permits focused analysis of identified sets of variants, in addition to unbiased analysis over the whole genome. Furthermore, by testing only a small subset of variants, we can ameliorate the effect of multiple hypothesis testing, thus increasing the power of statistical analysis. We conducted targeted analysis of common and rare variants on a set of 294 genes implicated in neurotransmitter transport and release (Table S5). These genes (SYNAPTIC) were identified based on their Gene Ontology[72] terms and they included *SV2A* and its paralogs, *SV2B* and *SV2C*. In addition, we tested all 402 high-confidence (i.e. "green") genes (EPILEPSY) in the genetic epilepsy syndromes panel v1.35 provided by Genomics England[73] (Table S5).

In the case of SYNAPTIC genes, which harboured ~51K common variants, we found evidence of association with response to LEV in the previously identified gene *NSG1* at a FWER<5% or <10%, depending on the methodology used for calculating the effective number of independent tests (Table S5). Furthermore, when examining the ~2.6K rare variants found in SYNAPTIC genes, *DLG2* was found associated with response to LEV at a FWER<0.1% (Table S5). This is not surprising, since *DLG2* was also found associated with LEV response in the previously conducted gene-based tests. We did not find any evidence of association between LEV response and *SV2A* or its paralogs, *SV2B* and *SV2C*, using either SNP- or gene-based tests. Furthermore, we did not find any evidence of association between ~66K common variants in EPILEPSY genes and response to LEV or between ~3.7K rare variants in the same genes and response to LEV at a FWER<10% (Table S5).

### DISCUSSION

Although the anticonvulsant properties of the prominent antiepileptic drug LEV have been linked to the activity levels of its molecular target, the synaptic glycoprotein SV2A[21,22], targeted sequencing did not reveal any associations between common[24] or rare[23] variation in this gene and LEV efficacy. This leaves open the question as to whether genetic variation is a component of response variability and, if so, the identity of the genomic variants underlying clinical response to LEV. In the present study, we followed a whole genome sequencing approach in an unbiased search of genomic polymorphisms that underlie clinical response to this drug.

According to one possible hypothesis for explaining variability in drug response, one or more common polymorphisms occur with different frequencies between responders and non-responders. Our analysis indicates that common polymorphisms in genes *NSG1*, *HDC, MDGA2*, *RASGEF1C* and *SPNS3* collectively predict clinical response to LEV in our cohort with overall accuracy ~91%. These genes are attractive candidates, since the first four are potentially implicated in synaptic neurotransmission, while the fourth is a transmembrane transporter protein homologous to *SV2A*.

A second hypothesis asserts that multiple rare variants act synergistically to influence a patient's response to the drug. Our analysis showed that groups of low-frequency variants in genes *PRKCB*, *DLG2*, *FILIP1* and *SEMA6D*, and in pathways involving *PRKCB* (and *LRP6*) demonstrate significant associations with the response/non-response phenotype.

From a neurophysiological perspective, there are three major, not mutually exclusive hypotheses for explaining pharmaco-resistant epilepsy[74]. First, the drug target hypothesis postulates that alterations in the activity of the molecular target of the drug (e.g. due to genomic polymorphisms coding for the drug-target binding site) result in reduced drug efficacy. Our analysis did not provide any evidence that *SV2A* or its paralogs (*SV2B* and *SV2C*) are associated with response to LEV, in agreement with previous studies[23,24]

Second, the drug transporter hypothesis states that reduced efficacy of antiepileptic drugs are due to low concentration of the drug at its target site due to over-active efflux drug transporters. A common intronic polymorphism in *SPNS3*, a gene homologous to *SV2A*, may be of interest in relation to this hypothesis. Both *SPNS3* and *SV2A* (and its paralogs) are structurally similar to the solute carrier family 22 (SLC22), a large family of transmembrane drug transporters. It should, however, be emphasised that homology (as established through structural similarity) is not definitive proof of biological relevance.

Finally, the intrinsic severity hypothesis postulates that severe epilepsy (manifested, for example, as high-frequency seizures) is linked to reduced response to antiepileptic drugs. Neurophysiological processes that are proposed to underlie the severity of epilepsy include neuroinflammation, aberrations in synaptic neurotransmission, and restructuring of neural networks[75]. Our analysis has identified common and low-frequency polymorphisms in genes and pathways, which are putatively related to these processes; for example, genes *HDC*, *NSG1*, *MDGA2*, *RASGEF1C*, *PRKCB* and *DLG2* (synaptic neurotransmission) and genes *FILIP1* and *SEMA6D* (restructuring of neural networks).

Whilst highlighting the approaches now available through the advent of NGS technologies, the findings in the present study need independent replication and potentially functional validation to confirm their role in determining response to LEV. Furthermore, we expect that the rapidly decreasing cost of WGS will allow conducting similar studies with a larger sample size in the near future. Nevertheless, our approach of using extremes of response is a pragmatic way to derive hypotheses for experimental testing. It is interesting to postulate what the remaining factors are that determine response to LEV. Drug response is likely to be a complex interaction of many factors, including interacting genetic factors, which should be explored through polygenic risk score analysis and integrative analysis of multiple data modalities utilizing machine learning approaches.

In summary, we have identified common and low-frequency variants in genes and pathways, which may influence clinical response to LEV in a cohort of 99 patients with epilepsy. We conclude that whole genome sequencing can be a useful approach for investigating the genomic correlates of pharmaco-resistant epilepsy.

# FIGURES

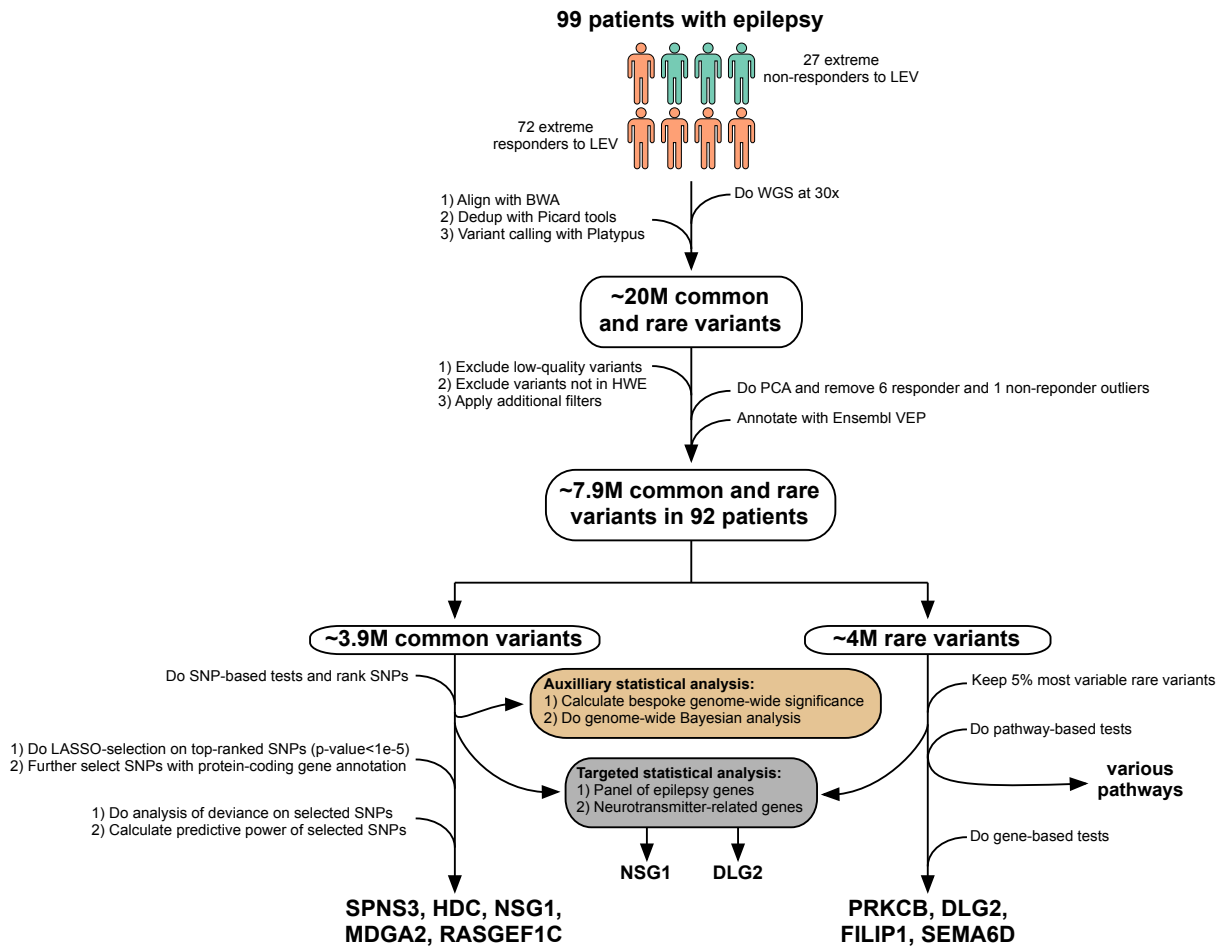

**Figure 1:** Overview of the study. We recruited 99 patients with epilepsy (72 extreme responders and 27 extreme non-responders to LEV). After performing WGS, alignment and variant calling, we identified ~20M unfiltered variants. After filtering across variants and samples, we ended up with ~3.9M common (MAF>5%) variants and ~4M low-frequency and rare (MAF<5%) variants across 92 patients. Subsequently, we calculated p-values for each common variant using a two-tailed Fisher's exact test. In the next step, we performed penalised logistic regression (LASSO) on all common variants with p-value less than the suggestive genome-wide significance threshold of $10^{-5}$ (n=23 variants; Supplementary Table S1). This was followed by further selecting variants with protein-coding gene annotation. In the last step, we performed analysis of deviance on the finally selected variants (n=5 variants) and we calculated their collective predictive accuracy using a cross-validation approach. For completeness, we also conducted additional auxiliary statistical analyses on the common variants (see Supplementary Material). In the case of low-frequency and rare variants, we focused on the top 5% most variable variants in our cohort and, by performing gene- and pathway-based tests on these, we identified associations between several genes or pathways and clinical response to LEV. Finally, for both common and low-frequency/rare variants, we conducted targeted analysis on a panel of epilepsy genes and on genes related to neurotransmitter transport and release.

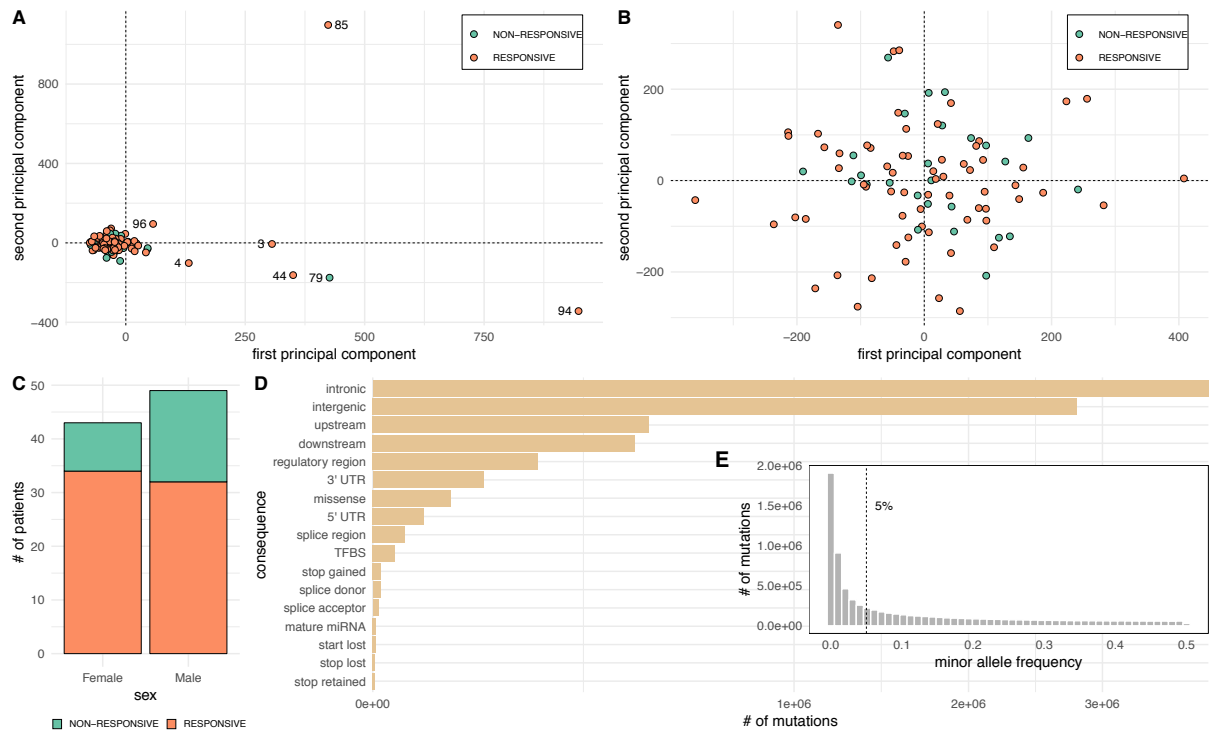

**Figure 2:** Overview of the WGS data from 99 extreme responders and non-responders to LEV. A) Principal component analysis (PCA) of the matrix of genotypes across all samples and variants. The first two principal components are illustrated. Seven samples appear as outliers. B) Repeating the PCA after removing the seven outliers identified in (A) indicates lack of any stratification (e.g. due to population structure) in the data. C) Number of male and female subjects among responders and non-responders to LEV. There are almost twice as many non-responders among 49 males (n=17), as among 43 females (n=9) in the data. A two-tailed Fisher's exact test of independence indicates that this difference is not statistically significant (odds ratio: 1.99; 95% CI: 0.72-5.86; p-value: 0.17). D) Consequences of all variants identified by WGS. Most variants are intronic, intergenic, or located immediately upstream or downstream of protein-coding genes. E) Minor allele frequencies (MAF) of all variants identified by WGS. A cut-off of 5% was chosen to discriminate between common and low-frequency or rare variants.

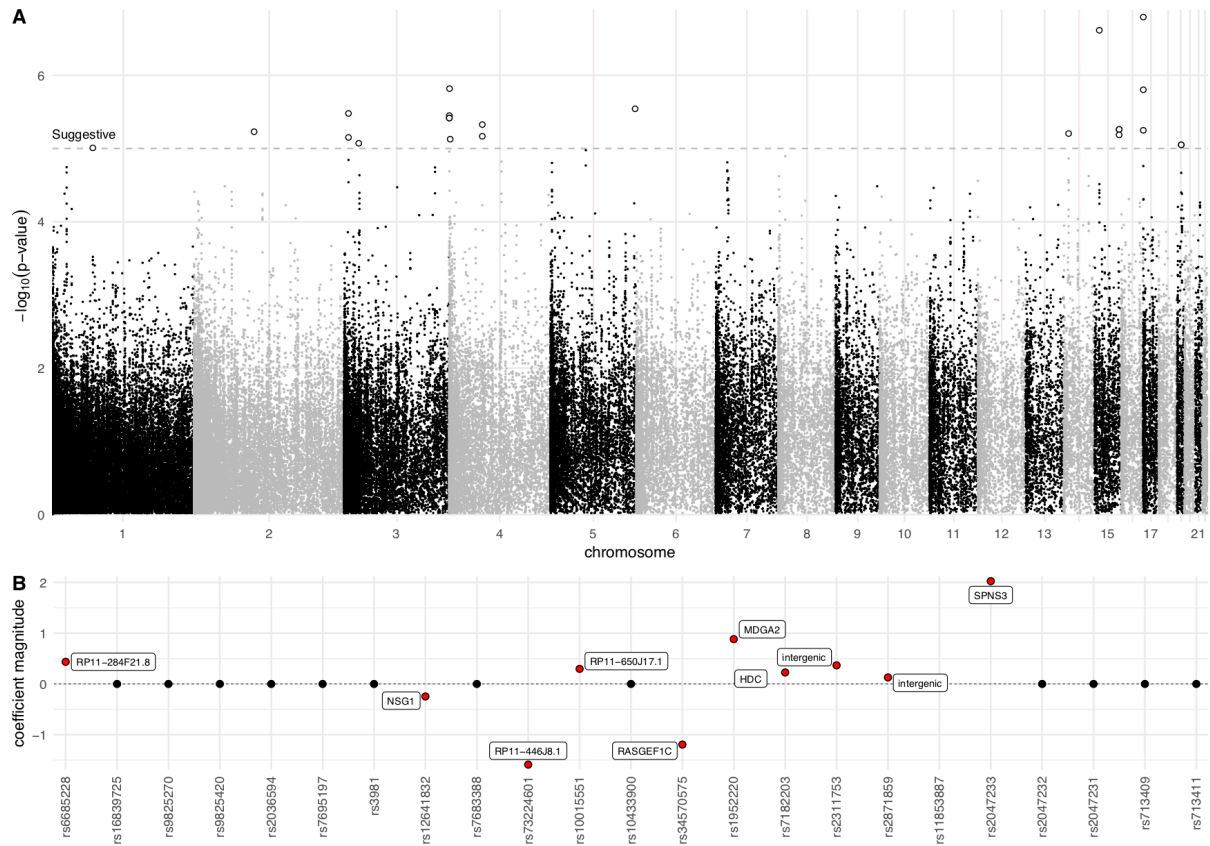

**Figure 3:** Genome-wide selection of a minimal set of common (MAF>5%) variants with maximal predictive power. A) Manhattan plot summarising SNP-based tests using a two-tailed Fisher's exact test of independence. All variants with p-values below a suggestive significance threshold of $10^{-5}$ are indicated with white circles (n=23). B) Summary of variable selection using penalised logistic regression (LASSO). All SNPs crossing the suggestive genome-wide significance threshold in (A) were used as predictors. Variants selected through this process have non-zero regression coefficients (red dots). Among these, the variants with protein-coding gene annotation (i.e. *SPNS3*, *HDC*, *MDGA2*, *NSG1* and *RASGEF1C*) were selected for further analysis.

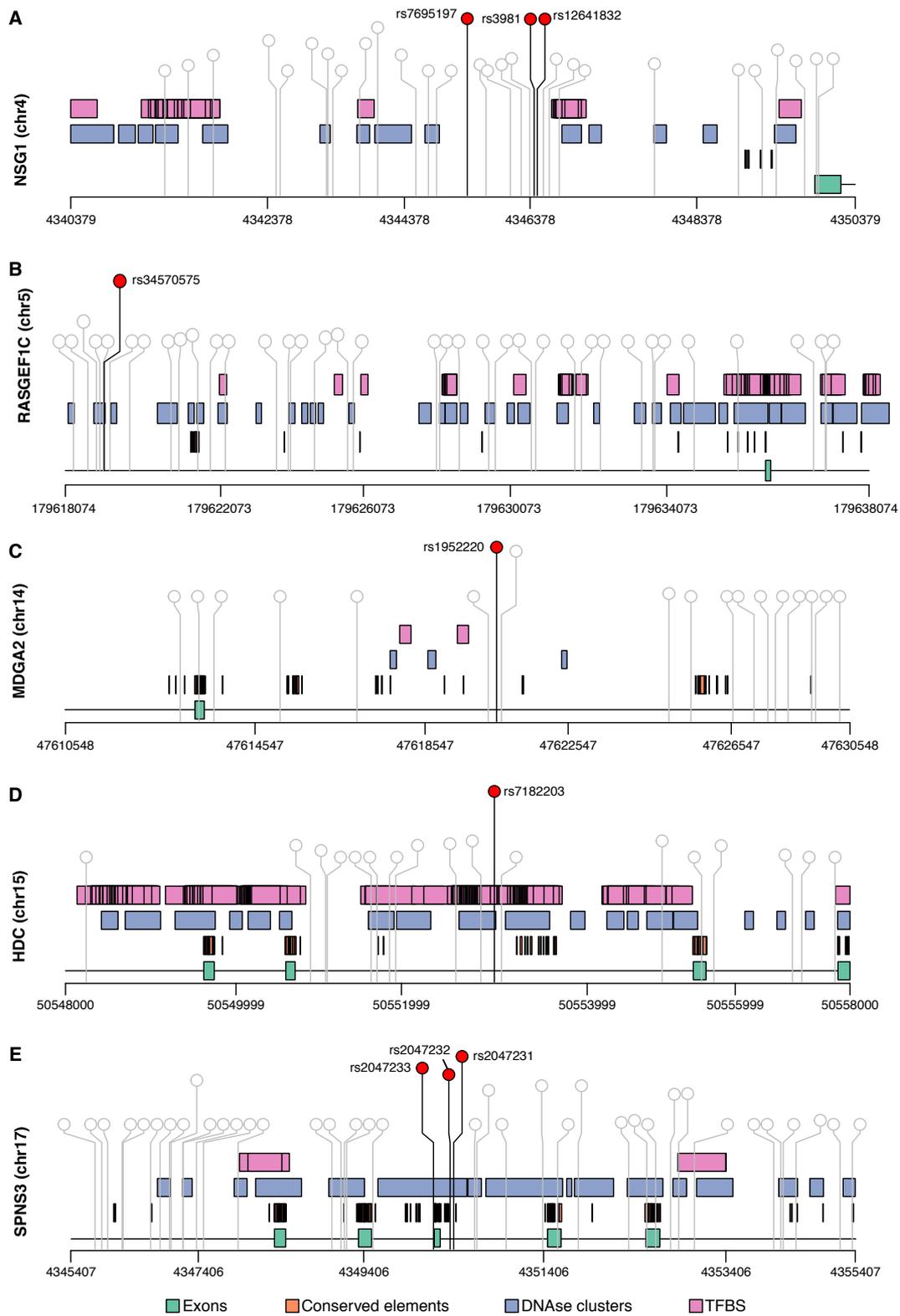

**Figure 4:** Local genomic structure near the most significant SNPs in genes *NSG1*, *RASGEF1C*, *MDGA2*, *HDC* and *SPNS3*. Common variants in these genes are strong predictors of clinical response to LEV in our cohort. SNPs crossing the suggestive genome-wide significance threshold of $10^{-5}$ are indicated in red.

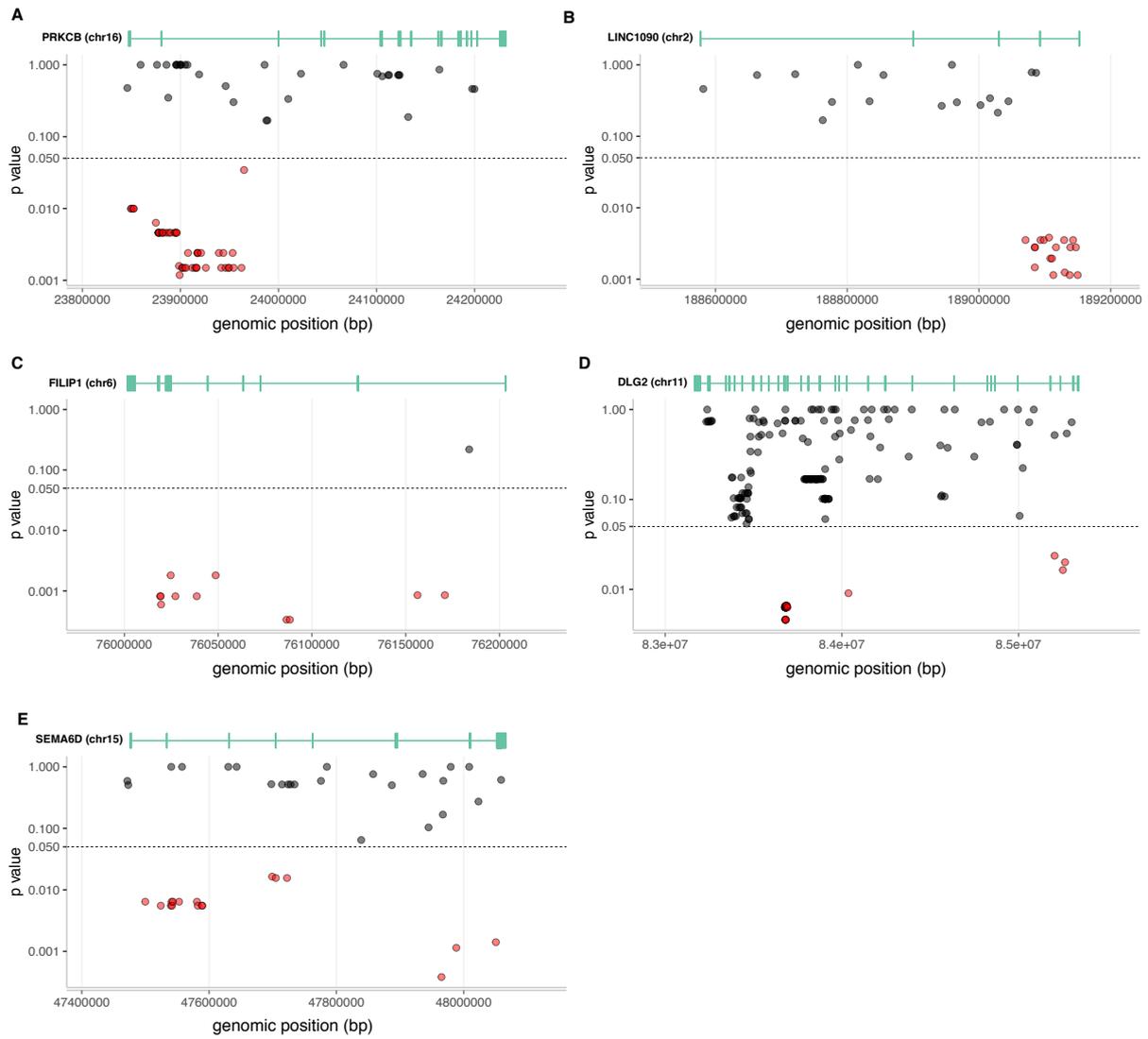

**Figure 5:** Distribution of rare variants along genes *PRKCB*, *LINC01090*, *FILIP1*, *DLG2* and *SEMA6D*. Based on gene-based tests, these genes are significantly associated with response to LEV at a FWER<10%. Rare variants with p-values below 5% (calculated using a two-tailed Fisher's exact test of independence) are indicated in red.

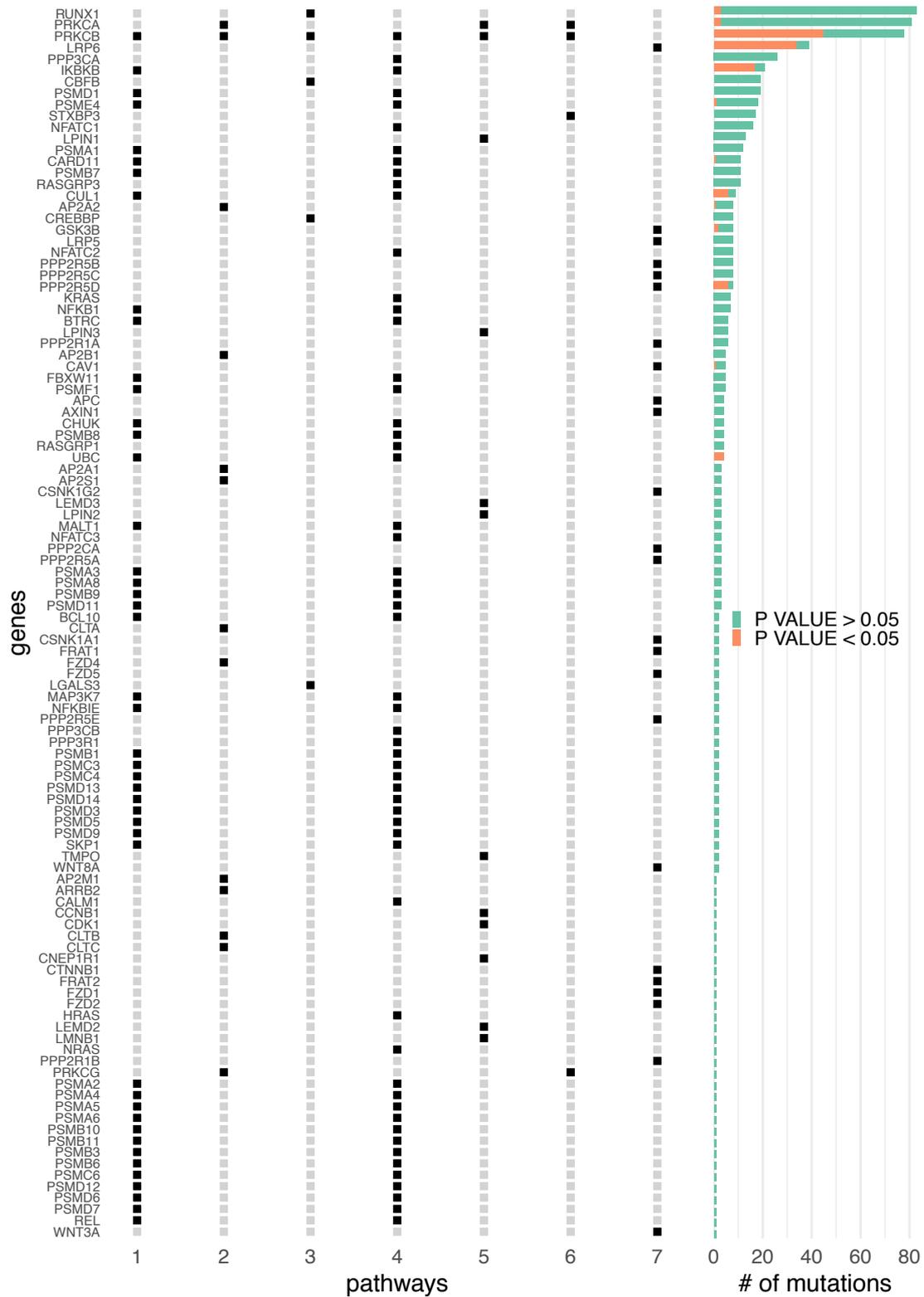

**Figure 6:** Summary of the results from Reactome pathway-based tests. In the left panel, pathways are as follows: *1-Activation of NF-kappaB in B cells; 2-WNT5A-dependent internalization of FZD4; 3-RUNX1 regulates transcription of genes involved in differentiation of myeloid cells; 4-Downstream signalling events of B Cell Receptor (BCR); 5-Depolymerisation of the Nuclear Lamina; 6-Disinhibition of SNARE formation; 7-Disassembly of the*

*destruction complex and recruitment of AXIN to the membrane*. Among all genes harbouring the highest number of rare variants (*RUNX1*, *PRKCA*, *PRKCB*, *LRP6*), *PRKCB* and *LRP6* have the highest proportion of rare variants with p-values less than 5% (right panel). *PRKCB* is participating in all but one pathway, with the remaining genes participating in only 1, 2 or 3 pathways (left panel).

# SUPPLEMENTARY MATERIAL
## Whole genome sequencing identifies putative associations between genomic polymorphisms and clinical response to the antiepileptic drug *levetiracetam*

Vavoulis DV[1*], Pagnamenta AT[1], Knight SJL[1], Pentony MM[1], Armstrong M[2], Galizia EC[3], Balestrini S[3,4], Sisodiya SM[3,4] & Taylor JC[1*]

1. NIHR Oxford Biomedical Research Centre, Wellcome Centre for Human Genetics, University of Oxford, Oxford, UK
2. New Medicines, UCB Pharma, Slough, UK
3. Department of Clinical and Experimental Epilepsy, UCL Institute of Neurology, London, UK
4. Chalfont Centre for Epilepsy, London, UK.

*corresponding authors

## STATISTICAL METHODS

### Statistical tests on single variants

For each common variant (and without assuming a particular genetic model), we constructed a 2x3 contingency table of counts, where the rows correspond to responders and non-responders (i.e. controls and cases, respectively), and the columns correspond to Ref/Ref, Ref/Alt and Alt/Alt genotypes. We analysed each such SNP-specific table using a two-sided Fisher's exact test of independence, which unlike the $\chi^2$ test or the Cohran-Armitage test for trend under an additive genetic model, is applicable even when the counts in any cell of the contingency matrix become very small (e.g. <5).

### Auxiliary statistical analysis of single variants

The genome-wide significance threshold for Type 1 error control due to multiple hypothesis testing is usually set at $5\times10^{-8}$, although it has been suggested[1] that this can be relaxed to the less stringent value of $10^{-7}$. In this study, we have calculated a bespoke significance threshold using the Sidak correction[2], in order to keep the family-wise error rate (FWER) below $\alpha=10\%$. Specifically, we calculated the significance threshold as $p_{thr} = 1-(1-\alpha)^{1/Neff}$, where *Neff* is the effective number of independent tests. For the calculation of *Neff*, we split each chromosome in a large number of blocks each containing 1000 variants, we calculate *Neff* in each block, and we sum the block-specific *Neff* values across all chromosomes to obtain a genome-wide estimate of *Neff*. For the actual calculation of *Neff* in each block, we applied four different methods[3–7], all of which operate on the eigenvalues of the correlation matrix of the SNP allele counts. Depending on which method is used, the estimated effective number of independent tests/SNPs was approximately equal to 3.6M[3,4], 358K[5], 259K[6] or 161K[7], leading to genome-wide significance thresholds equal to $3.0\times10^{-8}$, $3.0\times10^{-7}$, $4.1\times10^{-7}$ and $6.6\times10^{-7}$, respectively (Figure S1A; Table S1). Two SNPs (rs7182203 in gene *HDC* and rs2047231 in gene *SPNS3*, with p-values $2.43\times10^{-7}$ and $1.60\times10^{-7}$, respectively) are above the three least conservative thresholds[5–7], while 21 additional SNPs are above a suggestive significance threshold of $10^{-5}$ (Figure S1A; Table S6).

We have also applied a Bayesian approach to single-variant association testing, as previously described[8] (Figure S1B; Table S6). Specifically, we calculated the retrospective posterior probability of association (*rPPA*) for each SNP, which can be directly interpreted as the probability (given the data) that a particular SNP genotype is associated with the phenotype[8]. For each variant, we write: $rPPA = PO \times (1-PO)^{-1}$, where $PO = BF \times \pi \times (1-\pi)^{-1}$. In the previous expression, $PO$ are the posterior odds, $BF$ is the Bayes factor, and $\pi$ is the prior probability of association for the variant. This prior probability typically falls between $10^{-4}$ and $10^{-6}$, according to Stephens & Balding[8]. The Bayes factor is equal to:

$$BF = \frac{DirMult(n_{N0}, n_{N1}, n_{N2}|a,b,c) \times DirMult(n_{R0}, n_{R1}, n_{R2}|a,b,c)}{DirMult(n_0, n_1, n_2|a,b,c)}$$

where $DirMult(...)$ is the Dirichlet-Multinomial distribution; $n_{X0}$, $n_{X1}$, $n_{X2}$ are the number of responders ($X=R$) and non-responders ($X=N$) with Ref/Ref, Ref/Alt and Alt/Alt genotypes, respectively; $n_0$, $n_1$, $n_2$ are the total number of patients in each of the above genotype categories; and $a, b, c$ are the prior values of the concentration parameters of the Dirichlet-Multinomial distribution.

Although Bayesian approaches are, in principle, applicable regardless of the sample size or the number of variants entering the analysis, the calculation of $rPPA$ requires an explicit choice of priors $\pi$, $a$, $b$ and $c$. Here, we have assumed that $\pi=10^{-4}$, i.e. we expect a priori that 1 out of every 10K SNPs is truly associated with the phenotype (values of $rPPA$ for $\pi$ equal to $10^{-5}$ or $10^{-6}$ are also given in Table S6). The priors $a$, $b$ and $c$ were set either equal to 1 (flat prior), or equal to the empirical values $(a, b, c) = \left(\frac{n_0}{n}, \frac{n_1}{n}, \frac{n_2}{n}\right)$, where $n=n_0+n_1+n_2$ is the total number of patients. Both options return similar results, although the empirical prior is less conservative (see Table S6).

Under these explicit prior assumptions, we see that the previously identified variants rs7182203 (*HDC*) and rs2047231 (*SPNS3*) have *rPPA* values above 50%, i.e. they are more likely than not to be associated with drug response (Figure S1B; Table S6). In addition, the following variants also have an *rPPA* of more than 50%: rs7683388 (*HTRA3*), rs7695197 (*NSG1*), rs3981 (*NSG1*) and rs12641832 (*NSG1*).

**Statistical tests on groups of low-frequency and rare variants**

There are various approaches for testing groups of low-frequency and rare variants[9–11]. In this study, we followed a two-stage approach to group-based tests: a) first, we calculated for each variant in a group a p-value using a two-sided Fisher's exact test, as we did for the common variants, and b) we combined all variant-specific p-values in the same group into a group-specific statistic using an appropriately scaled Fisher's product method, which accounts for correlations between variants in the group[7]:

$$T = -2\frac{Meff}{M}\sum_{i=1}^{M} \log p_i$$

where $M$ is the total number of variants in the group, *Meff* is the effective number of uncorrelated variants in the group, and $p_i$ is the variant-specific p-value from Fisher's test. This statistic can be used to measure the strength of evidence that a particular group is associated with the response phenotype, and its distribution is approximated by a $\chi^2$ distribution with $2\times Meff$ degrees of freedom[7], which is used to derive a group-specific p-value, $p_j$. Assuming $N$ total groups, we use the Sidak correction to calculate an adjusted-for-multiplicity p-value, $p_{adj,j} = 1-(1-p_j)^N$. Groups are defined as either genes or pathways.

**SUPPLEMENTARY FIGURES**

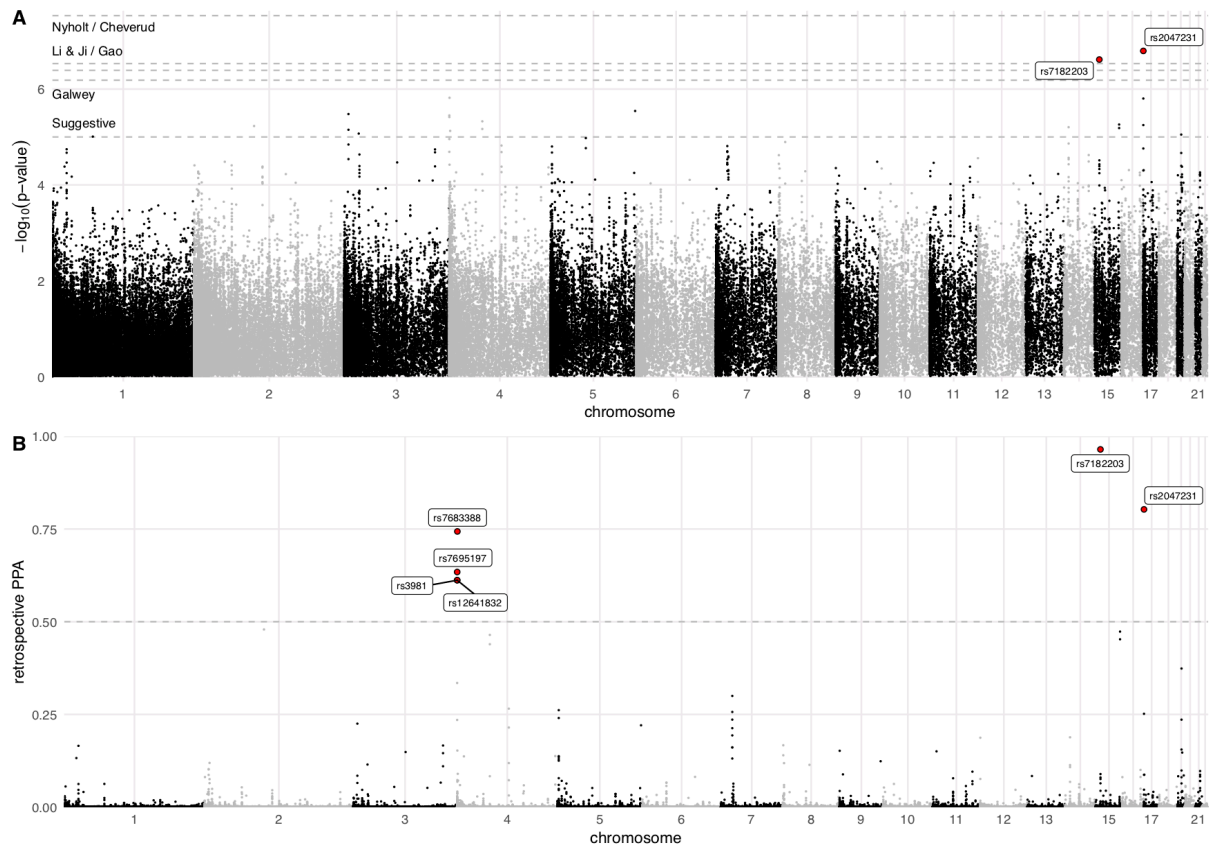

**Figure S1:** Auxiliary statistical analysis of common (MAF>5%) variants across the whole genome. A) Manhattan plot summarising SNP-based tests using a two-tailed Fisher's exact test of independence. Genome-wide significance thresholds (calculated using four different methodologies) are indicated, including a suggestive significance threshold of $10^{-5}$. Two variants (in red) cross the three least conservative thresholds and they could be considered statistically significant with respect to these thresholds. B) Manhattan plot summarising the Bayesian analysis of single SNPs. We assumed a prior retrospective probability of association (rPPA) equal to $\pi=10^{-4}$. Variants with rPPA values above 50% are more likely than not to be associated with response to the drug.

# SUPPLEMENTARY TABLES

**Table S1:** Summary of all common variants with p-values less than a suggestive genome-wide significance threshold of $10^{-5}$ (n=23). Among these, those selected by the LASSO are indicated in green and orange (n=10; also see Figure 3B). The variants indicated in green (n=5) have protein-coding genes annotations and they were selected for further analysis. Those indicated in orange (n=5) have non-protein-coding gene annotations or are annotated as *intergenic* and they were not selected for further analysis. In order to avoid division by zero in the calculation of odds ratios, the matrix of counts for each variant was pre-processed using Lidstone smoothing with pseudo-count parameter equal to 1 (i.e. add-one smoothing).

**Table S2:** Analysis of deviance using seven different logistic regression models of increasing complexity. The BASIC model includes only sex and the intercept as predictors. The FULL model includes in addition five SNPs from genes *SPNS3*, *HDC*, *NSG1, RASGEF1C* and *MDGA2*, which were previously selected using penalised logistic regression. Intermediate models include only sex, the intercept and the SNP harboured by the indicated gene. DF: degrees of freedom

**Table S3:** Summary of gene-based tests. Only genes with FWER<10% are shown. The low-frequency variants harboured by these genes are listed in the second spreadsheet.

**Table S4:** Summary of Reactome pathway-based tests. Only pathways with FWER<10% are shown. The rare variants harboured by genes in these pathways are listed in the second spreadsheet.

**Table S5:** Summary of results from the targeted analysis. We list the results from SNP- and gene-based tests on SYNAPTIC and EPILEPSY genes. A complete list of genes in each of these two groups is also provided in the last spreadsheet.

**Table S6:** Summary of results from the auxiliary statistical analysis on common variants. Only variants with a p-value below a suggestive genome-wide significance threshold of $10^{-5}$ are presented. P-values were corrected for multiplicity using Sidak's method and four different estimates of the effective number of independent tests, as indicated. The results of Bayesian analysis for different choices of the priors $\pi$ ($10^{-4}$, $10^{-5}$ and $10^{-6}$) and ($a$, $b$, $c$) (flat or empirical) are also given. rPPA: retrospective probability of association. FWER: family-wise error rate.

Table S1

| CHROM | POS | REF | ALT | MINOR ALLELE | ID | CONSEQUENCE | AF | MAF | SYMBOL | P VALUE | # RESPONDERS | | | # NON-RESPONDERS | | | % RESPONDERS | | | % NON-RESPONDERS | | | ODDS RATIO | |
|---|---|---|---|---|---|---|---|---|---|---|---|---|---|---|---|---|---|---|---|---|---|---|---|---|
| | | | | | | | | | | | Ref/Ref | Ref/Alt | Alt/Alt | Ref/Ref | Ref/Alt | Alt/Alt | Ref/Ref | Ref/Alt | Alt/Alt | Ref/Ref | Ref/Alt | Alt/Alt | Dominant model | Recessive model |
| 17 | 4350407 | G | C | G | rs2047231 | intronic | 0.82 | 0.18 | SPNS3 | 1.60E-07 | 1 | 4 | 60 | 0 | 15 | 10 | 2.94 | 7.35 | 89.71 | 3.57 | 57.14 | 39.29 | 0.82 | 0.07 |
| 15 | 50553000 | G | A | A | rs7182203 | intronic | 0.26 | 0.26 | HDC | 2.43E-07 | 21 | 35 | 10 | 22 | 0 | 4 | 31.88 | 52.17 | 15.94 | 79.31 | 3.45 | 17.24 | 0.12 | 1.10 |
| 4 | 8291102 | C | T | T | rs7683388 | intronic | 0.39 | 0.39 | HTRA3 | 1.52E-06 | 26 | 40 | 0 | 13 | 5 | 8 | 39.13 | 59.42 | 1.45 | 48.28 | 20.69 | 31.03 | 0.69 | 30.60 |
| 17 | 4350182 | T | C | T | rs2047233 | intronic | 0.88 | 0.12 | SPNS3 | 1.58E-06 | 0 | 3 | 63 | 0 | 13 | 13 | 1.45 | 5.80 | 92.75 | 3.45 | 48.28 | 48.28 | 0.41 | 0.07 |
| 5 | 179619074 | T | C | C | rs34570575 | intronic | 0.10 | 0.10 | RASGEF1C | 2.87E-06 | 57 | 9 | 0 | 9 | 14 | 2 | 84.06 | 14.49 | 1.45 | 35.71 | 53.57 | 10.71 | 9.49 | 8.16 |
| 3 | 37604012 | T | G | G | rs9825420 | intronic | 0.18 | 0.18 | ITGA9 | 3.32E-06 | 58 | 6 | 2 | 11 | 15 | 0 | 85.51 | 10.14 | 4.35 | 41.38 | 55.17 | 3.45 | 8.36 | 0.79 |
| 4 | 4345379 | T | C | C | rs7695197 | upstream | 0.31 | 0.31 | NSG1 | 3.55E-06 | 34 | 31 | 1 | 7 | 8 | 11 | 50.72 | 46.38 | 2.90 | 27.59 | 31.03 | 41.38 | 2.70 | 23.65 |
| 4 | 4346427 | G | A | A | rs3981 | upstream | 0.32 | 0.32 | NSG1 | 3.84E-06 | 34 | 30 | 1 | 7 | 8 | 11 | 51.47 | 45.59 | 2.94 | 27.59 | 31.03 | 41.38 | 2.78 | 23.29 |
| 4 | 4346465 | T | C | C | rs12641832 | upstream | 0.33 | 0.33 | NSG1 | 3.84E-06 | 34 | 30 | 1 | 7 | 8 | 11 | 51.47 | 45.59 | 2.94 | 27.59 | 31.03 | 41.38 | 2.78 | 23.29 |
| 4 | 113961039 | G | T | G | rs10015551 | intronic, non-coding transcript | 0.57 | 0.43 | RP11-650J17.1 | 4.72E-06 | 6 | 29 | 31 | 2 | 23 | 0 | 10.14 | 43.48 | 46.38 | 10.71 | 85.71 | 3.57 | 0.94 | 0.04 |
| 15 | 99100580 | A | G | A | rs2311753 | intergenic | 0.79 | 0.21 | | 5.49E-06 | 0 | 24 | 42 | 9 | 8 | 9 | 1.45 | 36.23 | 62.32 | 34.48 | 31.03 | 34.48 | 0.03 | 0.32 |
| 15 | 99101462 | A | G | A | rs2871859 | intergenic | 0.79 | 0.21 | | 5.49E-06 | 0 | 24 | 42 | 9 | 8 | 9 | 1.45 | 36.23 | 62.32 | 34.48 | 31.03 | 34.48 | 0.03 | 0.32 |
| 17 | 4350367 | A | G | A | rs2047232 | intronic | 0.88 | 0.12 | SPNS3 | 5.64E-06 | 0 | 4 | 62 | 0 | 13 | 13 | 1.45 | 7.25 | 91.30 | 3.45 | 48.28 | 48.28 | 0.41 | 0.09 |
| 2 | 156765630 | T | C | C | rs16839725 | intergenic | 0.29 | 0.29 | | 5.90E-06 | 40 | 15 | 11 | 3 | 19 | 4 | 59.42 | 23.19 | 17.39 | 13.79 | 68.97 | 17.24 | 9.15 | 0.99 |
| 14 | 47620548 | C | T | C | rs1952220 | intronic | 0.87 | 0.13 | MDGA2 | 6.25E-06 | 2 | 17 | 47 | 1 | 20 | 5 | 4.35 | 26.09 | 69.57 | 6.90 | 72.41 | 20.69 | 0.61 | 0.11 |
| 15 | 99102128 | A | G | A | rs11853887 | intergenic | 0.76 | 0.24 | | 6.52E-06 | 0 | 25 | 41 | 9 | 8 | 9 | 1.45 | 37.68 | 60.87 | 34.48 | 31.03 | 34.48 | 0.03 | 0.34 |
| 4 | 113961811 | C | T | C | rs10433900 | intronic, non-coding transcript | 0.52 | 0.48 | RP11-650J17.1 | 6.81E-06 | 9 | 27 | 30 | 3 | 22 | 0 | 14.49 | 40.58 | 44.93 | 14.29 | 82.14 | 3.57 | 1.02 | 0.05 |
| 3 | 37603952 | T | A | A | rs9825270 | intronic | 0.19 | 0.19 | ITGA9 | 7.07E-06 | 58 | 7 | 1 | 11 | 15 | 0 | 85.51 | 11.59 | 2.90 | 41.38 | 55.17 | 3.45 | 8.36 | 1.20 |
| 4 | 16428073 | G | A | A | rs73224601 | intronic, non-coding transcript | 0.10 | 0.10 | RP11-446J8.1 | 7.46E-06 | 64 | 1 | 0 | 16 | 9 | 1 | 95.59 | 2.94 | 1.47 | 58.62 | 34.48 | 6.90 | 15.29 | 4.96 |
| 3 | 72015276 | C | A | A | rs2036594 | intergenic | 0.17 | 0.17 | | 8.48E-06 | 61 | 2 | 3 | 15 | 11 | 0 | 89.86 | 4.35 | 5.80 | 55.17 | 41.38 | 3.45 | 7.20 | 0.58 |
| 19 | 46569619 | A | G | A | rs713409 | intronic | 0.53 | 0.47 | IGFL4 | 8.91E-06 | 18 | 24 | 23 | 3 | 22 | 0 | 27.94 | 36.76 | 35.29 | 14.29 | 82.14 | 3.57 | 2.33 | 0.07 |
| 19 | 46569630 | A | G | A | rs713411 | intronic | 0.53 | 0.47 | IGFL4 | 8.91E-06 | 18 | 24 | 23 | 3 | 22 | 0 | 27.94 | 36.76 | 35.29 | 14.29 | 82.14 | 3.57 | 2.33 | 0.07 |
| 1 | 156486009 | G | A | G | rs6685228 | downstream | 0.76 | 0.24 | RP11-284F21.8 | 9.82E-06 | 1 | 9 | 56 | 0 | 16 | 10 | 2.90 | 14.49 | 82.61 | 3.45 | 58.62 | 37.93 | 0.84 | 0.13 |

Table S2

| MODEL | DF | | DEVIANCE | | P VALUE | D2 | D2 (adj.) |
|---|---|---|---|---|---|---|---|
| | Residual | Null | Residual | Null | | | |
| BASIC | 90 | 91 | 107.38 | 109.55 | - | 0.02 | 0.01 |
| BASIC + HDC | 89 | 91 | 95.43 | 109.55 | 5.44E-04 | 0.13 | 0.11 |
| BASIC + NSG1 | 89 | 91 | 91.83 | 109.55 | 8.03E-05 | 0.16 | 0.14 |
| BASIC + MDGA2 | 89 | 91 | 89.68 | 109.55 | 2.58E-05 | 0.18 | 0.16 |
| BASIC + RASGEF1C | 89 | 91 | 84.62 | 109.55 | 1.84E-06 | 0.23 | 0.21 |
| BASIC + SPNS3 | 89 | 91 | 83.49 | 109.55 | 1.02E-06 | 0.24 | 0.22 |
| FULL | 86 | 91 | 28.04 | 109.55 | 1.15E-15 | 0.74 | 0.73 |

Table S3 (A): Significant results from gene-based tests

| SYMBOL | # MUTATIONS All | # MUTATIONS p value < 0.05 | P VALUE (GENE-BASED) | FWER | GENE ONTOLOGY | REACTOME |
|---|---|---|---|---|---|---|
| PRKCB | 78 | 45 | 8.24E-08 | 0.0016 | transferase activity, transferring phosphorus-containing groups, protein tyrosine kinase activity | glutamate binding, activation of AMPA receptors, synaptic plasticity |
| LINC01090 | 35 | 18 | 1.45E-07 | 0.0029 | - | - |
| DLG2 | 208 | 53 | 6.25E-07 | 0.0123 | kinase binding, guanylate kinase activity | protein-protein interactions at synapses |
| FILIP1 | 12 | 11 | 7.79E-07 | 0.0153 | cytoplasm, nucleolus, plasma membrane, actin cytoskeleton | - |
| SEMA6D | 41 | 17 | 4.87E-06 | 0.0921 | semaphorin receptor binding | axon guidance, semaphorin interactions |

## Table S3 (B): Variants with MAF<5% in significant genes

| CHROM | POS | REF | ALT | MINOR ALLELE | ID | CONSEQUENCE | AF | MAF | SYMBOL | P VALUE | # RESPONDERS | | | # NON-RESPONDERS | | |
|---|---|---|---|---|---|---|---|---|---|---|---|---|---|---|---|---|
| | | | | | | | | | | | Ref/Ref | Ref/Alt | Alt/Alt | Ref/Ref | Ref/Alt | Alt/Alt |
| 2 | 188581504 | C | T | T | rs10497687 | intronic,non_coding_transcript | 0.0495 | 0.0495 | LINC01090 | 0.46031 | 60 | 6 | 0 | 22 | 4 | 0 |
| 2 | 188663115 | C | G | G | rs116771909 | intronic,non_coding_transcript | 0.0371 | 0.0371 | LINC01090 | 0.72009 | 57 | 8 | 0 | 23 | 2 | 0 |
| 2 | 188721064 | T | C | C | rs7557550 | intronic,non_coding_transcript | 0.0361 | 0.0361 | LINC01090 | 0.74016 | 57 | 9 | 0 | 20 | 4 | 0 |
| 2 | 188762907 | A | T | T | rs116655447 | intronic,non_coding_transcript | 0.0293 | 0.0293 | LINC01090 | 0.16792 | 55 | 11 | 0 | 25 | 1 | 0 |
| 2 | 188776942 | T | C | C | rs75411057 | intronic,non_coding_transcript | 0.0254 | 0.0254 | LINC01090 | 0.30238 | 58 | 7 | 0 | 20 | 5 | 0 |
| 2 | 188816239 | C | G | G | rs112295437 | intronic,non_coding_transcript | 0.0367 | 0.0367 | LINC01090 | 1.00000 | 58 | 8 | 0 | 23 | 3 | 0 |
| 2 | 188833988 | C | A | A | rs115319205 | intronic,non_coding_transcript | 0.0176 | 0.0176 | LINC01090 | 0.30912 | 59 | 7 | 0 | 21 | 5 | 0 |
| 2 | 188855089 | C | T | T | rs17262471 | intronic,non_coding_transcript | 0.0277 | 0.0277 | LINC01090 | 0.72009 | 57 | 8 | 0 | 23 | 2 | 0 |
| 2 | 188943459 | A | C | C | rs77663937 | intronic,non_coding_transcript | 0.0499 | 0.0499 | LINC01090 | 0.26662 | 53 | 13 | 0 | 17 | 8 | 0 |
| 2 | 188959003 | T | C | C | rs7349373 | intronic,non_coding_transcript | 0.0434 | 0.0434 | LINC01090 | 1.00000 | 58 | 8 | 0 | 22 | 2 | 0 |
| 2 | 188966511 | C | T | T | rs76695614 | intronic,non_coding_transcript | 0.0174 | 0.0174 | LINC01090 | 0.29927 | 59 | 7 | 0 | 20 | 5 | 0 |
| 2 | 189002372 | G | A | A | rs77231981 | intronic,non_coding_transcript | 0.0404 | 0.0404 | LINC01090 | 0.27341 | 58 | 7 | 1 | 20 | 6 | 0 |
| 2 | 189016852 | A | G | G | rs1439880 | intronic,non_coding_transcript | 0.0387 | 0.0387 | LINC01090 | 0.34130 | 57 | 9 | 0 | 19 | 6 | 0 |
| 2 | 189028697 | A | G | G | rs62181870 | intronic,non_coding_transcript | 0.0386 | 0.0386 | LINC01090 | 0.21415 | 57 | 9 | 0 | 18 | 6 | 0 |
| 2 | 189044856 | A | C | C | rs75344621 | intronic,non_coding_transcript | 0.0231 | 0.0231 | LINC01090 | 0.30912 | 59 | 7 | 0 | 21 | 5 | 0 |
| 2 | 189070772 | C | T | T | rs73041656 | intronic,non_coding_transcript | 0.0370 | 0.0370 | LINC01090 | 0.00353 | 62 | 4 | 0 | 18 | 8 | 0 |
| 2 | 189080123 | A | G | G | rs72897997 | intronic,non_coding_transcript | 0.0295 | 0.0295 | LINC01090 | 0.78552 | 59 | 6 | 1 | 22 | 3 | 0 |
| 2 | 189084759 | T | G | G | rs73043732 | intronic,non_coding_transcript | 0.0377 | 0.0377 | LINC01090 | 0.00279 | 62 | 4 | 0 | 17 | 8 | 0 |
| 2 | 189084858 | G | A | A | rs16829913 | intronic,non_coding_transcript | 0.0311 | 0.0311 | LINC01090 | 0.00146 | 63 | 3 | 0 | 18 | 8 | 0 |
| 2 | 189085132 | T | C | C | rs16829916 | intronic,non_coding_transcript | 0.0421 | 0.0421 | LINC01090 | 0.00279 | 62 | 4 | 0 | 17 | 8 | 0 |
| 2 | 189086909 | A | G | G | rs17235639 | intronic,non_coding_transcript | 0.0486 | 0.0486 | LINC01090 | 0.77339 | 53 | 13 | 0 | 19 | 6 | 0 |
| 2 | 189093576 | A | G | G | rs139110719 | intronic,non_coding_transcript | 0.0345 | 0.0345 | LINC01090 | 0.00353 | 62 | 4 | 0 | 18 | 8 | 0 |
| 2 | 189098538 | T | C | C | rs77518981 | intronic,non_coding_transcript | 0.0346 | 0.0346 | LINC01090 | 0.00353 | 62 | 4 | 0 | 18 | 8 | 0 |
| 2 | 189106416 | C | T | T | rs4536617 | intronic,non_coding_transcript | 0.0346 | 0.0346 | LINC01090 | 0.00382 | 61 | 4 | 0 | 18 | 8 | 0 |
| 2 | 189108567 | A | G | G | rs34477786 | intronic,non_coding_transcript | 0.0489 | 0.0489 | LINC01090 | 0.00195 | 61 | 5 | 0 | 16 | 9 | 0 |
| 2 | 189110995 | A | C | C | rs72903517 | intronic,non_coding_transcript | 0.0493 | 0.0493 | LINC01090 | 0.00195 | 61 | 5 | 0 | 16 | 9 | 0 |
| 2 | 189112992 | A | G | G | rs76427030 | intronic,non_coding_transcript | 0.0304 | 0.0304 | LINC01090 | 0.00114 | 63 | 3 | 0 | 17 | 8 | 0 |
| 2 | 189117009 | A | G | G | rs74596536 | intronic,non_coding_transcript | 0.0380 | 0.0380 | LINC01090 | 0.00279 | 62 | 4 | 0 | 17 | 8 | 0 |
| 2 | 189129357 | G | A | A | rs79376882 | intronic,non_coding_transcript | 0.0339 | 0.0339 | LINC01090 | 0.00353 | 62 | 4 | 0 | 18 | 8 | 0 |
| 2 | 189130332 | T | A | A | rs78784062 | intronic,non_coding_transcript | 0.0306 | 0.0306 | LINC01090 | 0.00124 | 62 | 3 | 0 | 17 | 8 | 0 |
| 2 | 189137825 | G | T | T | rs78533331 | intronic,non_coding_transcript | 0.0305 | 0.0305 | LINC01090 | 0.00114 | 63 | 3 | 0 | 17 | 8 | 0 |
| 2 | 189138858 | A | T | T | rs76319142 | intronic,non_coding_transcript | 0.0339 | 0.0339 | LINC01090 | 0.00279 | 62 | 4 | 0 | 17 | 8 | 0 |
| 2 | 189142943 | A | T | T | rs60688387 | intronic,non_coding_transcript | 0.0378 | 0.0378 | LINC01090 | 0.00353 | 62 | 4 | 0 | 18 | 8 | 0 |
| 2 | 189147226 | C | T | T | rs75927445 | intronic,non_coding_transcript | 0.0377 | 0.0377 | LINC01090 | 0.00279 | 62 | 4 | 0 | 17 | 8 | 0 |
| 2 | 189149922 | C | T | T | rs4490147 | intronic,non_coding_transcript | 0.0311 | 0.0311 | LINC01090 | 0.00114 | 63 | 3 | 0 | 17 | 8 | 0 |
| 6 | 76019039 | G | A | A | rs13217015 | intronic | 0.0499 | 0.0499 | FILIP1 | 0.00082 | 61 | 5 | 0 | 16 | 9 | 1 |
| 6 | 76019395 | C | G | G | rs13217241 | intronic | 0.0495 | 0.0495 | FILIP1 | 0.00082 | 61 | 5 | 0 | 16 | 9 | 1 |
| 6 | 76019584 | A | T | T | rs13217398 | intronic | 0.0495 | 0.0495 | FILIP1 | 0.00060 | 61 | 5 | 0 | 15 | 9 | 1 |
| 6 | 76024704 | G | A | A | rs62415695 | missense | 0.0184 | 0.0184 | FILIP1 | 0.00181 | 65 | 1 | 0 | 20 | 5 | 1 |
| 6 | 76027221 | A | G | G | rs13198000 | intronic | 0.0485 | 0.0485 | FILIP1 | 0.00082 | 61 | 5 | 0 | 16 | 9 | 1 |
| 6 | 76038587 | G | C | C | rs13193201 | intronic | 0.0490 | 0.0490 | FILIP1 | 0.00082 | 61 | 5 | 0 | 16 | 9 | 1 |
| 6 | 76048681 | G | T | T | rs78980779 | intronic | 0.0192 | 0.0192 | FILIP1 | 0.00181 | 65 | 1 | 0 | 20 | 5 | 1 |
| 6 | 76086411 | G | A | A | rs13203246 | intronic | 0.0499 | 0.0499 | FILIP1 | 0.00033 | 62 | 4 | 0 | 16 | 9 | 1 |
| 6 | 76088207 | A | C | C | rs13211248 | intronic | 0.0499 | 0.0499 | FILIP1 | 0.00033 | 62 | 4 | 0 | 16 | 9 | 1 |
| 6 | 76156294 | G | A | A | rs34747139 | intronic | 0.0454 | 0.0454 | FILIP1 | 0.00085 | 62 | 4 | 0 | 16 | 8 | 1 |
| 6 | 76170874 | C | A | A | rs9443176 | intronic | 0.0416 | 0.0416 | FILIP1 | 0.00085 | 62 | 4 | 0 | 16 | 8 | 1 |
| 6 | 76183844 | G | A | A | rs72877517 | intronic | 0.0447 | 0.0447 | FILIP1 | 0.21916 | 52 | 13 | 0 | 23 | 2 | 0 |
| 11 | 83232915 | G | A | A | rs114513645 | intronic | 0.0474 | 0.0474 | DLG2 | 0.73157 | 57 | 8 | 0 | 21 | 4 | 0 |
| 11 | 83237817 | C | G | G | rs1155272 | intronic | 0.0455 | 0.0455 | DLG2 | 1.00000 | 57 | 9 | 0 | 22 | 4 | 0 |
| 11 | 83242900 | G | A | A | rs117159938 | intronic | 0.0424 | 0.0424 | DLG2 | 0.73017 | 58 | 8 | 0 | 21 | 4 | 0 |
| 11 | 83244370 | T | C | C | rs996925 | intronic | 0.0456 | 0.0456 | DLG2 | 0.74690 | 57 | 9 | 0 | 21 | 4 | 0 |
| 11 | 83255284 | T | C | C | rs116983205 | intronic | 0.0428 | 0.0428 | DLG2 | 0.73017 | 58 | 8 | 0 | 21 | 4 | 0 |
| 11 | 83259388 | A | C | C | rs118015302 | intronic | 0.0464 | 0.0464 | DLG2 | 0.74690 | 57 | 9 | 0 | 21 | 4 | 0 |
| 11 | 83264917 | A | G | G | rs17463588 | intronic | 0.0256 | 0.0256 | DLG2 | 0.74690 | 57 | 9 | 0 | 21 | 4 | 0 |
| 11 | 83375180 | A | T | T | rs12791719 | intronic | 0.0398 | 0.0398 | DLG2 | 0.06279 | 58 | 8 | 0 | 18 | 8 | 0 |
| 11 | 83377015 | C | A | A | rs977822 | intronic | 0.0335 | 0.0335 | DLG2 | 0.17547 | 59 | 7 | 0 | 19 | 6 | 0 |
| 11 | 83380474 | G | A | A | rs34718216 | intronic | 0.0336 | 0.0336 | DLG2 | 0.17547 | 59 | 7 | 0 | 19 | 6 | 0 |
| 11 | 83387738 | C | T | T | rs76751610 | intronic | 0.0333 | 0.0333 | DLG2 | 0.10333 | 59 | 7 | 0 | 19 | 5 | 1 |
| 11 | 83389547 | A | T | T | rs12803579 | intronic | 0.0343 | 0.0343 | DLG2 | 0.06500 | 58 | 8 | 0 | 18 | 6 | 1 |
| 11 | 83389577 | C | A | A | rs34839973 | intronic | 0.0398 | 0.0398 | DLG2 | 0.06500 | 58 | 8 | 0 | 18 | 6 | 1 |
| 11 | 83400031 | C | T | T | rs34843650 | intronic | 0.0345 | 0.0345 | DLG2 | 0.06500 | 58 | 8 | 0 | 18 | 6 | 1 |
| 11 | 83405196 | A | G | G | rs12806068 | intronic | 0.0319 | 0.0319 | DLG2 | 0.08178 | 59 | 7 | 0 | 19 | 6 | 1 |
| 11 | 83413022 | A | G | G | rs1892949 | intronic | 0.0320 | 0.0320 | DLG2 | 0.10333 | 59 | 7 | 0 | 19 | 5 | 1 |
| 11 | 83417779 | C | T | T | rs71465553 | intronic | 0.0317 | 0.0317 | DLG2 | 0.10333 | 59 | 7 | 0 | 19 | 5 | 1 |
| 11 | 83418812 | C | T | T | rs12787861 | intronic | 0.0329 | 0.0329 | DLG2 | 0.08178 | 59 | 7 | 0 | 19 | 6 | 1 |
| 11 | 83424657 | A | G | G | rs34514322 | intronic | 0.0320 | 0.0320 | DLG2 | 0.10333 | 59 | 7 | 0 | 19 | 5 | 1 |
| 11 | 83424825 | T | C | C | rs34994208 | intronic | 0.0330 | 0.0330 | DLG2 | 0.10333 | 59 | 7 | 0 | 19 | 5 | 1 |
| 11 | 83427143 | A | C | C | rs35954003 | intronic | 0.0327 | 0.0327 | DLG2 | 0.08178 | 59 | 7 | 0 | 19 | 6 | 1 |
| 11 | 83427549 | C | T | T | rs118106162 | intronic | 0.0327 | 0.0327 | DLG2 | 0.10333 | 59 | 7 | 0 | 19 | 5 | 1 |
| 11 | 83427586 | G | A | A | rs117792962 | intronic | 0.0326 | 0.0326 | DLG2 | 0.10333 | 59 | 7 | 0 | 19 | 5 | 1 |
| 11 | 83429731 | G | A | A | rs71465554 | intronic | 0.0317 | 0.0317 | DLG2 | 0.17547 | 59 | 7 | 0 | 19 | 6 | 0 |
| 11 | 83430048 | G | A | A | rs116872257 | intronic | 0.0316 | 0.0316 | DLG2 | 0.08178 | 59 | 7 | 0 | 19 | 6 | 1 |
| 11 | 83436606 | G | T | T | rs35301519 | intronic | 0.0327 | 0.0327 | DLG2 | 0.07030 | 58 | 8 | 0 | 19 | 6 | 1 |
| 11 | 83436879 | C | T | T | rs17481934 | intronic | 0.0328 | 0.0328 | DLG2 | 0.11742 | 58 | 8 | 0 | 19 | 5 | 1 |
| 11 | 83451753 | T | C | C | rs36110284 | intronic | 0.0344 | 0.0344 | DLG2 | 0.11742 | 58 | 8 | 0 | 19 | 5 | 1 |
| 11 | 83454878 | G | A | A | rs12792825 | intronic | 0.0331 | 0.0331 | DLG2 | 0.07030 | 58 | 8 | 0 | 19 | 6 | 1 |
| 11 | 83460859 | T | C | C | rs12786565 | intronic | 0.0405 | 0.0405 | DLG2 | 0.05388 | 57 | 8 | 0 | 18 | 7 | 1 |
| 11 | 83461366 | T | C | C | rs71465556 | intronic | 0.0345 | 0.0345 | DLG2 | 0.07030 | 58 | 8 | 0 | 19 | 6 | 1 |

| | | | | | | | | | | | | | | | |
|---|---|---|---|---|---|---|---|---|---|---|---|---|---|---|---|
| 11 | 83462110 | A | C | C | rs71465557 | intronic | 0.0344 | 0.0344 | DLG2 | 0.11742 | 58 | 8 | 0 | 19 | 5 | 1 |
| 11 | 83462245 | T | C | C | rs34873133 | intronic | 0.0403 | 0.0403 | DLG2 | 0.10134 | 57 | 9 | 0 | 18 | 6 | 1 |
| 11 | 83469530 | C | T | T | rs4456272 | intronic | 0.0342 | 0.0342 | DLG2 | 0.11742 | 58 | 8 | 0 | 19 | 5 | 1 |
| 11 | 83469764 | A | T | T | rs71465559 | intronic | 0.0365 | 0.0365 | DLG2 | 0.11742 | 58 | 8 | 0 | 19 | 5 | 1 |
| 11 | 83471488 | C | G | G | rs148662532 | intronic | 0.0124 | 0.0124 | DLG2 | 0.13742 | 62 | 3 | 0 | 22 | 3 | 1 |
| 11 | 83473103 | C | A | A | rs34520201 | intronic | 0.0445 | 0.0445 | DLG2 | 0.06038 | 57 | 9 | 0 | 18 | 7 | 1 |
| 11 | 83473291 | C | T | T | rs35047639 | intronic | 0.0441 | 0.0441 | DLG2 | 0.06038 | 57 | 9 | 0 | 18 | 7 | 1 |
| 11 | 83478153 | C | T | T | rs76663974 | intronic | 0.0409 | 0.0409 | DLG2 | 0.79699 | 57 | 8 | 1 | 23 | 2 | 0 |
| 11 | 83481286 | G | A | A | rs79417402 | intronic | 0.0346 | 0.0346 | DLG2 | 0.20777 | 58 | 8 | 0 | 20 | 6 | 0 |
| 11 | 83482085 | C | T | T | rs61900028 | intronic | 0.0374 | 0.0374 | DLG2 | 0.49988 | 56 | 10 | 0 | 23 | 2 | 0 |
| 11 | 83482647 | A | C | C | rs12802155 | intronic | 0.0322 | 0.0322 | DLG2 | 0.34130 | 57 | 9 | 0 | 19 | 6 | 0 |
| 11 | 83485395 | T | C | C | rs71468352 | intronic | 0.0307 | 0.0307 | DLG2 | 0.19653 | 58 | 8 | 0 | 19 | 6 | 0 |
| 11 | 83499391 | C | T | T | rs77713322 | intronic | 0.0304 | 0.0304 | DLG2 | 0.79699 | 57 | 8 | 1 | 23 | 2 | 0 |
| 11 | 83510016 | G | A | A | rs80098460 | intronic | 0.0215 | 0.0215 | DLG2 | 1.00000 | 57 | 9 | 0 | 22 | 3 | 0 |
| 11 | 83524658 | C | T | T | rs12362201 | intronic | 0.0449 | 0.0449 | DLG2 | 0.33478 | 54 | 12 | 0 | 23 | 2 | 0 |
| 11 | 83529668 | A | T | T | rs78005836 | intronic | 0.0224 | 0.0224 | DLG2 | 0.49839 | 59 | 7 | 0 | 22 | 4 | 0 |
| 11 | 83530139 | T | C | C | rs117322638 | intronic | 0.0227 | 0.0227 | DLG2 | 0.72340 | 57 | 7 | 0 | 22 | 4 | 0 |
| 11 | 83543747 | C | T | T | rs80100935 | intronic | 0.0355 | 0.0355 | DLG2 | 0.52285 | 62 | 3 | 1 | 22 | 3 | 0 |
| 11 | 83555201 | C | A | A | rs76144509 | intronic | 0.0461 | 0.0461 | DLG2 | 0.75217 | 56 | 10 | 0 | 23 | 3 | 0 |
| 11 | 83558415 | C | T | T | rs72958255 | intronic | 0.0260 | 0.0260 | DLG2 | 0.72086 | 57 | 9 | 0 | 23 | 2 | 0 |
| 11 | 83589911 | C | T | T | rs17146314 | intronic | 0.0462 | 0.0462 | DLG2 | 0.52280 | 54 | 11 | 1 | 24 | 2 | 0 |
| 11 | 83636021 | C | T | T | rs11233813 | intronic | 0.0453 | 0.0453 | DLG2 | 0.70191 | 60 | 6 | 0 | 22 | 3 | 0 |
| 11 | 83664354 | G | A | A | rs79392602 | intronic | 0.0490 | 0.0490 | DLG2 | 0.54308 | 54 | 12 | 0 | 23 | 3 | 0 |
| 11 | 83677047 | T | C | C | rs34747709 | intronic | 0.0464 | 0.0464 | DLG2 | 0.00633 | 52 | 14 | 0 | 24 | 0 | 1 |
| 11 | 83677711 | A | G | G | rs34310426 | intronic | 0.0462 | 0.0462 | DLG2 | 0.00633 | 52 | 14 | 0 | 24 | 0 | 1 |
| 11 | 83677783 | T | G | G | rs34826061 | intronic | 0.0463 | 0.0463 | DLG2 | 0.00633 | 52 | 14 | 0 | 24 | 0 | 1 |
| 11 | 83678621 | C | G | G | rs35494684 | intronic | 0.0464 | 0.0464 | DLG2 | 0.00633 | 52 | 14 | 0 | 24 | 0 | 1 |
| 11 | 83678640 | C | A | A | rs34197413 | intronic | 0.0464 | 0.0464 | DLG2 | 0.00633 | 52 | 14 | 0 | 24 | 0 | 1 |
| 11 | 83678730 | T | C | C | rs12282675 | intronic | 0.0484 | 0.0484 | DLG2 | 0.75015 | 55 | 11 | 0 | 22 | 3 | 0 |
| 11 | 83678734 | T | A | A | rs71465567 | intronic | 0.0464 | 0.0464 | DLG2 | 0.00633 | 52 | 14 | 0 | 24 | 0 | 1 |
| 11 | 83678887 | T | C | C | rs4133461 | intronic | 0.0464 | 0.0464 | DLG2 | 0.00633 | 52 | 14 | 0 | 24 | 0 | 1 |
| 11 | 83678920 | A | C | C | rs963090 | intronic | 0.0464 | 0.0464 | DLG2 | 0.00657 | 51 | 14 | 0 | 24 | 0 | 1 |
| 11 | 83679015 | A | C | C | rs12290311 | intronic | 0.0497 | 0.0497 | DLG2 | 0.75015 | 55 | 11 | 0 | 22 | 3 | 0 |
| 11 | 83679413 | T | C | C | rs7482523 | intronic | 0.0463 | 0.0463 | DLG2 | 0.00633 | 52 | 14 | 0 | 24 | 0 | 1 |
| 11 | 83679979 | T | C | C | rs71465568 | intronic | 0.0464 | 0.0464 | DLG2 | 0.00460 | 52 | 14 | 0 | 25 | 0 | 1 |
| 11 | 83680070 | G | A | A | rs71465569 | intronic | 0.0464 | 0.0464 | DLG2 | 0.00633 | 52 | 14 | 0 | 24 | 0 | 1 |
| 11 | 83680280 | A | G | G | rs35693607 | intronic | 0.0464 | 0.0464 | DLG2 | 0.00633 | 52 | 14 | 0 | 24 | 0 | 1 |
| 11 | 83680391 | G | C | C | rs12796172 | intronic | 0.0464 | 0.0464 | DLG2 | 0.00460 | 52 | 14 | 0 | 25 | 0 | 1 |
| 11 | 83680397 | T | C | C | rs11233866 | intronic | 0.0373 | 0.0373 | DLG2 | 1.00000 | 56 | 9 | 1 | 23 | 3 | 0 |
| 11 | 83680596 | T | C | C | rs34716185 | intronic | 0.0463 | 0.0463 | DLG2 | 0.00633 | 52 | 14 | 0 | 24 | 0 | 1 |

| | | | | | | | | | | | | | | | |
|---|---|---|---|---|---|---|---|---|---|---|---|---|---|---|---|
| 11 | 83680906 | A | G | G | rs35240755 | intronic | 0.0461 | 0.0461 | DLG2 | 0.00460 | 52 | 14 | 0 | 25 | 0 | 1 |
| 11 | 83681204 | C | T | T | rs12790276 | intronic | 0.0463 | 0.0463 | DLG2 | 0.00460 | 52 | 14 | 0 | 25 | 0 | 1 |
| 11 | 83681379 | C | T | T | rs35307147 | intronic | 0.0464 | 0.0464 | DLG2 | 0.00633 | 52 | 14 | 0 | 24 | 0 | 1 |
| 11 | 83681457 | G | C | C | rs35268360 | intronic | 0.0462 | 0.0462 | DLG2 | 0.00460 | 52 | 14 | 0 | 25 | 0 | 1 |
| 11 | 83681470 | G | A | A | rs34655432 | intronic | 0.0462 | 0.0462 | DLG2 | 0.00633 | 52 | 14 | 0 | 24 | 0 | 1 |
| 11 | 83681548 | G | A | A | rs34949067 | intronic | 0.0463 | 0.0463 | DLG2 | 0.00633 | 52 | 14 | 0 | 24 | 0 | 1 |
| 11 | 83682356 | A | G | G | rs35540097 | intronic | 0.0463 | 0.0463 | DLG2 | 0.00633 | 52 | 14 | 0 | 24 | 0 | 1 |
| 11 | 83683244 | G | T | T | rs71465571 | intronic | 0.0473 | 0.0473 | DLG2 | 0.00633 | 52 | 14 | 0 | 24 | 0 | 1 |
| 11 | 83683312 | A | G | G | rs71465572 | intronic | 0.0474 | 0.0474 | DLG2 | 0.00633 | 52 | 14 | 0 | 24 | 0 | 1 |
| 11 | 83683445 | G | A | A | rs12804617 | intronic | 0.0474 | 0.0474 | DLG2 | 0.00633 | 52 | 14 | 0 | 24 | 0 | 1 |
| 11 | 83683640 | A | G | G | rs12804532 | intronic | 0.0473 | 0.0473 | DLG2 | 0.00633 | 52 | 14 | 0 | 24 | 0 | 1 |
| 11 | 83683790 | G | C | C | rs12805402 | intronic | 0.0473 | 0.0473 | DLG2 | 0.00633 | 52 | 14 | 0 | 24 | 0 | 1 |
| 11 | 83683815 | T | A | A | rs34912232 | intronic | 0.0474 | 0.0474 | DLG2 | 0.00633 | 52 | 14 | 0 | 24 | 0 | 1 |
| 11 | 83683854 | C | T | T | rs35024645 | intronic | 0.0474 | 0.0474 | DLG2 | 0.00633 | 52 | 14 | 0 | 24 | 0 | 1 |
| 11 | 83683932 | G | A | A | rs36114419 | intronic | 0.0473 | 0.0473 | DLG2 | 0.00633 | 52 | 14 | 0 | 24 | 0 | 1 |
| 11 | 83684393 | G | C | C | rs34272592 | intronic | 0.0474 | 0.0474 | DLG2 | 0.00633 | 52 | 14 | 0 | 24 | 0 | 1 |
| 11 | 83684792 | A | T | T | rs35812794 | intronic | 0.0470 | 0.0470 | DLG2 | 0.00633 | 52 | 14 | 0 | 24 | 0 | 1 |
| 11 | 83684934 | A | G | G | rs71468355 | intronic | 0.0471 | 0.0471 | DLG2 | 0.00633 | 52 | 14 | 0 | 24 | 0 | 1 |
| 11 | 83684949 | C | T | T | rs71468356 | intronic | 0.0472 | 0.0472 | DLG2 | 0.00633 | 52 | 14 | 0 | 24 | 0 | 1 |
| 11 | 83685111 | T | C | C | rs77093335 | intronic | 0.0463 | 0.0463 | DLG2 | 0.00633 | 52 | 14 | 0 | 24 | 0 | 1 |
| 11 | 83685208 | A | T | T | rs77985485 | intronic | 0.0463 | 0.0463 | DLG2 | 0.00657 | 51 | 14 | 0 | 24 | 0 | 1 |
| 11 | 83685270 | A | G | G | rs80349643 | intronic | 0.0464 | 0.0464 | DLG2 | 0.00633 | 52 | 14 | 0 | 24 | 0 | 1 |
| 11 | 83685317 | G | A | A | rs79892044 | intronic | 0.0464 | 0.0464 | DLG2 | 0.00657 | 51 | 14 | 0 | 24 | 0 | 1 |
| 11 | 83685451 | T | G | G | rs71465573 | intronic | 0.0462 | 0.0462 | DLG2 | 0.00657 | 51 | 14 | 0 | 24 | 0 | 1 |
| 11 | 83685540 | T | G | G | rs71465574 | intronic | 0.0463 | 0.0463 | DLG2 | 0.00633 | 52 | 14 | 0 | 24 | 0 | 1 |
| 11 | 83685725 | A | C | C | rs12794213 | intronic | 0.0462 | 0.0462 | DLG2 | 0.00633 | 52 | 14 | 0 | 24 | 0 | 1 |
| 11 | 83685924 | C | T | T | rs34436974 | intronic | 0.0463 | 0.0463 | DLG2 | 0.00633 | 52 | 14 | 0 | 24 | 0 | 1 |
| 11 | 83686159 | T | C | C | rs71465575 | intronic | 0.0464 | 0.0464 | DLG2 | 0.00633 | 52 | 14 | 0 | 24 | 0 | 1 |
| 11 | 83686254 | A | G | G | rs71465576 | intronic | 0.0463 | 0.0463 | DLG2 | 0.00633 | 52 | 14 | 0 | 24 | 0 | 1 |
| 11 | 83686428 | A | G | G | rs71465577 | intronic | 0.0462 | 0.0462 | DLG2 | 0.00657 | 51 | 14 | 0 | 24 | 0 | 1 |
| 11 | 83686440 | A | T | T | rs71465578 | intronic | 0.0464 | 0.0464 | DLG2 | 0.00657 | 51 | 14 | 0 | 24 | 0 | 1 |
| 11 | 83686572 | A | G | G | rs71465579 | intronic | 0.0463 | 0.0463 | DLG2 | 0.00657 | 51 | 14 | 0 | 24 | 0 | 1 |
| 11 | 83686609 | A | G | G | rs71465580 | intronic | 0.0463 | 0.0463 | DLG2 | 0.00657 | 51 | 14 | 0 | 24 | 0 | 1 |
| 11 | 83686635 | A | G | G | rs71465581 | intronic | 0.0463 | 0.0463 | DLG2 | 0.00657 | 51 | 14 | 0 | 24 | 0 | 1 |
| 11 | 83692232 | T | G | G | rs34756786 | intronic | 0.0464 | 0.0464 | DLG2 | 0.00633 | 52 | 14 | 0 | 24 | 0 | 1 |
| 11 | 83735515 | G | A | A | rs80196056 | intronic | 0.0475 | 0.0475 | DLG2 | 0.75015 | 55 | 11 | 0 | 22 | 3 | 0 |
| 11 | 83737618 | G | A | A | rs17146592 | intronic | 0.0491 | 0.0491 | DLG2 | 0.74966 | 55 | 11 | 0 | 23 | 3 | 0 |
| 11 | 83767440 | G | A | A | rs11233892 | intronic | 0.0480 | 0.0480 | DLG2 | 0.74966 | 55 | 11 | 0 | 23 | 3 | 0 |
| 11 | 83778678 | A | G | G | rs35239124 | intronic | 0.0441 | 0.0441 | DLG2 | 0.47692 | 55 | 10 | 1 | 24 | 1 | 0 |
| 11 | 83786559 | T | C | C | rs34012933 | intronic | 0.0388 | 0.0388 | DLG2 | 0.16812 | 55 | 11 | 0 | 24 | 1 | 0 |

| | | | | | | | | | | | | | | | |
|---|---|---|---|---|---|---|---|---|---|---|---|---|---|---|---|
| 11 | 83791441 | T | C | C | rs17561231 | intronic | 0.0388 | 0.0388 | DLG2 | 0.16812 | 55 | 11 | 0 | 24 | 1 | 0 |
| 11 | 83796163 | T | C | C | rs2187553 | intronic | 0.0388 | 0.0388 | DLG2 | 0.16812 | 55 | 11 | 0 | 24 | 1 | 0 |
| 11 | 83798382 | G | A | A | rs34309181 | intronic | 0.0389 | 0.0389 | DLG2 | 0.16792 | 55 | 11 | 0 | 25 | 1 | 0 |
| 11 | 83798496 | T | C | C | rs34846175 | intronic | 0.0388 | 0.0388 | DLG2 | 0.16812 | 55 | 11 | 0 | 24 | 1 | 0 |
| 11 | 83798802 | T | C | C | rs34045362 | intronic | 0.0388 | 0.0388 | DLG2 | 0.16812 | 55 | 11 | 0 | 24 | 1 | 0 |
| 11 | 83802683 | G | T | T | rs34899787 | intronic | 0.0388 | 0.0388 | DLG2 | 0.16812 | 55 | 11 | 0 | 24 | 1 | 0 |
| 11 | 83803464 | A | G | G | rs71469619 | intronic | 0.0388 | 0.0388 | DLG2 | 0.16812 | 55 | 11 | 0 | 24 | 1 | 0 |
| 11 | 83807535 | T | C | C | rs17495791 | intronic | 0.0142 | 0.0142 | DLG2 | 0.43524 | 58 | 7 | 0 | 24 | 1 | 0 |
| 11 | 83813177 | T | C | C | rs75925390 | intronic | 0.0387 | 0.0387 | DLG2 | 0.16792 | 55 | 11 | 0 | 25 | 1 | 0 |
| 11 | 83814433 | G | C | C | rs71469623 | intronic | 0.0387 | 0.0387 | DLG2 | 0.16784 | 54 | 11 | 0 | 25 | 1 | 0 |
| 11 | 83821405 | C | T | T | rs12808867 | intronic | 0.0387 | 0.0387 | DLG2 | 0.16792 | 55 | 11 | 0 | 25 | 1 | 0 |
| 11 | 83821870 | C | A | A | rs12785867 | intronic | 0.0386 | 0.0386 | DLG2 | 0.16812 | 55 | 11 | 0 | 24 | 1 | 0 |
| 11 | 83827128 | G | T | T | rs75687645 | intronic | 0.0439 | 0.0439 | DLG2 | 1.00000 | 55 | 9 | 2 | 21 | 4 | 0 |
| 11 | 83828197 | G | C | C | rs12808314 | intronic | 0.0378 | 0.0378 | DLG2 | 0.16792 | 55 | 11 | 0 | 25 | 1 | 0 |
| 11 | 83828665 | T | C | C | rs71469624 | intronic | 0.0379 | 0.0379 | DLG2 | 0.16792 | 55 | 11 | 0 | 25 | 1 | 0 |
| 11 | 83830087 | C | T | T | rs35992468 | intronic | 0.0378 | 0.0378 | DLG2 | 0.16812 | 55 | 11 | 0 | 24 | 1 | 0 |
| 11 | 83839007 | G | A | A | rs116913890 | intronic | 0.0445 | 0.0445 | DLG2 | 1.00000 | 55 | 9 | 2 | 21 | 4 | 0 |
| 11 | 83839703 | C | A | A | rs71469625 | intronic | 0.0387 | 0.0387 | DLG2 | 0.16812 | 55 | 11 | 0 | 24 | 1 | 0 |
| 11 | 83844499 | T | C | C | rs34770791 | intronic | 0.0388 | 0.0388 | DLG2 | 0.16812 | 55 | 11 | 0 | 24 | 1 | 0 |
| 11 | 83846946 | G | A | A | rs35615041 | intronic | 0.0388 | 0.0388 | DLG2 | 0.16792 | 55 | 11 | 0 | 25 | 1 | 0 |
| 11 | 83850610 | T | C | C | rs71469626 | intronic | 0.0387 | 0.0387 | DLG2 | 0.16812 | 55 | 11 | 0 | 24 | 1 | 0 |
| 11 | 83851228 | G | A | A | rs71469627 | intronic | 0.0387 | 0.0387 | DLG2 | 0.16812 | 55 | 11 | 0 | 24 | 1 | 0 |
| 11 | 83855558 | A | C | C | rs12788130 | intronic | 0.0385 | 0.0385 | DLG2 | 0.16792 | 55 | 11 | 0 | 25 | 1 | 0 |
| 11 | 83856459 | T | C | C | rs34337231 | intronic | 0.0385 | 0.0385 | DLG2 | 0.16812 | 55 | 11 | 0 | 24 | 1 | 0 |
| 11 | 83857766 | T | C | C | rs35969270 | intronic | 0.0385 | 0.0385 | DLG2 | 0.16733 | 54 | 11 | 0 | 24 | 1 | 0 |
| 11 | 83860894 | G | A | A | rs71469629 | intronic | 0.0386 | 0.0386 | DLG2 | 0.16812 | 55 | 11 | 0 | 24 | 1 | 0 |
| 11 | 83862352 | T | G | G | rs71469630 | intronic | 0.0384 | 0.0384 | DLG2 | 0.16812 | 55 | 11 | 0 | 24 | 1 | 0 |
| 11 | 83866720 | A | T | T | rs11233929 | intronic | 0.0467 | 0.0467 | DLG2 | 1.00000 | 55 | 9 | 2 | 21 | 4 | 0 |
| 11 | 83867612 | T | C | C | rs71469631 | intronic | 0.0385 | 0.0385 | DLG2 | 0.16812 | 55 | 11 | 0 | 24 | 1 | 0 |
| 11 | 83875056 | A | C | C | rs35910823 | intronic | 0.0385 | 0.0385 | DLG2 | 0.16812 | 55 | 11 | 0 | 24 | 1 | 0 |
| 11 | 83878183 | T | C | C | rs12801364 | intronic | 0.0381 | 0.0381 | DLG2 | 0.16812 | 55 | 11 | 0 | 24 | 1 | 0 |
| 11 | 83880139 | A | T | T | rs373152600 | intronic | 0.0378 | 0.0378 | DLG2 | 0.17020 | 55 | 11 | 0 | 23 | 1 | 0 |
| 11 | 83882227 | C | T | T | rs78328809 | intronic | 0.0450 | 0.0450 | DLG2 | 1.00000 | 55 | 9 | 2 | 22 | 4 | 0 |
| 11 | 83888454 | C | G | G | rs12808048 | intronic | 0.0381 | 0.0381 | DLG2 | 0.16812 | 55 | 11 | 0 | 24 | 1 | 0 |
| 11 | 83890296 | T | C | C | rs12799079 | intronic | 0.0409 | 0.0409 | DLG2 | 0.10154 | 53 | 13 | 0 | 23 | 1 | 0 |
| 11 | 83892801 | C | T | T | rs34386657 | intronic | 0.0360 | 0.0360 | DLG2 | 0.16812 | 55 | 11 | 0 | 24 | 1 | 0 |
| 11 | 83899234 | A | G | G | rs11233934 | intronic | 0.0486 | 0.0486 | DLG2 | 0.75113 | 55 | 9 | 2 | 20 | 5 | 0 |
| 11 | 83900060 | A | G | G | rs34042104 | intronic | 0.0407 | 0.0407 | DLG2 | 0.10130 | 53 | 13 | 0 | 24 | 1 | 0 |
| 11 | 83903212 | A | T | T | rs73517787 | intronic | 0.0439 | 0.0439 | DLG2 | 0.10130 | 53 | 13 | 0 | 24 | 1 | 0 |
| 11 | 83903240 | A | G | G | rs74448633 | intronic | 0.0406 | 0.0406 | DLG2 | 0.10130 | 53 | 13 | 0 | 24 | 1 | 0 |
| 11 | 83903511 | C | T | T | rs35502448 | intronic | 0.0403 | 0.0403 | DLG2 | 0.10282 | 53 | 13 | 0 | 25 | 1 | 0 |
| 11 | 83905317 | G | A | A | rs73517792 | intronic | 0.0491 | 0.0491 | DLG2 | 0.21764 | 53 | 13 | 0 | 24 | 2 | 0 |
| 11 | 83906010 | C | G | G | rs12800271 | intronic | 0.0402 | 0.0402 | DLG2 | 0.06055 | 52 | 13 | 0 | 25 | 1 | 0 |
| 11 | 83907441 | G | A | A | rs77892201 | intronic | 0.0482 | 0.0482 | DLG2 | 0.10130 | 53 | 13 | 0 | 24 | 1 | 0 |
| 11 | 83907455 | C | T | T | rs79602990 | intronic | 0.0408 | 0.0408 | DLG2 | 0.10130 | 53 | 13 | 0 | 24 | 1 | 0 |
| 11 | 83908287 | T | C | C | rs59681368 | intronic | 0.0474 | 0.0474 | DLG2 | 0.10130 | 53 | 13 | 0 | 24 | 1 | 0 |
| 11 | 83908345 | G | A | A | rs75344440 | intronic | 0.0407 | 0.0407 | DLG2 | 0.10121 | 52 | 13 | 0 | 24 | 1 | 0 |
| 11 | 83922479 | G | C | C | rs71469640 | intronic | 0.0409 | 0.0409 | DLG2 | 0.10130 | 53 | 13 | 0 | 24 | 1 | 0 |
| 11 | 83924357 | T | C | C | rs75384284 | intronic | 0.0425 | 0.0425 | DLG2 | 0.10130 | 53 | 13 | 0 | 24 | 1 | 0 |
| 11 | 83926691 | A | G | G | rs34664733 | intronic | 0.0427 | 0.0427 | DLG2 | 0.10130 | 53 | 13 | 0 | 24 | 1 | 0 |
| 11 | 83942171 | T | C | C | rs11233957 | intronic | 0.0324 | 0.0324 | DLG2 | 1.00000 | 56 | 9 | 1 | 22 | 3 | 0 |
| 11 | 83954261 | T | C | C | rs74818442 | intronic | 0.0199 | 0.0199 | DLG2 | 1.00000 | 60 | 5 | 1 | 22 | 2 | 0 |
| 11 | 83961660 | C | G | G | rs17159833 | intronic | 0.0290 | 0.0290 | DLG2 | 0.49839 | 59 | 7 | 0 | 22 | 4 | 0 |
| 11 | 83965933 | A | T | T | rs11233966 | intronic | 0.0402 | 0.0402 | DLG2 | 1.00000 | 55 | 10 | 1 | 22 | 4 | 0 |
| 11 | 83979308 | C | A | A | rs11233971 | intronic | 0.0214 | 0.0214 | DLG2 | 0.75945 | 61 | 4 | 1 | 23 | 2 | 0 |
| 11 | 83985299 | C | G | G | rs17500125 | intronic | 0.0475 | 0.0475 | DLG2 | 0.27806 | 56 | 10 | 0 | 24 | 1 | 0 |
| 11 | 83987664 | C | T | T | rs80140888 | intronic | 0.0442 | 0.0442 | DLG2 | 0.54181 | 53 | 13 | 0 | 22 | 3 | 0 |
| 11 | 84037134 | C | T | T | rs72959544 | intronic | 0.0356 | 0.0356 | DLG2 | 0.00907 | 60 | 4 | 1 | 18 | 7 | 0 |
| 11 | 84051729 | G | A | A | rs61899215 | intronic | 0.0246 | 0.0246 | DLG2 | 0.59212 | 58 | 7 | 1 | 25 | 1 | 0 |
| 11 | 84073212 | A | G | G | rs116901444 | intronic | 0.0215 | 0.0215 | DLG2 | 0.75945 | 61 | 4 | 1 | 23 | 2 | 0 |
| 11 | 84124557 | A | T | T | rs11234016 | intronic | 0.0224 | 0.0224 | DLG2 | 1.00000 | 60 | 5 | 1 | 23 | 2 | 0 |
| 11 | 84148078 | C | T | T | rs12793229 | intronic | 0.0264 | 0.0264 | DLG2 | 0.76318 | 60 | 5 | 1 | 25 | 1 | 0 |
| 11 | 84157083 | A | G | G | rs71468372 | intronic | 0.0382 | 0.0382 | DLG2 | 0.16938 | 62 | 3 | 1 | 21 | 4 | 0 |
| 11 | 84161506 | A | C | C | rs79104196 | intronic | 0.0439 | 0.0439 | DLG2 | 0.50269 | 55 | 11 | 0 | 23 | 2 | 0 |
| 11 | 84168666 | C | T | T | rs11234035 | intronic | 0.0493 | 0.0493 | DLG2 | 1.00000 | 53 | 11 | 2 | 21 | 4 | 0 |
| 11 | 84203688 | C | T | T | rs11234045 | intronic | 0.0406 | 0.0406 | DLG2 | 0.16812 | 55 | 11 | 0 | 24 | 1 | 0 |
| 11 | 84216892 | T | C | C | rs17808372 | intronic | 0.0481 | 0.0481 | DLG2 | 0.37817 | 51 | 15 | 0 | 22 | 3 | 0 |
| 11 | 84238326 | T | C | C | rs1367538 | intronic | 0.0237 | 0.0237 | DLG2 | 1.00000 | 59 | 6 | 1 | 24 | 2 | 0 |
| 11 | 84257549 | C | T | T | rs11234071 | intronic | 0.0237 | 0.0237 | DLG2 | 1.00000 | 59 | 6 | 1 | 23 | 2 | 0 |
| 11 | 84265771 | G | A | A | rs111685275 | intronic | 0.0452 | 0.0452 | DLG2 | 0.77969 | 54 | 9 | 3 | 21 | 4 | 0 |
| 11 | 84298759 | G | A | A | rs77693614 | intronic | 0.0238 | 0.0238 | DLG2 | 1.00000 | 59 | 6 | 1 | 23 | 2 | 0 |
| 11 | 84378128 | T | C | C | rs71469652 | intronic | 0.0463 | 0.0463 | DLG2 | 0.29927 | 59 | 7 | 0 | 20 | 5 | 0 |
| 11 | 84395708 | C | A | A | rs118128180 | intronic | 0.0434 | 0.0434 | DLG2 | 1.00000 | 56 | 10 | 0 | 22 | 4 | 0 |
| 11 | 84557455 | T | G | G | rs113083358 | intronic | 0.0281 | 0.0281 | DLG2 | 0.39891 | 58 | 8 | 0 | 21 | 3 | 1 |
| 11 | 84561099 | T | G | G | rs117287296 | intronic | 0.0452 | 0.0452 | DLG2 | 0.10783 | 62 | 4 | 0 | 20 | 5 | 0 |
| 11 | 84565540 | C | A | A | rs1943717 | intronic | 0.0408 | 0.0408 | DLG2 | 0.11105 | 62 | 4 | 0 | 21 | 5 | 0 |
| 11 | 84580040 | A | C | C | rs76074207 | intronic | 0.0387 | 0.0387 | DLG2 | 1.00000 | 57 | 9 | 0 | 23 | 3 | 0 |
| 11 | 84581522 | T | C | C | rs7125471 | intronic | 0.0493 | 0.0493 | DLG2 | 0.10783 | 62 | 4 | 0 | 20 | 5 | 0 |
| 11 | 84598257 | T | G | G | rs76429934 | intronic | 0.0278 | 0.0278 | DLG2 | 0.37598 | 58 | 7 | 0 | 21 | 3 | 1 |
| 11 | 84639183 | G | C | C | rs72955531 | intronic | 0.0437 | 0.0437 | DLG2 | 1.00000 | 62 | 3 | 1 | 24 | 1 | 0 |

| | | | | | | | | | | | | | | | |
|---|---|---|---|---|---|---|---|---|---|---|---|---|---|---|---|
| 11 | 84749449 | C | T | T | rs74775461 | intronic | 0.0437 | 0.0437 | DLG2 | 0.29927 | 59 | 7 | 0 | 20 | 5 | 0 |
| 11 | 84789864 | A | G | G | rs111410955 | intronic | 0.0255 | 0.0255 | DLG2 | 0.72086 | 57 | 9 | 0 | 23 | 2 | 0 |
| 11 | 84837883 | G | A | A | rs35003694 | intronic | 0.0397 | 0.0397 | DLG2 | 0.73017 | 58 | 8 | 0 | 21 | 4 | 0 |
| 11 | 84918461 | T | C | C | rs112844595 | intronic | 0.0206 | 0.0206 | DLG2 | 1.00000 | 57 | 9 | 0 | 22 | 3 | 0 |
| 11 | 84991145 | C | T | T | rs189136848 | intronic | 0.0096 | 0.0096 | DLG2 | 0.40582 | 59 | 6 | 1 | 25 | 0 | 0 |
| 11 | 84991169 | C | A | A | rs192808445 | intronic | 0.0096 | 0.0096 | DLG2 | 0.40582 | 59 | 6 | 1 | 25 | 0 | 0 |
| 11 | 84996976 | A | G | G | rs35695258 | intronic | 0.0145 | 0.0145 | DLG2 | 1.00000 | 58 | 7 | 0 | 23 | 2 | 0 |
| 11 | 85006195 | A | T | T | rs1016636 | intronic | 0.0388 | 0.0388 | DLG2 | 0.06574 | 60 | 5 | 0 | 19 | 6 | 0 |
| 11 | 85024079 | T | C | C | rs185755081 | intronic | 0.0097 | 0.0097 | DLG2 | 0.22343 | 59 | 6 | 1 | 26 | 0 | 0 |
| 11 | 85060360 | C | G | G | rs12794044 | intronic | 0.0157 | 0.0157 | DLG2 | 0.72129 | 58 | 8 | 0 | 23 | 2 | 0 |
| 11 | 85086043 | A | T | T | rs35395221 | intronic | 0.0142 | 0.0142 | DLG2 | 1.00000 | 59 | 7 | 0 | 23 | 2 | 0 |
| 11 | 85203814 | C | T | T | rs77687788 | intronic | 0.0399 | 0.0399 | DLG2 | 0.02367 | 62 | 4 | 0 | 19 | 6 | 0 |
| 11 | 85203882 | G | A | A | rs75993784 | intronic | 0.0431 | 0.0431 | DLG2 | 0.51888 | 57 | 9 | 0 | 20 | 5 | 0 |
| 11 | 85250747 | G | A | A | rs7939547 | intronic | 0.0393 | 0.0393 | DLG2 | 0.01641 | 61 | 5 | 0 | 18 | 7 | 0 |
| 11 | 85262722 | T | C | C | rs7931722 | intronic | 0.0394 | 0.0394 | DLG2 | 0.02002 | 62 | 4 | 0 | 18 | 6 | 0 |
| 11 | 85272815 | T | G | G | rs505238 | intronic | 0.0310 | 0.0310 | DLG2 | 0.54181 | 53 | 13 | 0 | 22 | 3 | 0 |
| 11 | 85300458 | G | A | A | rs576722 | intronic | 0.0191 | 0.0191 | DLG2 | 0.72218 | 57 | 9 | 0 | 24 | 2 | 0 |
| 15 | 47471564 | A | G | G | rs146546695 | upstream | 0.0428 | 0.0428 | SEMA6D | 0.59042 | 57 | 8 | 1 | 24 | 1 | 0 |
| 15 | 47472960 | G | A | A | rs79135353 | upstream | 0.0407 | 0.0407 | SEMA6D | 0.50649 | 58 | 8 | 0 | 21 | 5 | 0 |
| 15 | 47499732 | G | A | A | rs77819625 | intronic | 0.0189 | 0.0189 | SEMA6D | 0.00641 | 64 | 1 | 1 | 21 | 5 | 0 |
| 15 | 47524094 | T | C | C | rs80085470 | intronic | 0.0192 | 0.0192 | SEMA6D | 0.00553 | 64 | 1 | 1 | 20 | 5 | 0 |
| 15 | 47540002 | A | G | G | rs78106502 | intronic | 0.0408 | 0.0408 | SEMA6D | 0.00553 | 64 | 1 | 1 | 20 | 5 | 0 |
| 15 | 47540559 | A | G | G | rs11856688 | intronic | 0.0462 | 0.0462 | SEMA6D | 1.00000 | 58 | 8 | 0 | 22 | 3 | 0 |
| 15 | 47541152 | C | T | T | rs117884763 | intronic | 0.0211 | 0.0211 | SEMA6D | 0.00641 | 64 | 1 | 1 | 21 | 5 | 0 |
| 15 | 47541811 | C | T | T | rs76035674 | intronic | 0.0210 | 0.0210 | SEMA6D | 0.00553 | 64 | 1 | 1 | 20 | 5 | 0 |
| 15 | 47543232 | T | C | C | rs76999103 | intronic | 0.0192 | 0.0192 | SEMA6D | 0.00641 | 64 | 1 | 1 | 21 | 5 | 0 |
| 15 | 47552814 | A | G | G | rs78673643 | intronic | 0.0216 | 0.0216 | SEMA6D | 0.00641 | 64 | 1 | 1 | 21 | 5 | 0 |
| 15 | 47557518 | C | T | T | rs72731766 | intronic | 0.0276 | 0.0276 | SEMA6D | 1.00000 | 57 | 8 | 1 | 23 | 3 | 0 |
| 15 | 47580975 | A | G | G | rs7174561 | intronic | 0.0320 | 0.0320 | SEMA6D | 0.00641 | 64 | 1 | 1 | 21 | 5 | 0 |
| 15 | 47582288 | C | G | G | rs78142733 | intronic | 0.0217 | 0.0217 | SEMA6D | 0.00553 | 64 | 1 | 1 | 20 | 5 | 0 |
| 15 | 47589031 | T | G | G | rs17358151 | intronic | 0.0211 | 0.0211 | SEMA6D | 0.00553 | 64 | 1 | 1 | 20 | 5 | 0 |
| 15 | 47589185 | G | T | T | rs75474152 | intronic | 0.0259 | 0.0259 | SEMA6D | 0.00553 | 64 | 1 | 1 | 20 | 5 | 0 |
| 15 | 47630156 | C | T | T | rs77144407 | intronic | 0.0389 | 0.0389 | SEMA6D | 1.00000 | 58 | 8 | 0 | 23 | 3 | 0 |
| 15 | 47643236 | C | T | T | rs117401545 | intronic | 0.0375 | 0.0375 | SEMA6D | 1.00000 | 58 | 8 | 0 | 22 | 3 | 0 |
| 15 | 47697796 | G | C | C | rs61998609 | intronic | 0.0429 | 0.0429 | SEMA6D | 0.52276 | 53 | 12 | 1 | 24 | 2 | 0 |
| 15 | 47699041 | T | C | C | rs76187847 | intronic | 0.0220 | 0.0220 | SEMA6D | 0.01640 | 63 | 2 | 0 | 20 | 4 | 1 |
| 15 | 47704791 | G | A | A | rs79372522 | intronic | 0.0268 | 0.0268 | SEMA6D | 0.01558 | 64 | 2 | 0 | 20 | 4 | 1 |
| 15 | 47714563 | C | A | A | rs16959446 | intronic | 0.0437 | 0.0437 | SEMA6D | 0.51682 | 53 | 12 | 1 | 23 | 2 | 0 |
| 15 | 47722435 | A | G | G | rs111589399 | intronic | 0.0238 | 0.0238 | SEMA6D | 0.01558 | 64 | 2 | 0 | 20 | 4 | 1 |
| 15 | 47724040 | G | A | A | rs4775685 | intronic | 0.0437 | 0.0437 | SEMA6D | 0.51682 | 53 | 12 | 1 | 23 | 2 | 0 |
| 15 | 47727933 | T | C | C | rs74787996 | intronic | 0.0492 | 0.0492 | SEMA6D | 0.51682 | 53 | 12 | 1 | 23 | 2 | 0 |
| 15 | 47734225 | T | A | A | rs76552436 | intronic | 0.0498 | 0.0498 | SEMA6D | 0.51682 | 53 | 12 | 1 | 23 | 2 | 0 |
| 15 | 47775809 | G | A | A | rs61998616 | intronic | 0.0283 | 0.0283 | SEMA6D | 0.59042 | 57 | 8 | 1 | 24 | 1 | 0 |
| 15 | 47785052 | T | G | G | rs75701467 | intronic | 0.0200 | 0.0200 | SEMA6D | 1.00000 | 55 | 11 | 0 | 22 | 4 | 0 |
| 15 | 47839035 | A | G | G | rs72733864 | intronic | 0.0498 | 0.0498 | SEMA6D | 0.06460 | 61 | 5 | 0 | 19 | 6 | 0 |
| 15 | 47857783 | G | C | C | rs61998619 | intronic | 0.0297 | 0.0297 | SEMA6D | 0.75945 | 59 | 6 | 1 | 24 | 1 | 0 |
| 15 | 47887164 | G | A | A | rs117833852 | intronic | 0.0268 | 0.0268 | SEMA6D | 0.50269 | 55 | 11 | 0 | 23 | 2 | 0 |
| 15 | 47935535 | T | C | C | rs76732631 | intronic | 0.0290 | 0.0290 | SEMA6D | 0.76310 | 59 | 5 | 1 | 25 | 1 | 0 |
| 15 | 47944782 | T | C | C | rs1369639 | intronic | 0.0472 | 0.0472 | SEMA6D | 0.10353 | 54 | 12 | 0 | 24 | 1 | 0 |
| 15 | 47965075 | A | G | G | rs72735833 | intronic | 0.0394 | 0.0394 | SEMA6D | 0.00038 | 64 | 2 | 0 | 17 | 8 | 0 |
| 15 | 47967262 | C | T | T | rs17386817 | intronic | 0.0191 | 0.0191 | SEMA6D | 0.16733 | 54 | 11 | 0 | 24 | 1 | 0 |
| 15 | 47967971 | C | T | T | rs79207249 | intronic | 0.0329 | 0.0329 | SEMA6D | 0.59212 | 58 | 7 | 1 | 25 | 1 | 0 |
| 15 | 47979578 | G | A | A | rs17327778 | intronic | 0.0463 | 0.0463 | SEMA6D | 1.00000 | 58 | 7 | 0 | 22 | 3 | 0 |
| 15 | 47988326 | G | A | A | rs72735842 | intronic | 0.0386 | 0.0386 | SEMA6D | 0.00114 | 63 | 3 | 0 | 17 | 8 | 0 |
| 15 | 48008619 | A | G | G | rs72735850 | upstream | 0.0478 | 0.0478 | SEMA6D | 1.00000 | 60 | 5 | 1 | 23 | 2 | 0 |
| 15 | 48023229 | A | G | G | rs150484536 | intronic | 0.0274 | 0.0274 | SEMA6D | 0.27275 | 56 | 9 | 0 | 24 | 1 | 0 |
| 15 | 48050564 | C | T | T | rs56376571 | intronic | 0.0209 | 0.0209 | SEMA6D | 0.00141 | 64 | 2 | 0 | 18 | 7 | 0 |
| 15 | 48058769 | T | A | A | rs16960071 | intronic | 0.0263 | 0.0263 | SEMA6D | 0.61255 | 63 | 3 | 0 | 23 | 2 | 0 |
| 16 | 23845860 | G | T | T | rs72777910 | upstream | 0.0300 | 0.0300 | PRKCB | 0.47621 | 60 | 5 | 1 | 22 | 4 | 0 |
| 16 | 23849482 | T | C | T | rs2023670 | intronic | 0.9513 | 0.0487 | PRKCB | 0.00994 | 0 | 11 | 55 | 1 | 0 | 25 |
| 16 | 23850240 | A | G | A | rs11074581 | intronic | 0.9663 | 0.0337 | PRKCB | 0.00994 | 0 | 11 | 55 | 1 | 0 | 25 |
| 16 | 23851956 | T | C | T | rs7189210 | intronic | 0.9663 | 0.0337 | PRKCB | 0.00994 | 0 | 11 | 55 | 1 | 0 | 25 |
| 16 | 23852415 | A | T | A | rs2188359 | intronic | 0.9528 | 0.0472 | PRKCB | 0.00994 | 0 | 11 | 55 | 1 | 0 | 25 |
| 16 | 23859391 | A | G | G | rs62030647 | intronic | 0.0226 | 0.0226 | PRKCB | 1.00000 | 59 | 7 | 0 | 23 | 3 | 0 |
| 16 | 23874933 | A | C | A | rs6497691 | intronic | 0.9663 | 0.0337 | PRKCB | 0.00633 | 0 | 14 | 52 | 1 | 0 | 24 |
| 16 | 23876099 | C | T | T | rs79131874 | intronic | 0.0303 | 0.0303 | PRKCB | 1.00000 | 59 | 7 | 0 | 23 | 3 | 0 |
| 16 | 23877500 | A | G | A | rs8059885 | intronic | 0.9663 | 0.0337 | PRKCB | 0.00460 | 0 | 14 | 52 | 1 | 0 | 25 |
| 16 | 23877606 | A | G | A | rs8060048 | intronic | 0.9644 | 0.0356 | PRKCB | 0.00460 | 0 | 14 | 52 | 1 | 0 | 25 |
| 16 | 23877781 | G | A | G | rs8060718 | intronic | 0.9664 | 0.0336 | PRKCB | 0.00460 | 0 | 14 | 52 | 1 | 0 | 25 |
| 16 | 23878470 | C | T | C | rs12935004 | intronic | 0.9657 | 0.0343 | PRKCB | 0.00460 | 0 | 14 | 52 | 1 | 0 | 25 |
| 16 | 23880851 | C | T | C | rs8061523 | intronic | 0.9664 | 0.0336 | PRKCB | 0.00460 | 0 | 14 | 52 | 1 | 0 | 25 |
| 16 | 23881930 | G | A | G | rs8047121 | intronic | 0.9662 | 0.0338 | PRKCB | 0.00460 | 0 | 14 | 52 | 1 | 0 | 25 |
| 16 | 23882469 | T | C | T | rs1468129 | intronic | 0.9663 | 0.0337 | PRKCB | 0.00460 | 0 | 14 | 52 | 1 | 0 | 25 |
| 16 | 23885608 | A | T | A | rs8044732 | intronic | 0.9664 | 0.0336 | PRKCB | 0.00460 | 0 | 14 | 52 | 1 | 0 | 25 |
| 16 | 23885751 | A | G | G | rs62031692 | intronic | 0.0253 | 0.0253 | PRKCB | 1.00000 | 59 | 7 | 0 | 23 | 3 | 0 |
| 16 | 23887574 | G | T | T | rs79034087 | intronic | 0.0290 | 0.0290 | PRKCB | 0.34868 | 61 | 4 | 1 | 22 | 4 | 0 |
| 16 | 23888354 | C | T | C | rs7404417 | intronic | 0.9664 | 0.0336 | PRKCB | 0.00460 | 0 | 14 | 52 | 1 | 0 | 25 |
| 16 | 23889896 | T | C | T | rs8063823 | intronic | 0.9665 | 0.0335 | PRKCB | 0.00460 | 0 | 14 | 52 | 1 | 0 | 25 |
| 16 | 23893893 | G | A | G | rs11647359 | intronic | 0.9664 | 0.0336 | PRKCB | 0.00460 | 0 | 14 | 52 | 1 | 0 | 25 |
| 16 | 23895034 | A | G | A | rs6497695 | intronic | 0.9665 | 0.0335 | PRKCB | 0.00460 | 0 | 14 | 52 | 1 | 0 | 25 |

| | | | | | | | | | | | | | | | |
|---|---|---|---|---|---|---|---|---|---|---|---|---|---|---|---|
| 16 | 23895443 | A | G | G | rs62028075 | intronic | 0.0253 | 0.0253 | PRKCB | 1.00000 | 59 | 7 | 0 | 23 | 3 | 0 |
| 16 | 23895884 | T | C | T | rs9944348 | intronic | 0.9665 | 0.0335 | PRKCB | 0.00460 | 0 | 14 | 52 | 1 | 0 | 25 |
| 16 | 23896089 | T | C | C | rs74572166 | intronic | 0.0245 | 0.0245 | PRKCB | 1.00000 | 59 | 7 | 0 | 23 | 3 | 0 |
| 16 | 23896209 | C | A | C | rs9302418 | intronic | 0.9664 | 0.0336 | PRKCB | 0.00460 | 0 | 14 | 52 | 1 | 0 | 25 |
| 16 | 23896438 | G | T | T | rs62028076 | intronic | 0.0252 | 0.0252 | PRKCB | 1.00000 | 59 | 7 | 0 | 22 | 3 | 0 |
| 16 | 23898605 | A | T | A | rs933290 | intronic | 0.9632 | 0.0368 | PRKCB | 0.00159 | 0 | 17 | 49 | 1 | 0 | 24 |
| 16 | 23899211 | A | T | A | rs12926245 | intronic | 0.9632 | 0.0368 | PRKCB | 0.00118 | 0 | 17 | 49 | 1 | 0 | 25 |
| 16 | 23899610 | G | A | A | rs17753246 | intronic | 0.0252 | 0.0252 | PRKCB | 1.00000 | 59 | 7 | 0 | 23 | 3 | 0 |
| 16 | 23899951 | G | A | A | rs62028077 | intronic | 0.0254 | 0.0254 | PRKCB | 1.00000 | 59 | 7 | 0 | 23 | 3 | 0 |
| 16 | 23900716 | T | C | C | rs62028078 | intronic | 0.0252 | 0.0252 | PRKCB | 1.00000 | 59 | 7 | 0 | 23 | 3 | 0 |
| 16 | 23901896 | C | T | C | rs6497696 | intronic | 0.9632 | 0.0368 | PRKCB | 0.00149 | 0 | 16 | 50 | 1 | 0 | 25 |
| 16 | 23901948 | A | C | A | rs6497697 | intronic | 0.9630 | 0.0370 | PRKCB | 0.00149 | 0 | 16 | 49 | 1 | 0 | 25 |
| 16 | 23904058 | A | G | A | rs886115 | intronic | 0.9632 | 0.0368 | PRKCB | 0.00149 | 0 | 16 | 50 | 1 | 0 | 25 |
| 16 | 23904781 | G | A | A | rs17753509 | intronic | 0.0253 | 0.0253 | PRKCB | 1.00000 | 59 | 7 | 0 | 23 | 3 | 0 |
| 16 | 23905676 | C | T | C | rs7200610 | intronic | 0.9631 | 0.0369 | PRKCB | 0.00149 | 0 | 16 | 50 | 1 | 0 | 25 |
| 16 | 23907177 | A | C | C | rs17810011 | intronic | 0.0251 | 0.0251 | PRKCB | 1.00000 | 59 | 7 | 0 | 23 | 3 | 0 |
| 16 | 23907765 | C | T | C | rs9925890 | intronic | 0.9632 | 0.0368 | PRKCB | 0.00240 | 0 | 16 | 50 | 1 | 0 | 24 |
| 16 | 23912174 | A | G | A | rs12448249 | intronic | 0.9519 | 0.0481 | PRKCB | 0.00149 | 0 | 16 | 50 | 1 | 0 | 25 |
| 16 | 23914915 | C | A | C | rs1004186 | intronic | 0.9632 | 0.0368 | PRKCB | 0.00149 | 0 | 16 | 50 | 1 | 0 | 25 |
| 16 | 23916258 | G | A | G | rs1004187 | intronic | 0.9632 | 0.0368 | PRKCB | 0.00149 | 0 | 16 | 50 | 1 | 0 | 25 |
| 16 | 23916521 | G | C | G | rs1008654 | intronic | 0.9633 | 0.0367 | PRKCB | 0.00149 | 0 | 16 | 50 | 1 | 0 | 25 |
| 16 | 23917335 | G | A | G | rs6497699 | intronic | 0.9645 | 0.0355 | PRKCB | 0.00240 | 0 | 16 | 50 | 1 | 0 | 24 |
| 16 | 23917465 | C | G | C | rs7186538 | intronic | 0.9645 | 0.0355 | PRKCB | 0.00240 | 0 | 16 | 50 | 1 | 0 | 24 |
| 16 | 23917700 | C | A | C | rs7187091 | intronic | 0.9646 | 0.0354 | PRKCB | 0.00240 | 0 | 16 | 50 | 1 | 0 | 24 |
| 16 | 23919088 | C | T | T | rs78322646 | intronic | 0.0278 | 0.0278 | PRKCB | 0.73476 | 58 | 8 | 0 | 22 | 4 | 0 |
| 16 | 23921083 | C | T | C | rs6497702 | intronic | 0.9647 | 0.0353 | PRKCB | 0.00240 | 0 | 16 | 50 | 1 | 0 | 24 |
| 16 | 23925936 | C | G | C | rs11074588 | intronic | 0.9649 | 0.0351 | PRKCB | 0.00149 | 0 | 16 | 50 | 1 | 0 | 25 |
| 16 | 23939212 | G | A | G | rs11074590 | intronic | 0.9650 | 0.0350 | PRKCB | 0.00240 | 0 | 16 | 50 | 1 | 0 | 24 |
| 16 | 23941628 | C | A | C | rs2005671 | intronic | 0.9647 | 0.0353 | PRKCB | 0.00149 | 0 | 16 | 49 | 1 | 0 | 25 |
| 16 | 23943749 | T | C | T | rs9302420 | intronic | 0.9649 | 0.0351 | PRKCB | 0.00240 | 0 | 16 | 50 | 1 | 0 | 24 |
| 16 | 23945985 | T | G | T | rs195989 | intronic | 0.9651 | 0.0349 | PRKCB | 0.00149 | 0 | 16 | 50 | 1 | 0 | 25 |
| 16 | 23946157 | G | A | A | rs76973283 | intronic | 0.0302 | 0.0302 | PRKCB | 0.50649 | 58 | 8 | 0 | 21 | 5 | 0 |
| 16 | 23949175 | G | C | G | rs2560403 | intronic | 0.9657 | 0.0343 | PRKCB | 0.00149 | 0 | 16 | 50 | 1 | 0 | 25 |
| 16 | 23949438 | A | G | A | rs195985 | intronic | 0.9658 | 0.0342 | PRKCB | 0.00149 | 0 | 16 | 50 | 1 | 0 | 25 |
| 16 | 23953265 | T | C | T | rs2560404 | intronic | 0.9656 | 0.0344 | PRKCB | 0.00240 | 0 | 16 | 50 | 1 | 0 | 24 |
| 16 | 23954128 | T | C | C | rs17810486 | intronic | 0.0308 | 0.0308 | PRKCB | 0.30238 | 58 | 7 | 0 | 20 | 5 | 0 |
| 16 | 23954253 | G | A | G | rs195994 | intronic | 0.9653 | 0.0347 | PRKCB | 0.00149 | 0 | 16 | 50 | 1 | 0 | 25 |
| 16 | 23962258 | G | C | G | rs196000 | intronic | 0.9659 | 0.0341 | PRKCB | 0.00149 | 0 | 16 | 50 | 1 | 0 | 25 |
| 16 | 23964858 | T | A | T | rs196003 | intronic | 0.9647 | 0.0353 | PRKCB | 0.03433 | 0 | 15 | 51 | 0 | 1 | 25 |
| 16 | 23985814 | C | T | T | rs72779914 | intronic | 0.0487 | 0.0487 | PRKCB | 1.00000 | 53 | 12 | 0 | 21 | 4 | 0 |

| | | | | | | | | | | | | | | | |
|---|---|---|---|---|---|---|---|---|---|---|---|---|---|---|---|
| 16 | 23987552 | A | G | A | rs169030 | intronic | 0.9709 | 0.0291 | PRKCB | 0.16792 | 0 | 11 | 55 | 0 | 1 | 25 |
| 16 | 23988755 | T | C | T | rs196013 | intronic | 0.9681 | 0.0319 | PRKCB | 0.16812 | 0 | 11 | 55 | 0 | 1 | 24 |
| 16 | 24009919 | A | G | G | rs75622923 | intronic | 0.0319 | 0.0319 | PRKCB | 0.33476 | 54 | 12 | 0 | 24 | 2 | 0 |
| 16 | 24022944 | C | T | T | rs111746132 | intronic | 0.0229 | 0.0229 | PRKCB | 0.75217 | 56 | 10 | 0 | 23 | 3 | 0 |
| 16 | 24066378 | G | A | A | rs113426570 | intronic | 0.0216 | 0.0216 | PRKCB | 1.00000 | 57 | 9 | 0 | 23 | 3 | 0 |
| 16 | 24100759 | T | A | A | rs11643939 | intronic | 0.0294 | 0.0294 | PRKCB | 0.75217 | 56 | 10 | 0 | 23 | 3 | 0 |
| 16 | 24105816 | G | A | A | rs56316329 | intronic | 0.0251 | 0.0251 | PRKCB | 0.69190 | 60 | 4 | 1 | 25 | 0 | 0 |
| 16 | 24111853 | T | C | C | rs55959083 | intronic | 0.0431 | 0.0431 | PRKCB | 0.72129 | 58 | 8 | 0 | 23 | 2 | 0 |
| 16 | 24112768 | G | A | A | rs117056307 | intronic | 0.0430 | 0.0430 | PRKCB | 0.71919 | 58 | 8 | 0 | 24 | 2 | 0 |
| 16 | 24122052 | G | A | A | rs117467859 | intronic | 0.0433 | 0.0433 | PRKCB | 0.72129 | 58 | 8 | 0 | 23 | 2 | 0 |
| 16 | 24122492 | C | T | T | rs72779977 | intronic | 0.0457 | 0.0457 | PRKCB | 0.72009 | 57 | 8 | 0 | 23 | 2 | 0 |
| 16 | 24123560 | G | A | A | rs60261043 | intronic | 0.0457 | 0.0457 | PRKCB | 0.72009 | 57 | 8 | 0 | 23 | 2 | 0 |
| 16 | 24132273 | G | A | A | rs62027458 | intronic | 0.0232 | 0.0232 | PRKCB | 0.18770 | 59 | 7 | 0 | 21 | 4 | 1 |
| 16 | 24164042 | G | T | T | rs72779989 | intronic | 0.0487 | 0.0487 | PRKCB | 0.85826 | 57 | 7 | 2 | 22 | 4 | 0 |
| 16 | 24197496 | A | T | T | rs79699525 | intronic | 0.0261 | 0.0261 | PRKCB | 0.46312 | 59 | 6 | 0 | 22 | 4 | 0 |
| 16 | 24199852 | C | T | T | rs78424166 | intronic | 0.0359 | 0.0359 | PRKCB | 0.46031 | 60 | 6 | 0 | 22 | 4 | 0 |

Table S4 (A): Significant results from pathway-based tests

| REACTOME PATHWAY | # MUTATIONS All | # MUTATIONS p value < 0.05 | P VALUE (P/WAY-BASED) | FWER |
|---|---|---|---|---|
| Activation of NF-kappaB in B cells | 264 | 74 | 4.89E-09 | 9.68E-06 |
| WNT5A-dependent internalization of FZD4 | 187 | 49 | 1.31E-06 | 2.59E-03 |
| RUNX1 regulates transcription of genes involved in differentiation of myeloid cells | 190 | 48 | 1.33E-06 | 2.64E-03 |
| Downstream signaling events of B Cell Receptor (BCR) | 345 | 74 | 4.60E-06 | 9.05E-03 |
| Depolymerisation of the Nuclear Lamina | 191 | 48 | 5.88E-06 | 1.16E-02 |
| Disinhibition of SNARE formation | 177 | 48 | 6.89E-06 | 1.35E-02 |
| Disassembly of the destruction complex and recruitment of AXIN to the membrane | 123 | 43 | 3.67E-05 | 7.00E-02 |

## Table S4 (B): Variants with MAF<5% in signifiant pathways

| REACTOME PATHWAY | CHROM | POS | REF | ALT | MINOR ALLELE | ID | CONSEQUENCE | AF | MAF | SYMBOL | P VALUE | # RESPONDERS Ref/Ref | # RESPONDERS Ref/Alt | # RESPONDERS Alt/Alt | # NON-RESPONDERS Ref/Ref | # NON-RESPONDERS Ref/Alt | # NON-RESPONDERS Alt/Alt |
|---|---|---|---|---|---|---|---|---|---|---|---|---|---|---|---|---|---|
| Activation of NF-kappaB in B cells | 1 | 85738304 | T | G | G | rs141532778 | intronic | 0.0265 | 0.0265 | BCL10 | 0.1949 | 59 | 7 | 0 | 24 | 1 | 1 |
| Activation of NF-kappaB in B cells | 1 | 85742155 | G | A | A | rs78416998 | 5_prime_UTR | 0.0349 | 0.0349 | BCL10 | 0.7734 | 60 | 5 | 1 | 23 | 3 | 0 |
| Activation of NF-kappaB in B cells | 1 | 109959684 | G | A | A | rs12119154 | intronic | 0.0463 | 0.0463 | PSMA5 | 0.3842 | 59 | 7 | 0 | 22 | 3 | 1 |
| Activation of NF-kappaB in B cells | 2 | 54106726 | G | A | A | rs78542544 | intronic | 0.0466 | 0.0466 | PSME4 | 0.0565 | 56 | 10 | 0 | 25 | 0 | 0 |
| Activation of NF-kappaB in B cells | 2 | 54109699 | G | A | A | rs75679248 | intronic | 0.0345 | 0.0345 | PSME4 | 1.0000 | 57 | 9 | 0 | 22 | 3 | 0 |
| Activation of NF-kappaB in B cells | 2 | 54112612 | T | G | G | rs115678984 | intronic | 0.0340 | 0.0340 | PSME4 | 0.7302 | 58 | 8 | 0 | 21 | 4 | 0 |
| Activation of NF-kappaB in B cells | 2 | 54116463 | C | T | T | rs805403 | intronic | 0.0376 | 0.0376 | PSME4 | 0.2083 | 62 | 4 | 0 | 21 | 4 | 0 |
| Activation of NF-kappaB in B cells | 2 | 54119870 | T | C | C | rs77215245 | intronic | 0.0487 | 0.0487 | PSME4 | 0.6362 | 55 | 10 | 1 | 23 | 2 | 0 |
| Activation of NF-kappaB in B cells | 2 | 54121956 | T | C | C | rs79436269 | intronic | 0.0498 | 0.0498 | PSME4 | 0.6362 | 55 | 10 | 1 | 23 | 2 | 0 |
| Activation of NF-kappaB in B cells | 2 | 54131546 | T | C | C | rs78887562 | intronic | 0.0467 | 0.0467 | PSME4 | 0.0565 | 56 | 10 | 0 | 25 | 0 | 0 |
| Activation of NF-kappaB in B cells | 2 | 54132737 | G | A | A | rs75125553 | intronic | 0.0493 | 0.0493 | PSME4 | 0.6407 | 54 | 10 | 1 | 24 | 2 | 0 |
| Activation of NF-kappaB in B cells | 2 | 54148310 | G | C | C | rs74545963 | intronic | 0.0348 | 0.0348 | PSME4 | 1.0000 | 57 | 9 | 0 | 23 | 3 | 0 |
| Activation of NF-kappaB in B cells | 2 | 54153284 | A | G | G | rs62139281 | intronic | 0.0226 | 0.0226 | PSME4 | 0.0237 | 62 | 4 | 0 | 19 | 6 | 0 |
| Activation of NF-kappaB in B cells | 2 | 54161109 | A | C | C | rs74627832 | intronic | 0.0468 | 0.0468 | PSME4 | 0.0565 | 56 | 10 | 0 | 25 | 0 | 0 |
| Activation of NF-kappaB in B cells | 2 | 54166671 | C | T | T | rs79435818 | intronic | 0.0317 | 0.0317 | PSME4 | 1.0000 | 58 | 8 | 0 | 23 | 3 | 0 |
| Activation of NF-kappaB in B cells | 2 | 54166800 | A | G | G | rs76811723 | intronic | 0.0317 | 0.0317 | PSME4 | 1.0000 | 58 | 8 | 0 | 22 | 3 | 0 |
| Activation of NF-kappaB in B cells | 2 | 54176733 | C | A | A | rs115660490 | intronic | 0.0366 | 0.0366 | PSME4 | 0.7213 | 58 | 8 | 0 | 23 | 2 | 0 |
| Activation of NF-kappaB in B cells | 2 | 54182973 | T | A | A | rs114515761 | intronic | 0.0346 | 0.0346 | PSME4 | 0.7192 | 58 | 8 | 0 | 24 | 2 | 0 |
| Activation of NF-kappaB in B cells | 2 | 54188394 | T | A | A | rs11692784 | intronic | 0.0346 | 0.0346 | PSME4 | 0.7192 | 58 | 8 | 0 | 24 | 2 | 0 |
| Activation of NF-kappaB in B cells | 2 | 54198172 | G | A | A | rs141622297 | upstream | 0.0360 | 0.0360 | PSME4 | 0.7192 | 58 | 8 | 0 | 24 | 2 | 0 |
| Activation of NF-kappaB in B cells | 2 | 54201823 | G | A | A | rs74336645 | upstream | 0.0362 | 0.0362 | PSME4 | 0.7213 | 58 | 8 | 0 | 23 | 2 | 0 |
| Activation of NF-kappaB in B cells | 2 | 61155012 | C | G | G | rs79263888 | 3_prime_UTR | 0.0229 | 0.0229 | REL | 0.7522 | 56 | 10 | 0 | 23 | 3 | 0 |
| Activation of NF-kappaB in B cells | 2 | 162169482 | G | A | A | rs7605885 | intronic | 0.0425 | 0.0425 | PSMD14 | 0.8183 | 56 | 9 | 1 | 21 | 4 | 0 |
| Activation of NF-kappaB in B cells | 2 | 162215335 | C | T | T | rs6722186 | intronic | 0.0427 | 0.0427 | PSMD14 | 1.0000 | 56 | 9 | 1 | 22 | 4 | 0 |
| Activation of NF-kappaB in B cells | 2 | 231921109 | G | A | A | rs3754982 | upstream | 0.0430 | 0.0430 | PSMD1 | 0.0899 | 60 | 6 | 0 | 25 | 0 | 1 |
| Activation of NF-kappaB in B cells | 2 | 231926692 | G | A | A | rs80031661 | intronic | 0.0429 | 0.0429 | PSMD1 | 0.0899 | 60 | 6 | 0 | 25 | 0 | 1 |
| Activation of NF-kappaB in B cells | 2 | 231932134 | T | C | C | rs78964764 | intronic | 0.0433 | 0.0433 | PSMD1 | 0.0863 | 60 | 6 | 0 | 24 | 0 | 1 |
| Activation of NF-kappaB in B cells | 2 | 231933408 | C | T | T | rs13424110 | intronic | 0.0476 | 0.0476 | PSMD1 | 0.1782 | 59 | 7 | 0 | 20 | 4 | 1 |
| Activation of NF-kappaB in B cells | 2 | 231937539 | T | C | C | rs2288148 | intronic | 0.0434 | 0.0434 | PSMD1 | 0.0863 | 60 | 6 | 0 | 24 | 0 | 1 |
| Activation of NF-kappaB in B cells | 2 | 231939526 | A | G | G | rs12620983 | intronic | 0.0434 | 0.0434 | PSMD1 | 0.0863 | 60 | 6 | 0 | 24 | 0 | 1 |
| Activation of NF-kappaB in B cells | 2 | 231947467 | A | G | G | rs11694724 | intronic | 0.0411 | 0.0411 | PSMD1 | 0.2744 | 57 | 9 | 0 | 24 | 1 | 0 |
| Activation of NF-kappaB in B cells | 2 | 231949569 | T | C | C | rs2303354 | intronic | 0.0433 | 0.0433 | PSMD1 | 0.0876 | 59 | 6 | 0 | 24 | 0 | 1 |
| Activation of NF-kappaB in B cells | 2 | 231950847 | T | G | G | rs78363488 | intronic | 0.0407 | 0.0407 | PSMD1 | 0.4353 | 57 | 8 | 0 | 24 | 1 | 0 |
| Activation of NF-kappaB in B cells | 2 | 231967886 | T | C | C | rs12616914 | intronic | 0.0436 | 0.0436 | PSMD1 | 0.0863 | 60 | 6 | 0 | 24 | 0 | 1 |
| Activation of NF-kappaB in B cells | 2 | 231979258 | A | G | G | rs17586405 | intronic | 0.0434 | 0.0434 | PSMD1 | 0.0863 | 60 | 6 | 0 | 24 | 0 | 1 |
| Activation of NF-kappaB in B cells | 2 | 231982950 | T | C | C | rs60707561 | intronic | 0.0434 | 0.0434 | PSMD1 | 0.0863 | 60 | 6 | 0 | 24 | 0 | 1 |
| Activation of NF-kappaB in B cells | 2 | 231983382 | T | G | G | rs17619636 | intronic | 0.0478 | 0.0478 | PSMD1 | 0.0899 | 60 | 6 | 0 | 25 | 0 | 1 |
| Activation of NF-kappaB in B cells | 2 | 231998010 | G | A | A | rs77199363 | intronic | 0.0431 | 0.0431 | PSMD1 | 0.0899 | 60 | 6 | 0 | 25 | 0 | 1 |
| Activation of NF-kappaB in B cells | 2 | 232003189 | C | T | T | rs80121410 | intronic | 0.0408 | 0.0408 | PSMD1 | 0.2744 | 57 | 9 | 0 | 24 | 1 | 0 |
| Activation of NF-kappaB in B cells | 2 | 232021354 | T | C | C | rs115196328 | intronic | 0.0402 | 0.0402 | PSMD1 | 0.5418 | 53 | 13 | 0 | 22 | 3 | 0 |
| Activation of NF-kappaB in B cells | 2 | 232025880 | C | T | T | rs76901853 | intronic | 0.0472 | 0.0472 | PSMD1 | 0.0863 | 60 | 6 | 0 | 24 | 0 | 1 |
| Activation of NF-kappaB in B cells | 2 | 232036567 | G | A | A | rs11674175 | intronic | 0.0409 | 0.0409 | PSMD1 | 0.2744 | 57 | 9 | 0 | 24 | 1 | 0 |
| Activation of NF-kappaB in B cells | 2 | 232041640 | G | A | A | rs111612792 | downstream | 0.0413 | 0.0413 | PSMD1 | 0.2719 | 57 | 9 | 0 | 25 | 1 | 0 |
| Activation of NF-kappaB in B cells | 3 | 63994801 | G | A | A | rs62252370 | downstream | 0.0360 | 0.0360 | PSMD6 | 1.0000 | 58 | 8 | 0 | 23 | 3 | 0 |
| Activation of NF-kappaB in B cells | 4 | 103436737 | T | C | C | rs79590323 | intronic | 0.0393 | 0.0393 | NFKB1 | 0.4981 | 56 | 10 | 0 | 24 | 2 | 0 |
| Activation of NF-kappaB in B cells | 4 | 103441152 | G | A | A | rs74833382 | intronic | 0.0318 | 0.0318 | NFKB1 | 0.7621 | 59 | 6 | 1 | 25 | 1 | 0 |
| Activation of NF-kappaB in B cells | 4 | 103441742 | G | A | A | rs78900265 | intronic | 0.0393 | 0.0393 | NFKB1 | 0.4999 | 56 | 10 | 0 | 23 | 2 | 0 |
| Activation of NF-kappaB in B cells | 4 | 103465229 | C | T | T | rs76016852 | intronic | 0.0414 | 0.0414 | NFKB1 | 0.4999 | 56 | 10 | 0 | 23 | 2 | 0 |
| Activation of NF-kappaB in B cells | 4 | 103534557 | C | T | T | rs4648117 | intronic | 0.0416 | 0.0416 | NFKB1 | 0.7621 | 59 | 6 | 1 | 25 | 1 | 0 |
| Activation of NF-kappaB in B cells | 4 | 103535905 | C | T | T | rs4648127 | intronic | 0.0389 | 0.0389 | NFKB1 | 0.4974 | 55 | 10 | 0 | 24 | 2 | 0 |
| Activation of NF-kappaB in B cells | 4 | 103542005 | G | T | T | rs997476 | downstream | 0.0397 | 0.0397 | NFKB1 | 0.4981 | 56 | 10 | 0 | 24 | 2 | 0 |
| Activation of NF-kappaB in B cells | 5 | 133505097 | T | C | C | rs34751193 | intronic | 0.0303 | 0.0303 | SKP1 | 0.7497 | 55 | 11 | 0 | 23 | 3 | 0 |
| Activation of NF-kappaB in B cells | 5 | 133512605 | G | T | T | rs11538030 | 5_prime_UTR | 0.0356 | 0.0356 | SKP1 | 0.6295 | 55 | 9 | 2 | 24 | 2 | 0 |
| Activation of NF-kappaB in B cells | 5 | 171287624 | C | G | C | rs702109 | downstream | 0.9605 | 0.0395 | FBXW11 | 0.1356 | 0 | 13 | 52 | 1 | 2 | 22 |
| Activation of NF-kappaB in B cells | 5 | 171319929 | T | C | C | rs17569783 | intronic | 0.0456 | 0.0456 | FBXW11 | 1.0000 | 59 | 6 | 1 | 23 | 2 | 0 |
| Activation of NF-kappaB in B cells | 5 | 171387634 | G | A | A | rs72835279 | intronic | 0.0450 | 0.0450 | FBXW11 | 1.0000 | 59 | 6 | 1 | 24 | 2 | 0 |
| Activation of NF-kappaB in B cells | 5 | 171420105 | G | C | C | rs72835287 | intronic | 0.0441 | 0.0441 | FBXW11 | 0.2719 | 57 | 9 | 0 | 25 | 1 | 0 |
| Activation of NF-kappaB in B cells | 5 | 171424137 | T | G | G | rs72835288 | intronic | 0.0465 | 0.0465 | FBXW11 | 1.0000 | 59 | 6 | 1 | 24 | 2 | 0 |
| Activation of NF-kappaB in B cells | 6 | 32810147 | G | T | T | rs41270496 | intronic | 0.0323 | 0.0323 | PSMB8 | 0.4984 | 59 | 7 | 0 | 22 | 4 | 0 |
| Activation of NF-kappaB in B cells | 6 | 32811224 | A | T | T | rs72854938 | intronic | 0.0322 | 0.0322 | PSMB8 | 0.4984 | 59 | 7 | 0 | 22 | 4 | 0 |
| Activation of NF-kappaB in B cells | 6 | 32811366 | G | T | T | rs72854939 | intronic | 0.0322 | 0.0322 | PSMB8 | 0.4984 | 59 | 7 | 0 | 22 | 4 | 0 |
| Activation of NF-kappaB in B cells | 6 | 32826688 | G | T | T | rs115353581 | intronic | 0.0219 | 0.0219 | PSMB9 | 0.4984 | 59 | 7 | 0 | 22 | 4 | 0 |
| Activation of NF-kappaB in B cells | 6 | 32832218 | A | T | T | rs116481206 | downstream | 0.0219 | 0.0219 | PSMB9 | 0.4862 | 59 | 7 | 0 | 21 | 4 | 0 |
| Activation of NF-kappaB in B cells | 6 | 44227224 | C | T | T | rs28362860 | intronic | 0.0269 | 0.0269 | NFKBIE | 0.4603 | 60 | 6 | 0 | 22 | 4 | 0 |
| Activation of NF-kappaB in B cells | 6 | 44233216 | G | C | C | rs28362857 | missense | 0.0296 | 0.0296 | NFKBIE | 0.0685 | 61 | 5 | 0 | 20 | 6 | 0 |
| Activation of NF-kappaB in B cells | 6 | 91239702 | T | A | A | rs62409064 | intronic | 0.0407 | 0.0407 | MAP3K7 | 0.4940 | 58 | 7 | 1 | 20 | 5 | 0 |
| Activation of NF-kappaB in B cells | 6 | 91296810 | C | T | T | rs34016005 | upstream | 0.0303 | 0.0303 | MAP3K7 | 0.4762 | 60 | 5 | 1 | 22 | 4 | 0 |
| Activation of NF-kappaB in B cells | 6 | 170851525 | G | A | A | rs12200064 | intronic | 0.0431 | 0.0431 | PSMB1 | 0.3694 | 59 | 5 | 2 | 26 | 0 | 0 |
| Activation of NF-kappaB in B cells | 6 | 170858332 | G | A | A | rs17860779 | intronic | 0.0420 | 0.0420 | PSMB1 | 1.0000 | 59 | 6 | 1 | 23 | 2 | 0 |
| Activation of NF-kappaB in B cells | 7 | 2955476 | G | C | C | rs41336352 | intronic | 0.0400 | 0.0400 | CARD11 | 1.0000 | 58 | 8 | 0 | 22 | 3 | 0 |
| Activation of NF-kappaB in B cells | 7 | 2966334 | G | A | A | rs41448444 | intronic | 0.0454 | 0.0454 | CARD11 | 0.7213 | 58 | 8 | 0 | 23 | 2 | 0 |
| Activation of NF-kappaB in B cells | 7 | 2968105 | C | A | A | rs71527417 | intronic | 0.0472 | 0.0472 | CARD11 | 0.0436 | 49 | 16 | 1 | 25 | 1 | 0 |
| Activation of NF-kappaB in B cells | 7 | 2970297 | C | T | T | rs41386651 | intronic | 0.0410 | 0.0410 | CARD11 | 0.7386 | 56 | 8 | 0 | 22 | 4 | 0 |
| Activation of NF-kappaB in B cells | 7 | 2970866 | C | T | T | rs9648301 | intronic | 0.0444 | 0.0444 | CARD11 | 1.0000 | 57 | 9 | 0 | 22 | 4 | 0 |
| Activation of NF-kappaB in B cells | 7 | 3029796 | G | C | C | rs12700500 | intronic | 0.0454 | 0.0454 | CARD11 | 1.0000 | 59 | 7 | 0 | 22 | 3 | 0 |
| Activation of NF-kappaB in B cells | 7 | 3042802 | G | A | A | rs35579453 | intronic | 0.0363 | 0.0363 | CARD11 | 1.0000 | 57 | 9 | 0 | 22 | 4 | 0 |
| Activation of NF-kappaB in B cells | 7 | 3052087 | G | A | A | rs75468256 | intronic | 0.0407 | 0.0407 | CARD11 | 0.2020 | 62 | 4 | 0 | 20 | 4 | 0 |
| Activation of NF-kappaB in B cells | 7 | 3072512 | T | C | C | rs6975176 | intronic | 0.0452 | 0.0452 | CARD11 | 1.0000 | 61 | 4 | 1 | 25 | 1 | 0 |
| Activation of NF-kappaB in B cells | 7 | 3073243 | G | A | A | rs78353391 | intronic | 0.0427 | 0.0427 | CARD11 | 0.0650 | 56 | 8 | 0 | 18 | 8 | 0 |
| Activation of NF-kappaB in B cells | 7 | 3079195 | G | T | T | rs62439353 | intronic | 0.0400 | 0.0400 | CARD11 | 0.7348 | 58 | 8 | 0 | 22 | 4 | 0 |
| Activation of NF-kappaB in B cells | 7 | 42971385 | G | A | A | rs76256519 | intronic | 0.0432 | 0.0432 | PSMA2 | 0.1923 | 61 | 5 | 0 | 21 | 3 | 1 |
| Activation of NF-kappaB in B cells | 7 | 148411091 | C | T | T | rs11764941 | intronic | 0.0416 | 0.0416 | CUL1 | 0.0082 | 58 | 8 | 0 | 16 | 9 | 1 |
| Activation of NF-kappaB in B cells | 7 | 148422581 | A | G | G | rs11760399 | intronic | 0.0409 | 0.0409 | CUL1 | 0.0158 | 57 | 9 | 0 | 16 | 9 | 1 |
| Activation of NF-kappaB in B cells | 7 | 148427470 | A | T | T | rs17625893 | intronic | 0.0188 | 0.0188 | CUL1 | 0.3091 | 59 | 7 | 0 | 21 | 5 | 0 |
| Activation of NF-kappaB in B cells | 7 | 148453031 | C | G | G | rs4726991 | intronic | 0.0427 | 0.0427 | CUL1 | 0.0164 | 61 | 5 | 0 | 18 | 7 | 0 |
| Activation of NF-kappaB in B cells | 7 | 148455710 | A | G | G | rs73158225 | intronic | 0.0426 | 0.0426 | CUL1 | 0.0333 | 61 | 5 | 0 | 19 | 7 | 0 |
| Activation of NF-kappaB in B cells | 7 | 148456174 | T | G | G | rs17537343 | intronic | 0.0430 | 0.0430 | CUL1 | 0.0164 | 61 | 5 | 0 | 18 | 7 | 0 |
| Activation of NF-kappaB in B cells | 7 | 148458141 | C | T | T | rs73158226 | intronic | 0.0426 | 0.0426 | CUL1 | 0.0616 | 61 | 5 | 0 | 18 | 6 | 0 |
| Activation of NF-kappaB in B cells | 7 | 148470413 | T | C | C | rs73158234 | intronic | 0.0423 | 0.0423 | CUL1 | 0.0333 | 61 | 5 | 0 | 19 | 7 | 0 |
| Activation of NF-kappaB in B cells | 7 | 148502252 | T | C | C | rs77948593 | downstream | 0.0395 | 0.0395 | CUL1 | 1.0000 | 59 | 6 | 1 | 23 | 2 | 0 |
| Activation of NF-kappaB in B cells | 8 | 42125339 | T | C | C | rs76361874 | upstream | 0.0303 | 0.0303 | IKBKB | 0.0653 | 59 | 7 | 0 | 25 | 0 | 1 |
| Activation of NF-kappaB in B cells | 8 | 42127643 | A | C | C | rs17875739 | upstream | 0.0241 | 0.0241 | IKBKB | 0.0653 | 59 | 7 | 0 | 25 | 0 | 1 |
| Activation of NF-kappaB in B cells | 8 | 42128072 | T | A | A | rs62507976 | upstream | 0.0296 | 0.0296 | IKBKB | 0.4157 | 61 | 5 | 0 | 23 | 2 | 1 |
| Activation of NF-kappaB in B cells | 8 | 42135678 | T | G | G | rs17875740 | intronic | 0.0335 | 0.0335 | IKBKB | 0.0330 | 58 | 8 | 0 | 25 | 0 | 1 |
| Activation of NF-kappaB in B cells | 8 | 42138781 | G | C | C | rs80313154 | intronic | 0.0274 | 0.0274 | IKBKB | 0.0330 | 58 | 8 | 0 | 25 | 0 | 1 |
| Activation of NF-kappaB in B cells | 8 | 42142316 | G | A | A | rs79881854 | intronic | 0.0326 | 0.0326 | IKBKB | 0.0321 | 58 | 8 | 0 | 24 | 0 | 1 |
| Activation of NF-kappaB in B cells | 8 | 42145446 | A | G | G | rs17875744 | intronic | 0.0327 | 0.0327 | IKBKB | 0.0272 | 57 | 9 | 0 | 25 | 0 | 1 |
| Activation of NF-kappaB in B cells | 8 | 42150794 | C | T | T | rs79123247 | intronic | 0.0326 | 0.0326 | IKBKB | 0.0330 | 58 | 8 | 0 | 25 | 0 | 1 |
| Activation of NF-kappaB in B cells | 8 | 42154866 | A | G | G | rs78342373 | intronic | 0.0326 | 0.0326 | IKBKB | 0.0330 | 58 | 8 | 0 | 25 | 0 | 1 |
| Activation of NF-kappaB in B cells | 8 | 42155029 | G | A | A | rs112141582 | intronic | 0.0267 | 0.0267 | IKBKB | 0.0321 | 58 | 8 | 0 | 24 | 0 | 1 |
| Activation of NF-kappaB in B cells | 8 | 42158881 | A | G | G | rs75230171 | intronic | 0.0327 | 0.0327 | IKBKB | 0.0321 | 58 | 8 | 0 | 24 | 0 | 1 |
| Activation of NF-kappaB in B cells | 8 | 42159022 | C | T | T | rs79460550 | intronic | 0.0326 | 0.0326 | IKBKB | 0.0330 | 58 | 8 | 0 | 25 | 0 | 1 |
| Activation of NF-kappaB in B cells | 8 | 42161317 | G | A | A | rs113993388 | intronic | 0.0280 | 0.0280 | IKBKB | 0.0330 | 58 | 8 | 0 | 25 | 0 | 1 |
| Activation of NF-kappaB in B cells | 8 | 42162675 | C | T | T | rs17875700 | intronic | 0.0338 | 0.0338 | IKBKB | 0.0321 | 58 | 8 | 0 | 24 | 0 | 1 |
| Activation of NF-kappaB in B cells | 8 | 42170928 | C | T | T | rs76891399 | intronic | 0.0325 | 0.0325 | IKBKB | 0.0321 | 58 | 8 | 0 | 24 | 0 | 1 |
| Activation of NF-kappaB in B cells | 8 | 42173677 | G | T | T | rs17875746 | intronic | 0.0322 | 0.0322 | IKBKB | 0.0330 | 58 | 8 | 0 | 25 | 0 | 1 |
| Activation of NF-kappaB in B cells | 8 | 42180084 | G | A | A | rs17875721 | intronic | 0.0326 | 0.0326 | IKBKB | 0.0330 | 58 | 8 | 0 | 25 | 0 | 1 |
| Activation of NF-kappaB in B cells | 8 | 42180328 | G | T | T | rs17875751 | intronic | 0.0327 | 0.0327 | IKBKB | 0.0330 | 58 | 8 | 0 | 25 | 0 | 1 |
| Activation of NF-kappaB in B cells | 8 | 42180679 | G | A | A | rs78599265 | intronic | 0.0333 | 0.0333 | IKBKB | 0.0330 | 58 | 8 | 0 | 25 | 0 | 1 |
| Activation of NF-kappaB in B cells | 8 | 42186989 | C | T | T | rs17875731 | intronic | 0.0324 | 0.0324 | IKBKB | 0.0330 | 58 | 8 | 0 | 25 | 0 | 1 |
| Activation of NF-kappaB in B cells | 8 | 42188334 | G | T | T | rs79200457 | intronic | 0.0325 | 0.0325 | IKBKB | 0.0666 | 58 | 7 | 0 | 25 | 0 | 1 |
| Activation of NF-kappaB in B cells | 9 | 123578733 | G | A | A | rs10384 | 3_prime_UTR | 0.0186 | 0.0186 | PSMD5 | 0.3096 | 61 | 5 | 0 | 23 | 1 | 1 |
| Activation of NF-kappaB in B cells | 9 | 123598623 | A | G | G | rs62581704 | intronic | 0.0476 | 0.0476 | PSMD5 | 0.3820 | 62 | 4 | 0 | 24 | 1 | 1 |
| Activation of NF-kappaB in B cells | 9 | 127118932 | G | A | A | rs117300833 | intronic | 0.0270 | 0.0270 | PSMB7 | 0.3024 | 58 | 7 | 0 | 20 | 5 | 0 |
| Activation of NF-kappaB in B cells | 9 | 127119442 | G | A | A | rs16927388 | intronic | 0.0435 | 0.0435 | PSMB7 | 0.4984 | 59 | 7 | 0 | 22 | 4 | 0 |
| Activation of NF-kappaB in B cells | 9 | 127119799 | G | A | A | rs56058032 | intronic | 0.0172 | 0.0172 | PSMB7 | 0.4984 | 59 | 7 | 0 | 22 | 4 | 0 |
| Activation of NF-kappaB in B cells | 9 | 127141846 | G | C | C | rs79069085 | intronic | 0.0178 | 0.0178 | PSMB7 | 0.4984 | 59 | 7 | 0 | 22 | 4 | 0 |
| Activation of NF-kappaB in B cells | 9 | 127143001 | T | A | A | rs76699962 | intronic | 0.0448 | 0.0448 | PSMB7 | 0.4984 | 59 | 7 | 0 | 22 | 4 | 0 |
| Activation of NF-kappaB in B cells | 9 | 127144906 | A | C | C | rs79321634 | intronic | 0.0239 | 0.0239 | PSMB7 | 0.2833 | 59 | 6 | 0 | 21 | 5 | 0 |
| Activation of NF-kappaB in B cells | 9 | 127145687 | T | C | C | rs77214542 | intronic | 0.0335 | 0.0335 | PSMB7 | 1.0000 | 58 | 8 | 0 | 23 | 3 | 0 |

| Pathway | Chr | Position | Ref | Alt | Allele | rsID | Region | Freq1 | Freq2 | Gene | P-value | N1 | N2 | N3 | N4 | N5 | N6 |
|---|---|---|---|---|---|---|---|---|---|---|---|---|---|---|---|---|---|
| Activation of NF-kappaB in B cells | 9 | 127145878 | G | C | C | rs41274376 | intronic | 0.0237 | 0.0237 | PSMB7 | 0.3091 | 59 | 7 | 0 | 21 | 5 | 0 |
| Activation of NF-kappaB in B cells | 9 | 127151807 | T | G | G | rs114451602 | intronic | 0.0301 | 0.0301 | PSMB7 | 0.5095 | 60 | 5 | 1 | 26 | 0 | 0 |
| Activation of NF-kappaB in B cells | 9 | 127172515 | T | C | C | rs140504275 | intronic | 0.0204 | 0.0204 | PSMB7 | 0.1666 | 60 | 6 | 0 | 20 | 5 | 0 |
| Activation of NF-kappaB in B cells | 9 | 127177053 | C | T | T | rs73588260 | intronic | 0.0344 | 0.0344 | PSMB7 | 0.7366 | 57 | 8 | 0 | 22 | 4 | 0 |
| Activation of NF-kappaB in B cells | 10 | 101958770 | C | T | T | rs12764732 | intronic | 0.0464 | 0.0464 | CHUK | 0.7209 | 57 | 9 | 0 | 23 | 2 | 0 |
| Activation of NF-kappaB in B cells | 10 | 101958858 | T | G | G | rs17883365 | intronic | 0.0229 | 0.0229 | CHUK | 1.0000 | 61 | 5 | 0 | 24 | 2 | 0 |
| Activation of NF-kappaB in B cells | 10 | 101961171 | G | A | A | rs17885986 | intronic | 0.0496 | 0.0496 | CHUK | 0.6322 | 57 | 8 | 1 | 21 | 4 | 1 |
| Activation of NF-kappaB in B cells | 10 | 101979482 | T | C | C | rs12764370 | intronic | 0.0302 | 0.0302 | CHUK | 0.7209 | 57 | 9 | 0 | 23 | 2 | 0 |
| Activation of NF-kappaB in B cells | 10 | 103121589 | C | T | T | rs11190960 | intronic | 0.0325 | 0.0325 | BTRC | 0.2234 | 59 | 6 | 1 | 26 | 0 | 0 |
| Activation of NF-kappaB in B cells | 10 | 103165858 | G | A | A | rs145847638 | intronic | 0.0278 | 0.0278 | BTRC | 0.4058 | 59 | 6 | 1 | 25 | 0 | 0 |
| Activation of NF-kappaB in B cells | 10 | 103180336 | G | A | A | rs11191003 | intronic | 0.0274 | 0.0274 | BTRC | 0.2234 | 59 | 6 | 1 | 26 | 0 | 0 |
| Activation of NF-kappaB in B cells | 10 | 103231195 | T | C | C | rs34711120 | intronic | 0.0435 | 0.0435 | BTRC | 0.7302 | 58 | 8 | 0 | 21 | 4 | 0 |
| Activation of NF-kappaB in B cells | 10 | 103272221 | C | T | T | rs12774622 | intronic | 0.0257 | 0.0257 | BTRC | 0.7970 | 57 | 8 | 1 | 23 | 2 | 0 |
| Activation of NF-kappaB in B cells | 10 | 103298099 | G | T | T | rs4151060 | missense | 0.0289 | 0.0289 | BTRC | 0.6362 | 55 | 10 | 1 | 23 | 2 | 0 |
| Activation of NF-kappaB in B cells | 11 | 242014 | T | A | A | rs17727753 | intronic | 0.0189 | 0.0189 | PSMD13 | 1.0000 | 60 | 5 | 1 | 24 | 1 | 0 |
| Activation of NF-kappaB in B cells | 11 | 249105 | G | A | A | rs11601352 | intronic | 0.0465 | 0.0465 | PSMD13 | 0.7522 | 56 | 10 | 0 | 23 | 3 | 0 |
| Activation of NF-kappaB in B cells | 11 | 14528592 | T | A | A | rs11023241 | intronic | 0.0417 | 0.0417 | PSMA1 | 0.6117 | 57 | 8 | 1 | 23 | 2 | 1 |
| Activation of NF-kappaB in B cells | 11 | 14531031 | C | T | T | rs78398913 | intronic | 0.0273 | 0.0273 | PSMA1 | 1.0000 | 56 | 9 | 0 | 22 | 4 | 0 |
| Activation of NF-kappaB in B cells | 11 | 14537004 | A | G | G | rs74589503 | intronic | 0.0418 | 0.0418 | PSMA1 | 0.4981 | 56 | 10 | 0 | 24 | 2 | 0 |
| Activation of NF-kappaB in B cells | 11 | 14540827 | C | T | T | rs79966935 | intronic | 0.0233 | 0.0233 | PSMA1 | 0.7222 | 57 | 9 | 0 | 24 | 2 | 0 |
| Activation of NF-kappaB in B cells | 11 | 14541179 | A | T | T | rs61883612 | intronic | 0.0441 | 0.0441 | PSMA1 | 0.7666 | 55 | 11 | 0 | 21 | 5 | 0 |
| Activation of NF-kappaB in B cells | 11 | 14556220 | C | T | T | rs34162548 | intronic | 0.0376 | 0.0376 | PSMA1 | 1.0000 | 55 | 11 | 0 | 21 | 4 | 0 |
| Activation of NF-kappaB in B cells | 11 | 14588324 | G | A | A | rs78854818 | intronic | 0.0413 | 0.0413 | PSMA1 | 1.0000 | 59 | 6 | 1 | 24 | 2 | 0 |
| Activation of NF-kappaB in B cells | 11 | 14602698 | T | G | G | rs17567703 | intronic | 0.0418 | 0.0418 | PSMA1 | 1.0000 | 58 | 8 | 0 | 23 | 3 | 0 |
| Activation of NF-kappaB in B cells | 11 | 14622982 | G | A | A | rs16930367 | intronic | 0.0250 | 0.0250 | PSMA1 | 0.7192 | 58 | 8 | 0 | 24 | 2 | 0 |
| Activation of NF-kappaB in B cells | 11 | 14627135 | G | A | A | rs55760529 | intronic | 0.0485 | 0.0485 | PSMA1 | 1.0000 | 56 | 9 | 1 | 23 | 3 | 0 |
| Activation of NF-kappaB in B cells | 11 | 14633490 | G | A | A | rs11023274 | intronic | 0.0317 | 0.0317 | PSMA1 | 1.0000 | 56 | 9 | 1 | 23 | 3 | 0 |
| Activation of NF-kappaB in B cells | 11 | 14648387 | G | A | A | rs79123458 | intronic | 0.0412 | 0.0412 | PSMA1 | 1.0000 | 59 | 6 | 1 | 24 | 2 | 0 |
| Activation of NF-kappaB in B cells | 11 | 47441683 | C | T | T | rs72903900 | intronic | 0.0375 | 0.0375 | PSMC3 | 0.8389 | 59 | 5 | 2 | 23 | 3 | 0 |
| Activation of NF-kappaB in B cells | 11 | 47449591 | G | A | A | rs116930066 | upstream | 0.0311 | 0.0311 | PSMC3 | 0.5028 | 56 | 9 | 0 | 24 | 2 | 0 |
| Activation of NF-kappaB in B cells | 12 | 122349404 | T | G | G | rs113810917 | intronic | 0.0229 | 0.0229 | PSMD9 | 0.1282 | 57 | 8 | 1 | 25 | 0 | 0 |
| Activation of NF-kappaB in B cells | 12 | 122351293 | A | C | C | rs73229956 | intronic | 0.0288 | 0.0288 | PSMD9 | 0.2597 | 54 | 11 | 1 | 24 | 1 | 0 |
| Activation of NF-kappaB in B cells | 12 | 125395161 | G | C | C | rs113660988 | downstream | 0.0428 | 0.0428 | UBC | 0.0064 | 64 | 1 | 1 | 21 | 5 | 0 |
| Activation of NF-kappaB in B cells | 12 | 125395728 | C | T | T | rs112205208 | downstream | 0.0429 | 0.0429 | UBC | 0.0064 | 64 | 1 | 1 | 21 | 5 | 0 |
| Activation of NF-kappaB in B cells | 12 | 125398911 | C | T | T | rs112043091 | 5_prime_UTR | 0.0430 | 0.0430 | UBC | 0.0064 | 64 | 1 | 1 | 21 | 5 | 0 |
| Activation of NF-kappaB in B cells | 12 | 125399133 | C | T | T | rs41276688 | 5_prime_UTR | 0.0434 | 0.0434 | UBC | 0.0064 | 64 | 1 | 1 | 21 | 5 | 0 |
| Activation of NF-kappaB in B cells | 14 | 23512430 | C | T | T | rs78162644 | 3_prime_UTR | 0.0459 | 0.0459 | PSMB11 | 1.0000 | 57 | 9 | 0 | 22 | 4 | 0 |
| Activation of NF-kappaB in B cells | 14 | 35787993 | A | T | T | rs12890150 | downstream | 0.0324 | 0.0324 | PSMA6 | 0.5321 | 62 | 3 | 1 | 23 | 3 | 0 |
| Activation of NF-kappaB in B cells | 14 | 53178728 | T | C | C | rs117516552 | intronic | 0.0276 | 0.0276 | PSMC6 | 0.7209 | 57 | 9 | 0 | 23 | 2 | 0 |
| Activation of NF-kappaB in B cells | 14 | 58710711 | C | G | G | rs117328560 | upstream | 0.0280 | 0.0280 | PSMA3 | 0.4999 | 56 | 10 | 0 | 23 | 2 | 0 |
| Activation of NF-kappaB in B cells | 14 | 58729996 | C | G | G | rs111364917 | intronic | 0.0397 | 0.0397 | PSMA3 | 0.4984 | 59 | 7 | 0 | 22 | 4 | 0 |
| Activation of NF-kappaB in B cells | 14 | 58731398 | T | G | G | rs117955444 | intronic | 0.0210 | 0.0210 | PSMA3 | 1.0000 | 59 | 7 | 0 | 22 | 3 | 0 |
| Activation of NF-kappaB in B cells | 15 | 78834476 | A | G | G | rs41280046 | intronic | 0.0393 | 0.0393 | PSMA4 | 0.7087 | 60 | 6 | 0 | 23 | 3 | 0 |
| Activation of NF-kappaB in B cells | 16 | 23845860 | G | T | T | rs72777910 | upstream | 0.0300 | 0.0300 | PRKCB | 0.4762 | 60 | 5 | 1 | 22 | 4 | 0 |
| Activation of NF-kappaB in B cells | 16 | 23849482 | T | C | T | rs2023670 | intronic | 0.9513 | 0.0487 | PRKCB | 0.0099 | 0 | 11 | 55 | 1 | 0 | 25 |
| Activation of NF-kappaB in B cells | 16 | 23850240 | A | G | A | rs11074581 | intronic | 0.9663 | 0.0337 | PRKCB | 0.0099 | 0 | 11 | 55 | 1 | 0 | 25 |
| Activation of NF-kappaB in B cells | 16 | 23851956 | T | C | T | rs7189210 | intronic | 0.9663 | 0.0337 | PRKCB | 0.0099 | 0 | 11 | 55 | 1 | 0 | 25 |
| Activation of NF-kappaB in B cells | 16 | 23852415 | A | T | A | rs2188359 | intronic | 0.9528 | 0.0472 | PRKCB | 0.0099 | 0 | 11 | 55 | 1 | 0 | 25 |
| Activation of NF-kappaB in B cells | 16 | 23859391 | A | G | G | rs62030647 | intronic | 0.0226 | 0.0226 | PRKCB | 1.0000 | 59 | 7 | 0 | 23 | 3 | 0 |
| Activation of NF-kappaB in B cells | 16 | 23874933 | A | C | A | rs6497691 | intronic | 0.9663 | 0.0337 | PRKCB | 0.0063 | 0 | 14 | 52 | 1 | 0 | 24 |
| Activation of NF-kappaB in B cells | 16 | 23876099 | C | T | T | rs79131874 | intronic | 0.0303 | 0.0303 | PRKCB | 1.0000 | 59 | 7 | 0 | 23 | 3 | 0 |
| Activation of NF-kappaB in B cells | 16 | 23877500 | A | G | A | rs8059885 | intronic | 0.9663 | 0.0337 | PRKCB | 0.0046 | 0 | 14 | 52 | 1 | 0 | 25 |
| Activation of NF-kappaB in B cells | 16 | 23877606 | A | G | A | rs8060048 | intronic | 0.9644 | 0.0356 | PRKCB | 0.0046 | 0 | 14 | 52 | 1 | 0 | 25 |
| Activation of NF-kappaB in B cells | 16 | 23877781 | G | A | G | rs8060718 | intronic | 0.9664 | 0.0336 | PRKCB | 0.0046 | 0 | 14 | 52 | 1 | 0 | 25 |
| Activation of NF-kappaB in B cells | 16 | 23878470 | C | T | C | rs12935004 | intronic | 0.9657 | 0.0343 | PRKCB | 0.0046 | 0 | 14 | 52 | 1 | 0 | 25 |
| Activation of NF-kappaB in B cells | 16 | 23880851 | C | T | C | rs8061523 | intronic | 0.9664 | 0.0336 | PRKCB | 0.0046 | 0 | 14 | 52 | 1 | 0 | 25 |
| Activation of NF-kappaB in B cells | 16 | 23881930 | G | A | G | rs8047121 | intronic | 0.9662 | 0.0338 | PRKCB | 0.0046 | 0 | 14 | 52 | 1 | 0 | 25 |
| Activation of NF-kappaB in B cells | 16 | 23882469 | T | C | T | rs1468129 | intronic | 0.9663 | 0.0337 | PRKCB | 0.0046 | 0 | 14 | 52 | 1 | 0 | 25 |
| Activation of NF-kappaB in B cells | 16 | 23885608 | A | T | A | rs8044732 | intronic | 0.9664 | 0.0336 | PRKCB | 0.0046 | 0 | 14 | 52 | 1 | 0 | 25 |
| Activation of NF-kappaB in B cells | 16 | 23885751 | A | G | G | rs62031692 | intronic | 0.0253 | 0.0253 | PRKCB | 1.0000 | 59 | 7 | 0 | 23 | 3 | 0 |
| Activation of NF-kappaB in B cells | 16 | 23887574 | G | T | T | rs79034087 | intronic | 0.0290 | 0.0290 | PRKCB | 0.3487 | 61 | 4 | 1 | 22 | 4 | 0 |

| Pathway | Chr | Position | Ref | Alt | Allele | rsID | Region | Freq1 | Freq2 | Gene | P-value | c1 | c2 | c3 | c4 | c5 | c6 |
|---|---|---|---|---|---|---|---|---|---|---|---|---|---|---|---|---|---|
| Activation of NF-kappaB in B cells | 16 | 23888354 | C | T | C | rs7404417 | intronic | 0.9664 | 0.0336 | PRKCB | 0.0046 | 0 | 14 | 52 | 1 | 0 | 25 |
| Activation of NF-kappaB in B cells | 16 | 23889896 | T | C | T | rs8063823 | intronic | 0.9665 | 0.0335 | PRKCB | 0.0046 | 0 | 14 | 52 | 1 | 0 | 25 |
| Activation of NF-kappaB in B cells | 16 | 23893893 | G | A | G | rs11647359 | intronic | 0.9664 | 0.0336 | PRKCB | 0.0046 | 0 | 14 | 52 | 1 | 0 | 25 |
| Activation of NF-kappaB in B cells | 16 | 23895034 | A | G | A | rs6497695 | intronic | 0.9665 | 0.0335 | PRKCB | 0.0046 | 0 | 14 | 52 | 1 | 0 | 25 |
| Activation of NF-kappaB in B cells | 16 | 23895443 | A | G | G | rs62028075 | intronic | 0.0253 | 0.0253 | PRKCB | 1.0000 | 59 | 7 | 0 | 23 | 3 | 0 |
| Activation of NF-kappaB in B cells | 16 | 23895884 | T | C | T | rs9944348 | intronic | 0.9665 | 0.0335 | PRKCB | 0.0046 | 0 | 14 | 52 | 1 | 0 | 25 |
| Activation of NF-kappaB in B cells | 16 | 23896089 | T | C | C | rs74572166 | intronic | 0.0245 | 0.0245 | PRKCB | 1.0000 | 59 | 7 | 0 | 23 | 3 | 0 |
| Activation of NF-kappaB in B cells | 16 | 23896209 | C | A | C | rs9302418 | intronic | 0.9664 | 0.0336 | PRKCB | 0.0046 | 0 | 14 | 52 | 1 | 0 | 25 |
| Activation of NF-kappaB in B cells | 16 | 23896438 | G | T | T | rs62028076 | intronic | 0.0252 | 0.0252 | PRKCB | 1.0000 | 59 | 7 | 0 | 22 | 3 | 0 |
| Activation of NF-kappaB in B cells | 16 | 23898605 | A | T | A | rs933290 | intronic | 0.9632 | 0.0368 | PRKCB | 0.0016 | 0 | 17 | 49 | 1 | 0 | 24 |
| Activation of NF-kappaB in B cells | 16 | 23899211 | A | T | A | rs12926245 | intronic | 0.9632 | 0.0368 | PRKCB | 0.0012 | 0 | 17 | 49 | 1 | 0 | 25 |
| Activation of NF-kappaB in B cells | 16 | 23899610 | G | A | A | rs17753246 | intronic | 0.0252 | 0.0252 | PRKCB | 1.0000 | 59 | 7 | 0 | 23 | 3 | 0 |
| Activation of NF-kappaB in B cells | 16 | 23899951 | G | A | A | rs62028077 | intronic | 0.0254 | 0.0254 | PRKCB | 1.0000 | 59 | 7 | 0 | 23 | 3 | 0 |
| Activation of NF-kappaB in B cells | 16 | 23900716 | T | C | C | rs62028078 | intronic | 0.0252 | 0.0252 | PRKCB | 1.0000 | 59 | 7 | 0 | 23 | 3 | 0 |
| Activation of NF-kappaB in B cells | 16 | 23901896 | C | T | C | rs6497696 | intronic | 0.9632 | 0.0368 | PRKCB | 0.0015 | 0 | 16 | 50 | 1 | 0 | 25 |
| Activation of NF-kappaB in B cells | 16 | 23901948 | A | C | A | rs6497697 | intronic | 0.9630 | 0.0370 | PRKCB | 0.0015 | 0 | 16 | 49 | 1 | 0 | 25 |
| Activation of NF-kappaB in B cells | 16 | 23904058 | A | G | A | rs886115 | intronic | 0.9632 | 0.0368 | PRKCB | 0.0015 | 0 | 16 | 50 | 1 | 0 | 25 |
| Activation of NF-kappaB in B cells | 16 | 23904781 | G | A | A | rs17753509 | intronic | 0.0253 | 0.0253 | PRKCB | 1.0000 | 59 | 7 | 0 | 23 | 3 | 0 |
| Activation of NF-kappaB in B cells | 16 | 23905676 | C | T | C | rs7200610 | intronic | 0.9631 | 0.0369 | PRKCB | 0.0015 | 0 | 16 | 50 | 1 | 0 | 25 |
| Activation of NF-kappaB in B cells | 16 | 23907177 | A | C | C | rs17810011 | intronic | 0.0251 | 0.0251 | PRKCB | 1.0000 | 59 | 7 | 0 | 23 | 3 | 0 |
| Activation of NF-kappaB in B cells | 16 | 23907765 | C | T | C | rs9925890 | intronic | 0.9632 | 0.0368 | PRKCB | 0.0024 | 0 | 16 | 50 | 1 | 0 | 24 |
| Activation of NF-kappaB in B cells | 16 | 23912174 | A | G | A | rs12448249 | intronic | 0.9519 | 0.0481 | PRKCB | 0.0015 | 0 | 16 | 50 | 1 | 0 | 25 |
| Activation of NF-kappaB in B cells | 16 | 23914915 | C | A | C | rs1004186 | intronic | 0.9632 | 0.0368 | PRKCB | 0.0015 | 0 | 16 | 50 | 1 | 0 | 25 |
| Activation of NF-kappaB in B cells | 16 | 23916258 | G | A | G | rs1004187 | intronic | 0.9632 | 0.0368 | PRKCB | 0.0015 | 0 | 16 | 50 | 1 | 0 | 25 |
| Activation of NF-kappaB in B cells | 16 | 23916521 | G | C | G | rs1008654 | intronic | 0.9633 | 0.0367 | PRKCB | 0.0015 | 0 | 16 | 50 | 1 | 0 | 25 |
| Activation of NF-kappaB in B cells | 16 | 23917335 | G | A | G | rs6497699 | intronic | 0.9645 | 0.0355 | PRKCB | 0.0024 | 0 | 16 | 50 | 1 | 0 | 24 |
| Activation of NF-kappaB in B cells | 16 | 23917465 | C | G | G | rs7186538 | intronic | 0.9645 | 0.0355 | PRKCB | 0.0024 | 0 | 16 | 50 | 1 | 0 | 24 |
| Activation of NF-kappaB in B cells | 16 | 23917700 | C | A | C | rs7187091 | intronic | 0.9646 | 0.0354 | PRKCB | 0.0024 | 0 | 16 | 50 | 1 | 0 | 24 |
| Activation of NF-kappaB in B cells | 16 | 23919088 | C | T | T | rs78322646 | intronic | 0.0278 | 0.0278 | PRKCB | 0.7348 | 58 | 8 | 0 | 22 | 4 | 0 |
| Activation of NF-kappaB in B cells | 16 | 23921083 | C | T | C | rs6497702 | intronic | 0.9647 | 0.0353 | PRKCB | 0.0024 | 0 | 16 | 50 | 1 | 0 | 24 |
| Activation of NF-kappaB in B cells | 16 | 23925936 | C | G | C | rs11074588 | intronic | 0.9649 | 0.0351 | PRKCB | 0.0015 | 0 | 16 | 50 | 1 | 0 | 25 |
| Activation of NF-kappaB in B cells | 16 | 23939212 | G | A | G | rs11074590 | intronic | 0.9650 | 0.0350 | PRKCB | 0.0024 | 0 | 16 | 50 | 1 | 0 | 24 |
| Activation of NF-kappaB in B cells | 16 | 23941628 | C | A | C | rs2005671 | intronic | 0.9647 | 0.0353 | PRKCB | 0.0015 | 0 | 16 | 49 | 1 | 0 | 25 |
| Activation of NF-kappaB in B cells | 16 | 23943749 | T | C | T | rs9302420 | intronic | 0.9649 | 0.0351 | PRKCB | 0.0024 | 0 | 16 | 50 | 1 | 0 | 24 |
| Activation of NF-kappaB in B cells | 16 | 23945985 | T | G | T | rs195989 | intronic | 0.9651 | 0.0349 | PRKCB | 0.0015 | 0 | 16 | 50 | 1 | 0 | 25 |
| Activation of NF-kappaB in B cells | 16 | 23946157 | G | A | A | rs76973283 | intronic | 0.0302 | 0.0302 | PRKCB | 0.5065 | 58 | 8 | 0 | 21 | 5 | 0 |
| Activation of NF-kappaB in B cells | 16 | 23949175 | G | C | G | rs2560403 | intronic | 0.9657 | 0.0343 | PRKCB | 0.0015 | 0 | 16 | 50 | 1 | 0 | 25 |
| Activation of NF-kappaB in B cells | 16 | 23949438 | A | G | A | rs195985 | intronic | 0.9658 | 0.0342 | PRKCB | 0.0015 | 0 | 16 | 50 | 1 | 0 | 25 |
| Activation of NF-kappaB in B cells | 16 | 23953265 | T | C | T | rs2560404 | intronic | 0.9656 | 0.0344 | PRKCB | 0.0024 | 0 | 16 | 50 | 1 | 0 | 24 |
| Activation of NF-kappaB in B cells | 16 | 23954128 | T | C | C | rs17810486 | intronic | 0.0308 | 0.0308 | PRKCB | 0.3024 | 58 | 7 | 0 | 20 | 5 | 0 |
| Activation of NF-kappaB in B cells | 16 | 23954253 | G | A | G | rs195994 | intronic | 0.9653 | 0.0347 | PRKCB | 0.0015 | 0 | 16 | 50 | 1 | 0 | 25 |
| Activation of NF-kappaB in B cells | 16 | 23962258 | G | C | G | rs196000 | intronic | 0.9659 | 0.0341 | PRKCB | 0.0015 | 0 | 16 | 50 | 1 | 0 | 25 |
| Activation of NF-kappaB in B cells | 16 | 23964858 | T | A | T | rs196003 | intronic | 0.9647 | 0.0353 | PRKCB | 0.0343 | 0 | 15 | 51 | 0 | 1 | 25 |
| Activation of NF-kappaB in B cells | 16 | 23985814 | C | T | T | rs72779914 | intronic | 0.0487 | 0.0487 | PRKCB | 1.0000 | 53 | 12 | 0 | 21 | 4 | 0 |
| Activation of NF-kappaB in B cells | 16 | 23987552 | A | G | A | rs169030 | intronic | 0.9709 | 0.0291 | PRKCB | 0.1679 | 0 | 11 | 55 | 0 | 1 | 25 |
| Activation of NF-kappaB in B cells | 16 | 23988755 | T | C | T | rs196013 | intronic | 0.9681 | 0.0319 | PRKCB | 0.1681 | 0 | 11 | 55 | 0 | 1 | 24 |
| Activation of NF-kappaB in B cells | 16 | 24009919 | A | G | G | rs75622923 | intronic | 0.0319 | 0.0319 | PRKCB | 0.3348 | 54 | 12 | 0 | 24 | 2 | 0 |
| Activation of NF-kappaB in B cells | 16 | 24022944 | C | T | T | rs111746132 | intronic | 0.0229 | 0.0229 | PRKCB | 0.7522 | 56 | 10 | 0 | 23 | 3 | 0 |
| Activation of NF-kappaB in B cells | 16 | 24066378 | G | A | A | rs113426570 | intronic | 0.0216 | 0.0216 | PRKCB | 1.0000 | 57 | 9 | 0 | 23 | 3 | 0 |
| Activation of NF-kappaB in B cells | 16 | 24100759 | T | A | A | rs11643939 | intronic | 0.0294 | 0.0294 | PRKCB | 0.7522 | 56 | 10 | 0 | 23 | 3 | 0 |
| Activation of NF-kappaB in B cells | 16 | 24105816 | G | A | A | rs56316329 | intronic | 0.0251 | 0.0251 | PRKCB | 0.6919 | 60 | 4 | 1 | 25 | 0 | 0 |
| Activation of NF-kappaB in B cells | 16 | 24111853 | T | C | C | rs55959083 | intronic | 0.0431 | 0.0431 | PRKCB | 0.7213 | 58 | 8 | 0 | 23 | 2 | 0 |
| Activation of NF-kappaB in B cells | 16 | 24112768 | G | A | A | rs117056307 | intronic | 0.0430 | 0.0430 | PRKCB | 0.7192 | 58 | 8 | 0 | 24 | 2 | 0 |
| Activation of NF-kappaB in B cells | 16 | 24122052 | G | A | A | rs117467859 | intronic | 0.0433 | 0.0433 | PRKCB | 0.7213 | 58 | 8 | 0 | 23 | 2 | 0 |
| Activation of NF-kappaB in B cells | 16 | 24122492 | C | T | T | rs72779977 | intronic | 0.0457 | 0.0457 | PRKCB | 0.7201 | 57 | 8 | 0 | 23 | 2 | 0 |
| Activation of NF-kappaB in B cells | 16 | 24123560 | G | A | A | rs60261043 | intronic | 0.0457 | 0.0457 | PRKCB | 0.7201 | 57 | 8 | 0 | 23 | 2 | 0 |
| Activation of NF-kappaB in B cells | 16 | 24132273 | G | A | A | rs62027458 | intronic | 0.0232 | 0.0232 | PRKCB | 0.1877 | 59 | 7 | 0 | 21 | 4 | 1 |
| Activation of NF-kappaB in B cells | 16 | 24164042 | G | T | T | rs72779989 | intronic | 0.0487 | 0.0487 | PRKCB | 0.8583 | 57 | 7 | 2 | 22 | 4 | 0 |
| Activation of NF-kappaB in B cells | 16 | 24197496 | A | T | T | rs79699525 | intronic | 0.0261 | 0.0261 | PRKCB | 0.4631 | 59 | 6 | 0 | 22 | 4 | 0 |
| Activation of NF-kappaB in B cells | 16 | 24199852 | C | T | T | rs78424166 | intronic | 0.0359 | 0.0359 | PRKCB | 0.4603 | 60 | 6 | 0 | 22 | 4 | 0 |
| Activation of NF-kappaB in B cells | 16 | 67973569 | C | T | T | rs17240392 | upstream | 0.0497 | 0.0497 | PSMB10 | 0.4984 | 59 | 7 | 0 | 22 | 4 | 0 |

| Pathway | Chr | Position | Ref | Alt | Alt2 | rsID | Region | Freq1 | Freq2 | Gene | P | N1 | N2 | N3 | N4 | N5 | N6 |
|---|---|---|---|---|---|---|---|---|---|---|---|---|---|---|---|---|---|
| Activation of NF-kappaB in B cells | 16 | 74334545 | G | T | T | rs149977784 | intronic | 0.0196 | 0.0196 | PSMD7 | 1.0000 | 62 | 2 | 2 | 26 | 0 | 0 |
| Activation of NF-kappaB in B cells | 17 | 4702206 | C | A | A | rs71368518 | downstream | 0.0416 | 0.0416 | PSMB6 | 1.0000 | 56 | 9 | 1 | 23 | 3 | 0 |
| Activation of NF-kappaB in B cells | 17 | 30778483 | C | T | T | rs117117721 | intronic | 0.0190 | 0.0190 | PSMD11 | 1.0000 | 59 | 7 | 0 | 22 | 3 | 0 |
| Activation of NF-kappaB in B cells | 17 | 30791889 | C | T | T | rs35225085 | intronic | 0.0427 | 0.0427 | PSMD11 | 0.4999 | 56 | 10 | 0 | 23 | 2 | 0 |
| Activation of NF-kappaB in B cells | 17 | 30794889 | C | G | G | rs35177842 | intronic | 0.0430 | 0.0430 | PSMD11 | 0.4999 | 56 | 10 | 0 | 23 | 2 | 0 |
| Activation of NF-kappaB in B cells | 17 | 36920050 | C | T | T | rs118080693 | intronic | 0.0415 | 0.0415 | PSMB3 | 0.1732 | 63 | 3 | 0 | 25 | 0 | 1 |
| Activation of NF-kappaB in B cells | 17 | 38144079 | G | T | T | rs118009374 | intronic | 0.0189 | 0.0189 | PSMD3 | 0.8767 | 53 | 11 | 2 | 20 | 5 | 0 |
| Activation of NF-kappaB in B cells | 17 | 38154396 | C | T | T | rs118034841 | downstream | 0.0176 | 0.0176 | PSMD3 | 0.0569 | 56 | 10 | 0 | 26 | 0 | 0 |
| Activation of NF-kappaB in B cells | 17 | 65334270 | T | A | A | rs146515782 | 3_prime_UTR | 0.0418 | 0.0418 | PSMD12 | 1.0000 | 58 | 8 | 0 | 23 | 3 | 0 |
| Activation of NF-kappaB in B cells | 18 | 23711373 | G | T | T | rs79820119 | upstream | 0.0407 | 0.0407 | PSMA8 | 0.2218 | 53 | 13 | 0 | 23 | 2 | 0 |
| Activation of NF-kappaB in B cells | 18 | 23715815 | T | G | G | rs4800242 | intronic | 0.0460 | 0.0460 | PSMA8 | 0.7043 | 59 | 6 | 0 | 22 | 3 | 0 |
| Activation of NF-kappaB in B cells | 18 | 23774138 | C | A | A | rs79452515 | downstream | 0.0424 | 0.0424 | PSMA8 | 0.2993 | 59 | 7 | 0 | 20 | 5 | 0 |
| Activation of NF-kappaB in B cells | 18 | 56333866 | G | A | A | rs72958690 | upstream | 0.0368 | 0.0368 | MALT1 | 0.8183 | 56 | 9 | 1 | 21 | 4 | 0 |
| Activation of NF-kappaB in B cells | 18 | 56338792 | G | A | A | rs56142402 | 5_prime_UTR | 0.0327 | 0.0327 | MALT1 | 0.8110 | 57 | 8 | 1 | 22 | 4 | 0 |
| Activation of NF-kappaB in B cells | 18 | 56363534 | A | C | C | rs55825071 | intronic | 0.0377 | 0.0377 | MALT1 | 0.6531 | 56 | 9 | 1 | 20 | 5 | 0 |
| Activation of NF-kappaB in B cells | 19 | 40472450 | A | C | C | rs147915270 | upstream | 0.0382 | 0.0382 | PSMC4 | 1.0000 | 58 | 7 | 1 | 22 | 3 | 0 |
| Activation of NF-kappaB in B cells | 19 | 40475070 | T | G | G | rs139876278 | upstream | 0.0147 | 0.0147 | PSMC4 | 0.2725 | 63 | 2 | 1 | 23 | 3 | 0 |
| Activation of NF-kappaB in B cells | 20 | 1128622 | T | C | C | rs74871431 | intronic | 0.0320 | 0.0320 | PSMF1 | 0.7087 | 60 | 6 | 0 | 23 | 3 | 0 |
| Activation of NF-kappaB in B cells | 20 | 1146048 | C | T | T | rs17716261 | 3_prime_UTR | 0.0279 | 0.0279 | PSMF1 | 0.4532 | 60 | 6 | 0 | 21 | 4 | 0 |
| Activation of NF-kappaB in B cells | 20 | 1149980 | C | T | T | rs77625408 | downstream | 0.0254 | 0.0254 | PSMF1 | 0.7495 | 54 | 11 | 0 | 23 | 3 | 0 |
| Activation of NF-kappaB in B cells | 20 | 1152323 | C | T | T | rs34552580 | downstream | 0.0323 | 0.0323 | PSMF1 | 1.0000 | 59 | 5 | 1 | 24 | 2 | 0 |
| Activation of NF-kappaB in B cells | 20 | 1155154 | G | A | A | rs78313102 | intronic,non_coding_transcript | 0.0258 | 0.0258 | PSMF1 | 1.0000 | 58 | 7 | 0 | 23 | 3 | 0 |
| WNT5A-dependent internalization of FZD4 | 3 | 183896966 | G | A | A | rs2231217 | intronic | 0.0406 | 0.0406 | AP2M1 | 0.7980 | 57 | 8 | 1 | 24 | 2 | 0 |
| WNT5A-dependent internalization of FZD4 | 5 | 175838749 | T | A | A | rs72807224 | intronic | 0.0466 | 0.0466 | CLTB | 0.0974 | 60 | 5 | 1 | 20 | 6 | 0 |
| WNT5A-dependent internalization of FZD4 | 9 | 36189647 | T | C | C | rs3739608 | upstream | 0.0486 | 0.0486 | CLTA | 0.4858 | 59 | 6 | 1 | 21 | 5 | 0 |
| WNT5A-dependent internalization of FZD4 | 9 | 36201064 | G | A | A | rs10972788 | intronic | 0.0274 | 0.0274 | CLTA | 0.2818 | 60 | 6 | 0 | 21 | 5 | 0 |
| WNT5A-dependent internalization of FZD4 | 11 | 941844 | C | T | T | rs10902236 | intronic | 0.0339 | 0.0339 | AP2A2 | 1.0000 | 59 | 7 | 0 | 23 | 3 | 0 |
| WNT5A-dependent internalization of FZD4 | 11 | 965671 | C | T | T | rs74045443 | intronic | 0.0489 | 0.0489 | AP2A2 | 0.4662 | 58 | 6 | 0 | 22 | 4 | 0 |
| WNT5A-dependent internalization of FZD4 | 11 | 966609 | C | T | T | rs7394613 | intronic | 0.0457 | 0.0457 | AP2A2 | 0.0005 | 64 | 2 | 0 | 18 | 7 | 1 |
| WNT5A-dependent internalization of FZD4 | 11 | 967822 | C | T | T | rs117367662 | intronic | 0.0338 | 0.0338 | AP2A2 | 1.0000 | 59 | 7 | 0 | 23 | 3 | 0 |
| WNT5A-dependent internalization of FZD4 | 11 | 970818 | C | T | T | rs74045446 | intronic | 0.0491 | 0.0491 | AP2A2 | 0.4474 | 60 | 6 | 0 | 20 | 4 | 0 |
| WNT5A-dependent internalization of FZD4 | 11 | 971884 | A | G | G | rs7945582 | intronic | 0.0470 | 0.0470 | AP2A2 | 0.6640 | 56 | 9 | 1 | 21 | 5 | 0 |
| WNT5A-dependent internalization of FZD4 | 11 | 978921 | C | T | T | rs7122686 | intronic | 0.0468 | 0.0468 | AP2A2 | 0.8183 | 56 | 9 | 1 | 21 | 4 | 0 |
| WNT5A-dependent internalization of FZD4 | 11 | 986568 | G | C | C | rs112772664 | intronic | 0.0457 | 0.0457 | AP2A2 | 0.4532 | 60 | 6 | 0 | 21 | 4 | 0 |
| WNT5A-dependent internalization of FZD4 | 11 | 86653253 | G | A | A | rs79378895 | downstream | 0.0478 | 0.0478 | FZD4 | 0.7222 | 57 | 9 | 0 | 24 | 2 | 0 |
| WNT5A-dependent internalization of FZD4 | 11 | 86657419 | A | G | G | rs72963441 | 3_prime_UTR | 0.0393 | 0.0393 | FZD4 | 0.5065 | 58 | 8 | 0 | 21 | 5 | 0 |
| WNT5A-dependent internalization of FZD4 | 16 | 23845860 | G | T | T | rs72777910 | upstream | 0.0300 | 0.0300 | PRKCB | 0.4762 | 60 | 5 | 1 | 22 | 4 | 0 |
| WNT5A-dependent internalization of FZD4 | 16 | 23849482 | T | C | T | rs2023670 | intronic | 0.9513 | 0.0487 | PRKCB | 0.0099 | 0 | 11 | 55 | 1 | 0 | 25 |
| WNT5A-dependent internalization of FZD4 | 16 | 23850240 | A | G | A | rs11074581 | intronic | 0.9663 | 0.0337 | PRKCB | 0.0099 | 0 | 11 | 55 | 1 | 0 | 25 |
| WNT5A-dependent internalization of FZD4 | 16 | 23851956 | T | C | T | rs7189210 | intronic | 0.9663 | 0.0337 | PRKCB | 0.0099 | 0 | 11 | 55 | 1 | 0 | 25 |
| WNT5A-dependent internalization of FZD4 | 16 | 23852415 | A | T | A | rs2188359 | intronic | 0.9528 | 0.0472 | PRKCB | 0.0099 | 0 | 11 | 55 | 1 | 0 | 25 |
| WNT5A-dependent internalization of FZD4 | 16 | 23859391 | A | G | G | rs62030647 | intronic | 0.0226 | 0.0226 | PRKCB | 1.0000 | 59 | 7 | 0 | 23 | 3 | 0 |
| WNT5A-dependent internalization of FZD4 | 16 | 23874933 | A | C | A | rs6497691 | intronic | 0.9663 | 0.0337 | PRKCB | 0.0063 | 0 | 14 | 52 | 1 | 0 | 24 |
| WNT5A-dependent internalization of FZD4 | 16 | 23876099 | C | T | T | rs79131874 | intronic | 0.0303 | 0.0303 | PRKCB | 1.0000 | 59 | 7 | 0 | 23 | 3 | 0 |
| WNT5A-dependent internalization of FZD4 | 16 | 23877500 | A | G | A | rs8059885 | intronic | 0.9663 | 0.0337 | PRKCB | 0.0046 | 0 | 14 | 52 | 1 | 0 | 25 |
| WNT5A-dependent internalization of FZD4 | 16 | 23877606 | A | G | A | rs8060048 | intronic | 0.9644 | 0.0356 | PRKCB | 0.0046 | 0 | 14 | 52 | 1 | 0 | 25 |
| WNT5A-dependent internalization of FZD4 | 16 | 23877781 | G | A | G | rs8060718 | intronic | 0.9664 | 0.0336 | PRKCB | 0.0046 | 0 | 14 | 52 | 1 | 0 | 25 |
| WNT5A-dependent internalization of FZD4 | 16 | 23878470 | C | T | C | rs12935004 | intronic | 0.9657 | 0.0343 | PRKCB | 0.0046 | 0 | 14 | 52 | 1 | 0 | 25 |
| WNT5A-dependent internalization of FZD4 | 16 | 23880851 | C | T | C | rs8061523 | intronic | 0.9664 | 0.0336 | PRKCB | 0.0046 | 0 | 14 | 52 | 1 | 0 | 25 |
| WNT5A-dependent internalization of FZD4 | 16 | 23881930 | G | A | G | rs8047121 | intronic | 0.9662 | 0.0338 | PRKCB | 0.0046 | 0 | 14 | 52 | 1 | 0 | 25 |
| WNT5A-dependent internalization of FZD4 | 16 | 23882469 | T | C | T | rs1468129 | intronic | 0.9663 | 0.0337 | PRKCB | 0.0046 | 0 | 14 | 52 | 1 | 0 | 25 |
| WNT5A-dependent internalization of FZD4 | 16 | 23885608 | A | T | A | rs8044732 | intronic | 0.9664 | 0.0336 | PRKCB | 0.0046 | 0 | 14 | 52 | 1 | 0 | 25 |
| WNT5A-dependent internalization of FZD4 | 16 | 23885751 | A | G | G | rs62031692 | intronic | 0.0253 | 0.0253 | PRKCB | 1.0000 | 59 | 7 | 0 | 23 | 3 | 0 |
| WNT5A-dependent internalization of FZD4 | 16 | 23887574 | G | T | T | rs79034087 | intronic | 0.0290 | 0.0290 | PRKCB | 0.3487 | 61 | 4 | 1 | 22 | 4 | 0 |
| WNT5A-dependent internalization of FZD4 | 16 | 23888354 | C | T | C | rs7404417 | intronic | 0.9664 | 0.0336 | PRKCB | 0.0046 | 0 | 14 | 52 | 1 | 0 | 25 |
| WNT5A-dependent internalization of FZD4 | 16 | 23889896 | T | C | T | rs8063823 | intronic | 0.9665 | 0.0335 | PRKCB | 0.0046 | 0 | 14 | 52 | 1 | 0 | 25 |
| WNT5A-dependent internalization of FZD4 | 16 | 23893893 | G | A | G | rs11647359 | intronic | 0.9664 | 0.0336 | PRKCB | 0.0046 | 0 | 14 | 52 | 1 | 0 | 25 |
| WNT5A-dependent internalization of FZD4 | 16 | 23895034 | A | G | A | rs6497695 | intronic | 0.9665 | 0.0335 | PRKCB | 0.0046 | 0 | 14 | 52 | 1 | 0 | 25 |
| WNT5A-dependent internalization of FZD4 | 16 | 23895443 | A | G | G | rs62028075 | intronic | 0.0253 | 0.0253 | PRKCB | 1.0000 | 59 | 7 | 0 | 23 | 3 | 0 |
| WNT5A-dependent internalization of FZD4 | 16 | 23895884 | T | C | T | rs9944348 | intronic | 0.9665 | 0.0335 | PRKCB | 0.0046 | 0 | 14 | 52 | 1 | 0 | 25 |
| WNT5A-dependent internalization of FZD4 | 16 | 23896089 | T | C | C | rs74572166 | intronic | 0.0245 | 0.0245 | PRKCB | 1.0000 | 59 | 7 | 0 | 23 | 3 | 0 |
| WNT5A-dependent internalization of FZD4 | 16 | 23896209 | C | A | C | rs9302418 | intronic | 0.9664 | 0.0336 | PRKCB | 0.0046 | 0 | 14 | 52 | 1 | 0 | 25 |
| WNT5A-dependent internalization of FZD4 | 16 | 23896438 | G | T | T | rs62028076 | intronic | 0.0252 | 0.0252 | PRKCB | 1.0000 | 59 | 7 | 0 | 22 | 3 | 0 |
| WNT5A-dependent internalization of FZD4 | 16 | 23898605 | A | T | A | rs933290 | intronic | 0.9632 | 0.0368 | PRKCB | 0.0016 | 0 | 17 | 49 | 1 | 0 | 24 |
| WNT5A-dependent internalization of FZD4 | 16 | 23899211 | A | T | A | rs12926245 | intronic | 0.9632 | 0.0368 | PRKCB | 0.0012 | 0 | 17 | 49 | 1 | 0 | 25 |
| WNT5A-dependent internalization of FZD4 | 16 | 23899610 | G | A | A | rs17753246 | intronic | 0.0252 | 0.0252 | PRKCB | 1.0000 | 59 | 7 | 0 | 23 | 3 | 0 |
| WNT5A-dependent internalization of FZD4 | 16 | 23899951 | G | A | A | rs62028077 | intronic | 0.0254 | 0.0254 | PRKCB | 1.0000 | 59 | 7 | 0 | 23 | 3 | 0 |
| WNT5A-dependent internalization of FZD4 | 16 | 23900716 | T | C | C | rs62028078 | intronic | 0.0252 | 0.0252 | PRKCB | 1.0000 | 59 | 7 | 0 | 23 | 3 | 0 |
| WNT5A-dependent internalization of FZD4 | 16 | 23901896 | C | T | C | rs6497696 | intronic | 0.9632 | 0.0368 | PRKCB | 0.0015 | 0 | 16 | 50 | 1 | 0 | 25 |
| WNT5A-dependent internalization of FZD4 | 16 | 23901948 | A | C | A | rs6497697 | intronic | 0.9630 | 0.0370 | PRKCB | 0.0015 | 0 | 16 | 49 | 1 | 0 | 25 |
| WNT5A-dependent internalization of FZD4 | 16 | 23904058 | A | G | A | rs886115 | intronic | 0.9632 | 0.0368 | PRKCB | 0.0015 | 0 | 16 | 50 | 1 | 0 | 25 |
| WNT5A-dependent internalization of FZD4 | 16 | 23904781 | G | A | A | rs17753509 | intronic | 0.0253 | 0.0253 | PRKCB | 1.0000 | 59 | 7 | 0 | 23 | 3 | 0 |
| WNT5A-dependent internalization of FZD4 | 16 | 23905676 | C | T | C | rs7200610 | intronic | 0.9631 | 0.0369 | PRKCB | 0.0015 | 0 | 16 | 50 | 1 | 0 | 25 |
| WNT5A-dependent internalization of FZD4 | 16 | 23907177 | A | C | C | rs17810011 | intronic | 0.0251 | 0.0251 | PRKCB | 1.0000 | 59 | 7 | 0 | 23 | 3 | 0 |
| WNT5A-dependent internalization of FZD4 | 16 | 23907765 | C | T | C | rs9925890 | intronic | 0.9632 | 0.0368 | PRKCB | 0.0024 | 0 | 16 | 50 | 1 | 0 | 24 |
| WNT5A-dependent internalization of FZD4 | 16 | 23912174 | A | G | A | rs12448249 | intronic | 0.9519 | 0.0481 | PRKCB | 0.0015 | 0 | 16 | 50 | 1 | 0 | 25 |
| WNT5A-dependent internalization of FZD4 | 16 | 23914915 | C | A | C | rs1004186 | intronic | 0.9632 | 0.0368 | PRKCB | 0.0015 | 0 | 16 | 50 | 1 | 0 | 25 |
| WNT5A-dependent internalization of FZD4 | 16 | 23916258 | G | A | G | rs1004187 | intronic | 0.9632 | 0.0368 | PRKCB | 0.0015 | 0 | 16 | 50 | 1 | 0 | 25 |
| WNT5A-dependent internalization of FZD4 | 16 | 23916521 | G | C | G | rs1008654 | intronic | 0.9633 | 0.0367 | PRKCB | 0.0015 | 0 | 16 | 50 | 1 | 0 | 25 |
| WNT5A-dependent internalization of FZD4 | 16 | 23917335 | G | A | G | rs6497699 | intronic | 0.9645 | 0.0355 | PRKCB | 0.0024 | 0 | 16 | 50 | 1 | 0 | 24 |
| WNT5A-dependent internalization of FZD4 | 16 | 23917465 | C | G | C | rs7186538 | intronic | 0.9645 | 0.0355 | PRKCB | 0.0024 | 0 | 16 | 50 | 1 | 0 | 24 |
| WNT5A-dependent internalization of FZD4 | 16 | 23917700 | C | A | C | rs7187091 | intronic | 0.9646 | 0.0354 | PRKCB | 0.0024 | 0 | 16 | 50 | 1 | 0 | 24 |
| WNT5A-dependent internalization of FZD4 | 16 | 23919088 | C | T | T | rs78322646 | intronic | 0.0278 | 0.0278 | PRKCB | 0.7348 | 58 | 8 | 0 | 22 | 4 | 0 |
| WNT5A-dependent internalization of FZD4 | 16 | 23921083 | C | T | C | rs6497702 | intronic | 0.9647 | 0.0353 | PRKCB | 0.0024 | 0 | 16 | 50 | 1 | 0 | 24 |
| WNT5A-dependent internalization of FZD4 | 16 | 23925936 | C | G | C | rs11074588 | intronic | 0.9649 | 0.0351 | PRKCB | 0.0015 | 0 | 16 | 50 | 1 | 0 | 25 |
| WNT5A-dependent internalization of FZD4 | 16 | 23939212 | G | A | G | rs11074590 | intronic | 0.9650 | 0.0350 | PRKCB | 0.0024 | 0 | 16 | 50 | 1 | 0 | 24 |
| WNT5A-dependent internalization of FZD4 | 16 | 23941628 | C | A | C | rs2005671 | intronic | 0.9647 | 0.0353 | PRKCB | 0.0015 | 0 | 16 | 49 | 1 | 0 | 25 |
| WNT5A-dependent internalization of FZD4 | 16 | 23943749 | T | C | T | rs9302420 | intronic | 0.9649 | 0.0351 | PRKCB | 0.0024 | 0 | 16 | 50 | 1 | 0 | 24 |
| WNT5A-dependent internalization of FZD4 | 16 | 23945985 | T | G | T | rs195989 | intronic | 0.9651 | 0.0349 | PRKCB | 0.0015 | 0 | 16 | 50 | 1 | 0 | 25 |
| WNT5A-dependent internalization of FZD4 | 16 | 23946157 | G | A | A | rs76973283 | intronic | 0.0302 | 0.0302 | PRKCB | 0.5065 | 58 | 8 | 0 | 21 | 5 | 0 |
| WNT5A-dependent internalization of FZD4 | 16 | 23949175 | G | C | G | rs2560403 | intronic | 0.9657 | 0.0343 | PRKCB | 0.0015 | 0 | 16 | 50 | 1 | 0 | 25 |
| WNT5A-dependent internalization of FZD4 | 16 | 23949438 | A | G | A | rs195985 | intronic | 0.9658 | 0.0342 | PRKCB | 0.0015 | 0 | 16 | 50 | 1 | 0 | 25 |
| WNT5A-dependent internalization of FZD4 | 16 | 23953265 | T | C | T | rs2560404 | intronic | 0.9656 | 0.0344 | PRKCB | 0.0024 | 0 | 16 | 50 | 1 | 0 | 24 |
| WNT5A-dependent internalization of FZD4 | 16 | 23954128 | T | C | C | rs17810486 | intronic | 0.0308 | 0.0308 | PRKCB | 0.3024 | 58 | 7 | 0 | 20 | 5 | 0 |
| WNT5A-dependent internalization of FZD4 | 16 | 23954253 | G | A | G | rs195994 | intronic | 0.9653 | 0.0347 | PRKCB | 0.0015 | 0 | 16 | 50 | 1 | 0 | 25 |
| WNT5A-dependent internalization of FZD4 | 16 | 23962258 | G | C | G | rs196000 | intronic | 0.9659 | 0.0341 | PRKCB | 0.0015 | 0 | 16 | 50 | 1 | 0 | 25 |
| WNT5A-dependent internalization of FZD4 | 16 | 23964858 | T | A | T | rs196003 | intronic | 0.9647 | 0.0353 | PRKCB | 0.0343 | 0 | 15 | 51 | 0 | 1 | 25 |
| WNT5A-dependent internalization of FZD4 | 16 | 23985814 | C | T | T | rs72779914 | intronic | 0.0487 | 0.0487 | PRKCB | 1.0000 | 53 | 12 | 0 | 21 | 4 | 0 |
| WNT5A-dependent internalization of FZD4 | 16 | 23987552 | A | G | A | rs169030 | intronic | 0.9709 | 0.0291 | PRKCB | 0.1679 | 0 | 11 | 55 | 0 | 1 | 25 |
| WNT5A-dependent internalization of FZD4 | 16 | 23988755 | T | C | T | rs196013 | intronic | 0.9681 | 0.0319 | PRKCB | 0.1681 | 0 | 11 | 55 | 0 | 1 | 24 |
| WNT5A-dependent internalization of FZD4 | 16 | 24009919 | A | G | G | rs75622923 | intronic | 0.0319 | 0.0319 | PRKCB | 0.3348 | 54 | 12 | 0 | 24 | 2 | 0 |
| WNT5A-dependent internalization of FZD4 | 16 | 24022944 | C | T | T | rs111746132 | intronic | 0.0229 | 0.0229 | PRKCB | 0.7522 | 56 | 10 | 0 | 23 | 3 | 0 |
| WNT5A-dependent internalization of FZD4 | 16 | 24066378 | G | A | A | rs113426570 | intronic | 0.0216 | 0.0216 | PRKCB | 1.0000 | 57 | 9 | 0 | 23 | 3 | 0 |
| WNT5A-dependent internalization of FZD4 | 16 | 24100759 | T | A | A | rs11643939 | intronic | 0.0294 | 0.0294 | PRKCB | 0.7522 | 56 | 10 | 0 | 23 | 3 | 0 |
| WNT5A-dependent internalization of FZD4 | 16 | 24105816 | G | A | A | rs56316329 | intronic | 0.0251 | 0.0251 | PRKCB | 0.6919 | 60 | 4 | 1 | 25 | 0 | 0 |
| WNT5A-dependent internalization of FZD4 | 16 | 24111853 | T | C | C | rs55959083 | intronic | 0.0431 | 0.0431 | PRKCB | 0.7213 | 58 | 8 | 0 | 23 | 2 | 0 |
| WNT5A-dependent internalization of FZD4 | 16 | 24112768 | G | A | A | rs117056307 | intronic | 0.0430 | 0.0430 | PRKCB | 0.7192 | 58 | 8 | 0 | 24 | 2 | 0 |
| WNT5A-dependent internalization of FZD4 | 16 | 24122052 | G | A | A | rs117467859 | intronic | 0.0433 | 0.0433 | PRKCB | 0.7213 | 58 | 8 | 0 | 23 | 2 | 0 |
| WNT5A-dependent internalization of FZD4 | 16 | 24122492 | C | T | T | rs72779977 | intronic | 0.0457 | 0.0457 | PRKCB | 0.7201 | 57 | 8 | 0 | 23 | 2 | 0 |
| WNT5A-dependent internalization of FZD4 | 16 | 24123560 | G | A | A | rs60261043 | intronic | 0.0457 | 0.0457 | PRKCB | 0.7201 | 57 | 8 | 0 | 23 | 2 | 0 |
| WNT5A-dependent internalization of FZD4 | 16 | 24132273 | G | A | A | rs62027458 | intronic | 0.0232 | 0.0232 | PRKCB | 0.1877 | 59 | 7 | 0 | 21 | 4 | 1 |
| WNT5A-dependent internalization of FZD4 | 16 | 24164042 | G | T | T | rs72779989 | intronic | 0.0487 | 0.0487 | PRKCB | 0.8583 | 57 | 7 | 2 | 22 | 4 | 0 |
| WNT5A-dependent internalization of FZD4 | 16 | 24197496 | A | T | T | rs79699525 | intronic | 0.0261 | 0.0261 | PRKCB | 0.4631 | 59 | 6 | 0 | 22 | 4 | 0 |
| WNT5A-dependent internalization of FZD4 | 16 | 24199852 | C | T | T | rs78424166 | intronic | 0.0359 | 0.0359 | PRKCB | 0.4603 | 60 | 6 | 0 | 22 | 4 | 0 |
| WNT5A-dependent internalization of FZD4 | 17 | 4626173 | A | G | G | rs78082978 | downstream | 0.0315 | 0.0315 | ARRB2 | 0.8212 | 55 | 10 | 1 | 23 | 3 | 0 |
| WNT5A-dependent internalization of FZD4 | 17 | 33953109 | G | A | A | rs225253 | intronic | 0.0178 | 0.0178 | AP2B1 | 0.2428 | 58 | 7 | 1 | 26 | 0 | 0 |
| WNT5A-dependent internalization of FZD4 | 17 | 33990042 | T | C | C | rs75950451 | intronic | 0.0469 | 0.0469 | AP2B1 | 0.8528 | 56 | 8 | 2 | 23 | 2 | 0 |
| WNT5A-dependent internalization of FZD4 | 17 | 34003398 | C | T | T | rs79010821 | intronic | 0.0386 | 0.0386 | AP2B1 | 0.1965 | 58 | 8 | 0 | 19 | 6 | 0 |
| WNT5A-dependent internalization of FZD4 | 17 | 34006300 | T | A | A | rs79688427 | intronic | 0.0389 | 0.0389 | AP2B1 | 0.2078 | 58 | 8 | 0 | 20 | 6 | 0 |
| WNT5A-dependent internalization of FZD4 | 17 | 34051124 | T | G | G | rs11653803 | 3_prime_UTR | 0.0341 | 0.0341 | AP2B1 | 0.7192 | 58 | 8 | 0 | 24 | 2 | 0 |
| WNT5A-dependent internalization of FZD4 | 17 | 57746013 | C | T | T | rs117419555 | intronic | 0.0437 | 0.0437 | CLTC | 1.0000 | 57 | 9 | 0 | 23 | 3 | 0 |
| WNT5A-dependent internalization of FZD4 | 17 | 64305051 | A | G | G | rs78357146 | intronic | 0.0186 | 0.0186 | PRKCA | 0.7209 | 57 | 9 | 0 | 23 | 2 | 0 |
| WNT5A-dependent internalization of FZD4 | 17 | 64315409 | T | C | C | rs80130647 | intronic | 0.0184 | 0.0184 | PRKCA | 0.7043 | 59 | 6 | 0 | 22 | 3 | 0 |
| WNT5A-dependent internalization of FZD4 | 17 | 64318385 | G | A | A | rs12150623 | intronic | 0.0343 | 0.0343 | PRKCA | 0.7595 | 61 | 4 | 1 | 23 | 2 | 0 |
| WNT5A-dependent internalization of FZD4 | 17 | 64320040 | C | T | T | rs139317720 | intronic | 0.0229 | 0.0229 | PRKCA | 1.0000 | 62 | 3 | 1 | 24 | 1 | 0 |
| WNT5A-dependent internalization of FZD4 | 17 | 64326068 | T | G | G | rs12951126 | intronic | 0.0260 | 0.0260 | PRKCA | 1.0000 | 58 | 8 | 0 | 23 | 3 | 0 |
| WNT5A-dependent internalization of FZD4 | 17 | 64343295 | A | C | C | rs72843901 | intronic | 0.0457 | 0.0457 | PRKCA | 0.4725 | 56 | 9 | 1 | 24 | 1 | 0 |
| WNT5A-dependent internalization of FZD4 | 17 | 64344650 | C | T | T | rs72846606 | intronic | 0.0454 | 0.0454 | PRKCA | 0.4725 | 56 | 9 | 1 | 24 | 1 | 0 |
| WNT5A-dependent internalization of FZD4 | 17 | 64344788 | C | T | T | rs72846607 | intronic | 0.0457 | 0.0457 | PRKCA | 0.4725 | 56 | 9 | 1 | 24 | 1 | 0 |
| WNT5A-dependent internalization of FZD4 | 17 | 64346204 | A | G | G | rs12150367 | intronic | 0.0452 | 0.0452 | PRKCA | 0.4770 | 56 | 9 | 1 | 25 | 1 | 0 |
| WNT5A-dependent internalization of FZD4 | 17 | 64349104 | C | T | T | rs72846609 | intronic | 0.0497 | 0.0497 | PRKCA | 0.4770 | 56 | 9 | 1 | 25 | 1 | 0 |
| WNT5A-dependent internalization of FZD4 | 17 | 64351542 | G | C | C | rs28592028 | intronic | 0.0453 | 0.0453 | PRKCA | 0.4770 | 56 | 9 | 1 | 25 | 1 | 0 |
| WNT5A-dependent internalization of FZD4 | 17 | 64352431 | T | G | G | rs544435459 | intronic | 0.0451 | 0.0451 | PRKCA | 0.4725 | 56 | 9 | 1 | 24 | 1 | 0 |
| WNT5A-dependent internalization of FZD4 | 17 | 64354597 | T | C | C | rs72846612 | intronic | 0.0451 | 0.0451 | PRKCA | 0.4725 | 56 | 9 | 1 | 24 | 1 | 0 |
| WNT5A-dependent internalization of FZD4 | 17 | 64358121 | C | G | G | rs72846614 | intronic | 0.0451 | 0.0451 | PRKCA | 0.4770 | 56 | 9 | 1 | 25 | 1 | 0 |
| WNT5A-dependent internalization of FZD4 | 17 | 64359354 | T | C | C | rs72846615 | intronic | 0.0452 | 0.0452 | PRKCA | 0.4725 | 56 | 9 | 1 | 24 | 1 | 0 |
| WNT5A-dependent internalization of FZD4 | 17 | 64365593 | C | T | T | rs72846677 | intronic | 0.0452 | 0.0452 | PRKCA | 0.4770 | 56 | 9 | 1 | 25 | 1 | 0 |
| WNT5A-dependent internalization of FZD4 | 17 | 64367534 | C | T | T | rs72846678 | intronic | 0.0452 | 0.0452 | PRKCA | 0.4725 | 56 | 9 | 1 | 24 | 1 | 0 |
| WNT5A-dependent internalization of FZD4 | 17 | 64369822 | A | G | G | rs72846681 | intronic | 0.0451 | 0.0451 | PRKCA | 0.4770 | 56 | 9 | 1 | 25 | 1 | 0 |
| WNT5A-dependent internalization of FZD4 | 17 | 64375194 | T | C | C | rs72846695 | intronic | 0.0452 | 0.0452 | PRKCA | 0.4770 | 56 | 9 | 1 | 25 | 1 | 0 |
| WNT5A-dependent internalization of FZD4 | 17 | 64377301 | C | A | A | rs113134992 | intronic | 0.0448 | 0.0448 | PRKCA | 0.4770 | 56 | 9 | 1 | 25 | 1 | 0 |
| WNT5A-dependent internalization of FZD4 | 17 | 64382507 | C | G | G | rs79461368 | intronic | 0.0453 | 0.0453 | PRKCA | 0.4725 | 56 | 9 | 1 | 24 | 1 | 0 |
| WNT5A-dependent internalization of FZD4 | 17 | 64389524 | C | T | T | rs77682324 | intronic | 0.0403 | 0.0403 | PRKCA | 0.7621 | 59 | 6 | 1 | 25 | 1 | 0 |
| WNT5A-dependent internalization of FZD4 | 17 | 64395773 | T | C | C | rs78584531 | intronic | 0.0452 | 0.0452 | PRKCA | 0.4770 | 56 | 9 | 1 | 25 | 1 | 0 |
| WNT5A-dependent internalization of FZD4 | 17 | 64396570 | T | G | G | rs72838208 | intronic | 0.0452 | 0.0452 | PRKCA | 0.4725 | 56 | 9 | 1 | 24 | 1 | 0 |
| WNT5A-dependent internalization of FZD4 | 17 | 64396695 | A | G | G | rs74329211 | intronic | 0.0432 | 0.0432 | PRKCA | 0.7302 | 58 | 8 | 0 | 21 | 4 | 0 |
| WNT5A-dependent internalization of FZD4 | 17 | 64396883 | G | A | A | rs72838209 | intronic | 0.0452 | 0.0452 | PRKCA | 0.4770 | 56 | 9 | 1 | 25 | 1 | 0 |
| WNT5A-dependent internalization of FZD4 | 17 | 64398013 | G | A | A | rs75125285 | intronic | 0.0295 | 0.0295 | PRKCA | 0.7632 | 60 | 5 | 1 | 25 | 1 | 0 |
| WNT5A-dependent internalization of FZD4 | 17 | 64400648 | C | T | T | rs72838214 | intronic | 0.0452 | 0.0452 | PRKCA | 0.4788 | 55 | 9 | 1 | 25 | 1 | 0 |
| WNT5A-dependent internalization of FZD4 | 17 | 64402141 | G | T | T | rs10221238 | intronic | 0.0452 | 0.0452 | PRKCA | 0.4770 | 56 | 9 | 1 | 25 | 1 | 0 |
| WNT5A-dependent internalization of FZD4 | 17 | 64405430 | T | A | A | rs72838216 | intronic | 0.0453 | 0.0453 | PRKCA | 0.4725 | 56 | 9 | 1 | 24 | 1 | 0 |
| WNT5A-dependent internalization of FZD4 | 17 | 64410564 | C | T | T | rs9972974 | intronic | 0.0390 | 0.0390 | PRKCA | 0.7632 | 60 | 5 | 1 | 25 | 1 | 0 |
| WNT5A-dependent internalization of FZD4 | 17 | 64412921 | G | A | A | rs62070391 | intronic | 0.0426 | 0.0426 | PRKCA | 0.7621 | 59 | 6 | 1 | 25 | 1 | 0 |
| WNT5A-dependent internalization of FZD4 | 17 | 64424382 | C | T | T | rs62070395 | intronic | 0.0425 | 0.0425 | PRKCA | 0.7595 | 59 | 6 | 1 | 24 | 1 | 0 |
| WNT5A-dependent internalization of FZD4 | 17 | 64429928 | G | A | A | rs117729211 | intronic | 0.0272 | 0.0272 | PRKCA | 0.0477 | 60 | 6 | 0 | 19 | 6 | 1 |
| WNT5A-dependent internalization of FZD4 | 17 | 64431874 | G | A | A | rs72838278 | intronic | 0.0426 | 0.0426 | PRKCA | 0.7642 | 57 | 6 | 1 | 25 | 1 | 0 |
| WNT5A-dependent internalization of FZD4 | 17 | 64432202 | A | G | G | rs62070397 | intronic | 0.0427 | 0.0427 | PRKCA | 0.7621 | 59 | 6 | 1 | 25 | 1 | 0 |
| WNT5A-dependent internalization of FZD4 | 17 | 64432881 | A | G | G | rs62070398 | intronic | 0.0426 | 0.0426 | PRKCA | 0.7595 | 59 | 6 | 1 | 24 | 1 | 0 |
| WNT5A-dependent internalization of FZD4 | 17 | 64438732 | A | T | T | rs113684166 | intronic | 0.0445 | 0.0445 | PRKCA | 0.2744 | 57 | 9 | 0 | 24 | 1 | 0 |
| WNT5A-dependent internalization of FZD4 | 17 | 64441204 | T | C | C | rs75005068 | intronic | 0.0469 | 0.0469 | PRKCA | 0.2719 | 57 | 9 | 0 | 25 | 1 | 0 |
| WNT5A-dependent internalization of FZD4 | 17 | 64441759 | C | A | A | rs78330327 | intronic | 0.0465 | 0.0465 | PRKCA | 0.2710 | 56 | 9 | 0 | 25 | 1 | 0 |
| WNT5A-dependent internalization of FZD4 | 17 | 64445337 | A | C | C | rs118090701 | intronic | 0.0387 | 0.0387 | PRKCA | 1.0000 | 58 | 6 | 1 | 23 | 2 | 0 |
| WNT5A-dependent internalization of FZD4 | 17 | 64445856 | A | G | G | rs12451388 | intronic | 0.0449 | 0.0449 | PRKCA | 0.2744 | 57 | 9 | 0 | 24 | 1 | 0 |
| WNT5A-dependent internalization of FZD4 | 17 | 64447721 | G | A | A | rs12452749 | intronic | 0.0462 | 0.0462 | PRKCA | 0.2786 | 57 | 9 | 0 | 23 | 1 | 0 |
| WNT5A-dependent internalization of FZD4 | 17 | 64450113 | G | C | C | rs80162292 | intronic | 0.0461 | 0.0461 | PRKCA | 0.2710 | 56 | 9 | 0 | 25 | 1 | 0 |
| WNT5A-dependent internalization of FZD4 | 17 | 64453194 | A | G | G | rs111776777 | intronic | 0.0442 | 0.0442 | PRKCA | 0.8716 | 54 | 10 | 2 | 23 | 3 | 0 |
| WNT5A-dependent internalization of FZD4 | 17 | 64458036 | C | T | T | rs113153197 | intronic | 0.0339 | 0.0339 | PRKCA | 0.1678 | 54 | 11 | 0 | 25 | 1 | 0 |
| WNT5A-dependent internalization of FZD4 | 17 | 64462111 | T | G | G | rs79239451 | intronic | 0.0464 | 0.0464 | PRKCA | 0.2744 | 57 | 9 | 0 | 24 | 1 | 0 |
| WNT5A-dependent internalization of FZD4 | 17 | 64462288 | G | A | A | rs80080003 | intronic | 0.0305 | 0.0305 | PRKCA | 0.1706 | 56 | 10 | 0 | 25 | 1 | 0 |
| WNT5A-dependent internalization of FZD4 | 17 | 64483589 | G | T | T | rs79070174 | intronic | 0.0391 | 0.0391 | PRKCA | 1.0000 | 59 | 6 | 1 | 23 | 2 | 0 |
| WNT5A-dependent internalization of FZD4 | 17 | 64487077 | T | G | G | rs78149509 | intronic | 0.0200 | 0.0200 | PRKCA | 0.0381 | 63 | 3 | 0 | 21 | 4 | 1 |
| WNT5A-dependent internalization of FZD4 | 17 | 64494906 | A | G | G | rs117168126 | intronic | 0.0300 | 0.0300 | PRKCA | 1.0000 | 59 | 6 | 1 | 23 | 2 | 0 |
| WNT5A-dependent internalization of FZD4 | 17 | 64501943 | C | T | T | rs7217954 | intronic | 0.0422 | 0.0422 | PRKCA | 1.0000 | 59 | 7 | 0 | 23 | 3 | 0 |
| WNT5A-dependent internalization of FZD4 | 17 | 64519790 | G | A | A | rs11659067 | intronic | 0.0498 | 0.0498 | PRKCA | 0.0125 | 62 | 3 | 0 | 19 | 6 | 0 |
| WNT5A-dependent internalization of FZD4 | 17 | 64525002 | A | T | T | rs77462363 | intronic | 0.0380 | 0.0380 | PRKCA | 1.0000 | 59 | 6 | 1 | 23 | 2 | 0 |
| WNT5A-dependent internalization of FZD4 | 17 | 64544723 | C | T | T | rs116879811 | intronic | 0.0274 | 0.0274 | PRKCA | 1.0000 | 60 | 6 | 0 | 24 | 2 | 0 |
| WNT5A-dependent internalization of FZD4 | 17 | 64559535 | A | G | G | rs227907 | intronic | 0.0478 | 0.0478 | PRKCA | 0.2800 | 55 | 11 | 0 | 21 | 3 | 1 |
| WNT5A-dependent internalization of FZD4 | 17 | 64561055 | G | A | A | rs62071706 | intronic | 0.0238 | 0.0238 | PRKCA | 0.1546 | 60 | 5 | 0 | 21 | 4 | 1 |
| WNT5A-dependent internalization of FZD4 | 17 | 64590951 | A | T | T | rs11867591 | intronic | 0.0449 | 0.0449 | PRKCA | 0.2122 | 57 | 8 | 0 | 20 | 6 | 0 |
| WNT5A-dependent internalization of FZD4 | 17 | 64604776 | C | T | T | rs117539643 | intronic | 0.0479 | 0.0479 | PRKCA | 0.6653 | 51 | 11 | 2 | 23 | 3 | 0 |
| WNT5A-dependent internalization of FZD4 | 17 | 64608923 | T | C | C | rs72845947 | intronic | 0.0440 | 0.0440 | PRKCA | 0.7355 | 57 | 7 | 2 | 21 | 4 | 0 |
| WNT5A-dependent internalization of FZD4 | 17 | 64610285 | T | C | C | rs72845948 | intronic | 0.0445 | 0.0445 | PRKCA | 0.8583 | 57 | 7 | 2 | 22 | 4 | 0 |
| WNT5A-dependent internalization of FZD4 | 17 | 64610480 | A | C | C | rs17759657 | intronic | 0.0444 | 0.0444 | PRKCA | 0.8583 | 57 | 7 | 2 | 22 | 4 | 0 |

| Pathway | Chr | Position | Ref | Alt | Obs | rsID | Region | Freq1 | Freq2 | Gene | P | N1 | N2 | N3 | N4 | N5 | N6 |
|---|---|---|---|---|---|---|---|---|---|---|---|---|---|---|---|---|---|
| WNT5A-dependent internalization of FZD4 | 17 | 64612838 | C | T | T | rs16959714 | intronic | 0.0338 | 0.0338 | PRKCA | 0.4353 | 57 | 8 | 0 | 24 | 1 | 0 |
| WNT5A-dependent internalization of FZD4 | 17 | 64614717 | T | C | C | rs17686540 | intronic | 0.0442 | 0.0442 | PRKCA | 0.7355 | 57 | 7 | 2 | 21 | 4 | 0 |
| WNT5A-dependent internalization of FZD4 | 17 | 64626385 | G | A | A | rs74352723 | intronic | 0.0460 | 0.0460 | PRKCA | 0.2744 | 57 | 9 | 0 | 24 | 1 | 0 |
| WNT5A-dependent internalization of FZD4 | 17 | 64628634 | A | G | G | rs117353888 | intronic | 0.0437 | 0.0437 | PRKCA | 0.2521 | 54 | 11 | 0 | 18 | 7 | 0 |
| WNT5A-dependent internalization of FZD4 | 17 | 64660047 | G | C | C | rs146141011 | intronic | 0.0484 | 0.0484 | PRKCA | 1.0000 | 53 | 9 | 2 | 22 | 4 | 0 |
| WNT5A-dependent internalization of FZD4 | 17 | 64671047 | C | T | T | rs16959942 | intronic | 0.0496 | 0.0496 | PRKCA | 0.1706 | 56 | 10 | 0 | 25 | 1 | 0 |
| WNT5A-dependent internalization of FZD4 | 17 | 64712634 | T | C | C | rs78121420 | intronic | 0.0465 | 0.0465 | PRKCA | 1.0000 | 57 | 7 | 2 | 23 | 3 | 0 |
| WNT5A-dependent internalization of FZD4 | 17 | 64713847 | G | T | T | rs112934229 | intronic | 0.0464 | 0.0464 | PRKCA | 1.0000 | 57 | 7 | 2 | 23 | 3 | 0 |
| WNT5A-dependent internalization of FZD4 | 17 | 64723880 | G | A | A | rs79547774 | intronic | 0.0264 | 0.0264 | PRKCA | 1.0000 | 57 | 9 | 0 | 23 | 3 | 0 |
| WNT5A-dependent internalization of FZD4 | 17 | 64738427 | A | C | C | rs117138620 | intronic | 0.0400 | 0.0400 | PRKCA | 0.3055 | 58 | 8 | 0 | 23 | 2 | 1 |
| WNT5A-dependent internalization of FZD4 | 17 | 64748431 | G | T | T | rs141177250 | intronic | 0.0153 | 0.0153 | PRKCA | 0.7595 | 59 | 6 | 1 | 24 | 1 | 0 |
| WNT5A-dependent internalization of FZD4 | 17 | 64755018 | T | C | C | rs74831470 | intronic | 0.0226 | 0.0226 | PRKCA | 0.7976 | 56 | 9 | 1 | 23 | 2 | 0 |
| WNT5A-dependent internalization of FZD4 | 17 | 64760870 | A | G | G | rs77904275 | intronic | 0.0307 | 0.0307 | PRKCA | 0.5233 | 53 | 11 | 1 | 24 | 2 | 0 |
| WNT5A-dependent internalization of FZD4 | 17 | 64762410 | C | G | G | rs113025478 | intronic | 0.0307 | 0.0307 | PRKCA | 0.6398 | 54 | 11 | 1 | 23 | 2 | 0 |
| WNT5A-dependent internalization of FZD4 | 17 | 64763235 | T | C | C | rs77635068 | intronic | 0.0306 | 0.0306 | PRKCA | 0.6398 | 54 | 11 | 1 | 23 | 2 | 0 |
| WNT5A-dependent internalization of FZD4 | 17 | 64771614 | C | T | T | rs80238933 | intronic | 0.0297 | 0.0297 | PRKCA | 0.6398 | 54 | 11 | 1 | 23 | 2 | 0 |
| WNT5A-dependent internalization of FZD4 | 17 | 64776847 | G | A | A | rs113542727 | intronic | 0.0297 | 0.0297 | PRKCA | 0.5228 | 54 | 11 | 1 | 24 | 2 | 0 |
| WNT5A-dependent internalization of FZD4 | 17 | 64791836 | G | C | C | rs56884788 | intronic | 0.0390 | 0.0390 | PRKCA | 0.6362 | 55 | 10 | 1 | 23 | 2 | 0 |
| WNT5A-dependent internalization of FZD4 | 17 | 64792863 | G | C | C | rs72838636 | intronic | 0.0484 | 0.0484 | PRKCA | 0.6470 | 57 | 8 | 1 | 21 | 5 | 0 |
| WNT5A-dependent internalization of FZD4 | 19 | 47342929 | G | A | A | rs17716171 | intronic | 0.0468 | 0.0468 | AP2S1 | 0.3121 | 57 | 8 | 1 | 20 | 5 | 1 |
| WNT5A-dependent internalization of FZD4 | 19 | 47352639 | A | G | G | rs78730097 | intronic | 0.0144 | 0.0144 | AP2S1 | 1.0000 | 60 | 5 | 1 | 24 | 2 | 0 |
| WNT5A-dependent internalization of FZD4 | 19 | 47353083 | T | A | A | rs117750880 | intronic | 0.0252 | 0.0252 | AP2S1 | 0.7632 | 60 | 5 | 1 | 25 | 1 | 0 |
| WNT5A-dependent internalization of FZD4 | 19 | 50279933 | G | T | T | rs79075745 | intronic | 0.0372 | 0.0372 | AP2A1 | 0.7560 | 55 | 10 | 0 | 21 | 5 | 0 |
| WNT5A-dependent internalization of FZD4 | 19 | 50282053 | G | A | A | rs62129179 | intronic | 0.0227 | 0.0227 | AP2A1 | 0.2039 | 62 | 4 | 0 | 21 | 2 | 1 |
| WNT5A-dependent internalization of FZD4 | 19 | 50297520 | T | C | C | rs62129182 | intronic | 0.0212 | 0.0212 | AP2A1 | 0.3151 | 62 | 4 | 0 | 23 | 2 | 1 |
| WNT5A-dependent internalization of FZD4 | 19 | 54401602 | G | C | C | rs41311973 | intronic | 0.0163 | 0.0163 | PRKCG | 0.7213 | 58 | 8 | 0 | 23 | 2 | 0 |
| RUNX1 regulates transcription of genes involved in differentiation of myeloid cells | 14 | 55593382 | G | A | A | rs149885215 | upstream | 0.0492 | 0.0492 | LGALS3 | 1.0000 | 56 | 9 | 1 | 22 | 4 | 0 |
| RUNX1 regulates transcription of genes involved in differentiation of myeloid cells | 14 | 55598617 | G | T | T | rs78793419 | intronic | 0.0375 | 0.0375 | LGALS3 | 0.4984 | 59 | 7 | 0 | 22 | 4 | 0 |
| RUNX1 regulates transcription of genes involved in differentiation of myeloid cells | 16 | 3782471 | G | C | C | rs75401475 | intronic | 0.0208 | 0.0208 | CREBBP | 1.0000 | 58 | 8 | 0 | 23 | 3 | 0 |
| RUNX1 regulates transcription of genes involved in differentiation of myeloid cells | 16 | 3796252 | T | C | C | rs116481265 | intronic | 0.0390 | 0.0390 | CREBBP | 0.7549 | 56 | 10 | 0 | 21 | 5 | 0 |
| RUNX1 regulates transcription of genes involved in differentiation of myeloid cells | 16 | 3857035 | T | G | G | rs130038 | intronic | 0.0341 | 0.0341 | CREBBP | 0.5065 | 58 | 8 | 0 | 21 | 5 | 0 |
| RUNX1 regulates transcription of genes involved in differentiation of myeloid cells | 16 | 3876864 | T | C | C | rs77292571 | intronic | 0.0215 | 0.0215 | CREBBP | 0.7348 | 58 | 8 | 0 | 22 | 4 | 0 |
| RUNX1 regulates transcription of genes involved in differentiation of myeloid cells | 16 | 3880659 | A | G | G | rs12447449 | intronic | 0.0347 | 0.0347 | CREBBP | 1.0000 | 57 | 8 | 0 | 22 | 3 | 0 |
| RUNX1 regulates transcription of genes involved in differentiation of myeloid cells | 16 | 3881494 | T | C | C | rs113198082 | intronic | 0.0497 | 0.0497 | CREBBP | 0.2977 | 54 | 12 | 0 | 22 | 3 | 1 |
| RUNX1 regulates transcription of genes involved in differentiation of myeloid cells | 16 | 3888283 | A | T | T | rs35581149 | intronic | 0.0268 | 0.0268 | CREBBP | 1.0000 | 57 | 9 | 0 | 23 | 3 | 0 |
| RUNX1 regulates transcription of genes involved in differentiation of myeloid cells | 16 | 3901493 | T | C | C | rs113653086 | intronic | 0.0332 | 0.0332 | CREBBP | 0.3346 | 58 | 8 | 0 | 20 | 5 | 0 |
| RUNX1 regulates transcription of genes involved in differentiation of myeloid cells | 16 | 23845860 | G | T | T | rs72777910 | upstream | 0.0300 | 0.0300 | PRKCB | 0.4762 | 60 | 5 | 1 | 22 | 4 | 0 |
| RUNX1 regulates transcription of genes involved in differentiation of myeloid cells | 16 | 23849482 | T | C | T | rs2023670 | intronic | 0.9513 | 0.0487 | PRKCB | 0.0099 | 0 | 11 | 55 | 1 | 0 | 25 |
| RUNX1 regulates transcription of genes involved in differentiation of myeloid cells | 16 | 23850240 | A | G | A | rs11074581 | intronic | 0.9663 | 0.0337 | PRKCB | 0.0099 | 0 | 11 | 55 | 1 | 0 | 25 |
| RUNX1 regulates transcription of genes involved in differentiation of myeloid cells | 16 | 23851956 | T | C | T | rs7189210 | intronic | 0.9663 | 0.0337 | PRKCB | 0.0099 | 0 | 11 | 55 | 1 | 0 | 25 |
| RUNX1 regulates transcription of genes involved in differentiation of myeloid cells | 16 | 23852415 | A | T | A | rs2188359 | intronic | 0.9528 | 0.0472 | PRKCB | 0.0099 | 0 | 11 | 55 | 1 | 0 | 25 |
| RUNX1 regulates transcription of genes involved in differentiation of myeloid cells | 16 | 23859391 | A | G | G | rs62030647 | intronic | 0.0226 | 0.0226 | PRKCB | 1.0000 | 59 | 7 | 0 | 23 | 3 | 0 |
| RUNX1 regulates transcription of genes involved in differentiation of myeloid cells | 16 | 23874933 | A | C | A | rs6497691 | intronic | 0.9663 | 0.0337 | PRKCB | 0.0063 | 0 | 14 | 52 | 1 | 0 | 24 |
| RUNX1 regulates transcription of genes involved in differentiation of myeloid cells | 16 | 23876099 | C | T | T | rs79131874 | intronic | 0.0303 | 0.0303 | PRKCB | 1.0000 | 59 | 7 | 0 | 23 | 3 | 0 |
| RUNX1 regulates transcription of genes involved in differentiation of myeloid cells | 16 | 23877500 | A | G | A | rs8059885 | intronic | 0.9663 | 0.0337 | PRKCB | 0.0046 | 0 | 14 | 52 | 1 | 0 | 25 |
| RUNX1 regulates transcription of genes involved in differentiation of myeloid cells | 16 | 23877606 | A | G | A | rs8060048 | intronic | 0.9644 | 0.0356 | PRKCB | 0.0046 | 0 | 14 | 52 | 1 | 0 | 25 |
| RUNX1 regulates transcription of genes involved in differentiation of myeloid cells | 16 | 23877781 | G | A | G | rs8060718 | intronic | 0.9664 | 0.0336 | PRKCB | 0.0046 | 0 | 14 | 52 | 1 | 0 | 25 |
| RUNX1 regulates transcription of genes involved in differentiation of myeloid cells | 16 | 23878470 | C | T | C | rs12935004 | intronic | 0.9657 | 0.0343 | PRKCB | 0.0046 | 0 | 14 | 52 | 1 | 0 | 25 |
| RUNX1 regulates transcription of genes involved in differentiation of myeloid cells | 16 | 23880851 | C | T | C | rs8061523 | intronic | 0.9664 | 0.0336 | PRKCB | 0.0046 | 0 | 14 | 52 | 1 | 0 | 25 |
| RUNX1 regulates transcription of genes involved in differentiation of myeloid cells | 16 | 23881930 | G | A | G | rs8047121 | intronic | 0.9662 | 0.0338 | PRKCB | 0.0046 | 0 | 14 | 52 | 1 | 0 | 25 |
| RUNX1 regulates transcription of genes involved in differentiation of myeloid cells | 16 | 23882469 | T | C | T | rs1468129 | intronic | 0.9663 | 0.0337 | PRKCB | 0.0046 | 0 | 14 | 52 | 1 | 0 | 25 |
| RUNX1 regulates transcription of genes involved in differentiation of myeloid cells | 16 | 23885608 | A | T | A | rs8044732 | intronic | 0.9664 | 0.0336 | PRKCB | 0.0046 | 0 | 14 | 52 | 1 | 0 | 25 |
| RUNX1 regulates transcription of genes involved in differentiation of myeloid cells | 16 | 23885751 | A | G | G | rs62031692 | intronic | 0.0253 | 0.0253 | PRKCB | 1.0000 | 59 | 7 | 0 | 23 | 3 | 0 |
| RUNX1 regulates transcription of genes involved in differentiation of myeloid cells | 16 | 23887574 | G | T | T | rs79034087 | intronic | 0.0290 | 0.0290 | PRKCB | 0.3487 | 61 | 4 | 1 | 22 | 4 | 0 |
| RUNX1 regulates transcription of genes involved in differentiation of myeloid cells | 16 | 23888354 | C | T | C | rs7404417 | intronic | 0.9664 | 0.0336 | PRKCB | 0.0046 | 0 | 14 | 52 | 1 | 0 | 25 |
| RUNX1 regulates transcription of genes involved in differentiation of myeloid cells | 16 | 23889896 | T | C | T | rs8063823 | intronic | 0.9665 | 0.0335 | PRKCB | 0.0046 | 0 | 14 | 52 | 1 | 0 | 25 |
| RUNX1 regulates transcription of genes involved in differentiation of myeloid cells | 16 | 23893893 | G | A | G | rs11647359 | intronic | 0.9664 | 0.0336 | PRKCB | 0.0046 | 0 | 14 | 52 | 1 | 0 | 25 |
| RUNX1 regulates transcription of genes involved in differentiation of myeloid cells | 16 | 23895034 | A | G | A | rs6497695 | intronic | 0.9665 | 0.0335 | PRKCB | 0.0046 | 0 | 14 | 52 | 1 | 0 | 25 |
| RUNX1 regulates transcription of genes involved in differentiation of myeloid cells | 16 | 23895443 | A | G | G | rs62028075 | intronic | 0.0253 | 0.0253 | PRKCB | 1.0000 | 59 | 7 | 0 | 23 | 3 | 0 |
| RUNX1 regulates transcription of genes involved in differentiation of myeloid cells | 16 | 23895884 | T | C | T | rs9944348 | intronic | 0.9665 | 0.0335 | PRKCB | 0.0046 | 0 | 14 | 52 | 1 | 0 | 25 |
| RUNX1 regulates transcription of genes involved in differentiation of myeloid cells | 16 | 23896089 | T | C | C | rs74572166 | intronic | 0.0245 | 0.0245 | PRKCB | 1.0000 | 59 | 7 | 0 | 23 | 3 | 0 |

| Pathway | Chr | Position | Ref | Alt | Minor | rsID | Region | MAF1 | MAF2 | Gene | P | N1 | N2 | N3 | N4 | N5 | N6 |
|---|---|---|---|---|---|---|---|---|---|---|---|---|---|---|---|---|---|
| RUNX1 regulates transcription of genes involved in differentiation of myeloid cells | 16 | 23896209 | C | A | C | rs9302418 | intronic | 0.9664 | 0.0336 | PRKCB | 0.0046 | 0 | 14 | 52 | 1 | 0 | 25 |
| RUNX1 regulates transcription of genes involved in differentiation of myeloid cells | 16 | 23896438 | G | T | T | rs62028076 | intronic | 0.0252 | 0.0252 | PRKCB | 1.0000 | 59 | 7 | 0 | 22 | 3 | 0 |
| RUNX1 regulates transcription of genes involved in differentiation of myeloid cells | 16 | 23898605 | A | T | A | rs933290 | intronic | 0.9632 | 0.0368 | PRKCB | 0.0016 | 0 | 17 | 49 | 1 | 0 | 24 |
| RUNX1 regulates transcription of genes involved in differentiation of myeloid cells | 16 | 23899211 | A | T | A | rs12926245 | intronic | 0.9632 | 0.0368 | PRKCB | 0.0012 | 0 | 17 | 49 | 1 | 0 | 25 |
| RUNX1 regulates transcription of genes involved in differentiation of myeloid cells | 16 | 23899610 | G | A | A | rs17753246 | intronic | 0.0252 | 0.0252 | PRKCB | 1.0000 | 59 | 7 | 0 | 23 | 3 | 0 |
| RUNX1 regulates transcription of genes involved in differentiation of myeloid cells | 16 | 23899951 | G | A | A | rs62028077 | intronic | 0.0254 | 0.0254 | PRKCB | 1.0000 | 59 | 7 | 0 | 23 | 3 | 0 |
| RUNX1 regulates transcription of genes involved in differentiation of myeloid cells | 16 | 23900716 | T | C | C | rs62028078 | intronic | 0.0252 | 0.0252 | PRKCB | 1.0000 | 59 | 7 | 0 | 23 | 3 | 0 |
| RUNX1 regulates transcription of genes involved in differentiation of myeloid cells | 16 | 23901896 | C | T | C | rs6497696 | intronic | 0.9632 | 0.0368 | PRKCB | 0.0015 | 0 | 16 | 50 | 1 | 0 | 25 |
| RUNX1 regulates transcription of genes involved in differentiation of myeloid cells | 16 | 23901948 | A | C | A | rs6497697 | intronic | 0.9630 | 0.0370 | PRKCB | 0.0015 | 0 | 16 | 49 | 1 | 0 | 25 |
| RUNX1 regulates transcription of genes involved in differentiation of myeloid cells | 16 | 23904058 | A | G | A | rs886115 | intronic | 0.9632 | 0.0368 | PRKCB | 0.0015 | 0 | 16 | 50 | 1 | 0 | 25 |
| RUNX1 regulates transcription of genes involved in differentiation of myeloid cells | 16 | 23904781 | G | A | A | rs17753509 | intronic | 0.0253 | 0.0253 | PRKCB | 1.0000 | 59 | 7 | 0 | 23 | 3 | 0 |
| RUNX1 regulates transcription of genes involved in differentiation of myeloid cells | 16 | 23905676 | C | T | C | rs7200610 | intronic | 0.9631 | 0.0369 | PRKCB | 0.0015 | 0 | 16 | 50 | 1 | 0 | 25 |
| RUNX1 regulates transcription of genes involved in differentiation of myeloid cells | 16 | 23907177 | A | C | C | rs17810011 | intronic | 0.0251 | 0.0251 | PRKCB | 1.0000 | 59 | 7 | 0 | 23 | 3 | 0 |
| RUNX1 regulates transcription of genes involved in differentiation of myeloid cells | 16 | 23907765 | C | T | C | rs9925890 | intronic | 0.9632 | 0.0368 | PRKCB | 0.0024 | 0 | 16 | 50 | 1 | 0 | 24 |
| RUNX1 regulates transcription of genes involved in differentiation of myeloid cells | 16 | 23912174 | A | G | A | rs12448249 | intronic | 0.9519 | 0.0481 | PRKCB | 0.0015 | 0 | 16 | 50 | 1 | 0 | 25 |
| RUNX1 regulates transcription of genes involved in differentiation of myeloid cells | 16 | 23914915 | C | A | C | rs1004186 | intronic | 0.9632 | 0.0368 | PRKCB | 0.0015 | 0 | 16 | 50 | 1 | 0 | 25 |
| RUNX1 regulates transcription of genes involved in differentiation of myeloid cells | 16 | 23916258 | G | A | G | rs1004187 | intronic | 0.9632 | 0.0368 | PRKCB | 0.0015 | 0 | 16 | 50 | 1 | 0 | 25 |
| RUNX1 regulates transcription of genes involved in differentiation of myeloid cells | 16 | 23916521 | G | C | G | rs1008654 | intronic | 0.9633 | 0.0367 | PRKCB | 0.0015 | 0 | 16 | 50 | 1 | 0 | 25 |
| RUNX1 regulates transcription of genes involved in differentiation of myeloid cells | 16 | 23917335 | G | A | G | rs6497699 | intronic | 0.9645 | 0.0355 | PRKCB | 0.0024 | 0 | 16 | 50 | 1 | 0 | 24 |
| RUNX1 regulates transcription of genes involved in differentiation of myeloid cells | 16 | 23917465 | C | G | C | rs7186538 | intronic | 0.9645 | 0.0355 | PRKCB | 0.0024 | 0 | 16 | 50 | 1 | 0 | 24 |
| RUNX1 regulates transcription of genes involved in differentiation of myeloid cells | 16 | 23917700 | C | A | C | rs7187091 | intronic | 0.9646 | 0.0354 | PRKCB | 0.0024 | 0 | 16 | 50 | 1 | 0 | 24 |
| RUNX1 regulates transcription of genes involved in differentiation of myeloid cells | 16 | 23919088 | C | T | T | rs78322646 | intronic | 0.0278 | 0.0278 | PRKCB | 0.7348 | 58 | 8 | 0 | 22 | 4 | 0 |
| RUNX1 regulates transcription of genes involved in differentiation of myeloid cells | 16 | 23921083 | C | T | C | rs6497702 | intronic | 0.9647 | 0.0353 | PRKCB | 0.0024 | 0 | 16 | 50 | 1 | 0 | 24 |
| RUNX1 regulates transcription of genes involved in differentiation of myeloid cells | 16 | 23925936 | C | G | C | rs11074588 | intronic | 0.9649 | 0.0351 | PRKCB | 0.0015 | 0 | 16 | 50 | 1 | 0 | 25 |
| RUNX1 regulates transcription of genes involved in differentiation of myeloid cells | 16 | 23939212 | G | A | G | rs11074590 | intronic | 0.9650 | 0.0350 | PRKCB | 0.0024 | 0 | 16 | 50 | 1 | 0 | 24 |
| RUNX1 regulates transcription of genes involved in differentiation of myeloid cells | 16 | 23941628 | C | A | C | rs2005671 | intronic | 0.9647 | 0.0353 | PRKCB | 0.0015 | 0 | 16 | 49 | 1 | 0 | 25 |
| RUNX1 regulates transcription of genes involved in differentiation of myeloid cells | 16 | 23943749 | T | C | T | rs9302420 | intronic | 0.9649 | 0.0351 | PRKCB | 0.0024 | 0 | 16 | 50 | 1 | 0 | 24 |
| RUNX1 regulates transcription of genes involved in differentiation of myeloid cells | 16 | 23945985 | T | G | T | rs195989 | intronic | 0.9651 | 0.0349 | PRKCB | 0.0015 | 0 | 16 | 50 | 1 | 0 | 25 |
| RUNX1 regulates transcription of genes involved in differentiation of myeloid cells | 16 | 23946157 | G | A | A | rs76973283 | intronic | 0.0302 | 0.0302 | PRKCB | 0.5065 | 58 | 8 | 0 | 21 | 5 | 0 |
| RUNX1 regulates transcription of genes involved in differentiation of myeloid cells | 16 | 23949175 | G | C | G | rs2560403 | intronic | 0.9657 | 0.0343 | PRKCB | 0.0015 | 0 | 16 | 50 | 1 | 0 | 25 |
| RUNX1 regulates transcription of genes involved in differentiation of myeloid cells | 16 | 23949438 | A | G | A | rs195985 | intronic | 0.9658 | 0.0342 | PRKCB | 0.0015 | 0 | 16 | 50 | 1 | 0 | 25 |
| RUNX1 regulates transcription of genes involved in differentiation of myeloid cells | 16 | 23953265 | T | C | T | rs2560404 | intronic | 0.9656 | 0.0344 | PRKCB | 0.0024 | 0 | 16 | 50 | 1 | 0 | 24 |
| RUNX1 regulates transcription of genes involved in differentiation of myeloid cells | 16 | 23954128 | T | C | C | rs17810486 | intronic | 0.0308 | 0.0308 | PRKCB | 0.3024 | 58 | 7 | 0 | 20 | 5 | 0 |
| RUNX1 regulates transcription of genes involved in differentiation of myeloid cells | 16 | 23954253 | G | A | G | rs195994 | intronic | 0.9653 | 0.0347 | PRKCB | 0.0015 | 0 | 16 | 50 | 1 | 0 | 25 |
| RUNX1 regulates transcription of genes involved in differentiation of myeloid cells | 16 | 23962258 | G | C | G | rs196000 | intronic | 0.9659 | 0.0341 | PRKCB | 0.0015 | 0 | 16 | 50 | 1 | 0 | 25 |
| RUNX1 regulates transcription of genes involved in differentiation of myeloid cells | 16 | 23964858 | T | A | T | rs196003 | intronic | 0.9647 | 0.0353 | PRKCB | 0.0343 | 0 | 15 | 51 | 0 | 1 | 25 |
| RUNX1 regulates transcription of genes involved in differentiation of myeloid cells | 16 | 23985814 | C | T | T | rs72779914 | intronic | 0.0487 | 0.0487 | PRKCB | 1.0000 | 53 | 12 | 0 | 21 | 4 | 0 |
| RUNX1 regulates transcription of genes involved in differentiation of myeloid cells | 16 | 23987552 | A | G | A | rs169030 | intronic | 0.9709 | 0.0291 | PRKCB | 0.1679 | 0 | 11 | 55 | 0 | 1 | 25 |
| RUNX1 regulates transcription of genes involved in differentiation of myeloid cells | 16 | 23988755 | T | C | T | rs196013 | intronic | 0.9681 | 0.0319 | PRKCB | 0.1681 | 0 | 11 | 55 | 0 | 1 | 24 |
| RUNX1 regulates transcription of genes involved in differentiation of myeloid cells | 16 | 24009919 | A | G | G | rs75622923 | intronic | 0.0319 | 0.0319 | PRKCB | 0.3348 | 54 | 12 | 0 | 24 | 2 | 0 |
| RUNX1 regulates transcription of genes involved in differentiation of myeloid cells | 16 | 24022944 | C | T | T | rs111746132 | intronic | 0.0229 | 0.0229 | PRKCB | 0.7522 | 56 | 10 | 0 | 23 | 3 | 0 |
| RUNX1 regulates transcription of genes involved in differentiation of myeloid cells | 16 | 24066378 | G | A | A | rs113426570 | intronic | 0.0216 | 0.0216 | PRKCB | 1.0000 | 57 | 9 | 0 | 23 | 3 | 0 |
| RUNX1 regulates transcription of genes involved in differentiation of myeloid cells | 16 | 24100759 | T | A | A | rs11643939 | intronic | 0.0294 | 0.0294 | PRKCB | 0.7522 | 56 | 10 | 0 | 23 | 3 | 0 |
| RUNX1 regulates transcription of genes involved in differentiation of myeloid cells | 16 | 24105816 | G | A | A | rs56316329 | intronic | 0.0251 | 0.0251 | PRKCB | 0.6919 | 60 | 4 | 1 | 25 | 0 | 0 |
| RUNX1 regulates transcription of genes involved in differentiation of myeloid cells | 16 | 24111853 | T | C | C | rs55959083 | intronic | 0.0431 | 0.0431 | PRKCB | 0.7213 | 58 | 8 | 0 | 23 | 2 | 0 |
| RUNX1 regulates transcription of genes involved in differentiation of myeloid cells | 16 | 24112768 | G | A | A | rs117056307 | intronic | 0.0430 | 0.0430 | PRKCB | 0.7192 | 58 | 8 | 0 | 24 | 2 | 0 |
| RUNX1 regulates transcription of genes involved in differentiation of myeloid cells | 16 | 24122052 | G | A | A | rs117467859 | intronic | 0.0433 | 0.0433 | PRKCB | 0.7213 | 58 | 8 | 0 | 23 | 2 | 0 |
| RUNX1 regulates transcription of genes involved in differentiation of myeloid cells | 16 | 24122492 | C | T | T | rs72779977 | intronic | 0.0457 | 0.0457 | PRKCB | 0.7201 | 57 | 8 | 0 | 23 | 2 | 0 |
| RUNX1 regulates transcription of genes involved in differentiation of myeloid cells | 16 | 24123560 | G | A | A | rs60261043 | intronic | 0.0457 | 0.0457 | PRKCB | 0.7201 | 57 | 8 | 0 | 23 | 2 | 0 |
| RUNX1 regulates transcription of genes involved in differentiation of myeloid cells | 16 | 24132273 | G | A | A | rs62027458 | intronic | 0.0232 | 0.0232 | PRKCB | 0.1877 | 59 | 7 | 0 | 21 | 4 | 1 |
| RUNX1 regulates transcription of genes involved in differentiation of myeloid cells | 16 | 24164042 | G | T | T | rs72779989 | intronic | 0.0487 | 0.0487 | PRKCB | 0.8583 | 57 | 7 | 2 | 22 | 4 | 0 |
| RUNX1 regulates transcription of genes involved in differentiation of myeloid cells | 16 | 24197496 | A | T | T | rs79699525 | intronic | 0.0261 | 0.0261 | PRKCB | 0.4631 | 59 | 6 | 0 | 22 | 4 | 0 |
| RUNX1 regulates transcription of genes involved in differentiation of myeloid cells | 16 | 24199852 | C | T | T | rs78424166 | intronic | 0.0359 | 0.0359 | PRKCB | 0.4603 | 60 | 6 | 0 | 22 | 4 | 0 |
| RUNX1 regulates transcription of genes involved in differentiation of myeloid cells | 16 | 67058614 | T | G | G | rs116975210 | upstream | 0.0256 | 0.0256 | CBFB | 0.2078 | 58 | 8 | 0 | 20 | 6 | 0 |
| RUNX1 regulates transcription of genes involved in differentiation of myeloid cells | 16 | 67059547 | T | C | C | rs118064296 | upstream | 0.0255 | 0.0255 | CBFB | 0.2078 | 58 | 8 | 0 | 20 | 6 | 0 |
| RUNX1 regulates transcription of genes involved in differentiation of myeloid cells | 16 | 67060518 | C | A | A | rs78843766 | upstream | 0.0407 | 0.0407 | CBFB | 1.0000 | 59 | 7 | 0 | 23 | 3 | 0 |
| RUNX1 regulates transcription of genes involved in differentiation of myeloid cells | 16 | 67064218 | C | T | T | rs115088567 | intronic | 0.0408 | 0.0408 | CBFB | 1.0000 | 59 | 7 | 0 | 23 | 3 | 0 |
| RUNX1 regulates transcription of genes involved in differentiation of myeloid cells | 16 | 67065005 | T | G | G | rs60915579 | intronic | 0.0410 | 0.0410 | CBFB | 1.0000 | 59 | 7 | 0 | 23 | 3 | 0 |
| RUNX1 regulates transcription of genes involved in differentiation of myeloid cells | 16 | 67077985 | A | T | T | rs16957128 | intronic | 0.0410 | 0.0410 | CBFB | 1.0000 | 59 | 7 | 0 | 23 | 3 | 0 |
| RUNX1 regulates transcription of genes involved in differentiation of myeloid cells | 16 | 67081133 | C | A | A | rs116271212 | intronic | 0.0409 | 0.0409 | CBFB | 1.0000 | 59 | 7 | 0 | 23 | 3 | 0 |
| RUNX1 regulates transcription of genes involved in differentiation of myeloid cells | 16 | 67087645 | T | C | C | rs34968486 | intronic | 0.0413 | 0.0413 | CBFB | 1.0000 | 59 | 7 | 0 | 23 | 3 | 0 |
| RUNX1 regulates transcription of genes involved in differentiation of myeloid cells | 16 | 67089554 | C | T | T | rs115947495 | intronic | 0.0409 | 0.0409 | CBFB | 1.0000 | 59 | 7 | 0 | 22 | 3 | 0 |
| RUNX1 regulates transcription of genes involved in differentiation of myeloid cells | 16 | 67092389 | T | C | C | rs58448340 | intronic | 0.0410 | 0.0410 | CBFB | 1.0000 | 59 | 7 | 0 | 23 | 3 | 0 |
| RUNX1 regulates transcription of genes involved in differentiation of myeloid cells | 16 | 67097563 | C | G | G | rs146692508 | intronic | 0.0409 | 0.0409 | CBFB | 1.0000 | 59 | 7 | 0 | 22 | 3 | 0 |
| RUNX1 regulates transcription of genes involved in differentiation of myeloid cells | 16 | 67110174 | G | A | A | rs115775129 | intronic | 0.0409 | 0.0409 | CBFB | 1.0000 | 59 | 7 | 0 | 23 | 3 | 0 |
| RUNX1 regulates transcription of genes involved in differentiation of myeloid cells | 16 | 67113178 | A | G | G | rs78817904 | intronic | 0.0407 | 0.0407 | CBFB | 1.0000 | 59 | 7 | 0 | 23 | 3 | 0 |
| RUNX1 regulates transcription of genes involved in differentiation of myeloid cells | 16 | 67121057 | G | C | C | rs114556591 | intronic | 0.0408 | 0.0408 | CBFB | 1.0000 | 59 | 7 | 0 | 23 | 3 | 0 |
| RUNX1 regulates transcription of genes involved in differentiation of myeloid cells | 16 | 67121931 | C | G | G | rs80188152 | intronic | 0.0408 | 0.0408 | CBFB | 1.0000 | 59 | 7 | 0 | 22 | 3 | 0 |
| RUNX1 regulates transcription of genes involved in differentiation of myeloid cells | 16 | 67122543 | A | G | G | rs2204708 | intronic | 0.0407 | 0.0407 | CBFB | 1.0000 | 59 | 7 | 0 | 23 | 3 | 0 |
| RUNX1 regulates transcription of genes involved in differentiation of myeloid cells | 16 | 67127104 | C | T | T | rs116154207 | intronic | 0.0408 | 0.0408 | CBFB | 1.0000 | 59 | 7 | 0 | 23 | 3 | 0 |
| RUNX1 regulates transcription of genes involved in differentiation of myeloid cells | 16 | 67128349 | G | C | C | rs115375898 | intronic | 0.0408 | 0.0408 | CBFB | 1.0000 | 59 | 7 | 0 | 22 | 3 | 0 |
| RUNX1 regulates transcription of genes involved in differentiation of myeloid cells | 16 | 67138195 | C | G | G | rs76658185 | downstream | 0.0409 | 0.0409 | CBFB | 1.0000 | 59 | 7 | 0 | 23 | 3 | 0 |
| RUNX1 regulates transcription of genes involved in differentiation of myeloid cells | 21 | 36161662 | G | C | C | rs75192893 | 3_prime_UTR | 0.0252 | 0.0252 | RUNX1 | 1.0000 | 61 | 4 | 1 | 24 | 2 | 0 |
| RUNX1 regulates transcription of genes involved in differentiation of myeloid cells | 21 | 36258223 | T | C | C | rs2734472 | intronic | 0.0207 | 0.0207 | RUNX1 | 0.0718 | 55 | 11 | 0 | 24 | 1 | 1 |
| RUNX1 regulates transcription of genes involved in differentiation of myeloid cells | 21 | 36279730 | T | C | C | rs79087516 | intronic | 0.0328 | 0.0328 | RUNX1 | 0.7285 | 62 | 3 | 1 | 24 | 2 | 0 |
| RUNX1 regulates transcription of genes involved in differentiation of myeloid cells | 21 | 36298914 | T | C | C | rs2051392 | intronic | 0.0378 | 0.0378 | RUNX1 | 0.0576 | 28 | 30 | 8 | 6 | 19 | 1 |
| RUNX1 regulates transcription of genes involved in differentiation of myeloid cells | 21 | 36360816 | C | T | T | rs118131631 | intronic | 0.0328 | 0.0328 | RUNX1 | 1.0000 | 58 | 8 | 0 | 22 | 3 | 0 |
| RUNX1 regulates transcription of genes involved in differentiation of myeloid cells | 21 | 36385772 | G | A | A | rs78097430 | intronic | 0.0312 | 0.0312 | RUNX1 | 1.0000 | 60 | 5 | 1 | 24 | 2 | 0 |
| RUNX1 regulates transcription of genes involved in differentiation of myeloid cells | 21 | 36389539 | C | T | T | rs77728098 | intronic | 0.0493 | 0.0493 | RUNX1 | 1.0000 | 58 | 7 | 1 | 23 | 3 | 0 |
| RUNX1 regulates transcription of genes involved in differentiation of myeloid cells | 21 | 36392564 | A | G | G | rs71329093 | intronic | 0.0314 | 0.0314 | RUNX1 | 1.0000 | 58 | 8 | 0 | 23 | 3 | 0 |
| RUNX1 regulates transcription of genes involved in differentiation of myeloid cells | 21 | 36400441 | C | G | G | rs75967349 | intronic | 0.0294 | 0.0294 | RUNX1 | 1.0000 | 59 | 7 | 0 | 23 | 3 | 0 |
| RUNX1 regulates transcription of genes involved in differentiation of myeloid cells | 21 | 36405472 | C | G | G | rs75102343 | intronic | 0.0365 | 0.0365 | RUNX1 | 0.0087 | 52 | 14 | 0 | 26 | 0 | 0 |
| RUNX1 regulates transcription of genes involved in differentiation of myeloid cells | 21 | 36420012 | T | C | C | rs55800826 | intronic | 0.0435 | 0.0435 | RUNX1 | 0.7065 | 51 | 14 | 1 | 19 | 7 | 0 |
| RUNX1 regulates transcription of genes involved in differentiation of myeloid cells | 21 | 36420200 | T | G | G | rs77398576 | intronic | 0.0289 | 0.0289 | RUNX1 | 0.2818 | 60 | 6 | 0 | 21 | 5 | 0 |
| RUNX1 regulates transcription of genes involved in differentiation of myeloid cells | 21 | 36445895 | T | C | C | rs78203858 | intronic | 0.0402 | 0.0402 | RUNX1 | 1.0000 | 57 | 9 | 0 | 23 | 3 | 0 |
| RUNX1 regulates transcription of genes involved in differentiation of myeloid cells | 21 | 36481863 | C | A | A | rs9305560 | intronic | 0.0454 | 0.0454 | RUNX1 | 0.8770 | 56 | 8 | 2 | 22 | 2 | 1 |
| RUNX1 regulates transcription of genes involved in differentiation of myeloid cells | 21 | 36525063 | G | T | T | rs79134632 | intronic | 0.0303 | 0.0303 | RUNX1 | 0.0320 | 54 | 12 | 0 | 25 | 0 | 0 |
| RUNX1 regulates transcription of genes involved in differentiation of myeloid cells | 21 | 36525900 | T | C | C | rs62216988 | intronic | 0.0414 | 0.0414 | RUNX1 | 0.1174 | 57 | 9 | 0 | 23 | 1 | 1 |
| RUNX1 regulates transcription of genes involved in differentiation of myeloid cells | 21 | 36531645 | A | C | C | rs2834807 | intronic | 0.0357 | 0.0357 | RUNX1 | 1.0000 | 58 | 8 | 0 | 23 | 3 | 0 |
| RUNX1 regulates transcription of genes involved in differentiation of myeloid cells | 21 | 36540870 | T | C | C | rs9984209 | intronic | 0.0153 | 0.0153 | RUNX1 | 0.7929 | 59 | 6 | 1 | 23 | 3 | 0 |
| RUNX1 regulates transcription of genes involved in differentiation of myeloid cells | 21 | 36549995 | T | C | C | rs76673048 | intronic | 0.0153 | 0.0153 | RUNX1 | 0.7855 | 59 | 6 | 1 | 22 | 3 | 0 |
| RUNX1 regulates transcription of genes involved in differentiation of myeloid cells | 21 | 36554527 | C | T | T | rs80172635 | intronic | 0.0341 | 0.0341 | RUNX1 | 0.1028 | 53 | 13 | 0 | 25 | 1 | 0 |
| RUNX1 regulates transcription of genes involved in differentiation of myeloid cells | 21 | 36578831 | T | A | T | rs8132151 | intronic | 0.9516 | 0.0484 | RUNX1 | 0.2734 | 1 | 7 | 58 | 0 | 6 | 20 |
| RUNX1 regulates transcription of genes involved in differentiation of myeloid cells | 21 | 36600490 | G | A | A | rs113393734 | intronic | 0.0338 | 0.0338 | RUNX1 | 0.5316 | 56 | 9 | 0 | 21 | 5 | 0 |
| RUNX1 regulates transcription of genes involved in differentiation of myeloid cells | 21 | 36615789 | T | G | G | rs76608041 | intronic | 0.0356 | 0.0356 | RUNX1 | 0.4862 | 59 | 7 | 0 | 21 | 4 | 0 |
| RUNX1 regulates transcription of genes involved in differentiation of myeloid cells | 21 | 36644806 | C | T | T | rs73192907 | intronic | 0.0329 | 0.0329 | RUNX1 | 0.7549 | 56 | 10 | 0 | 21 | 5 | 0 |
| RUNX1 regulates transcription of genes involved in differentiation of myeloid cells | 21 | 36651166 | C | T | T | rs78150243 | intronic | 0.0348 | 0.0348 | RUNX1 | 0.4984 | 59 | 7 | 0 | 22 | 4 | 0 |
| RUNX1 regulates transcription of genes involved in differentiation of myeloid cells | 21 | 36678254 | G | A | A | rs77468602 | intronic | 0.0288 | 0.0288 | RUNX1 | 1.0000 | 56 | 10 | 0 | 22 | 3 | 0 |
| RUNX1 regulates transcription of genes involved in differentiation of myeloid cells | 21 | 36679979 | A | G | G | rs75531812 | intronic | 0.0486 | 0.0486 | RUNX1 | 0.4353 | 58 | 8 | 0 | 24 | 1 | 0 |
| RUNX1 regulates transcription of genes involved in differentiation of myeloid cells | 21 | 36682912 | T | A | A | rs66699816 | intronic | 0.0476 | 0.0476 | RUNX1 | 0.3666 | 4 | 34 | 28 | 0 | 11 | 14 |
| RUNX1 regulates transcription of genes involved in differentiation of myeloid cells | 21 | 36714201 | C | A | A | rs2242721 | intronic | 0.0422 | 0.0422 | RUNX1 | 1.0000 | 57 | 9 | 0 | 23 | 3 | 0 |
| RUNX1 regulates transcription of genes involved in differentiation of myeloid cells | 21 | 36741936 | A | G | G | rs2242724 | intronic | 0.0446 | 0.0446 | RUNX1 | 0.5457 | 56 | 10 | 0 | 20 | 5 | 0 |
| RUNX1 regulates transcription of genes involved in differentiation of myeloid cells | 21 | 36777301 | C | T | T | rs113799213 | intronic | 0.0482 | 0.0482 | RUNX1 | 0.7348 | 58 | 8 | 0 | 22 | 4 | 0 |
| RUNX1 regulates transcription of genes involved in differentiation of myeloid cells | 21 | 36790637 | C | T | T | rs73197362 | intronic | 0.0471 | 0.0471 | RUNX1 | 0.1026 | 56 | 9 | 0 | 18 | 6 | 1 |
| RUNX1 regulates transcription of genes involved in differentiation of myeloid cells | 21 | 36817202 | G | T | T | rs76384559 | intronic | 0.0472 | 0.0472 | RUNX1 | 0.7302 | 58 | 8 | 0 | 21 | 4 | 0 |
| RUNX1 regulates transcription of genes involved in differentiation of myeloid cells | 21 | 36827331 | G | C | C | rs62217064 | intronic | 0.0304 | 0.0304 | RUNX1 | 0.4532 | 60 | 6 | 0 | 21 | 4 | 0 |
| RUNX1 regulates transcription of genes involved in differentiation of myeloid cells | 21 | 36887862 | G | A | A | rs73201472 | intronic | 0.0285 | 0.0285 | RUNX1 | 0.0370 | 61 | 4 | 1 | 20 | 6 | 0 |
| RUNX1 regulates transcription of genes involved in differentiation of myeloid cells | 21 | 36904158 | A | G | G | rs78159534 | intronic | 0.0382 | 0.0382 | RUNX1 | 1.0000 | 58 | 7 | 1 | 23 | 3 | 0 |
| RUNX1 regulates transcription of genes involved in differentiation of myeloid cells | 21 | 36978701 | C | T | T | rs141048458 | intronic | 0.0137 | 0.0137 | RUNX1 | 0.7043 | 59 | 6 | 0 | 22 | 3 | 0 |
| RUNX1 regulates transcription of genes involved in differentiation of myeloid cells | 21 | 36988765 | A | G | G | rs117334794 | intronic | 0.0137 | 0.0137 | RUNX1 | 1.0000 | 59 | 7 | 0 | 22 | 3 | 0 |
| RUNX1 regulates transcription of genes involved in differentiation of myeloid cells | 21 | 37001252 | G | A | A | rs147475775 | intronic | 0.0133 | 0.0133 | RUNX1 | 1.0000 | 59 | 7 | 0 | 23 | 3 | 0 |
| RUNX1 regulates transcription of genes involved in differentiation of myeloid cells | 21 | 37041204 | A | T | T | rs9978750 | intronic | 0.0430 | 0.0430 | RUNX1 | 1.0000 | 58 | 8 | 0 | 22 | 3 | 0 |
| RUNX1 regulates transcription of genes involved in differentiation of myeloid cells | 21 | 37111438 | C | T | T | rs2835112 | intronic | 0.0109 | 0.0109 | RUNX1 | 1.0000 | 58 | 8 | 0 | 23 | 3 | 0 |
| RUNX1 regulates transcription of genes involved in differentiation of myeloid cells | 21 | 37145469 | A | G | G | rs78766167 | intronic | 0.0338 | 0.0338 | RUNX1 | 0.8583 | 57 | 8 | 1 | 22 | 3 | 1 |
| RUNX1 regulates transcription of genes involved in differentiation of myeloid cells | 21 | 37146205 | T | C | C | rs2835139 | intronic | 0.0440 | 0.0440 | RUNX1 | 1.0000 | 56 | 10 | 0 | 22 | 3 | 0 |
| RUNX1 regulates transcription of genes involved in differentiation of myeloid cells | 21 | 37165262 | T | G | G | rs79838028 | intronic | 0.0208 | 0.0208 | RUNX1 | 0.7681 | 60 | 5 | 1 | 22 | 3 | 0 |
| RUNX1 regulates transcription of genes involved in differentiation of myeloid cells | 21 | 37168627 | C | A | A | rs9976429 | intronic | 0.0381 | 0.0381 | RUNX1 | 1.0000 | 59 | 7 | 0 | 23 | 3 | 0 |
| RUNX1 regulates transcription of genes involved in differentiation of myeloid cells | 21 | 37201847 | G | A | A | rs2835146 | intronic | 0.0109 | 0.0109 | RUNX1 | 1.0000 | 58 | 8 | 0 | 23 | 3 | 0 |
| RUNX1 regulates transcription of genes involved in differentiation of myeloid cells | 21 | 37220269 | T | A | A | rs141274106 | intronic | 0.0109 | 0.0109 | RUNX1 | 1.0000 | 58 | 8 | 0 | 23 | 3 | 0 |
| RUNX1 regulates transcription of genes involved in differentiation of myeloid cells | 21 | 37276929 | C | T | T | rs75561514 | intronic | 0.0138 | 0.0138 | RUNX1 | 1.0000 | 61 | 4 | 1 | 24 | 2 | 0 |
| RUNX1 regulates transcription of genes involved in differentiation of myeloid cells | 21 | 37283402 | C | T | T | rs79437318 | intronic | 0.0284 | 0.0284 | RUNX1 | 0.7681 | 60 | 5 | 1 | 22 | 3 | 0 |
| RUNX1 regulates transcription of genes involved in differentiation of myeloid cells | 21 | 37314583 | C | T | T | rs35256126 | intronic | 0.0349 | 0.0349 | RUNX1 | 0.1421 | 59 | 6 | 1 | 19 | 6 | 0 |
| RUNX1 regulates transcription of genes involved in differentiation of myeloid cells | 21 | 37316253 | T | C | C | rs35676489 | intronic | 0.0484 | 0.0484 | RUNX1 | 1.0000 | 57 | 9 | 0 | 23 | 3 | 0 |
| RUNX1 regulates transcription of genes involved in differentiation of myeloid cells | 21 | 37318065 | G | A | A | rs112248952 | intronic | 0.0352 | 0.0352 | RUNX1 | 1.0000 | 58 | 8 | 0 | 22 | 3 | 0 |
| RUNX1 regulates transcription of genes involved in differentiation of myeloid cells | 21 | 37323869 | A | G | G | rs17813817 | intronic | 0.0353 | 0.0353 | RUNX1 | 0.1603 | 57 | 6 | 1 | 20 | 6 | 0 |
| RUNX1 regulates transcription of genes involved in differentiation of myeloid cells | 21 | 37325386 | A | G | G | rs74524623 | intronic | 0.0186 | 0.0186 | RUNX1 | 0.3387 | 61 | 4 | 1 | 21 | 4 | 0 |
| RUNX1 regulates transcription of genes involved in differentiation of myeloid cells | 21 | 37327434 | G | C | C | rs35296624 | intronic | 0.0356 | 0.0356 | RUNX1 | 0.2734 | 58 | 7 | 1 | 20 | 6 | 0 |
| RUNX1 regulates transcription of genes involved in differentiation of myeloid cells | 21 | 37327713 | G | C | C | rs35137377 | intronic | 0.0355 | 0.0355 | RUNX1 | 0.2734 | 58 | 7 | 1 | 20 | 6 | 0 |
| RUNX1 regulates transcription of genes involved in differentiation of myeloid cells | 21 | 37331245 | G | C | C | rs13049450 | intronic | 0.0354 | 0.0354 | RUNX1 | 0.1521 | 59 | 6 | 1 | 20 | 6 | 0 |
| RUNX1 regulates transcription of genes involved in differentiation of myeloid cells | 21 | 37332139 | G | T | T | rs75887851 | intronic | 0.0458 | 0.0458 | RUNX1 | 0.4799 | 58 | 6 | 1 | 20 | 5 | 0 |
| RUNX1 regulates transcription of genes involved in differentiation of myeloid cells | 21 | 37332988 | C | T | T | rs79662622 | intronic | 0.0372 | 0.0372 | RUNX1 | 0.2734 | 58 | 7 | 1 | 20 | 6 | 0 |
| RUNX1 regulates transcription of genes involved in differentiation of myeloid cells | 21 | 37333615 | C | T | T | rs34303222 | intronic | 0.0353 | 0.0353 | RUNX1 | 0.2734 | 58 | 7 | 1 | 20 | 6 | 0 |
| RUNX1 regulates transcription of genes involved in differentiation of myeloid cells | 21 | 37335408 | T | A | A | rs35338421 | intronic | 0.0354 | 0.0354 | RUNX1 | 0.2734 | 58 | 7 | 1 | 20 | 6 | 0 |
| RUNX1 regulates transcription of genes involved in differentiation of myeloid cells | 21 | 37342778 | C | T | T | rs13047609 | intronic | 0.0320 | 0.0320 | RUNX1 | 0.6499 | 56 | 8 | 1 | 21 | 5 | 0 |
| RUNX1 regulates transcription of genes involved in differentiation of myeloid cells | 21 | 37343254 | G | T | T | rs13048435 | intronic | 0.0319 | 0.0319 | RUNX1 | 0.4348 | 57 | 8 | 1 | 20 | 6 | 0 |
| RUNX1 regulates transcription of genes involved in differentiation of myeloid cells | 21 | 37345637 | G | A | A | rs13052673 | intronic | 0.0474 | 0.0474 | RUNX1 | 0.1196 | 56 | 9 | 1 | 18 | 8 | 0 |
| RUNX1 regulates transcription of genes involved in differentiation of myeloid cells | 21 | 37346157 | A | G | G | rs915742 | intronic | 0.0473 | 0.0473 | RUNX1 | 0.1196 | 56 | 9 | 1 | 18 | 8 | 0 |
| RUNX1 regulates transcription of genes involved in differentiation of myeloid cells | 21 | 37347057 | C | T | T | rs2835194 | intronic | 0.0475 | 0.0475 | RUNX1 | 0.1196 | 56 | 9 | 1 | 18 | 8 | 0 |
| RUNX1 regulates transcription of genes involved in differentiation of myeloid cells | 21 | 37348386 | A | G | G | rs34544538 | intronic | 0.0472 | 0.0472 | RUNX1 | 0.1196 | 56 | 9 | 1 | 18 | 8 | 0 |
| RUNX1 regulates transcription of genes involved in differentiation of myeloid cells | 21 | 37350173 | C | T | T | rs71330629 | intronic | 0.0472 | 0.0472 | RUNX1 | 0.1227 | 55 | 9 | 1 | 18 | 8 | 0 |
| RUNX1 regulates transcription of genes involved in differentiation of myeloid cells | 21 | 37351306 | C | T | T | rs2835205 | intronic | 0.0476 | 0.0476 | RUNX1 | 0.1139 | 55 | 9 | 1 | 17 | 8 | 0 |
| RUNX1 regulates transcription of genes involved in differentiation of myeloid cells | 21 | 37351459 | T | C | C | rs56317271 | intronic | 0.0472 | 0.0472 | RUNX1 | 0.1117 | 56 | 9 | 1 | 17 | 8 | 0 |
| RUNX1 regulates transcription of genes involved in differentiation of myeloid cells | 21 | 37351802 | C | T | T | rs13048310 | intronic | 0.0473 | 0.0473 | RUNX1 | 0.1196 | 56 | 9 | 1 | 18 | 8 | 0 |
| RUNX1 regulates transcription of genes involved in differentiation of myeloid cells | 21 | 37352953 | T | C | C | rs13046307 | intronic | 0.0473 | 0.0473 | RUNX1 | 0.1117 | 56 | 9 | 1 | 17 | 8 | 0 |
| RUNX1 regulates transcription of genes involved in differentiation of myeloid cells | 21 | 37353681 | T | C | C | rs71330630 | intronic | 0.0473 | 0.0473 | RUNX1 | 0.1227 | 55 | 9 | 1 | 18 | 8 | 0 |
| RUNX1 regulates transcription of genes involved in differentiation of myeloid cells | 21 | 37357051 | C | T | T | rs28665789 | 5_prime_UTR | 0.0498 | 0.0498 | RUNX1 | 1.0000 | 56 | 9 | 0 | 23 | 3 | 0 |
| RUNX1 regulates transcription of genes involved in differentiation of myeloid cells | 21 | 37361832 | A | C | C | rs36028697 | intronic | 0.0476 | 0.0476 | RUNX1 | 0.1196 | 56 | 9 | 1 | 18 | 8 | 0 |
| RUNX1 regulates transcription of genes involved in differentiation of myeloid cells | 21 | 37362718 | C | A | A | rs2835215 | intronic | 0.0476 | 0.0476 | RUNX1 | 0.1117 | 56 | 9 | 1 | 17 | 8 | 0 |
| RUNX1 regulates transcription of genes involved in differentiation of myeloid cells | 21 | 37363385 | A | G | G | rs2835216 | intronic | 0.0477 | 0.0477 | RUNX1 | 0.1196 | 56 | 9 | 1 | 18 | 8 | 0 |
| RUNX1 regulates transcription of genes involved in differentiation of myeloid cells | 21 | 37364473 | A | G | G | rs13047560 | intronic | 0.0432 | 0.0432 | RUNX1 | 0.3739 | 56 | 10 | 0 | 20 | 6 | 0 |
| RUNX1 regulates transcription of genes involved in differentiation of myeloid cells | 21 | 37366389 | A | T | T | rs34327189 | intronic | 0.0459 | 0.0459 | RUNX1 | 0.7528 | 55 | 10 | 0 | 20 | 5 | 0 |
| RUNX1 regulates transcription of genes involved in differentiation of myeloid cells | 21 | 37368100 | C | G | G | rs55824509 | intronic | 0.0431 | 0.0431 | RUNX1 | 0.3739 | 56 | 10 | 0 | 20 | 6 | 0 |
| RUNX1 regulates transcription of genes involved in differentiation of myeloid cells | 21 | 37369628 | C | T | T | rs13049536 | intronic | 0.0431 | 0.0431 | RUNX1 | 0.3612 | 56 | 10 | 0 | 19 | 6 | 0 |
| RUNX1 regulates transcription of genes involved in differentiation of myeloid cells | 21 | 37370442 | G | A | A | rs13050761 | intronic | 0.0430 | 0.0430 | RUNX1 | 0.3739 | 56 | 10 | 0 | 20 | 6 | 0 |
| RUNX1 regulates transcription of genes involved in differentiation of myeloid cells | 21 | 37371181 | C | T | T | rs71330631 | intronic | 0.0433 | 0.0433 | RUNX1 | 0.3788 | 55 | 10 | 0 | 20 | 6 | 0 |
| Downstream signaling events of B Cell Receptor (BCR) | 1 | 85738304 | T | G | G | rs141532778 | intronic | 0.0265 | 0.0265 | BCL10 | 0.1949 | 59 | 7 | 0 | 24 | 1 | 1 |
| Downstream signaling events of B Cell Receptor (BCR) | 1 | 85742155 | G | A | A | rs78416998 | 5_prime_UTR | 0.0349 | 0.0349 | BCL10 | 0.7734 | 60 | 5 | 1 | 23 | 3 | 0 |
| Downstream signaling events of B Cell Receptor (BCR) | 1 | 109959684 | G | A | A | rs12119154 | intronic | 0.0463 | 0.0463 | PSMA5 | 0.3842 | 59 | 7 | 0 | 22 | 3 | 1 |
| Downstream signaling events of B Cell Receptor (BCR) | 1 | 115255717 | T | C | C | rs9724630 | intronic | 0.0490 | 0.0490 | NRAS | 0.0910 | 60 | 6 | 0 | 20 | 6 | 0 |
| Downstream signaling events of B Cell Receptor (BCR) | 2 | 33674504 | A | G | G | rs17013001 | intronic | 0.0406 | 0.0406 | RASGRP3 | 0.2744 | 57 | 9 | 0 | 24 | 1 | 0 |
| Downstream signaling events of B Cell Receptor (BCR) | 2 | 33707486 | A | G | G | rs17648515 | intronic | 0.0454 | 0.0454 | RASGRP3 | 0.7549 | 56 | 10 | 0 | 21 | 5 | 0 |
| Downstream signaling events of B Cell Receptor (BCR) | 2 | 33715019 | A | C | C | rs17595032 | intronic | 0.0422 | 0.0422 | RASGRP3 | 0.7549 | 56 | 10 | 0 | 21 | 5 | 0 |
| Downstream signaling events of B Cell Receptor (BCR) | 2 | 33717791 | T | A | A | rs12470632 | intronic | 0.0329 | 0.0329 | RASGRP3 | 0.4981 | 56 | 10 | 0 | 24 | 2 | 0 |
| Downstream signaling events of B Cell Receptor (BCR) | 2 | 33718749 | G | A | A | rs72804240 | intronic | 0.0274 | 0.0274 | RASGRP3 | 0.3739 | 56 | 10 | 0 | 20 | 6 | 0 |
| Downstream signaling events of B Cell Receptor (BCR) | 2 | 33723194 | T | C | C | rs34719322 | intronic | 0.0204 | 0.0204 | RASGRP3 | 0.5590 | 61 | 4 | 1 | 22 | 3 | 0 |
| Downstream signaling events of B Cell Receptor (BCR) | 2 | 33725279 | T | C | C | rs34958143 | intronic | 0.0231 | 0.0231 | RASGRP3 | 0.7348 | 58 | 8 | 0 | 22 | 4 | 0 |
| Downstream signaling events of B Cell Receptor (BCR) | 2 | 33725706 | A | T | T | rs72785908 | intronic | 0.0406 | 0.0406 | RASGRP3 | 0.1706 | 56 | 10 | 0 | 25 | 1 | 0 |
| Downstream signaling events of B Cell Receptor (BCR) | 2 | 33732446 | A | G | G | rs80205916 | intronic | 0.0361 | 0.0361 | RASGRP3 | 0.7632 | 60 | 5 | 1 | 25 | 1 | 0 |
| Downstream signaling events of B Cell Receptor (BCR) | 2 | 33758737 | C | T | T | rs56197103 | intronic | 0.0426 | 0.0426 | RASGRP3 | 0.7384 | 55 | 9 | 1 | 21 | 3 | 1 |
| Downstream signaling events of B Cell Receptor (BCR) | 2 | 33760191 | A | G | G | rs148641284 | intronic | 0.0360 | 0.0360 | RASGRP3 | 0.7469 | 57 | 9 | 0 | 21 | 4 | 0 |
| Downstream signaling events of B Cell Receptor (BCR) | 2 | 54106726 | G | A | A | rs78542544 | intronic | 0.0466 | 0.0466 | PSME4 | 0.0565 | 56 | 10 | 0 | 25 | 0 | 0 |
| Downstream signaling events of B Cell Receptor (BCR) | 2 | 54109699 | G | A | A | rs75679248 | intronic | 0.0345 | 0.0345 | PSME4 | 1.0000 | 57 | 9 | 0 | 22 | 3 | 0 |
| Downstream signaling events of B Cell Receptor (BCR) | 2 | 54112612 | T | G | G | rs115678984 | intronic | 0.0340 | 0.0340 | PSME4 | 0.7302 | 58 | 8 | 0 | 21 | 4 | 0 |
| Downstream signaling events of B Cell Receptor (BCR) | 2 | 54116463 | C | T | T | rs805403 | intronic | 0.0376 | 0.0376 | PSME4 | 0.2083 | 62 | 4 | 0 | 21 | 4 | 0 |
| Downstream signaling events of B Cell Receptor (BCR) | 2 | 54119870 | T | C | C | rs77215245 | intronic | 0.0487 | 0.0487 | PSME4 | 0.6362 | 55 | 10 | 1 | 23 | 2 | 0 |
| Downstream signaling events of B Cell Receptor (BCR) | 2 | 54121956 | T | C | C | rs79436269 | intronic | 0.0498 | 0.0498 | PSME4 | 0.6362 | 55 | 10 | 1 | 23 | 2 | 0 |
| Downstream signaling events of B Cell Receptor (BCR) | 2 | 54131546 | T | C | C | rs78887562 | intronic | 0.0467 | 0.0467 | PSME4 | 0.0565 | 56 | 10 | 0 | 25 | 0 | 0 |
| Downstream signaling events of B Cell Receptor (BCR) | 2 | 54132737 | G | A | A | rs75125553 | intronic | 0.0493 | 0.0493 | PSME4 | 0.6407 | 54 | 10 | 1 | 24 | 2 | 0 |
| Downstream signaling events of B Cell Receptor (BCR) | 2 | 54148310 | G | C | C | rs74545963 | intronic | 0.0348 | 0.0348 | PSME4 | 1.0000 | 57 | 9 | 0 | 23 | 3 | 0 |
| Downstream signaling events of B Cell Receptor (BCR) | 2 | 54153284 | A | G | G | rs62139281 | intronic | 0.0226 | 0.0226 | PSME4 | 0.0237 | 62 | 4 | 0 | 19 | 6 | 0 |
| Downstream signaling events of B Cell Receptor (BCR) | 2 | 54161109 | A | C | C | rs74627832 | intronic | 0.0468 | 0.0468 | PSME4 | 0.0565 | 56 | 10 | 0 | 25 | 0 | 0 |
| Downstream signaling events of B Cell Receptor (BCR) | 2 | 54166671 | C | T | T | rs79435818 | intronic | 0.0317 | 0.0317 | PSME4 | 1.0000 | 58 | 8 | 0 | 23 | 3 | 0 |
| Downstream signaling events of B Cell Receptor (BCR) | 2 | 54166800 | A | G | G | rs76811723 | intronic | 0.0317 | 0.0317 | PSME4 | 1.0000 | 58 | 8 | 0 | 22 | 3 | 0 |
| Downstream signaling events of B Cell Receptor (BCR) | 2 | 54176733 | C | A | A | rs115660490 | intronic | 0.0366 | 0.0366 | PSME4 | 0.7213 | 58 | 8 | 0 | 23 | 2 | 0 |
| Downstream signaling events of B Cell Receptor (BCR) | 2 | 54182973 | T | A | A | rs114515761 | intronic | 0.0346 | 0.0346 | PSME4 | 0.7192 | 58 | 8 | 0 | 24 | 2 | 0 |
| Downstream signaling events of B Cell Receptor (BCR) | 2 | 54188394 | T | A | A | rs11692784 | intronic | 0.0346 | 0.0346 | PSME4 | 0.7192 | 58 | 8 | 0 | 24 | 2 | 0 |
| Downstream signaling events of B Cell Receptor (BCR) | 2 | 54198172 | G | A | A | rs141622297 | upstream | 0.0360 | 0.0360 | PSME4 | 0.7192 | 58 | 8 | 0 | 24 | 2 | 0 |
| Downstream signaling events of B Cell Receptor (BCR) | 2 | 54201823 | G | A | A | rs74336645 | upstream | 0.0362 | 0.0362 | PSME4 | 0.7213 | 58 | 8 | 0 | 23 | 2 | 0 |
| Downstream signaling events of B Cell Receptor (BCR) | 2 | 61155012 | C | G | G | rs79263888 | 3_prime_UTR | 0.0229 | 0.0229 | REL | 0.7522 | 56 | 10 | 0 | 23 | 3 | 0 |
| Downstream signaling events of B Cell Receptor (BCR) | 2 | 68411185 | T | C | C | rs72832011 | intronic | 0.0451 | 0.0451 | PPP3R1 | 0.0653 | 59 | 7 | 0 | 25 | 0 | 1 |
| Downstream signaling events of B Cell Receptor (BCR) | 2 | 68413244 | T | A | A | rs72832013 | intronic | 0.0455 | 0.0455 | PPP3R1 | 0.0621 | 59 | 7 | 0 | 24 | 0 | 1 |
| Downstream signaling events of B Cell Receptor (BCR) | 2 | 162169482 | G | A | A | rs7605885 | intronic | 0.0425 | 0.0425 | PSMD14 | 0.8183 | 56 | 9 | 1 | 21 | 4 | 0 |
| Downstream signaling events of B Cell Receptor (BCR) | 2 | 162215335 | C | T | T | rs6722186 | intronic | 0.0427 | 0.0427 | PSMD14 | 1.0000 | 56 | 9 | 1 | 22 | 4 | 0 |
| Downstream signaling events of B Cell Receptor (BCR) | 2 | 231921109 | G | A | A | rs3754982 | upstream | 0.0430 | 0.0430 | PSMD1 | 0.0899 | 60 | 6 | 0 | 25 | 0 | 1 |
| Downstream signaling events of B Cell Receptor (BCR) | 2 | 231926692 | G | A | A | rs80031661 | intronic | 0.0429 | 0.0429 | PSMD1 | 0.0899 | 60 | 6 | 0 | 25 | 0 | 1 |
| Downstream signaling events of B Cell Receptor (BCR) | 2 | 231932134 | T | C | C | rs78964764 | intronic | 0.0433 | 0.0433 | PSMD1 | 0.0863 | 60 | 6 | 0 | 24 | 0 | 1 |
| Downstream signaling events of B Cell Receptor (BCR) | 2 | 231933408 | C | T | T | rs13424110 | intronic | 0.0476 | 0.0476 | PSMD1 | 0.1782 | 59 | 7 | 0 | 20 | 4 | 1 |
| Downstream signaling events of B Cell Receptor (BCR) | 2 | 231937539 | T | C | C | rs2288148 | intronic | 0.0434 | 0.0434 | PSMD1 | 0.0863 | 60 | 6 | 0 | 24 | 0 | 1 |
| Downstream signaling events of B Cell Receptor (BCR) | 2 | 231939526 | A | G | G | rs12620983 | intronic | 0.0434 | 0.0434 | PSMD1 | 0.0863 | 60 | 6 | 0 | 24 | 0 | 1 |
| Downstream signaling events of B Cell Receptor (BCR) | 2 | 231947467 | A | G | G | rs11694724 | intronic | 0.0411 | 0.0411 | PSMD1 | 0.2744 | 57 | 9 | 0 | 24 | 1 | 0 |
| Downstream signaling events of B Cell Receptor (BCR) | 2 | 231949569 | T | C | C | rs2303354 | intronic | 0.0433 | 0.0433 | PSMD1 | 0.0876 | 59 | 6 | 0 | 24 | 0 | 1 |
| Downstream signaling events of B Cell Receptor (BCR) | 2 | 231950847 | T | G | G | rs78363488 | intronic | 0.0407 | 0.0407 | PSMD1 | 0.4353 | 57 | 8 | 0 | 24 | 1 | 0 |
| Downstream signaling events of B Cell Receptor (BCR) | 2 | 231967886 | T | C | C | rs12616914 | intronic | 0.0436 | 0.0436 | PSMD1 | 0.0863 | 60 | 6 | 0 | 24 | 0 | 1 |
| Downstream signaling events of B Cell Receptor (BCR) | 2 | 231979258 | A | G | G | rs17586405 | intronic | 0.0434 | 0.0434 | PSMD1 | 0.0863 | 60 | 6 | 0 | 24 | 0 | 1 |
| Downstream signaling events of B Cell Receptor (BCR) | 2 | 231982950 | T | C | C | rs60707561 | intronic | 0.0434 | 0.0434 | PSMD1 | 0.0863 | 60 | 6 | 0 | 24 | 0 | 1 |
| Downstream signaling events of B Cell Receptor (BCR) | 2 | 231983382 | T | G | G | rs17619636 | intronic | 0.0478 | 0.0478 | PSMD1 | 0.0899 | 60 | 6 | 0 | 25 | 0 | 1 |
| Downstream signaling events of B Cell Receptor (BCR) | 2 | 231998010 | G | A | A | rs77199363 | intronic | 0.0431 | 0.0431 | PSMD1 | 0.0899 | 60 | 6 | 0 | 25 | 0 | 1 |
| Downstream signaling events of B Cell Receptor (BCR) | 2 | 232003189 | C | T | T | rs80121410 | intronic | 0.0408 | 0.0408 | PSMD1 | 0.2744 | 57 | 9 | 0 | 24 | 1 | 0 |
| Downstream signaling events of B Cell Receptor (BCR) | 2 | 232021354 | T | C | C | rs115196328 | intronic | 0.0402 | 0.0402 | PSMD1 | 0.5418 | 53 | 13 | 0 | 22 | 3 | 0 |
| Downstream signaling events of B Cell Receptor (BCR) | 2 | 232025880 | C | T | T | rs76901853 | intronic | 0.0472 | 0.0472 | PSMD1 | 0.0863 | 60 | 6 | 0 | 24 | 0 | 1 |
| Downstream signaling events of B Cell Receptor (BCR) | 2 | 232036567 | G | A | A | rs11674175 | intronic | 0.0409 | 0.0409 | PSMD1 | 0.2744 | 57 | 9 | 0 | 24 | 1 | 0 |
| Downstream signaling events of B Cell Receptor (BCR) | 2 | 232041640 | G | A | A | rs111612792 | downstream | 0.0413 | 0.0413 | PSMD1 | 0.2719 | 57 | 9 | 0 | 25 | 1 | 0 |
| Downstream signaling events of B Cell Receptor (BCR) | 3 | 63994801 | G | A | A | rs62252370 | downstream | 0.0360 | 0.0360 | PSMD6 | 1.0000 | 58 | 8 | 0 | 23 | 3 | 0 |
| Downstream signaling events of B Cell Receptor (BCR) | 4 | 102027455 | C | T | T | rs78657301 | intronic | 0.0019 | 0.0019 | PPP3CA | 0.7980 | 57 | 8 | 1 | 24 | 2 | 0 |
| Downstream signaling events of B Cell Receptor (BCR) | 4 | 102030909 | T | C | C | rs143246322 | intronic | 0.0331 | 0.0331 | PPP3CA | 0.4999 | 56 | 10 | 0 | 23 | 2 | 0 |
| Downstream signaling events of B Cell Receptor (BCR) | 4 | 102046408 | C | T | T | rs79507245 | intronic | 0.0394 | 0.0394 | PPP3CA | 0.4999 | 56 | 10 | 0 | 23 | 2 | 0 |
| Downstream signaling events of B Cell Receptor (BCR) | 4 | 102046629 | C | T | T | rs74675949 | intronic | 0.0393 | 0.0393 | PPP3CA | 0.4981 | 56 | 10 | 0 | 24 | 2 | 0 |
| Downstream signaling events of B Cell Receptor (BCR) | 4 | 102047770 | T | C | C | rs78209620 | intronic | 0.0395 | 0.0395 | PPP3CA | 0.4999 | 56 | 10 | 0 | 23 | 2 | 0 |
| Downstream signaling events of B Cell Receptor (BCR) | 4 | 102047812 | A | G | G | rs78888749 | intronic | 0.0394 | 0.0394 | PPP3CA | 0.4999 | 56 | 10 | 0 | 23 | 2 | 0 |
| Downstream signaling events of B Cell Receptor (BCR) | 4 | 102049484 | A | G | G | rs113609659 | intronic | 0.0394 | 0.0394 | PPP3CA | 0.4999 | 56 | 10 | 0 | 23 | 2 | 0 |
| Downstream signaling events of B Cell Receptor (BCR) | 4 | 102059408 | A | G | G | rs111885453 | intronic | 0.0393 | 0.0393 | PPP3CA | 0.4999 | 56 | 10 | 0 | 23 | 2 | 0 |
| Downstream signaling events of B Cell Receptor (BCR) | 4 | 102060085 | C | A | A | rs150119589 | intronic | 0.0392 | 0.0392 | PPP3CA | 0.4999 | 56 | 10 | 0 | 23 | 2 | 0 |
| Downstream signaling events of B Cell Receptor (BCR) | 4 | 102061144 | C | T | T | rs113496555 | intronic | 0.0392 | 0.0392 | PPP3CA | 0.4985 | 55 | 10 | 0 | 23 | 2 | 0 |
| Downstream signaling events of B Cell Receptor (BCR) | 4 | 102061766 | T | C | C | rs116229517 | intronic | 0.0392 | 0.0392 | PPP3CA | 0.4999 | 56 | 10 | 0 | 23 | 2 | 0 |
| Downstream signaling events of B Cell Receptor (BCR) | 4 | 102061822 | C | T | T | rs112122964 | intronic | 0.0393 | 0.0393 | PPP3CA | 0.4999 | 56 | 10 | 0 | 23 | 2 | 0 |
| Downstream signaling events of B Cell Receptor (BCR) | 4 | 102062069 | G | A | A | rs74631130 | intronic | 0.0392 | 0.0392 | PPP3CA | 0.4999 | 56 | 10 | 0 | 23 | 2 | 0 |
| Downstream signaling events of B Cell Receptor (BCR) | 4 | 102063270 | T | C | C | rs111752790 | intronic | 0.0421 | 0.0421 | PPP3CA | 0.4999 | 56 | 10 | 0 | 23 | 2 | 0 |
| Downstream signaling events of B Cell Receptor (BCR) | 4 | 102064126 | A | T | T | rs113744576 | intronic | 0.0439 | 0.0439 | PPP3CA | 0.4981 | 56 | 10 | 0 | 24 | 2 | 0 |
| Downstream signaling events of B Cell Receptor (BCR) | 4 | 102064899 | T | C | C | rs112973280 | intronic | 0.0500 | 0.0500 | PPP3CA | 0.4981 | 56 | 10 | 0 | 24 | 2 | 0 |
| Downstream signaling events of B Cell Receptor (BCR) | 4 | 102065144 | G | A | A | rs112695279 | intronic | 0.0497 | 0.0497 | PPP3CA | 0.4981 | 56 | 10 | 0 | 24 | 2 | 0 |
| Downstream signaling events of B Cell Receptor (BCR) | 4 | 102065533 | G | A | A | rs75372805 | intronic | 0.0392 | 0.0392 | PPP3CA | 0.4981 | 56 | 10 | 0 | 24 | 2 | 0 |
| Downstream signaling events of B Cell Receptor (BCR) | 4 | 102065719 | T | C | C | rs17240082 | intronic | 0.0496 | 0.0496 | PPP3CA | 0.4999 | 56 | 10 | 0 | 23 | 2 | 0 |
| Downstream signaling events of B Cell Receptor (BCR) | 4 | 102071777 | C | T | T | rs112522326 | intronic | 0.0439 | 0.0439 | PPP3CA | 0.4999 | 56 | 10 | 0 | 23 | 2 | 0 |
| Downstream signaling events of B Cell Receptor (BCR) | 4 | 102073273 | G | A | A | rs111792710 | intronic | 0.0391 | 0.0391 | PPP3CA | 0.4999 | 56 | 10 | 0 | 23 | 2 | 0 |
| Downstream signaling events of B Cell Receptor (BCR) | 4 | 102105924 | C | T | T | rs80191143 | intronic | 0.0153 | 0.0153 | PPP3CA | 1.0000 | 60 | 5 | 1 | 24 | 1 | 0 |
| Downstream signaling events of B Cell Receptor (BCR) | 4 | 102128007 | A | C | C | rs75423137 | intronic | 0.0447 | 0.0447 | PPP3CA | 0.6398 | 54 | 11 | 1 | 23 | 2 | 0 |
| Downstream signaling events of B Cell Receptor (BCR) | 4 | 102158097 | G | A | A | rs116097940 | intronic | 0.0226 | 0.0226 | PPP3CA | 0.3171 | 60 | 6 | 0 | 23 | 1 | 1 |
| Downstream signaling events of B Cell Receptor (BCR) | 4 | 102159289 | C | T | T | rs13133951 | intronic | 0.0215 | 0.0215 | PPP3CA | 0.3221 | 60 | 6 | 0 | 24 | 1 | 1 |
| Downstream signaling events of B Cell Receptor (BCR) | 4 | 102174349 | T | C | C | rs114543570 | intronic | 0.0466 | 0.0466 | PPP3CA | 0.7621 | 59 | 6 | 1 | 25 | 1 | 0 |
| Downstream signaling events of B Cell Receptor (BCR) | 4 | 103436737 | T | C | C | rs79590323 | intronic | 0.0393 | 0.0393 | NFKB1 | 0.4981 | 56 | 10 | 0 | 24 | 2 | 0 |
| Downstream signaling events of B Cell Receptor (BCR) | 4 | 103441152 | G | A | A | rs74833382 | intronic | 0.0318 | 0.0318 | NFKB1 | 0.7621 | 59 | 6 | 1 | 25 | 1 | 0 |
| Downstream signaling events of B Cell Receptor (BCR) | 4 | 103441742 | G | A | A | rs78900265 | intronic | 0.0393 | 0.0393 | NFKB1 | 0.4999 | 56 | 10 | 0 | 23 | 2 | 0 |
| Downstream signaling events of B Cell Receptor (BCR) | 4 | 103465229 | C | T | T | rs76016852 | intronic | 0.0414 | 0.0414 | NFKB1 | 0.4999 | 56 | 10 | 0 | 23 | 2 | 0 |
| Downstream signaling events of B Cell Receptor (BCR) | 4 | 103534557 | C | T | T | rs4648117 | intronic | 0.0416 | 0.0416 | NFKB1 | 0.7621 | 59 | 6 | 1 | 25 | 1 | 0 |

| Pathway | Chr | Position | Ref | Alt | Minor | rsID | Region | MAF1 | MAF2 | Gene | HWE | N00 | N01 | N02 | N10 | N11 | N12 |
|---|---|---|---|---|---|---|---|---|---|---|---|---|---|---|---|---|---|
| Downstream signaling events of B Cell Receptor (BCR) | 4 | 103535905 | C | T | T | rs4648127 | intronic | 0.0389 | 0.0389 | NFKB1 | 0.4974 | 55 | 10 | 0 | 24 | 2 | 0 |
| Downstream signaling events of B Cell Receptor (BCR) | 4 | 103542005 | G | T | T | rs997476 | downstream | 0.0397 | 0.0397 | NFKB1 | 0.4981 | 56 | 10 | 0 | 24 | 2 | 0 |
| Downstream signaling events of B Cell Receptor (BCR) | 5 | 133505097 | T | C | C | rs34751193 | intronic | 0.0303 | 0.0303 | SKP1 | 0.7497 | 55 | 11 | 0 | 23 | 3 | 0 |
| Downstream signaling events of B Cell Receptor (BCR) | 5 | 133512605 | G | T | T | rs11538030 | 5_prime_UTR | 0.0356 | 0.0356 | SKP1 | 0.6295 | 55 | 9 | 2 | 24 | 2 | 0 |
| Downstream signaling events of B Cell Receptor (BCR) | 5 | 171287624 | C | G | C | rs702109 | downstream | 0.9605 | 0.0395 | FBXW11 | 0.1356 | 0 | 13 | 52 | 1 | 2 | 22 |
| Downstream signaling events of B Cell Receptor (BCR) | 5 | 171319929 | T | C | C | rs17569783 | intronic | 0.0456 | 0.0456 | FBXW11 | 1.0000 | 59 | 6 | 1 | 23 | 2 | 0 |
| Downstream signaling events of B Cell Receptor (BCR) | 5 | 171387634 | G | A | A | rs72835279 | intronic | 0.0450 | 0.0450 | FBXW11 | 1.0000 | 59 | 6 | 1 | 24 | 2 | 0 |
| Downstream signaling events of B Cell Receptor (BCR) | 5 | 171420105 | G | C | C | rs72835287 | intronic | 0.0441 | 0.0441 | FBXW11 | 0.2719 | 57 | 9 | 0 | 25 | 1 | 0 |
| Downstream signaling events of B Cell Receptor (BCR) | 5 | 171424137 | T | G | G | rs72835288 | intronic | 0.0465 | 0.0465 | FBXW11 | 1.0000 | 59 | 6 | 1 | 24 | 2 | 0 |
| Downstream signaling events of B Cell Receptor (BCR) | 6 | 32810147 | G | T | T | rs41270496 | intronic | 0.0323 | 0.0323 | PSMB8 | 0.4984 | 59 | 7 | 0 | 22 | 4 | 0 |
| Downstream signaling events of B Cell Receptor (BCR) | 6 | 32811224 | A | T | T | rs72854938 | intronic | 0.0322 | 0.0322 | PSMB8 | 0.4984 | 59 | 7 | 0 | 22 | 4 | 0 |
| Downstream signaling events of B Cell Receptor (BCR) | 6 | 32811366 | G | T | T | rs72854939 | intronic | 0.0322 | 0.0322 | PSMB8 | 0.4984 | 59 | 7 | 0 | 22 | 4 | 0 |
| Downstream signaling events of B Cell Receptor (BCR) | 6 | 32826688 | G | T | T | rs115353581 | intronic | 0.0219 | 0.0219 | PSMB9 | 0.4984 | 59 | 7 | 0 | 22 | 4 | 0 |
| Downstream signaling events of B Cell Receptor (BCR) | 6 | 32832218 | A | T | T | rs116481206 | downstream | 0.0219 | 0.0219 | PSMB9 | 0.4862 | 59 | 7 | 0 | 21 | 4 | 0 |
| Downstream signaling events of B Cell Receptor (BCR) | 6 | 44227224 | C | T | T | rs28362860 | intronic | 0.0269 | 0.0269 | NFKBIE | 0.4603 | 60 | 6 | 0 | 22 | 4 | 0 |
| Downstream signaling events of B Cell Receptor (BCR) | 6 | 44233216 | G | C | C | rs28362857 | missense | 0.0296 | 0.0296 | NFKBIE | 0.0685 | 61 | 5 | 0 | 20 | 6 | 0 |
| Downstream signaling events of B Cell Receptor (BCR) | 6 | 91239702 | T | A | A | rs62409064 | intronic | 0.0407 | 0.0407 | MAP3K7 | 0.4940 | 58 | 7 | 1 | 20 | 5 | 0 |
| Downstream signaling events of B Cell Receptor (BCR) | 6 | 91296810 | C | T | T | rs34016005 | upstream | 0.0303 | 0.0303 | MAP3K7 | 0.4762 | 60 | 5 | 1 | 22 | 4 | 0 |
| Downstream signaling events of B Cell Receptor (BCR) | 6 | 170851525 | G | A | A | rs12200064 | intronic | 0.0431 | 0.0431 | PSMB1 | 0.3694 | 59 | 5 | 2 | 26 | 0 | 0 |
| Downstream signaling events of B Cell Receptor (BCR) | 6 | 170858332 | G | A | A | rs17860779 | intronic | 0.0420 | 0.0420 | PSMB1 | 1.0000 | 59 | 6 | 1 | 23 | 2 | 0 |
| Downstream signaling events of B Cell Receptor (BCR) | 7 | 2955476 | G | C | C | rs41336352 | intronic | 0.0400 | 0.0400 | CARD11 | 1.0000 | 58 | 8 | 0 | 22 | 3 | 0 |
| Downstream signaling events of B Cell Receptor (BCR) | 7 | 2966334 | G | A | A | rs41448444 | intronic | 0.0454 | 0.0454 | CARD11 | 0.7213 | 58 | 8 | 0 | 23 | 2 | 0 |
| Downstream signaling events of B Cell Receptor (BCR) | 7 | 2968105 | C | A | A | rs71527417 | intronic | 0.0472 | 0.0472 | CARD11 | 0.0436 | 49 | 16 | 1 | 25 | 1 | 0 |
| Downstream signaling events of B Cell Receptor (BCR) | 7 | 2970297 | C | T | T | rs41386651 | intronic | 0.0410 | 0.0410 | CARD11 | 0.7386 | 56 | 8 | 0 | 22 | 4 | 0 |
| Downstream signaling events of B Cell Receptor (BCR) | 7 | 2970866 | C | T | T | rs9648301 | intronic | 0.0444 | 0.0444 | CARD11 | 1.0000 | 57 | 9 | 0 | 22 | 4 | 0 |
| Downstream signaling events of B Cell Receptor (BCR) | 7 | 3029796 | G | C | C | rs12700500 | intronic | 0.0454 | 0.0454 | CARD11 | 1.0000 | 59 | 7 | 0 | 22 | 3 | 0 |
| Downstream signaling events of B Cell Receptor (BCR) | 7 | 3042802 | G | A | A | rs35579453 | intronic | 0.0363 | 0.0363 | CARD11 | 1.0000 | 57 | 9 | 0 | 22 | 4 | 0 |
| Downstream signaling events of B Cell Receptor (BCR) | 7 | 3052087 | G | A | A | rs75468256 | intronic | 0.0407 | 0.0407 | CARD11 | 0.2020 | 62 | 4 | 0 | 20 | 4 | 0 |
| Downstream signaling events of B Cell Receptor (BCR) | 7 | 3072512 | T | C | C | rs6975176 | intronic | 0.0452 | 0.0452 | CARD11 | 1.0000 | 61 | 4 | 1 | 25 | 1 | 0 |
| Downstream signaling events of B Cell Receptor (BCR) | 7 | 3073243 | G | A | A | rs78353391 | intronic | 0.0427 | 0.0427 | CARD11 | 0.0650 | 56 | 8 | 0 | 18 | 8 | 0 |
| Downstream signaling events of B Cell Receptor (BCR) | 7 | 3079195 | G | T | T | rs62439353 | intronic | 0.0400 | 0.0400 | CARD11 | 0.7348 | 58 | 8 | 0 | 22 | 4 | 0 |
| Downstream signaling events of B Cell Receptor (BCR) | 7 | 42971385 | G | A | A | rs76256519 | intronic | 0.0432 | 0.0432 | PSMA2 | 0.1923 | 61 | 5 | 0 | 21 | 3 | 1 |
| Downstream signaling events of B Cell Receptor (BCR) | 7 | 148411091 | C | T | T | rs11764941 | intronic | 0.0416 | 0.0416 | CUL1 | 0.0082 | 58 | 8 | 0 | 16 | 9 | 1 |
| Downstream signaling events of B Cell Receptor (BCR) | 7 | 148422581 | A | G | G | rs11760399 | intronic | 0.0409 | 0.0409 | CUL1 | 0.0158 | 57 | 9 | 0 | 16 | 9 | 1 |
| Downstream signaling events of B Cell Receptor (BCR) | 7 | 148427470 | A | T | T | rs17625893 | intronic | 0.0188 | 0.0188 | CUL1 | 0.3091 | 59 | 7 | 0 | 21 | 5 | 0 |
| Downstream signaling events of B Cell Receptor (BCR) | 7 | 148453031 | C | G | G | rs4726991 | intronic | 0.0427 | 0.0427 | CUL1 | 0.0164 | 61 | 5 | 0 | 18 | 7 | 0 |
| Downstream signaling events of B Cell Receptor (BCR) | 7 | 148455710 | A | G | G | rs73158225 | intronic | 0.0426 | 0.0426 | CUL1 | 0.0333 | 61 | 5 | 0 | 19 | 7 | 0 |
| Downstream signaling events of B Cell Receptor (BCR) | 7 | 148456174 | T | G | G | rs17537343 | intronic | 0.0430 | 0.0430 | CUL1 | 0.0164 | 61 | 5 | 0 | 18 | 7 | 0 |
| Downstream signaling events of B Cell Receptor (BCR) | 7 | 148458141 | C | T | T | rs73158226 | intronic | 0.0426 | 0.0426 | CUL1 | 0.0616 | 61 | 5 | 0 | 18 | 6 | 0 |
| Downstream signaling events of B Cell Receptor (BCR) | 7 | 148470413 | T | C | C | rs73158234 | intronic | 0.0423 | 0.0423 | CUL1 | 0.0333 | 61 | 5 | 0 | 19 | 7 | 0 |
| Downstream signaling events of B Cell Receptor (BCR) | 7 | 148502252 | T | C | C | rs77948593 | downstream | 0.0395 | 0.0395 | CUL1 | 1.0000 | 59 | 6 | 1 | 23 | 2 | 0 |
| Downstream signaling events of B Cell Receptor (BCR) | 8 | 42125339 | T | C | C | rs76361874 | upstream | 0.0303 | 0.0303 | IKBKB | 0.0653 | 59 | 7 | 0 | 25 | 0 | 1 |
| Downstream signaling events of B Cell Receptor (BCR) | 8 | 42127643 | A | C | C | rs17875739 | upstream | 0.0241 | 0.0241 | IKBKB | 0.0653 | 59 | 7 | 0 | 25 | 0 | 1 |
| Downstream signaling events of B Cell Receptor (BCR) | 8 | 42128072 | T | A | A | rs62507976 | upstream | 0.0296 | 0.0296 | IKBKB | 0.4157 | 61 | 5 | 0 | 23 | 2 | 1 |
| Downstream signaling events of B Cell Receptor (BCR) | 8 | 42135678 | T | G | G | rs17875740 | intronic | 0.0335 | 0.0335 | IKBKB | 0.0330 | 58 | 8 | 0 | 25 | 0 | 1 |
| Downstream signaling events of B Cell Receptor (BCR) | 8 | 42138781 | G | C | C | rs80313154 | intronic | 0.0274 | 0.0274 | IKBKB | 0.0330 | 58 | 8 | 0 | 25 | 0 | 1 |
| Downstream signaling events of B Cell Receptor (BCR) | 8 | 42142316 | G | A | A | rs79881854 | intronic | 0.0326 | 0.0326 | IKBKB | 0.0321 | 58 | 8 | 0 | 24 | 0 | 1 |
| Downstream signaling events of B Cell Receptor (BCR) | 8 | 42145446 | A | G | G | rs17875744 | intronic | 0.0327 | 0.0327 | IKBKB | 0.0272 | 57 | 9 | 0 | 25 | 0 | 1 |
| Downstream signaling events of B Cell Receptor (BCR) | 8 | 42150794 | C | T | T | rs79123247 | intronic | 0.0326 | 0.0326 | IKBKB | 0.0330 | 58 | 8 | 0 | 25 | 0 | 1 |
| Downstream signaling events of B Cell Receptor (BCR) | 8 | 42154866 | A | G | G | rs78342373 | intronic | 0.0326 | 0.0326 | IKBKB | 0.0330 | 58 | 8 | 0 | 25 | 0 | 1 |
| Downstream signaling events of B Cell Receptor (BCR) | 8 | 42155029 | G | A | A | rs112141582 | intronic | 0.0267 | 0.0267 | IKBKB | 0.0321 | 58 | 8 | 0 | 24 | 0 | 1 |
| Downstream signaling events of B Cell Receptor (BCR) | 8 | 42158881 | A | G | G | rs75230171 | intronic | 0.0327 | 0.0327 | IKBKB | 0.0321 | 58 | 8 | 0 | 24 | 0 | 1 |
| Downstream signaling events of B Cell Receptor (BCR) | 8 | 42159022 | C | T | T | rs79460550 | intronic | 0.0326 | 0.0326 | IKBKB | 0.0330 | 58 | 8 | 0 | 25 | 0 | 1 |
| Downstream signaling events of B Cell Receptor (BCR) | 8 | 42161317 | G | A | A | rs113993388 | intronic | 0.0280 | 0.0280 | IKBKB | 0.0330 | 58 | 8 | 0 | 25 | 0 | 1 |
| Downstream signaling events of B Cell Receptor (BCR) | 8 | 42162675 | C | T | T | rs17875700 | intronic | 0.0338 | 0.0338 | IKBKB | 0.0321 | 58 | 8 | 0 | 24 | 0 | 1 |
| Downstream signaling events of B Cell Receptor (BCR) | 8 | 42170928 | C | T | T | rs76891399 | intronic | 0.0325 | 0.0325 | IKBKB | 0.0321 | 58 | 8 | 0 | 24 | 0 | 1 |
| Downstream signaling events of B Cell Receptor (BCR) | 8 | 42173677 | G | T | T | rs17875746 | intronic | 0.0322 | 0.0322 | IKBKB | 0.0330 | 58 | 8 | 0 | 25 | 0 | 1 |
| Downstream signaling events of B Cell Receptor (BCR) | 8 | 42180084 | G | A | A | rs17875721 | intronic | 0.0326 | 0.0326 | IKBKB | 0.0330 | 58 | 8 | 0 | 25 | 0 | 1 |
| Downstream signaling events of B Cell Receptor (BCR) | 8 | 42180328 | G | T | T | rs17875751 | intronic | 0.0327 | 0.0327 | IKBKB | 0.0330 | 58 | 8 | 0 | 25 | 0 | 1 |
| Downstream signaling events of B Cell Receptor (BCR) | 8 | 42180679 | G | A | A | rs78599265 | intronic | 0.0333 | 0.0333 | IKBKB | 0.0330 | 58 | 8 | 0 | 25 | 0 | 1 |
| Downstream signaling events of B Cell Receptor (BCR) | 8 | 42186989 | C | T | T | rs17875731 | intronic | 0.0324 | 0.0324 | IKBKB | 0.0330 | 58 | 8 | 0 | 25 | 0 | 1 |
| Downstream signaling events of B Cell Receptor (BCR) | 8 | 42188334 | G | T | T | rs79200457 | intronic | 0.0325 | 0.0325 | IKBKB | 0.0666 | 58 | 7 | 0 | 25 | 0 | 1 |
| Downstream signaling events of B Cell Receptor (BCR) | 9 | 123578733 | G | A | A | rs10384 | 3_prime_UTR | 0.0186 | 0.0186 | PSMD5 | 0.3096 | 61 | 5 | 0 | 23 | 1 | 1 |
| Downstream signaling events of B Cell Receptor (BCR) | 9 | 123598623 | A | G | G | rs62581704 | intronic | 0.0476 | 0.0476 | PSMD5 | 0.3820 | 62 | 4 | 0 | 24 | 1 | 1 |
| Downstream signaling events of B Cell Receptor (BCR) | 9 | 127118932 | G | A | A | rs117300833 | intronic | 0.0270 | 0.0270 | PSMB7 | 0.3024 | 58 | 7 | 0 | 20 | 5 | 0 |
| Downstream signaling events of B Cell Receptor (BCR) | 9 | 127119442 | G | A | A | rs16927388 | intronic | 0.0435 | 0.0435 | PSMB7 | 0.4984 | 59 | 7 | 0 | 22 | 4 | 0 |
| Downstream signaling events of B Cell Receptor (BCR) | 9 | 127119799 | G | A | A | rs56058032 | intronic | 0.0172 | 0.0172 | PSMB7 | 0.4984 | 59 | 7 | 0 | 22 | 4 | 0 |
| Downstream signaling events of B Cell Receptor (BCR) | 9 | 127141846 | G | C | C | rs79069085 | intronic | 0.0178 | 0.0178 | PSMB7 | 0.4984 | 59 | 7 | 0 | 22 | 4 | 0 |
| Downstream signaling events of B Cell Receptor (BCR) | 9 | 127143001 | T | A | A | rs76699962 | intronic | 0.0448 | 0.0448 | PSMB7 | 0.4984 | 59 | 7 | 0 | 22 | 4 | 0 |
| Downstream signaling events of B Cell Receptor (BCR) | 9 | 127144906 | A | C | C | rs79321634 | intronic | 0.0239 | 0.0239 | PSMB7 | 0.2833 | 59 | 6 | 0 | 21 | 5 | 0 |
| Downstream signaling events of B Cell Receptor (BCR) | 9 | 127145687 | T | C | C | rs77214542 | intronic | 0.0335 | 0.0335 | PSMB7 | 1.0000 | 58 | 8 | 0 | 23 | 3 | 0 |
| Downstream signaling events of B Cell Receptor (BCR) | 9 | 127145878 | G | C | C | rs41274376 | intronic | 0.0237 | 0.0237 | PSMB7 | 0.3091 | 59 | 7 | 0 | 21 | 5 | 0 |
| Downstream signaling events of B Cell Receptor (BCR) | 9 | 127151807 | T | G | G | rs114451602 | intronic | 0.0301 | 0.0301 | PSMB7 | 0.5095 | 60 | 5 | 1 | 26 | 0 | 0 |
| Downstream signaling events of B Cell Receptor (BCR) | 9 | 127172515 | T | C | C | rs140504275 | intronic | 0.0204 | 0.0204 | PSMB7 | 0.1666 | 60 | 6 | 0 | 20 | 5 | 0 |
| Downstream signaling events of B Cell Receptor (BCR) | 9 | 127177053 | C | T | T | rs73588260 | intronic | 0.0344 | 0.0344 | PSMB7 | 0.7366 | 57 | 8 | 0 | 22 | 4 | 0 |
| Downstream signaling events of B Cell Receptor (BCR) | 10 | 75203331 | A | T | T | rs12775630 | intronic | 0.0464 | 0.0464 | PPP3CB | 0.8528 | 56 | 8 | 2 | 23 | 2 | 0 |
| Downstream signaling events of B Cell Receptor (BCR) | 10 | 75243476 | G | A | A | rs72814306 | intronic | 0.0163 | 0.0163 | PPP3CB | 0.2725 | 63 | 2 | 1 | 23 | 3 | 0 |
| Downstream signaling events of B Cell Receptor (BCR) | 10 | 101958770 | C | T | T | rs12764732 | intronic | 0.0464 | 0.0464 | CHUK | 0.7209 | 57 | 9 | 0 | 23 | 2 | 0 |
| Downstream signaling events of B Cell Receptor (BCR) | 10 | 101958858 | T | G | G | rs17883365 | intronic | 0.0229 | 0.0229 | CHUK | 1.0000 | 61 | 5 | 0 | 24 | 2 | 0 |
| Downstream signaling events of B Cell Receptor (BCR) | 10 | 101961171 | G | A | A | rs17885986 | intronic | 0.0496 | 0.0496 | CHUK | 0.6322 | 57 | 8 | 1 | 21 | 4 | 1 |
| Downstream signaling events of B Cell Receptor (BCR) | 10 | 101979482 | T | C | C | rs12764370 | intronic | 0.0302 | 0.0302 | CHUK | 0.7209 | 57 | 9 | 0 | 23 | 2 | 0 |
| Downstream signaling events of B Cell Receptor (BCR) | 10 | 103121589 | C | T | T | rs11190960 | intronic | 0.0325 | 0.0325 | BTRC | 0.2234 | 59 | 6 | 1 | 26 | 0 | 0 |
| Downstream signaling events of B Cell Receptor (BCR) | 10 | 103165858 | G | A | A | rs145847638 | intronic | 0.0278 | 0.0278 | BTRC | 0.4058 | 59 | 6 | 1 | 25 | 0 | 0 |
| Downstream signaling events of B Cell Receptor (BCR) | 10 | 103180336 | G | A | A | rs11191003 | intronic | 0.0274 | 0.0274 | BTRC | 0.2234 | 59 | 6 | 1 | 26 | 0 | 0 |
| Downstream signaling events of B Cell Receptor (BCR) | 10 | 103231195 | T | C | C | rs34711120 | intronic | 0.0435 | 0.0435 | BTRC | 0.7302 | 58 | 8 | 0 | 21 | 4 | 0 |
| Downstream signaling events of B Cell Receptor (BCR) | 10 | 103272221 | C | T | T | rs12774622 | intronic | 0.0257 | 0.0257 | BTRC | 0.7970 | 57 | 8 | 1 | 23 | 2 | 0 |
| Downstream signaling events of B Cell Receptor (BCR) | 10 | 103298099 | G | T | T | rs4151060 | missense | 0.0289 | 0.0289 | BTRC | 0.6362 | 55 | 10 | 1 | 23 | 2 | 0 |
| Downstream signaling events of B Cell Receptor (BCR) | 11 | 242014 | T | A | A | rs17727753 | intronic | 0.0189 | 0.0189 | PSMD13 | 1.0000 | 60 | 5 | 1 | 24 | 1 | 0 |
| Downstream signaling events of B Cell Receptor (BCR) | 11 | 249105 | G | A | A | rs11601352 | intronic | 0.0465 | 0.0465 | PSMD13 | 0.7522 | 56 | 10 | 0 | 23 | 3 | 0 |
| Downstream signaling events of B Cell Receptor (BCR) | 11 | 527311 | T | G | G | rs117011293 | downstream | 0.0497 | 0.0497 | HRAS | 0.4157 | 61 | 5 | 0 | 23 | 2 | 1 |
| Downstream signaling events of B Cell Receptor (BCR) | 11 | 14528592 | T | A | A | rs11023241 | intronic | 0.0417 | 0.0417 | PSMA1 | 0.6117 | 57 | 8 | 1 | 23 | 2 | 1 |
| Downstream signaling events of B Cell Receptor (BCR) | 11 | 14531031 | C | T | T | rs78398913 | intronic | 0.0273 | 0.0273 | PSMA1 | 1.0000 | 56 | 9 | 0 | 22 | 4 | 0 |
| Downstream signaling events of B Cell Receptor (BCR) | 11 | 14537004 | A | G | G | rs74589503 | intronic | 0.0418 | 0.0418 | PSMA1 | 0.4981 | 56 | 10 | 0 | 24 | 2 | 0 |
| Downstream signaling events of B Cell Receptor (BCR) | 11 | 14540827 | C | T | T | rs79966935 | intronic | 0.0233 | 0.0233 | PSMA1 | 0.7222 | 57 | 9 | 0 | 24 | 2 | 0 |
| Downstream signaling events of B Cell Receptor (BCR) | 11 | 14541179 | A | T | T | rs61883612 | intronic | 0.0441 | 0.0441 | PSMA1 | 0.7666 | 55 | 11 | 0 | 21 | 5 | 0 |
| Downstream signaling events of B Cell Receptor (BCR) | 11 | 14556220 | C | T | T | rs34162548 | intronic | 0.0376 | 0.0376 | PSMA1 | 1.0000 | 55 | 11 | 0 | 21 | 4 | 0 |
| Downstream signaling events of B Cell Receptor (BCR) | 11 | 14588324 | G | A | A | rs78854818 | intronic | 0.0413 | 0.0413 | PSMA1 | 1.0000 | 59 | 6 | 1 | 24 | 2 | 0 |
| Downstream signaling events of B Cell Receptor (BCR) | 11 | 14602698 | T | G | G | rs17567703 | intronic | 0.0418 | 0.0418 | PSMA1 | 1.0000 | 58 | 8 | 0 | 23 | 3 | 0 |
| Downstream signaling events of B Cell Receptor (BCR) | 11 | 14622982 | G | A | A | rs16930367 | intronic | 0.0250 | 0.0250 | PSMA1 | 0.7192 | 58 | 8 | 0 | 24 | 2 | 0 |
| Downstream signaling events of B Cell Receptor (BCR) | 11 | 14627135 | G | A | A | rs55760529 | intronic | 0.0485 | 0.0485 | PSMA1 | 1.0000 | 56 | 9 | 1 | 23 | 3 | 0 |
| Downstream signaling events of B Cell Receptor (BCR) | 11 | 14633490 | G | A | A | rs11023274 | intronic | 0.0317 | 0.0317 | PSMA1 | 1.0000 | 56 | 9 | 1 | 23 | 3 | 0 |
| Downstream signaling events of B Cell Receptor (BCR) | 11 | 14648387 | G | A | A | rs79123458 | intronic | 0.0412 | 0.0412 | PSMA1 | 1.0000 | 59 | 6 | 1 | 24 | 2 | 0 |
| Downstream signaling events of B Cell Receptor (BCR) | 11 | 47441683 | C | T | T | rs72903900 | intronic | 0.0375 | 0.0375 | PSMC3 | 0.8389 | 59 | 5 | 2 | 23 | 3 | 0 |
| Downstream signaling events of B Cell Receptor (BCR) | 11 | 47449591 | G | A | A | rs116930066 | upstream | 0.0311 | 0.0311 | PSMC3 | 0.5028 | 56 | 9 | 0 | 24 | 2 | 0 |
| Downstream signaling events of B Cell Receptor (BCR) | 12 | 25374046 | T | C | C | rs117991169 | intronic | 0.0070 | 0.0070 | KRAS | 1.0000 | 63 | 2 | 1 | 26 | 0 | 0 |
| Downstream signaling events of B Cell Receptor (BCR) | 12 | 25384938 | C | T | T | rs187995827 | intronic | 0.0218 | 0.0218 | KRAS | 0.6900 | 61 | 4 | 1 | 25 | 0 | 0 |
| Downstream signaling events of B Cell Receptor (BCR) | 12 | 25389886 | T | C | C | rs61761102 | intronic | 0.0354 | 0.0354 | KRAS | 1.0000 | 57 | 7 | 2 | 23 | 3 | 0 |
| Downstream signaling events of B Cell Receptor (BCR) | 12 | 25393865 | A | C | C | rs17388893 | intronic | 0.0338 | 0.0338 | KRAS | 1.0000 | 57 | 7 | 2 | 22 | 3 | 0 |
| Downstream signaling events of B Cell Receptor (BCR) | 12 | 25395953 | T | C | C | rs61761078 | intronic | 0.0309 | 0.0309 | KRAS | 0.5904 | 57 | 8 | 1 | 24 | 1 | 0 |
| Downstream signaling events of B Cell Receptor (BCR) | 12 | 25397593 | A | C | C | rs61761074 | intronic | 0.0336 | 0.0336 | KRAS | 0.3096 | 61 | 5 | 0 | 23 | 1 | 1 |
| Downstream signaling events of B Cell Receptor (BCR) | 12 | 25398972 | C | T | T | rs17329975 | intronic | 0.0340 | 0.0340 | KRAS | 1.0000 | 57 | 7 | 2 | 22 | 3 | 0 |
| Downstream signaling events of B Cell Receptor (BCR) | 12 | 122349404 | T | G | G | rs113810917 | intronic | 0.0229 | 0.0229 | PSMD9 | 0.1282 | 57 | 8 | 1 | 25 | 0 | 0 |
| Downstream signaling events of B Cell Receptor (BCR) | 12 | 122351293 | A | C | C | rs73229956 | intronic | 0.0288 | 0.0288 | PSMD9 | 0.2597 | 54 | 11 | 0 | 24 | 1 | 0 |
| Downstream signaling events of B Cell Receptor (BCR) | 12 | 125395161 | G | C | C | rs113660988 | downstream | 0.0428 | 0.0428 | UBC | 0.0064 | 64 | 1 | 1 | 21 | 5 | 0 |
| Downstream signaling events of B Cell Receptor (BCR) | 12 | 125395728 | C | T | T | rs112205208 | downstream | 0.0429 | 0.0429 | UBC | 0.0064 | 64 | 1 | 1 | 21 | 5 | 0 |
| Downstream signaling events of B Cell Receptor (BCR) | 12 | 125398911 | C | T | T | rs112043091 | 5_prime_UTR | 0.0430 | 0.0430 | UBC | 0.0064 | 64 | 1 | 1 | 21 | 5 | 0 |
| Downstream signaling events of B Cell Receptor (BCR) | 12 | 125399133 | C | T | T | rs41276688 | 5_prime_UTR | 0.0434 | 0.0434 | UBC | 0.0064 | 64 | 1 | 1 | 21 | 5 | 0 |
| Downstream signaling events of B Cell Receptor (BCR) | 14 | 23512430 | C | T | T | rs78162644 | 3_prime_UTR | 0.0459 | 0.0459 | PSMB11 | 1.0000 | 57 | 9 | 0 | 22 | 4 | 0 |
| Downstream signaling events of B Cell Receptor (BCR) | 14 | 35787993 | A | T | T | rs12890150 | downstream | 0.0324 | 0.0324 | PSMA6 | 0.5321 | 62 | 3 | 1 | 23 | 3 | 0 |
| Downstream signaling events of B Cell Receptor (BCR) | 14 | 53178728 | T | C | C | rs117516552 | intronic | 0.0276 | 0.0276 | PSMC6 | 0.7209 | 57 | 9 | 0 | 23 | 2 | 0 |
| Downstream signaling events of B Cell Receptor (BCR) | 14 | 58710711 | C | G | G | rs117328560 | upstream | 0.0280 | 0.0280 | PSMA3 | 0.4999 | 56 | 10 | 0 | 23 | 2 | 0 |
| Downstream signaling events of B Cell Receptor (BCR) | 14 | 58729996 | C | G | G | rs111364917 | intronic | 0.0397 | 0.0397 | PSMA3 | 0.4984 | 59 | 7 | 0 | 22 | 4 | 0 |
| Downstream signaling events of B Cell Receptor (BCR) | 14 | 58731398 | T | G | G | rs117955444 | intronic | 0.0210 | 0.0210 | PSMA3 | 1.0000 | 59 | 7 | 0 | 22 | 3 | 0 |
| Downstream signaling events of B Cell Receptor (BCR) | 14 | 90868665 | A | G | G | rs34825576 | intronic | 0.0382 | 0.0382 | CALM1 | 0.5425 | 53 | 13 | 0 | 23 | 3 | 0 |
| Downstream signaling events of B Cell Receptor (BCR) | 15 | 38801129 | C | T | T | rs12901652 | intronic | 0.0478 | 0.0478 | RASGRP1 | 0.0703 | 56 | 9 | 1 | 26 | 0 | 0 |
| Downstream signaling events of B Cell Receptor (BCR) | 15 | 38803689 | G | A | A | rs28710819 | intronic | 0.0461 | 0.0461 | RASGRP1 | 0.7703 | 53 | 13 | 0 | 22 | 4 | 0 |
| Downstream signaling events of B Cell Receptor (BCR) | 15 | 38805185 | A | C | C | rs55888286 | intronic | 0.0498 | 0.0498 | RASGRP1 | 0.7632 | 60 | 5 | 1 | 25 | 1 | 0 |
| Downstream signaling events of B Cell Receptor (BCR) | 15 | 38816519 | G | A | A | rs35925499 | intronic | 0.0463 | 0.0463 | RASGRP1 | 0.7302 | 58 | 8 | 0 | 21 | 4 | 0 |
| Downstream signaling events of B Cell Receptor (BCR) | 15 | 78834476 | A | G | G | rs41280046 | intronic | 0.0393 | 0.0393 | PSMA4 | 0.7087 | 60 | 6 | 0 | 23 | 3 | 0 |
| Downstream signaling events of B Cell Receptor (BCR) | 16 | 23845860 | G | T | T | rs72777910 | upstream | 0.0300 | 0.0300 | PRKCB | 0.4762 | 60 | 5 | 1 | 22 | 4 | 0 |
| Downstream signaling events of B Cell Receptor (BCR) | 16 | 23849482 | T | C | T | rs2023670 | intronic | 0.9513 | 0.0487 | PRKCB | 0.0099 | 0 | 11 | 55 | 1 | 0 | 25 |
| Downstream signaling events of B Cell Receptor (BCR) | 16 | 23850240 | A | G | A | rs11074581 | intronic | 0.9663 | 0.0337 | PRKCB | 0.0099 | 0 | 11 | 55 | 1 | 0 | 25 |
| Downstream signaling events of B Cell Receptor (BCR) | 16 | 23851956 | T | C | T | rs7189210 | intronic | 0.9663 | 0.0337 | PRKCB | 0.0099 | 0 | 11 | 55 | 1 | 0 | 25 |
| Downstream signaling events of B Cell Receptor (BCR) | 16 | 23852415 | A | T | A | rs2188359 | intronic | 0.9528 | 0.0472 | PRKCB | 0.0099 | 0 | 11 | 55 | 1 | 0 | 25 |
| Downstream signaling events of B Cell Receptor (BCR) | 16 | 23859391 | A | G | G | rs62030647 | intronic | 0.0226 | 0.0226 | PRKCB | 1.0000 | 59 | 7 | 0 | 23 | 3 | 0 |
| Downstream signaling events of B Cell Receptor (BCR) | 16 | 23874933 | A | C | A | rs6497691 | intronic | 0.9663 | 0.0337 | PRKCB | 0.0063 | 0 | 14 | 52 | 1 | 0 | 24 |
| Downstream signaling events of B Cell Receptor (BCR) | 16 | 23876099 | C | T | T | rs79131874 | intronic | 0.0303 | 0.0303 | PRKCB | 1.0000 | 59 | 7 | 0 | 23 | 3 | 0 |
| Downstream signaling events of B Cell Receptor (BCR) | 16 | 23877500 | A | G | A | rs8059885 | intronic | 0.9663 | 0.0337 | PRKCB | 0.0046 | 0 | 14 | 52 | 1 | 0 | 25 |
| Downstream signaling events of B Cell Receptor (BCR) | 16 | 23877606 | A | G | A | rs8060048 | intronic | 0.9644 | 0.0356 | PRKCB | 0.0046 | 0 | 14 | 52 | 1 | 0 | 25 |
| Downstream signaling events of B Cell Receptor (BCR) | 16 | 23877781 | G | A | G | rs8060718 | intronic | 0.9664 | 0.0336 | PRKCB | 0.0046 | 0 | 14 | 52 | 1 | 0 | 25 |
| Downstream signaling events of B Cell Receptor (BCR) | 16 | 23878470 | C | T | C | rs12935004 | intronic | 0.9657 | 0.0343 | PRKCB | 0.0046 | 0 | 14 | 52 | 1 | 0 | 25 |
| Downstream signaling events of B Cell Receptor (BCR) | 16 | 23880851 | C | T | C | rs8061523 | intronic | 0.9664 | 0.0336 | PRKCB | 0.0046 | 0 | 14 | 52 | 1 | 0 | 25 |
| Downstream signaling events of B Cell Receptor (BCR) | 16 | 23881930 | G | A | G | rs8047121 | intronic | 0.9662 | 0.0338 | PRKCB | 0.0046 | 0 | 14 | 52 | 1 | 0 | 25 |
| Downstream signaling events of B Cell Receptor (BCR) | 16 | 23882469 | T | C | T | rs1468129 | intronic | 0.9663 | 0.0337 | PRKCB | 0.0046 | 0 | 14 | 52 | 1 | 0 | 25 |
| Downstream signaling events of B Cell Receptor (BCR) | 16 | 23885608 | A | T | A | rs8044732 | intronic | 0.9664 | 0.0336 | PRKCB | 0.0046 | 0 | 14 | 52 | 1 | 0 | 25 |
| Downstream signaling events of B Cell Receptor (BCR) | 16 | 23885751 | A | G | G | rs62031692 | intronic | 0.0253 | 0.0253 | PRKCB | 1.0000 | 59 | 7 | 0 | 23 | 3 | 0 |
| Downstream signaling events of B Cell Receptor (BCR) | 16 | 23887574 | G | T | T | rs79034087 | intronic | 0.0290 | 0.0290 | PRKCB | 0.3487 | 61 | 4 | 1 | 22 | 4 | 0 |
| Downstream signaling events of B Cell Receptor (BCR) | 16 | 23888354 | C | T | C | rs7404417 | intronic | 0.9664 | 0.0336 | PRKCB | 0.0046 | 0 | 14 | 52 | 1 | 0 | 25 |
| Downstream signaling events of B Cell Receptor (BCR) | 16 | 23889896 | T | C | T | rs8063823 | intronic | 0.9665 | 0.0335 | PRKCB | 0.0046 | 0 | 14 | 52 | 1 | 0 | 25 |
| Downstream signaling events of B Cell Receptor (BCR) | 16 | 23893893 | G | A | G | rs11647359 | intronic | 0.9664 | 0.0336 | PRKCB | 0.0046 | 0 | 14 | 52 | 1 | 0 | 25 |
| Downstream signaling events of B Cell Receptor (BCR) | 16 | 23895034 | A | G | A | rs6497695 | intronic | 0.9665 | 0.0335 | PRKCB | 0.0046 | 0 | 14 | 52 | 1 | 0 | 25 |
| Downstream signaling events of B Cell Receptor (BCR) | 16 | 23895443 | A | G | G | rs62028075 | intronic | 0.0253 | 0.0253 | PRKCB | 1.0000 | 59 | 7 | 0 | 23 | 3 | 0 |
| Downstream signaling events of B Cell Receptor (BCR) | 16 | 23895884 | T | C | T | rs9944348 | intronic | 0.9665 | 0.0335 | PRKCB | 0.0046 | 0 | 14 | 52 | 1 | 0 | 25 |
| Downstream signaling events of B Cell Receptor (BCR) | 16 | 23896089 | T | C | C | rs74572166 | intronic | 0.0245 | 0.0245 | PRKCB | 1.0000 | 59 | 7 | 0 | 23 | 3 | 0 |
| Downstream signaling events of B Cell Receptor (BCR) | 16 | 23896209 | C | A | C | rs9302418 | intronic | 0.9664 | 0.0336 | PRKCB | 0.0046 | 0 | 14 | 52 | 1 | 0 | 25 |
| Downstream signaling events of B Cell Receptor (BCR) | 16 | 23896438 | G | T | T | rs62028076 | intronic | 0.0252 | 0.0252 | PRKCB | 1.0000 | 59 | 7 | 0 | 22 | 3 | 0 |
| Downstream signaling events of B Cell Receptor (BCR) | 16 | 23898605 | A | T | A | rs933290 | intronic | 0.9632 | 0.0368 | PRKCB | 0.0016 | 0 | 17 | 49 | 1 | 0 | 24 |
| Downstream signaling events of B Cell Receptor (BCR) | 16 | 23899211 | A | T | A | rs12926245 | intronic | 0.9632 | 0.0368 | PRKCB | 0.0012 | 0 | 17 | 49 | 1 | 0 | 25 |
| Downstream signaling events of B Cell Receptor (BCR) | 16 | 23899610 | G | A | A | rs17753246 | intronic | 0.0252 | 0.0252 | PRKCB | 1.0000 | 59 | 7 | 0 | 23 | 3 | 0 |
| Downstream signaling events of B Cell Receptor (BCR) | 16 | 23899951 | G | A | A | rs62028077 | intronic | 0.0254 | 0.0254 | PRKCB | 1.0000 | 59 | 7 | 0 | 23 | 3 | 0 |
| Downstream signaling events of B Cell Receptor (BCR) | 16 | 23900716 | T | C | C | rs62028078 | intronic | 0.0252 | 0.0252 | PRKCB | 1.0000 | 59 | 7 | 0 | 23 | 3 | 0 |
| Downstream signaling events of B Cell Receptor (BCR) | 16 | 23901896 | C | T | C | rs6497696 | intronic | 0.9632 | 0.0368 | PRKCB | 0.0015 | 0 | 16 | 50 | 1 | 0 | 25 |
| Downstream signaling events of B Cell Receptor (BCR) | 16 | 23901948 | A | C | A | rs6497697 | intronic | 0.9630 | 0.0370 | PRKCB | 0.0015 | 0 | 16 | 49 | 1 | 0 | 25 |
| Downstream signaling events of B Cell Receptor (BCR) | 16 | 23904058 | A | G | A | rs886115 | intronic | 0.9632 | 0.0368 | PRKCB | 0.0015 | 0 | 16 | 50 | 1 | 0 | 25 |
| Downstream signaling events of B Cell Receptor (BCR) | 16 | 23904781 | G | A | A | rs17753509 | intronic | 0.0253 | 0.0253 | PRKCB | 1.0000 | 59 | 7 | 0 | 23 | 3 | 0 |
| Downstream signaling events of B Cell Receptor (BCR) | 16 | 23905676 | C | T | C | rs7200610 | intronic | 0.9631 | 0.0369 | PRKCB | 0.0015 | 0 | 16 | 50 | 1 | 0 | 25 |
| Downstream signaling events of B Cell Receptor (BCR) | 16 | 23907177 | A | C | C | rs17810011 | intronic | 0.0251 | 0.0251 | PRKCB | 1.0000 | 59 | 7 | 0 | 23 | 3 | 0 |
| Downstream signaling events of B Cell Receptor (BCR) | 16 | 23907765 | C | T | C | rs9925890 | intronic | 0.9632 | 0.0368 | PRKCB | 0.0024 | 0 | 16 | 50 | 1 | 0 | 24 |
| Downstream signaling events of B Cell Receptor (BCR) | 16 | 23912174 | A | G | A | rs12448249 | intronic | 0.9519 | 0.0481 | PRKCB | 0.0015 | 0 | 16 | 50 | 1 | 0 | 25 |
| Downstream signaling events of B Cell Receptor (BCR) | 16 | 23914915 | C | A | C | rs1004186 | intronic | 0.9632 | 0.0368 | PRKCB | 0.0015 | 0 | 16 | 50 | 1 | 0 | 25 |
| Downstream signaling events of B Cell Receptor (BCR) | 16 | 23916258 | G | A | G | rs1004187 | intronic | 0.9632 | 0.0368 | PRKCB | 0.0015 | 0 | 16 | 50 | 1 | 0 | 25 |
| Downstream signaling events of B Cell Receptor (BCR) | 16 | 23916521 | G | C | G | rs1008654 | intronic | 0.9633 | 0.0367 | PRKCB | 0.0015 | 0 | 16 | 50 | 1 | 0 | 25 |
| Downstream signaling events of B Cell Receptor (BCR) | 16 | 23917335 | G | A | G | rs6497699 | intronic | 0.9645 | 0.0355 | PRKCB | 0.0024 | 0 | 16 | 50 | 1 | 0 | 24 |
| Downstream signaling events of B Cell Receptor (BCR) | 16 | 23917465 | C | G | C | rs7186538 | intronic | 0.9645 | 0.0355 | PRKCB | 0.0024 | 0 | 16 | 50 | 1 | 0 | 24 |
| Downstream signaling events of B Cell Receptor (BCR) | 16 | 23917700 | C | A | C | rs7187091 | intronic | 0.9646 | 0.0354 | PRKCB | 0.0024 | 0 | 16 | 50 | 1 | 0 | 24 |
| Downstream signaling events of B Cell Receptor (BCR) | 16 | 23919088 | C | T | T | rs78322646 | intronic | 0.0278 | 0.0278 | PRKCB | 0.7348 | 58 | 8 | 0 | 22 | 4 | 0 |
| Downstream signaling events of B Cell Receptor (BCR) | 16 | 23921083 | C | T | C | rs6497702 | intronic | 0.9647 | 0.0353 | PRKCB | 0.0024 | 0 | 16 | 50 | 1 | 0 | 24 |
| Downstream signaling events of B Cell Receptor (BCR) | 16 | 23925936 | C | G | C | rs11074588 | intronic | 0.9649 | 0.0351 | PRKCB | 0.0015 | 0 | 16 | 50 | 1 | 0 | 25 |
| Downstream signaling events of B Cell Receptor (BCR) | 16 | 23939212 | G | A | G | rs11074590 | intronic | 0.9650 | 0.0350 | PRKCB | 0.0024 | 0 | 16 | 50 | 1 | 0 | 24 |
| Downstream signaling events of B Cell Receptor (BCR) | 16 | 23941628 | C | A | C | rs2005671 | intronic | 0.9647 | 0.0353 | PRKCB | 0.0015 | 0 | 16 | 49 | 1 | 0 | 25 |
| Downstream signaling events of B Cell Receptor (BCR) | 16 | 23943749 | T | C | T | rs9302420 | intronic | 0.9649 | 0.0351 | PRKCB | 0.0024 | 0 | 16 | 50 | 1 | 0 | 24 |
| Downstream signaling events of B Cell Receptor (BCR) | 16 | 23945985 | T | G | T | rs195989 | intronic | 0.9651 | 0.0349 | PRKCB | 0.0015 | 0 | 16 | 50 | 1 | 0 | 25 |
| Downstream signaling events of B Cell Receptor (BCR) | 16 | 23946157 | G | A | A | rs76973283 | intronic | 0.0302 | 0.0302 | PRKCB | 0.5065 | 58 | 8 | 0 | 21 | 5 | 0 |

| Pathway | Chr | Position | Ref | Alt | Minor | rsID | Region | Freq1 | Freq2 | Gene | P | c1 | c2 | c3 | c4 | c5 | c6 |
|---|---|---|---|---|---|---|---|---|---|---|---|---|---|---|---|---|---|
| Downstream signaling events of B Cell Receptor (BCR) | 16 | 23949175 | G | C | G | rs2560403 | intronic | 0.9657 | 0.0343 | PRKCB | 0.0015 | 0 | 16 | 50 | 1 | 0 | 25 |
| Downstream signaling events of B Cell Receptor (BCR) | 16 | 23949438 | A | G | A | rs195985 | intronic | 0.9658 | 0.0342 | PRKCB | 0.0015 | 0 | 16 | 50 | 1 | 0 | 25 |
| Downstream signaling events of B Cell Receptor (BCR) | 16 | 23953265 | T | C | T | rs2560404 | intronic | 0.9656 | 0.0344 | PRKCB | 0.0024 | 0 | 16 | 50 | 1 | 0 | 24 |
| Downstream signaling events of B Cell Receptor (BCR) | 16 | 23954128 | T | C | C | rs17810486 | intronic | 0.0308 | 0.0308 | PRKCB | 0.3024 | 58 | 7 | 0 | 20 | 5 | 0 |
| Downstream signaling events of B Cell Receptor (BCR) | 16 | 23954253 | G | A | G | rs195994 | intronic | 0.9653 | 0.0347 | PRKCB | 0.0015 | 0 | 16 | 50 | 1 | 0 | 25 |
| Downstream signaling events of B Cell Receptor (BCR) | 16 | 23962258 | G | C | G | rs196000 | intronic | 0.9659 | 0.0341 | PRKCB | 0.0015 | 0 | 16 | 50 | 1 | 0 | 25 |
| Downstream signaling events of B Cell Receptor (BCR) | 16 | 23964858 | T | A | T | rs196003 | intronic | 0.9647 | 0.0353 | PRKCB | 0.0343 | 0 | 15 | 51 | 0 | 1 | 25 |
| Downstream signaling events of B Cell Receptor (BCR) | 16 | 23985814 | C | T | T | rs72779914 | intronic | 0.0487 | 0.0487 | PRKCB | 1.0000 | 53 | 12 | 0 | 21 | 4 | 0 |
| Downstream signaling events of B Cell Receptor (BCR) | 16 | 23987552 | A | G | A | rs169030 | intronic | 0.9709 | 0.0291 | PRKCB | 0.1679 | 0 | 11 | 55 | 0 | 1 | 25 |
| Downstream signaling events of B Cell Receptor (BCR) | 16 | 23988755 | T | C | T | rs196013 | intronic | 0.9681 | 0.0319 | PRKCB | 0.1681 | 0 | 11 | 55 | 0 | 1 | 24 |
| Downstream signaling events of B Cell Receptor (BCR) | 16 | 24009919 | A | G | G | rs75622923 | intronic | 0.0319 | 0.0319 | PRKCB | 0.3348 | 54 | 12 | 0 | 24 | 2 | 0 |
| Downstream signaling events of B Cell Receptor (BCR) | 16 | 24022944 | C | T | T | rs111746132 | intronic | 0.0229 | 0.0229 | PRKCB | 0.7522 | 56 | 10 | 0 | 23 | 3 | 0 |
| Downstream signaling events of B Cell Receptor (BCR) | 16 | 24066378 | G | A | A | rs113426570 | intronic | 0.0216 | 0.0216 | PRKCB | 1.0000 | 57 | 9 | 0 | 23 | 3 | 0 |
| Downstream signaling events of B Cell Receptor (BCR) | 16 | 24100759 | T | A | A | rs11643939 | intronic | 0.0294 | 0.0294 | PRKCB | 0.7522 | 56 | 10 | 0 | 23 | 3 | 0 |
| Downstream signaling events of B Cell Receptor (BCR) | 16 | 24105816 | G | A | A | rs56316329 | intronic | 0.0251 | 0.0251 | PRKCB | 0.6919 | 60 | 4 | 1 | 25 | 0 | 0 |
| Downstream signaling events of B Cell Receptor (BCR) | 16 | 24111853 | T | C | C | rs55959083 | intronic | 0.0431 | 0.0431 | PRKCB | 0.7213 | 58 | 8 | 0 | 23 | 2 | 0 |
| Downstream signaling events of B Cell Receptor (BCR) | 16 | 24112768 | G | A | A | rs117056307 | intronic | 0.0430 | 0.0430 | PRKCB | 0.7192 | 58 | 8 | 0 | 24 | 2 | 0 |
| Downstream signaling events of B Cell Receptor (BCR) | 16 | 24122052 | G | A | A | rs117467859 | intronic | 0.0433 | 0.0433 | PRKCB | 0.7213 | 58 | 8 | 0 | 23 | 2 | 0 |
| Downstream signaling events of B Cell Receptor (BCR) | 16 | 24122492 | C | T | T | rs72779977 | intronic | 0.0457 | 0.0457 | PRKCB | 0.7201 | 57 | 8 | 0 | 23 | 2 | 0 |
| Downstream signaling events of B Cell Receptor (BCR) | 16 | 24123560 | G | A | A | rs60261043 | intronic | 0.0457 | 0.0457 | PRKCB | 0.7201 | 57 | 8 | 0 | 23 | 2 | 0 |
| Downstream signaling events of B Cell Receptor (BCR) | 16 | 24132273 | G | A | A | rs62027458 | intronic | 0.0232 | 0.0232 | PRKCB | 0.1877 | 59 | 7 | 0 | 21 | 4 | 1 |
| Downstream signaling events of B Cell Receptor (BCR) | 16 | 24164042 | G | T | T | rs72779989 | intronic | 0.0487 | 0.0487 | PRKCB | 0.8583 | 57 | 7 | 2 | 22 | 4 | 0 |
| Downstream signaling events of B Cell Receptor (BCR) | 16 | 24197496 | A | T | T | rs79699525 | intronic | 0.0261 | 0.0261 | PRKCB | 0.4631 | 59 | 6 | 0 | 22 | 4 | 0 |
| Downstream signaling events of B Cell Receptor (BCR) | 16 | 24199852 | C | T | T | rs78424166 | intronic | 0.0359 | 0.0359 | PRKCB | 0.4603 | 60 | 6 | 0 | 22 | 4 | 0 |
| Downstream signaling events of B Cell Receptor (BCR) | 16 | 67973569 | C | T | T | rs17240392 | upstream | 0.0497 | 0.0497 | PSMB10 | 0.4984 | 59 | 7 | 0 | 22 | 4 | 0 |
| Downstream signaling events of B Cell Receptor (BCR) | 16 | 68137486 | G | T | T | rs78479045 | intronic | 0.0183 | 0.0183 | NFATC3 | 0.7734 | 60 | 5 | 1 | 23 | 3 | 0 |
| Downstream signaling events of B Cell Receptor (BCR) | 16 | 68165961 | G | C | C | rs75914405 | intronic | 0.0183 | 0.0183 | NFATC3 | 0.7734 | 60 | 5 | 1 | 23 | 3 | 0 |
| Downstream signaling events of B Cell Receptor (BCR) | 16 | 68170298 | C | T | T | rs117300105 | intronic | 0.0352 | 0.0352 | NFATC3 | 0.1078 | 62 | 4 | 0 | 20 | 5 | 0 |
| Downstream signaling events of B Cell Receptor (BCR) | 16 | 74334545 | G | T | T | rs149977784 | intronic | 0.0196 | 0.0196 | PSMD7 | 1.0000 | 62 | 2 | 2 | 26 | 0 | 0 |
| Downstream signaling events of B Cell Receptor (BCR) | 17 | 4702206 | C | A | A | rs71368518 | downstream | 0.0416 | 0.0416 | PSMB6 | 1.0000 | 56 | 9 | 1 | 23 | 3 | 0 |
| Downstream signaling events of B Cell Receptor (BCR) | 17 | 30778483 | C | T | T | rs117117721 | intronic | 0.0190 | 0.0190 | PSMD11 | 1.0000 | 59 | 7 | 0 | 22 | 3 | 0 |
| Downstream signaling events of B Cell Receptor (BCR) | 17 | 30791889 | C | T | T | rs35225085 | intronic | 0.0427 | 0.0427 | PSMD11 | 0.4999 | 56 | 10 | 0 | 23 | 2 | 0 |
| Downstream signaling events of B Cell Receptor (BCR) | 17 | 30794889 | C | G | G | rs35177842 | intronic | 0.0430 | 0.0430 | PSMD11 | 0.4999 | 56 | 10 | 0 | 23 | 2 | 0 |
| Downstream signaling events of B Cell Receptor (BCR) | 17 | 36920050 | C | T | T | rs118080693 | intronic | 0.0415 | 0.0415 | PSMB3 | 0.1732 | 63 | 3 | 0 | 25 | 0 | 1 |
| Downstream signaling events of B Cell Receptor (BCR) | 17 | 38144079 | G | T | T | rs118009374 | intronic | 0.0189 | 0.0189 | PSMD3 | 0.8767 | 53 | 11 | 2 | 20 | 5 | 0 |
| Downstream signaling events of B Cell Receptor (BCR) | 17 | 38154396 | C | T | T | rs118034841 | downstream | 0.0176 | 0.0176 | PSMD3 | 0.0569 | 56 | 10 | 0 | 26 | 0 | 0 |
| Downstream signaling events of B Cell Receptor (BCR) | 17 | 65334270 | T | A | A | rs146515782 | 3_prime_UTR | 0.0418 | 0.0418 | PSMD12 | 1.0000 | 58 | 8 | 0 | 23 | 3 | 0 |
| Downstream signaling events of B Cell Receptor (BCR) | 18 | 23711373 | G | T | T | rs79820119 | upstream | 0.0407 | 0.0407 | PSMA8 | 0.2218 | 53 | 13 | 0 | 23 | 2 | 0 |
| Downstream signaling events of B Cell Receptor (BCR) | 18 | 23715815 | T | G | G | rs4800242 | intronic | 0.0460 | 0.0460 | PSMA8 | 0.7043 | 59 | 6 | 0 | 22 | 3 | 0 |
| Downstream signaling events of B Cell Receptor (BCR) | 18 | 23774138 | C | A | A | rs79452515 | downstream | 0.0424 | 0.0424 | PSMA8 | 0.2993 | 59 | 7 | 0 | 20 | 5 | 0 |
| Downstream signaling events of B Cell Receptor (BCR) | 18 | 56333866 | G | A | A | rs72958690 | upstream | 0.0368 | 0.0368 | MALT1 | 0.8183 | 56 | 9 | 1 | 21 | 4 | 0 |
| Downstream signaling events of B Cell Receptor (BCR) | 18 | 56338792 | G | A | A | rs56142402 | 5_prime_UTR | 0.0327 | 0.0327 | MALT1 | 0.8110 | 57 | 8 | 1 | 22 | 4 | 0 |
| Downstream signaling events of B Cell Receptor (BCR) | 18 | 56363534 | A | C | C | rs55825071 | intronic | 0.0377 | 0.0377 | MALT1 | 0.6531 | 56 | 9 | 1 | 20 | 5 | 0 |
| Downstream signaling events of B Cell Receptor (BCR) | 18 | 77161434 | C | T | T | rs73007652 | intronic | 0.0439 | 0.0439 | NFATC1 | 1.0000 | 54 | 12 | 0 | 22 | 4 | 0 |
| Downstream signaling events of B Cell Receptor (BCR) | 18 | 77181200 | C | T | T | rs4799054 | intronic | 0.0312 | 0.0312 | NFATC1 | 0.3820 | 62 | 4 | 0 | 24 | 1 | 1 |
| Downstream signaling events of B Cell Receptor (BCR) | 18 | 77193040 | G | A | A | rs142796661 | intronic | 0.0417 | 0.0417 | NFATC1 | 0.4442 | 53 | 12 | 0 | 21 | 4 | 1 |
| Downstream signaling events of B Cell Receptor (BCR) | 18 | 77197508 | G | A | A | rs117121680 | intronic | 0.0416 | 0.0416 | NFATC1 | 0.3172 | 52 | 13 | 0 | 21 | 4 | 1 |
| Downstream signaling events of B Cell Receptor (BCR) | 18 | 77199322 | C | T | T | rs75588847 | intronic | 0.0455 | 0.0455 | NFATC1 | 0.3091 | 59 | 7 | 0 | 21 | 5 | 0 |
| Downstream signaling events of B Cell Receptor (BCR) | 18 | 77199460 | G | A | A | rs62096887 | intronic | 0.0211 | 0.0211 | NFATC1 | 1.0000 | 60 | 6 | 0 | 24 | 2 | 0 |
| Downstream signaling events of B Cell Receptor (BCR) | 18 | 77211908 | G | A | A | rs55953864 | intronic | 0.0480 | 0.0480 | NFATC1 | 0.3091 | 59 | 7 | 0 | 21 | 5 | 0 |
| Downstream signaling events of B Cell Receptor (BCR) | 18 | 77212989 | G | T | T | rs73007700 | intronic | 0.0438 | 0.0438 | NFATC1 | 0.5065 | 58 | 8 | 0 | 21 | 5 | 0 |
| Downstream signaling events of B Cell Receptor (BCR) | 18 | 77239775 | G | A | A | rs117383241 | intronic | 0.0246 | 0.0246 | NFATC1 | 0.7113 | 59 | 6 | 0 | 23 | 3 | 0 |
| Downstream signaling events of B Cell Receptor (BCR) | 18 | 77241715 | G | A | A | rs117246954 | intronic | 0.0175 | 0.0175 | NFATC1 | 0.7222 | 57 | 9 | 0 | 24 | 2 | 0 |
| Downstream signaling events of B Cell Receptor (BCR) | 18 | 77256499 | A | G | G | rs62096906 | intronic | 0.0466 | 0.0466 | NFATC1 | 1.0000 | 53 | 12 | 0 | 22 | 4 | 0 |
| Downstream signaling events of B Cell Receptor (BCR) | 18 | 77275049 | A | G | G | rs113507668 | intronic | 0.0393 | 0.0393 | NFATC1 | 1.0000 | 58 | 7 | 1 | 23 | 3 | 0 |
| Downstream signaling events of B Cell Receptor (BCR) | 18 | 77278071 | G | A | A | rs116844899 | intronic | 0.0231 | 0.0231 | NFATC1 | 1.0000 | 59 | 7 | 0 | 23 | 3 | 0 |
| Downstream signaling events of B Cell Receptor (BCR) | 18 | 77279921 | G | A | A | rs79426764 | intronic | 0.0497 | 0.0497 | NFATC1 | 0.2606 | 51 | 15 | 0 | 23 | 3 | 0 |
| Downstream signaling events of B Cell Receptor (BCR) | 18 | 77279961 | A | G | G | rs76389821 | intronic | 0.0306 | 0.0306 | NFATC1 | 0.2710 | 56 | 9 | 0 | 25 | 1 | 0 |
| Downstream signaling events of B Cell Receptor (BCR) | 19 | 40472450 | A | C | C | rs147915270 | upstream | 0.0382 | 0.0382 | PSMC4 | 1.0000 | 58 | 7 | 1 | 22 | 3 | 0 |
| Downstream signaling events of B Cell Receptor (BCR) | 19 | 40475070 | T | G | G | rs139876278 | upstream | 0.0147 | 0.0147 | PSMC4 | 0.2725 | 63 | 2 | 1 | 23 | 3 | 0 |
| Downstream signaling events of B Cell Receptor (BCR) | 20 | 1128622 | T | C | C | rs74871431 | intronic | 0.0320 | 0.0320 | PSMF1 | 0.7087 | 60 | 6 | 0 | 23 | 3 | 0 |
| Downstream signaling events of B Cell Receptor (BCR) | 20 | 1146048 | C | T | T | rs17716261 | 3_prime_UTR | 0.0279 | 0.0279 | PSMF1 | 0.4532 | 60 | 6 | 0 | 21 | 4 | 0 |
| Downstream signaling events of B Cell Receptor (BCR) | 20 | 1149980 | C | T | T | rs77625408 | downstream | 0.0254 | 0.0254 | PSMF1 | 0.7495 | 54 | 11 | 0 | 23 | 3 | 0 |
| Downstream signaling events of B Cell Receptor (BCR) | 20 | 1152323 | C | T | T | rs34552580 | downstream | 0.0323 | 0.0323 | PSMF1 | 1.0000 | 59 | 5 | 1 | 24 | 2 | 0 |
| Downstream signaling events of B Cell Receptor (BCR) | 20 | 1155154 | G | A | A | rs78313102 | intronic,non_coding_transcript | 0.0258 | 0.0258 | PSMF1 | 1.0000 | 58 | 7 | 0 | 23 | 3 | 0 |
| Downstream signaling events of B Cell Receptor (BCR) | 20 | 50001132 | C | T | T | rs75343958 | downstream | 0.0191 | 0.0191 | NFATC2 | 0.0863 | 60 | 6 | 0 | 24 | 0 | 1 |
| Downstream signaling events of B Cell Receptor (BCR) | 20 | 50014957 | G | A | A | rs77379100 | intronic | 0.0207 | 0.0207 | NFATC2 | 0.7222 | 57 | 9 | 0 | 24 | 2 | 0 |
| Downstream signaling events of B Cell Receptor (BCR) | 20 | 50113131 | A | C | C | rs73128899 | intronic | 0.0467 | 0.0467 | NFATC2 | 0.7632 | 60 | 5 | 1 | 25 | 1 | 0 |
| Downstream signaling events of B Cell Receptor (BCR) | 20 | 50113240 | A | G | G | rs73128900 | intronic | 0.0466 | 0.0466 | NFATC2 | 0.7632 | 60 | 5 | 1 | 25 | 1 | 0 |
| Downstream signaling events of B Cell Receptor (BCR) | 20 | 50124126 | C | T | T | rs6013204 | intronic | 0.0473 | 0.0473 | NFATC2 | 1.0000 | 60 | 6 | 0 | 24 | 2 | 0 |
| Downstream signaling events of B Cell Receptor (BCR) | 20 | 50129716 | T | C | C | rs17728960 | intronic | 0.0477 | 0.0477 | NFATC2 | 0.7621 | 59 | 6 | 1 | 25 | 1 | 0 |
| Downstream signaling events of B Cell Receptor (BCR) | 20 | 50143470 | C | T | T | rs77042341 | intronic | 0.0367 | 0.0367 | NFATC2 | 0.7631 | 59 | 5 | 1 | 25 | 1 | 0 |
| Downstream signaling events of B Cell Receptor (BCR) | 20 | 50170376 | T | G | G | rs80208372 | intronic | 0.0480 | 0.0480 | NFATC2 | 1.0000 | 55 | 10 | 0 | 22 | 4 | 0 |
| Depolymerisation of the Nuclear Lamina | 2 | 11819328 | G | A | A | rs140904588 | intronic | 0.0333 | 0.0333 | LPIN1 | 0.1086 | 59 | 5 | 0 | 25 | 0 | 1 |
| Depolymerisation of the Nuclear Lamina | 2 | 11848161 | C | G | G | rs77534915 | intronic | 0.0360 | 0.0360 | LPIN1 | 0.6828 | 61 | 5 | 0 | 23 | 3 | 0 |
| Depolymerisation of the Nuclear Lamina | 2 | 11882703 | A | C | C | rs80102993 | intronic | 0.0434 | 0.0434 | LPIN1 | 0.4984 | 59 | 7 | 0 | 22 | 4 | 0 |
| Depolymerisation of the Nuclear Lamina | 2 | 11885139 | C | T | T | rs74899248 | intronic | 0.0444 | 0.0444 | LPIN1 | 0.4984 | 59 | 7 | 0 | 22 | 4 | 0 |
| Depolymerisation of the Nuclear Lamina | 2 | 11885972 | C | G | G | rs78733123 | intronic | 0.0354 | 0.0354 | LPIN1 | 0.4393 | 60 | 6 | 0 | 23 | 2 | 1 |
| Depolymerisation of the Nuclear Lamina | 2 | 11894527 | C | G | G | rs75421235 | intronic | 0.0416 | 0.0416 | LPIN1 | 0.7222 | 57 | 9 | 0 | 24 | 2 | 0 |
| Depolymerisation of the Nuclear Lamina | 2 | 11895388 | C | T | T | rs11694928 | intronic | 0.0498 | 0.0498 | LPIN1 | 0.5065 | 58 | 8 | 0 | 21 | 5 | 0 |
| Depolymerisation of the Nuclear Lamina | 2 | 11911946 | G | A | A | rs111726318 | intronic | 0.0471 | 0.0471 | LPIN1 | 0.2818 | 60 | 6 | 0 | 21 | 5 | 0 |
| Depolymerisation of the Nuclear Lamina | 2 | 11943082 | C | T | T | rs4669781 | missense | 0.0342 | 0.0342 | LPIN1 | 0.1706 | 56 | 10 | 0 | 25 | 1 | 0 |
| Depolymerisation of the Nuclear Lamina | 2 | 11945495 | A | C | C | rs79025286 | intronic | 0.0282 | 0.0282 | LPIN1 | 0.2744 | 57 | 9 | 0 | 24 | 1 | 0 |
| Depolymerisation of the Nuclear Lamina | 2 | 11955603 | C | T | T | rs75560733 | intronic | 0.0410 | 0.0410 | LPIN1 | 0.5425 | 53 | 13 | 0 | 23 | 3 | 0 |
| Depolymerisation of the Nuclear Lamina | 2 | 11963271 | A | C | C | rs56163146 | intronic | 0.0449 | 0.0449 | LPIN1 | 0.4984 | 59 | 7 | 0 | 22 | 4 | 0 |
| Depolymerisation of the Nuclear Lamina | 2 | 11968029 | A | G | G | rs34857485 | downstream | 0.0304 | 0.0304 | LPIN1 | 0.4098 | 61 | 5 | 0 | 24 | 1 | 1 |
| Depolymerisation of the Nuclear Lamina | 5 | 68462801 | C | G | G | rs8192259 | upstream | 0.0444 | 0.0444 | CCNB1 | 0.7688 | 54 | 11 | 0 | 21 | 5 | 0 |
| Depolymerisation of the Nuclear Lamina | 5 | 126143504 | T | C | C | rs2973605 | intronic | 0.9641 | 0.0359 | LMNB1 | 1.0000 | 1 | 8 | 57 | 0 | 3 | 22 |
| Depolymerisation of the Nuclear Lamina | 6 | 33746837 | C | T | T | rs151266730 | intronic | 0.0122 | 0.0122 | LEMD2 | 0.5321 | 62 | 3 | 1 | 23 | 3 | 0 |
| Depolymerisation of the Nuclear Lamina | 10 | 62548390 | G | A | A | rs3213056 | intronic | 0.0326 | 0.0326 | CDK1 | 0.7469 | 57 | 9 | 0 | 21 | 4 | 0 |
| Depolymerisation of the Nuclear Lamina | 12 | 65578819 | T | G | G | rs75563741 | intronic | 0.0221 | 0.0221 | LEMD3 | 0.4862 | 59 | 7 | 0 | 21 | 4 | 0 |
| Depolymerisation of the Nuclear Lamina | 12 | 65586630 | A | G | G | rs78691095 | intronic | 0.0389 | 0.0389 | LEMD3 | 0.8053 | 57 | 8 | 1 | 21 | 4 | 0 |
| Depolymerisation of the Nuclear Lamina | 12 | 65594504 | A | G | G | rs74809372 | intronic | 0.0234 | 0.0234 | LEMD3 | 0.2781 | 56 | 10 | 0 | 24 | 1 | 0 |
| Depolymerisation of the Nuclear Lamina | 12 | 98930771 | G | A | A | rs57467331 | downstream | 0.0354 | 0.0354 | TMPO | 0.4862 | 59 | 7 | 0 | 21 | 4 | 0 |
| Depolymerisation of the Nuclear Lamina | 12 | 98932013 | G | C | C | rs76069997 | downstream | 0.0352 | 0.0352 | TMPO | 0.4984 | 59 | 7 | 0 | 22 | 4 | 0 |
| Depolymerisation of the Nuclear Lamina | 16 | 23845860 | G | T | T | rs72777910 | upstream | 0.0300 | 0.0300 | PRKCB | 0.4762 | 60 | 5 | 1 | 22 | 4 | 0 |
| Depolymerisation of the Nuclear Lamina | 16 | 23849482 | T | C | T | rs2023670 | intronic | 0.9513 | 0.0487 | PRKCB | 0.0099 | 0 | 11 | 55 | 1 | 0 | 25 |
| Depolymerisation of the Nuclear Lamina | 16 | 23850240 | A | G | A | rs11074581 | intronic | 0.9663 | 0.0337 | PRKCB | 0.0099 | 0 | 11 | 55 | 1 | 0 | 25 |
| Depolymerisation of the Nuclear Lamina | 16 | 23851956 | T | C | T | rs7189210 | intronic | 0.9663 | 0.0337 | PRKCB | 0.0099 | 0 | 11 | 55 | 1 | 0 | 25 |
| Depolymerisation of the Nuclear Lamina | 16 | 23852415 | A | T | A | rs2188359 | intronic | 0.9528 | 0.0472 | PRKCB | 0.0099 | 0 | 11 | 55 | 1 | 0 | 25 |
| Depolymerisation of the Nuclear Lamina | 16 | 23859391 | A | G | G | rs62030647 | intronic | 0.0226 | 0.0226 | PRKCB | 1.0000 | 59 | 7 | 0 | 23 | 3 | 0 |
| Depolymerisation of the Nuclear Lamina | 16 | 23874933 | A | C | A | rs6497691 | intronic | 0.9663 | 0.0337 | PRKCB | 0.0063 | 0 | 14 | 52 | 1 | 0 | 24 |
| Depolymerisation of the Nuclear Lamina | 16 | 23876099 | C | T | T | rs79131874 | intronic | 0.0303 | 0.0303 | PRKCB | 1.0000 | 59 | 7 | 0 | 23 | 3 | 0 |
| Depolymerisation of the Nuclear Lamina | 16 | 23877500 | A | G | A | rs8059885 | intronic | 0.9663 | 0.0337 | PRKCB | 0.0046 | 0 | 14 | 52 | 1 | 0 | 25 |
| Depolymerisation of the Nuclear Lamina | 16 | 23877606 | A | G | A | rs8060048 | intronic | 0.9644 | 0.0356 | PRKCB | 0.0046 | 0 | 14 | 52 | 1 | 0 | 25 |
| Depolymerisation of the Nuclear Lamina | 16 | 23877781 | G | A | G | rs8060718 | intronic | 0.9664 | 0.0336 | PRKCB | 0.0046 | 0 | 14 | 52 | 1 | 0 | 25 |
| Depolymerisation of the Nuclear Lamina | 16 | 23878470 | C | T | C | rs12935004 | intronic | 0.9657 | 0.0343 | PRKCB | 0.0046 | 0 | 14 | 52 | 1 | 0 | 25 |
| Depolymerisation of the Nuclear Lamina | 16 | 23880851 | C | T | C | rs8061523 | intronic | 0.9664 | 0.0336 | PRKCB | 0.0046 | 0 | 14 | 52 | 1 | 0 | 25 |
| Depolymerisation of the Nuclear Lamina | 16 | 23881930 | G | A | G | rs8047121 | intronic | 0.9662 | 0.0338 | PRKCB | 0.0046 | 0 | 14 | 52 | 1 | 0 | 25 |
| Depolymerisation of the Nuclear Lamina | 16 | 23882469 | T | C | T | rs1468129 | intronic | 0.9663 | 0.0337 | PRKCB | 0.0046 | 0 | 14 | 52 | 1 | 0 | 25 |
| Depolymerisation of the Nuclear Lamina | 16 | 23885608 | A | T | A | rs8044732 | intronic | 0.9664 | 0.0336 | PRKCB | 0.0046 | 0 | 14 | 52 | 1 | 0 | 25 |
| Depolymerisation of the Nuclear Lamina | 16 | 23885751 | A | G | G | rs62031692 | intronic | 0.0253 | 0.0253 | PRKCB | 1.0000 | 59 | 7 | 0 | 23 | 3 | 0 |
| Depolymerisation of the Nuclear Lamina | 16 | 23887574 | G | T | T | rs79034087 | intronic | 0.0290 | 0.0290 | PRKCB | 0.3487 | 61 | 4 | 1 | 22 | 4 | 0 |
| Depolymerisation of the Nuclear Lamina | 16 | 23888354 | C | T | C | rs7404417 | intronic | 0.9664 | 0.0336 | PRKCB | 0.0046 | 0 | 14 | 52 | 1 | 0 | 25 |
| Depolymerisation of the Nuclear Lamina | 16 | 23889896 | T | C | T | rs8063823 | intronic | 0.9665 | 0.0335 | PRKCB | 0.0046 | 0 | 14 | 52 | 1 | 0 | 25 |
| Depolymerisation of the Nuclear Lamina | 16 | 23893893 | G | A | G | rs11647359 | intronic | 0.9664 | 0.0336 | PRKCB | 0.0046 | 0 | 14 | 52 | 1 | 0 | 25 |
| Depolymerisation of the Nuclear Lamina | 16 | 23895034 | A | G | A | rs6497695 | intronic | 0.9665 | 0.0335 | PRKCB | 0.0046 | 0 | 14 | 52 | 1 | 0 | 25 |
| Depolymerisation of the Nuclear Lamina | 16 | 23895443 | A | G | G | rs62028075 | intronic | 0.0253 | 0.0253 | PRKCB | 1.0000 | 59 | 7 | 0 | 23 | 3 | 0 |
| Depolymerisation of the Nuclear Lamina | 16 | 23895884 | T | C | T | rs9944348 | intronic | 0.9665 | 0.0335 | PRKCB | 0.0046 | 0 | 14 | 52 | 1 | 0 | 25 |
| Depolymerisation of the Nuclear Lamina | 16 | 23896089 | T | C | C | rs74572166 | intronic | 0.0245 | 0.0245 | PRKCB | 1.0000 | 59 | 7 | 0 | 23 | 3 | 0 |
| Depolymerisation of the Nuclear Lamina | 16 | 23896209 | C | A | C | rs9302418 | intronic | 0.9664 | 0.0336 | PRKCB | 0.0046 | 0 | 14 | 52 | 1 | 0 | 25 |
| Depolymerisation of the Nuclear Lamina | 16 | 23896438 | G | T | T | rs62028076 | intronic | 0.0252 | 0.0252 | PRKCB | 1.0000 | 59 | 7 | 0 | 22 | 3 | 0 |
| Depolymerisation of the Nuclear Lamina | 16 | 23898605 | A | T | A | rs933290 | intronic | 0.9632 | 0.0368 | PRKCB | 0.0016 | 0 | 17 | 49 | 1 | 0 | 24 |
| Depolymerisation of the Nuclear Lamina | 16 | 23899211 | A | T | A | rs12926245 | intronic | 0.9632 | 0.0368 | PRKCB | 0.0012 | 0 | 17 | 49 | 1 | 0 | 25 |
| Depolymerisation of the Nuclear Lamina | 16 | 23899610 | G | A | A | rs17753246 | intronic | 0.0252 | 0.0252 | PRKCB | 1.0000 | 59 | 7 | 0 | 23 | 3 | 0 |
| Depolymerisation of the Nuclear Lamina | 16 | 23899951 | G | A | A | rs62028077 | intronic | 0.0254 | 0.0254 | PRKCB | 1.0000 | 59 | 7 | 0 | 23 | 3 | 0 |
| Depolymerisation of the Nuclear Lamina | 16 | 23900716 | T | C | C | rs62028078 | intronic | 0.0252 | 0.0252 | PRKCB | 1.0000 | 59 | 7 | 0 | 23 | 3 | 0 |
| Depolymerisation of the Nuclear Lamina | 16 | 23901896 | C | T | C | rs6497696 | intronic | 0.9632 | 0.0368 | PRKCB | 0.0015 | 0 | 16 | 50 | 1 | 0 | 25 |
| Depolymerisation of the Nuclear Lamina | 16 | 23901948 | A | C | A | rs6497697 | intronic | 0.9630 | 0.0370 | PRKCB | 0.0015 | 0 | 16 | 49 | 1 | 0 | 25 |
| Depolymerisation of the Nuclear Lamina | 16 | 23904058 | A | G | A | rs886115 | intronic | 0.9632 | 0.0368 | PRKCB | 0.0015 | 0 | 16 | 50 | 1 | 0 | 25 |
| Depolymerisation of the Nuclear Lamina | 16 | 23904781 | G | A | A | rs17753509 | intronic | 0.0253 | 0.0253 | PRKCB | 1.0000 | 59 | 7 | 0 | 23 | 3 | 0 |
| Depolymerisation of the Nuclear Lamina | 16 | 23905676 | C | T | C | rs7200610 | intronic | 0.9631 | 0.0369 | PRKCB | 0.0015 | 0 | 16 | 50 | 1 | 0 | 25 |
| Depolymerisation of the Nuclear Lamina | 16 | 23907177 | A | C | C | rs17810011 | intronic | 0.0251 | 0.0251 | PRKCB | 1.0000 | 59 | 7 | 0 | 23 | 3 | 0 |
| Depolymerisation of the Nuclear Lamina | 16 | 23907765 | C | T | C | rs9925890 | intronic | 0.9632 | 0.0368 | PRKCB | 0.0024 | 0 | 16 | 50 | 1 | 0 | 24 |
| Depolymerisation of the Nuclear Lamina | 16 | 23912174 | A | G | A | rs12448249 | intronic | 0.9519 | 0.0481 | PRKCB | 0.0015 | 0 | 16 | 50 | 1 | 0 | 25 |
| Depolymerisation of the Nuclear Lamina | 16 | 23914915 | C | A | C | rs1004186 | intronic | 0.9632 | 0.0368 | PRKCB | 0.0015 | 0 | 16 | 50 | 1 | 0 | 25 |
| Depolymerisation of the Nuclear Lamina | 16 | 23916258 | G | A | G | rs1004187 | intronic | 0.9632 | 0.0368 | PRKCB | 0.0015 | 0 | 16 | 50 | 1 | 0 | 25 |
| Depolymerisation of the Nuclear Lamina | 16 | 23916521 | G | C | G | rs1008654 | intronic | 0.9633 | 0.0367 | PRKCB | 0.0015 | 0 | 16 | 50 | 1 | 0 | 25 |
| Depolymerisation of the Nuclear Lamina | 16 | 23917335 | G | A | G | rs6497699 | intronic | 0.9645 | 0.0355 | PRKCB | 0.0024 | 0 | 16 | 50 | 1 | 0 | 24 |
| Depolymerisation of the Nuclear Lamina | 16 | 23917465 | C | G | C | rs7186538 | intronic | 0.9645 | 0.0355 | PRKCB | 0.0024 | 0 | 16 | 50 | 1 | 0 | 24 |
| Depolymerisation of the Nuclear Lamina | 16 | 23917700 | C | A | C | rs7187091 | intronic | 0.9646 | 0.0354 | PRKCB | 0.0024 | 0 | 16 | 50 | 1 | 0 | 24 |
| Depolymerisation of the Nuclear Lamina | 16 | 23919088 | C | T | T | rs78322646 | intronic | 0.0278 | 0.0278 | PRKCB | 0.7348 | 58 | 8 | 0 | 22 | 4 | 0 |
| Depolymerisation of the Nuclear Lamina | 16 | 23921083 | C | T | C | rs6497702 | intronic | 0.9647 | 0.0353 | PRKCB | 0.0024 | 0 | 16 | 50 | 1 | 0 | 24 |
| Depolymerisation of the Nuclear Lamina | 16 | 23925936 | C | G | C | rs11074588 | intronic | 0.9649 | 0.0351 | PRKCB | 0.0015 | 0 | 16 | 50 | 1 | 0 | 25 |
| Depolymerisation of the Nuclear Lamina | 16 | 23939212 | G | A | G | rs11074590 | intronic | 0.9650 | 0.0350 | PRKCB | 0.0024 | 0 | 16 | 50 | 1 | 0 | 24 |
| Depolymerisation of the Nuclear Lamina | 16 | 23941628 | C | A | C | rs2005671 | intronic | 0.9647 | 0.0353 | PRKCB | 0.0015 | 0 | 16 | 49 | 1 | 0 | 25 |
| Depolymerisation of the Nuclear Lamina | 16 | 23943749 | T | C | T | rs9302420 | intronic | 0.9649 | 0.0351 | PRKCB | 0.0024 | 0 | 16 | 50 | 1 | 0 | 24 |
| Depolymerisation of the Nuclear Lamina | 16 | 23945985 | T | G | T | rs195989 | intronic | 0.9651 | 0.0349 | PRKCB | 0.0015 | 0 | 16 | 50 | 1 | 0 | 25 |
| Depolymerisation of the Nuclear Lamina | 16 | 23946157 | G | A | A | rs76973283 | intronic | 0.0302 | 0.0302 | PRKCB | 0.5065 | 58 | 8 | 0 | 21 | 5 | 0 |
| Depolymerisation of the Nuclear Lamina | 16 | 23949175 | G | C | G | rs2560403 | intronic | 0.9657 | 0.0343 | PRKCB | 0.0015 | 0 | 16 | 50 | 1 | 0 | 25 |
| Depolymerisation of the Nuclear Lamina | 16 | 23949438 | A | G | A | rs195985 | intronic | 0.9658 | 0.0342 | PRKCB | 0.0015 | 0 | 16 | 50 | 1 | 0 | 25 |
| Depolymerisation of the Nuclear Lamina | 16 | 23953265 | T | C | T | rs2560404 | intronic | 0.9656 | 0.0344 | PRKCB | 0.0024 | 0 | 16 | 50 | 1 | 0 | 24 |
| Depolymerisation of the Nuclear Lamina | 16 | 23954128 | T | C | C | rs17810486 | intronic | 0.0308 | 0.0308 | PRKCB | 0.3024 | 58 | 7 | 0 | 20 | 5 | 0 |
| Depolymerisation of the Nuclear Lamina | 16 | 23954253 | G | A | G | rs195994 | intronic | 0.9653 | 0.0347 | PRKCB | 0.0015 | 0 | 16 | 50 | 1 | 0 | 25 |
| Depolymerisation of the Nuclear Lamina | 16 | 23962258 | G | C | G | rs196000 | intronic | 0.9659 | 0.0341 | PRKCB | 0.0015 | 0 | 16 | 50 | 1 | 0 | 25 |
| Depolymerisation of the Nuclear Lamina | 16 | 23964858 | T | A | T | rs196003 | intronic | 0.9647 | 0.0353 | PRKCB | 0.0343 | 0 | 15 | 51 | 0 | 1 | 25 |
| Depolymerisation of the Nuclear Lamina | 16 | 23985814 | C | T | T | rs72779914 | intronic | 0.0487 | 0.0487 | PRKCB | 1.0000 | 53 | 12 | 0 | 21 | 4 | 0 |
| Depolymerisation of the Nuclear Lamina | 16 | 23987552 | A | G | A | rs169030 | intronic | 0.9709 | 0.0291 | PRKCB | 0.1679 | 0 | 11 | 55 | 0 | 1 | 25 |
| Depolymerisation of the Nuclear Lamina | 16 | 23988755 | T | C | T | rs196013 | intronic | 0.9681 | 0.0319 | PRKCB | 0.1681 | 0 | 11 | 55 | 0 | 1 | 24 |
| Depolymerisation of the Nuclear Lamina | 16 | 24009919 | A | G | G | rs75622923 | intronic | 0.0319 | 0.0319 | PRKCB | 0.3348 | 54 | 12 | 0 | 24 | 2 | 0 |
| Depolymerisation of the Nuclear Lamina | 16 | 24022944 | C | T | T | rs111746132 | intronic | 0.0229 | 0.0229 | PRKCB | 0.7522 | 56 | 10 | 0 | 23 | 3 | 0 |
| Depolymerisation of the Nuclear Lamina | 16 | 24066378 | G | A | A | rs113426570 | intronic | 0.0216 | 0.0216 | PRKCB | 1.0000 | 57 | 9 | 0 | 23 | 3 | 0 |
| Depolymerisation of the Nuclear Lamina | 16 | 24100759 | T | A | A | rs11643939 | intronic | 0.0294 | 0.0294 | PRKCB | 0.7522 | 56 | 10 | 0 | 23 | 3 | 0 |
| Depolymerisation of the Nuclear Lamina | 16 | 24105816 | G | A | A | rs56316329 | intronic | 0.0251 | 0.0251 | PRKCB | 0.6919 | 60 | 4 | 1 | 25 | 0 | 0 |
| Depolymerisation of the Nuclear Lamina | 16 | 24111853 | T | C | C | rs55959083 | intronic | 0.0431 | 0.0431 | PRKCB | 0.7213 | 58 | 8 | 0 | 23 | 2 | 0 |
| Depolymerisation of the Nuclear Lamina | 16 | 24112768 | G | A | A | rs117056307 | intronic | 0.0430 | 0.0430 | PRKCB | 0.7192 | 58 | 8 | 0 | 24 | 2 | 0 |
| Depolymerisation of the Nuclear Lamina | 16 | 24122052 | G | A | A | rs117467859 | intronic | 0.0433 | 0.0433 | PRKCB | 0.7213 | 58 | 8 | 0 | 23 | 2 | 0 |
| Depolymerisation of the Nuclear Lamina | 16 | 24122492 | C | T | T | rs72779977 | intronic | 0.0457 | 0.0457 | PRKCB | 0.7201 | 57 | 8 | 0 | 23 | 2 | 0 |
| Depolymerisation of the Nuclear Lamina | 16 | 24123560 | G | A | A | rs60261043 | intronic | 0.0457 | 0.0457 | PRKCB | 0.7201 | 57 | 8 | 0 | 23 | 2 | 0 |
| Depolymerisation of the Nuclear Lamina | 16 | 24132273 | G | A | A | rs62027458 | intronic | 0.0232 | 0.0232 | PRKCB | 0.1877 | 59 | 7 | 0 | 21 | 4 | 1 |
| Depolymerisation of the Nuclear Lamina | 16 | 24164042 | G | T | T | rs72779989 | intronic | 0.0487 | 0.0487 | PRKCB | 0.8583 | 57 | 7 | 2 | 22 | 4 | 0 |
| Depolymerisation of the Nuclear Lamina | 16 | 24197496 | A | T | T | rs79699525 | intronic | 0.0261 | 0.0261 | PRKCB | 0.4631 | 59 | 6 | 0 | 22 | 4 | 0 |
| Depolymerisation of the Nuclear Lamina | 16 | 24199852 | C | T | T | rs78424166 | intronic | 0.0359 | 0.0359 | PRKCB | 0.4603 | 60 | 6 | 0 | 22 | 4 | 0 |
| Depolymerisation of the Nuclear Lamina | 16 | 50058523 | C | T | T | rs111253683 | 5_prime_UTR | 0.0302 | 0.0302 | CNEP1R1 | 0.5717 | 61 | 4 | 1 | 23 | 3 | 0 |
| Depolymerisation of the Nuclear Lamina | 17 | 64305051 | A | G | G | rs78357146 | intronic | 0.0186 | 0.0186 | PRKCA | 0.7209 | 57 | 9 | 0 | 23 | 2 | 0 |
| Depolymerisation of the Nuclear Lamina | 17 | 64315409 | T | C | C | rs80130647 | intronic | 0.0184 | 0.0184 | PRKCA | 0.7043 | 59 | 6 | 0 | 22 | 3 | 0 |
| Depolymerisation of the Nuclear Lamina | 17 | 64318385 | G | A | A | rs12150623 | intronic | 0.0343 | 0.0343 | PRKCA | 0.7595 | 61 | 4 | 1 | 23 | 2 | 0 |
| Depolymerisation of the Nuclear Lamina | 17 | 64320040 | C | T | T | rs139317720 | intronic | 0.0229 | 0.0229 | PRKCA | 1.0000 | 62 | 3 | 1 | 24 | 1 | 0 |
| Depolymerisation of the Nuclear Lamina | 17 | 64326068 | T | G | G | rs12951126 | intronic | 0.0260 | 0.0260 | PRKCA | 1.0000 | 58 | 8 | 0 | 23 | 3 | 0 |
| Depolymerisation of the Nuclear Lamina | 17 | 64343295 | A | C | C | rs72843901 | intronic | 0.0457 | 0.0457 | PRKCA | 0.4725 | 56 | 9 | 1 | 24 | 1 | 0 |
| Depolymerisation of the Nuclear Lamina | 17 | 64344650 | C | T | T | rs72846606 | intronic | 0.0454 | 0.0454 | PRKCA | 0.4725 | 56 | 9 | 1 | 24 | 1 | 0 |
| Depolymerisation of the Nuclear Lamina | 17 | 64344788 | C | T | T | rs72846607 | intronic | 0.0457 | 0.0457 | PRKCA | 0.4725 | 56 | 9 | 1 | 24 | 1 | 0 |
| Depolymerisation of the Nuclear Lamina | 17 | 64346204 | A | G | G | rs12150367 | intronic | 0.0452 | 0.0452 | PRKCA | 0.4770 | 56 | 9 | 1 | 25 | 1 | 0 |
| Depolymerisation of the Nuclear Lamina | 17 | 64349104 | C | T | T | rs72846609 | intronic | 0.0497 | 0.0497 | PRKCA | 0.4770 | 56 | 9 | 1 | 25 | 1 | 0 |
| Depolymerisation of the Nuclear Lamina | 17 | 64351542 | G | C | C | rs28592028 | intronic | 0.0453 | 0.0453 | PRKCA | 0.4770 | 56 | 9 | 1 | 25 | 1 | 0 |
| Depolymerisation of the Nuclear Lamina | 17 | 64352431 | T | G | G | rs544435459 | intronic | 0.0451 | 0.0451 | PRKCA | 0.4725 | 56 | 9 | 1 | 24 | 1 | 0 |
| Depolymerisation of the Nuclear Lamina | 17 | 64354597 | T | C | C | rs72846612 | intronic | 0.0451 | 0.0451 | PRKCA | 0.4725 | 56 | 9 | 1 | 24 | 1 | 0 |
| Depolymerisation of the Nuclear Lamina | 17 | 64358121 | C | G | G | rs72846614 | intronic | 0.0451 | 0.0451 | PRKCA | 0.4770 | 56 | 9 | 1 | 25 | 1 | 0 |
| Depolymerisation of the Nuclear Lamina | 17 | 64359354 | T | C | C | rs72846615 | intronic | 0.0452 | 0.0452 | PRKCA | 0.4725 | 56 | 9 | 1 | 24 | 1 | 0 |
| Depolymerisation of the Nuclear Lamina | 17 | 64365593 | C | T | T | rs72846677 | intronic | 0.0452 | 0.0452 | PRKCA | 0.4770 | 56 | 9 | 1 | 25 | 1 | 0 |
| Depolymerisation of the Nuclear Lamina | 17 | 64367534 | C | T | T | rs72846678 | intronic | 0.0452 | 0.0452 | PRKCA | 0.4725 | 56 | 9 | 1 | 24 | 1 | 0 |
| Depolymerisation of the Nuclear Lamina | 17 | 64369822 | A | G | G | rs72846681 | intronic | 0.0451 | 0.0451 | PRKCA | 0.4770 | 56 | 9 | 1 | 25 | 1 | 0 |
| Depolymerisation of the Nuclear Lamina | 17 | 64375194 | T | C | C | rs72846695 | intronic | 0.0452 | 0.0452 | PRKCA | 0.4770 | 56 | 9 | 1 | 25 | 1 | 0 |
| Depolymerisation of the Nuclear Lamina | 17 | 64377301 | C | A | A | rs113134992 | intronic | 0.0448 | 0.0448 | PRKCA | 0.4770 | 56 | 9 | 1 | 25 | 1 | 0 |
| Depolymerisation of the Nuclear Lamina | 17 | 64382507 | C | G | G | rs79461368 | intronic | 0.0453 | 0.0453 | PRKCA | 0.4725 | 56 | 9 | 1 | 24 | 1 | 0 |
| Depolymerisation of the Nuclear Lamina | 17 | 64389524 | C | T | T | rs77682324 | intronic | 0.0403 | 0.0403 | PRKCA | 0.7621 | 59 | 6 | 1 | 25 | 1 | 0 |
| Depolymerisation of the Nuclear Lamina | 17 | 64395773 | T | C | C | rs78584531 | intronic | 0.0452 | 0.0452 | PRKCA | 0.4770 | 56 | 9 | 1 | 25 | 1 | 0 |
| Depolymerisation of the Nuclear Lamina | 17 | 64396570 | T | G | G | rs72838208 | intronic | 0.0452 | 0.0452 | PRKCA | 0.4725 | 56 | 9 | 1 | 24 | 1 | 0 |
| Depolymerisation of the Nuclear Lamina | 17 | 64396695 | A | G | G | rs74329211 | intronic | 0.0432 | 0.0432 | PRKCA | 0.7302 | 58 | 8 | 0 | 21 | 4 | 0 |
| Depolymerisation of the Nuclear Lamina | 17 | 64396883 | G | A | A | rs72838209 | intronic | 0.0452 | 0.0452 | PRKCA | 0.4770 | 56 | 9 | 1 | 25 | 1 | 0 |
| Depolymerisation of the Nuclear Lamina | 17 | 64398013 | G | A | A | rs75125285 | intronic | 0.0295 | 0.0295 | PRKCA | 0.7632 | 60 | 5 | 1 | 25 | 1 | 0 |
| Depolymerisation of the Nuclear Lamina | 17 | 64400648 | C | T | T | rs72838214 | intronic | 0.0452 | 0.0452 | PRKCA | 0.4788 | 55 | 9 | 1 | 25 | 1 | 0 |
| Depolymerisation of the Nuclear Lamina | 17 | 64402141 | G | T | T | rs10221238 | intronic | 0.0452 | 0.0452 | PRKCA | 0.4770 | 56 | 9 | 1 | 25 | 1 | 0 |
| Depolymerisation of the Nuclear Lamina | 17 | 64405430 | T | A | A | rs72838216 | intronic | 0.0453 | 0.0453 | PRKCA | 0.4725 | 56 | 9 | 1 | 24 | 1 | 0 |
| Depolymerisation of the Nuclear Lamina | 17 | 64410564 | C | T | T | rs9972974 | intronic | 0.0390 | 0.0390 | PRKCA | 0.7632 | 60 | 5 | 1 | 25 | 1 | 0 |
| Depolymerisation of the Nuclear Lamina | 17 | 64412921 | G | A | A | rs62070391 | intronic | 0.0426 | 0.0426 | PRKCA | 0.7621 | 59 | 6 | 1 | 25 | 1 | 0 |
| Depolymerisation of the Nuclear Lamina | 17 | 64424382 | C | T | T | rs62070395 | intronic | 0.0425 | 0.0425 | PRKCA | 0.7595 | 59 | 6 | 1 | 24 | 1 | 0 |
| Depolymerisation of the Nuclear Lamina | 17 | 64429928 | G | A | A | rs117729211 | intronic | 0.0272 | 0.0272 | PRKCA | 0.0477 | 60 | 6 | 0 | 19 | 6 | 1 |
| Depolymerisation of the Nuclear Lamina | 17 | 64431874 | G | A | A | rs72838278 | intronic | 0.0426 | 0.0426 | PRKCA | 0.7642 | 57 | 6 | 1 | 25 | 1 | 0 |
| Depolymerisation of the Nuclear Lamina | 17 | 64432202 | A | G | G | rs62070397 | intronic | 0.0427 | 0.0427 | PRKCA | 0.7621 | 59 | 6 | 1 | 25 | 1 | 0 |
| Depolymerisation of the Nuclear Lamina | 17 | 64432881 | A | G | G | rs62070398 | intronic | 0.0426 | 0.0426 | PRKCA | 0.7595 | 59 | 6 | 1 | 24 | 1 | 0 |
| Depolymerisation of the Nuclear Lamina | 17 | 64438732 | A | T | T | rs113684166 | intronic | 0.0445 | 0.0445 | PRKCA | 0.2744 | 57 | 9 | 0 | 24 | 1 | 0 |
| Depolymerisation of the Nuclear Lamina | 17 | 64441204 | T | C | C | rs75005068 | intronic | 0.0469 | 0.0469 | PRKCA | 0.2719 | 57 | 9 | 0 | 25 | 1 | 0 |
| Depolymerisation of the Nuclear Lamina | 17 | 64441759 | C | A | A | rs78330327 | intronic | 0.0465 | 0.0465 | PRKCA | 0.2710 | 56 | 9 | 0 | 25 | 1 | 0 |
| Depolymerisation of the Nuclear Lamina | 17 | 64445337 | A | C | C | rs118090701 | intronic | 0.0387 | 0.0387 | PRKCA | 1.0000 | 58 | 6 | 1 | 23 | 2 | 0 |
| Depolymerisation of the Nuclear Lamina | 17 | 64445856 | A | G | G | rs12451388 | intronic | 0.0449 | 0.0449 | PRKCA | 0.2744 | 57 | 9 | 0 | 24 | 1 | 0 |
| Depolymerisation of the Nuclear Lamina | 17 | 64447721 | G | A | A | rs12452749 | intronic | 0.0462 | 0.0462 | PRKCA | 0.2786 | 57 | 9 | 0 | 23 | 1 | 0 |
| Depolymerisation of the Nuclear Lamina | 17 | 64450113 | G | C | C | rs80162292 | intronic | 0.0461 | 0.0461 | PRKCA | 0.2710 | 56 | 9 | 0 | 25 | 1 | 0 |
| Depolymerisation of the Nuclear Lamina | 17 | 64453194 | A | G | G | rs111776777 | intronic | 0.0442 | 0.0442 | PRKCA | 0.8716 | 54 | 10 | 2 | 23 | 3 | 0 |
| Depolymerisation of the Nuclear Lamina | 17 | 64458036 | C | T | T | rs113153197 | intronic | 0.0339 | 0.0339 | PRKCA | 0.1678 | 54 | 11 | 0 | 25 | 1 | 0 |
| Depolymerisation of the Nuclear Lamina | 17 | 64462111 | T | G | G | rs79239451 | intronic | 0.0464 | 0.0464 | PRKCA | 0.2744 | 57 | 9 | 0 | 24 | 1 | 0 |
| Depolymerisation of the Nuclear Lamina | 17 | 64462288 | G | A | A | rs80080003 | intronic | 0.0305 | 0.0305 | PRKCA | 0.1706 | 56 | 10 | 0 | 25 | 1 | 0 |
| Depolymerisation of the Nuclear Lamina | 17 | 64483589 | G | T | T | rs79070174 | intronic | 0.0391 | 0.0391 | PRKCA | 1.0000 | 59 | 6 | 1 | 23 | 2 | 0 |
| Depolymerisation of the Nuclear Lamina | 17 | 64487077 | T | G | G | rs78149509 | intronic | 0.0200 | 0.0200 | PRKCA | 0.0381 | 63 | 3 | 0 | 21 | 4 | 1 |
| Depolymerisation of the Nuclear Lamina | 17 | 64494906 | A | G | G | rs117168126 | intronic | 0.0300 | 0.0300 | PRKCA | 1.0000 | 59 | 6 | 1 | 23 | 2 | 0 |
| Depolymerisation of the Nuclear Lamina | 17 | 64501943 | C | T | T | rs7217954 | intronic | 0.0422 | 0.0422 | PRKCA | 1.0000 | 59 | 7 | 0 | 23 | 3 | 0 |
| Depolymerisation of the Nuclear Lamina | 17 | 64519790 | G | A | A | rs11659067 | intronic | 0.0498 | 0.0498 | PRKCA | 0.0125 | 62 | 3 | 0 | 19 | 6 | 1 |
| Depolymerisation of the Nuclear Lamina | 17 | 64525002 | A | T | T | rs77462363 | intronic | 0.0380 | 0.0380 | PRKCA | 1.0000 | 59 | 6 | 1 | 23 | 2 | 0 |
| Depolymerisation of the Nuclear Lamina | 17 | 64544723 | C | T | T | rs116879811 | intronic | 0.0274 | 0.0274 | PRKCA | 1.0000 | 60 | 6 | 0 | 24 | 2 | 0 |
| Depolymerisation of the Nuclear Lamina | 17 | 64559535 | A | G | G | rs227907 | intronic | 0.0478 | 0.0478 | PRKCA | 0.2800 | 55 | 11 | 0 | 21 | 3 | 1 |
| Depolymerisation of the Nuclear Lamina | 17 | 64561055 | G | A | A | rs62071706 | intronic | 0.0238 | 0.0238 | PRKCA | 0.1546 | 60 | 5 | 0 | 21 | 4 | 1 |
| Depolymerisation of the Nuclear Lamina | 17 | 64590951 | A | T | T | rs11867591 | intronic | 0.0449 | 0.0449 | PRKCA | 0.2122 | 57 | 8 | 0 | 20 | 6 | 0 |
| Depolymerisation of the Nuclear Lamina | 17 | 64604776 | C | T | T | rs117539643 | intronic | 0.0479 | 0.0479 | PRKCA | 0.6653 | 51 | 11 | 2 | 23 | 3 | 0 |
| Depolymerisation of the Nuclear Lamina | 17 | 64608923 | T | C | C | rs72845947 | intronic | 0.0440 | 0.0440 | PRKCA | 0.7355 | 57 | 7 | 2 | 21 | 4 | 0 |
| Depolymerisation of the Nuclear Lamina | 17 | 64610285 | T | C | C | rs72845948 | intronic | 0.0445 | 0.0445 | PRKCA | 0.8583 | 57 | 7 | 2 | 22 | 4 | 0 |
| Depolymerisation of the Nuclear Lamina | 17 | 64610480 | A | C | C | rs17759657 | intronic | 0.0444 | 0.0444 | PRKCA | 0.8583 | 57 | 7 | 2 | 22 | 4 | 0 |
| Depolymerisation of the Nuclear Lamina | 17 | 64612838 | C | T | T | rs16959714 | intronic | 0.0338 | 0.0338 | PRKCA | 0.4353 | 57 | 8 | 0 | 24 | 1 | 0 |
| Depolymerisation of the Nuclear Lamina | 17 | 64614717 | T | C | C | rs17686540 | intronic | 0.0442 | 0.0442 | PRKCA | 0.7355 | 57 | 7 | 2 | 21 | 4 | 0 |
| Depolymerisation of the Nuclear Lamina | 17 | 64626385 | G | A | A | rs74352723 | intronic | 0.0460 | 0.0460 | PRKCA | 0.2744 | 57 | 9 | 0 | 24 | 1 | 0 |
| Depolymerisation of the Nuclear Lamina | 17 | 64628634 | A | G | G | rs117353888 | intronic | 0.0437 | 0.0437 | PRKCA | 0.2521 | 54 | 11 | 0 | 18 | 7 | 0 |
| Depolymerisation of the Nuclear Lamina | 17 | 64660047 | G | C | C | rs146141011 | intronic | 0.0484 | 0.0484 | PRKCA | 1.0000 | 53 | 9 | 2 | 22 | 4 | 0 |
| Depolymerisation of the Nuclear Lamina | 17 | 64671047 | C | T | T | rs16959942 | intronic | 0.0496 | 0.0496 | PRKCA | 0.1706 | 56 | 10 | 0 | 25 | 1 | 0 |
| Depolymerisation of the Nuclear Lamina | 17 | 64712634 | T | C | C | rs78121420 | intronic | 0.0465 | 0.0465 | PRKCA | 1.0000 | 57 | 7 | 2 | 23 | 3 | 0 |
| Depolymerisation of the Nuclear Lamina | 17 | 64713847 | G | T | T | rs112934229 | intronic | 0.0464 | 0.0464 | PRKCA | 1.0000 | 57 | 7 | 2 | 23 | 3 | 0 |
| Depolymerisation of the Nuclear Lamina | 17 | 64723880 | G | A | A | rs79547774 | intronic | 0.0264 | 0.0264 | PRKCA | 1.0000 | 57 | 9 | 0 | 23 | 3 | 0 |
| Depolymerisation of the Nuclear Lamina | 17 | 64738427 | A | C | C | rs117138620 | intronic | 0.0400 | 0.0400 | PRKCA | 0.3055 | 58 | 8 | 0 | 23 | 2 | 1 |
| Depolymerisation of the Nuclear Lamina | 17 | 64748431 | G | T | T | rs141177250 | intronic | 0.0153 | 0.0153 | PRKCA | 0.7595 | 59 | 6 | 1 | 24 | 1 | 0 |
| Depolymerisation of the Nuclear Lamina | 17 | 64755018 | T | C | C | rs74831470 | intronic | 0.0226 | 0.0226 | PRKCA | 0.7976 | 56 | 9 | 1 | 23 | 2 | 0 |
| Depolymerisation of the Nuclear Lamina | 17 | 64760870 | A | G | G | rs77904275 | intronic | 0.0307 | 0.0307 | PRKCA | 0.5233 | 53 | 11 | 1 | 24 | 2 | 0 |
| Depolymerisation of the Nuclear Lamina | 17 | 64762410 | C | G | G | rs113025478 | intronic | 0.0307 | 0.0307 | PRKCA | 0.6398 | 54 | 11 | 1 | 23 | 2 | 0 |
| Depolymerisation of the Nuclear Lamina | 17 | 64763235 | T | C | C | rs77635068 | intronic | 0.0306 | 0.0306 | PRKCA | 0.6398 | 54 | 11 | 1 | 23 | 2 | 0 |
| Depolymerisation of the Nuclear Lamina | 17 | 64771614 | C | T | T | rs80238933 | intronic | 0.0297 | 0.0297 | PRKCA | 0.6398 | 54 | 11 | 1 | 23 | 2 | 0 |
| Depolymerisation of the Nuclear Lamina | 17 | 64776847 | G | A | A | rs113542727 | intronic | 0.0297 | 0.0297 | PRKCA | 0.5228 | 54 | 11 | 1 | 24 | 2 | 0 |
| Depolymerisation of the Nuclear Lamina | 17 | 64791836 | G | C | C | rs56884788 | intronic | 0.0390 | 0.0390 | PRKCA | 0.6362 | 55 | 10 | 1 | 23 | 2 | 0 |
| Depolymerisation of the Nuclear Lamina | 17 | 64792863 | G | C | C | rs72838636 | intronic | 0.0484 | 0.0484 | PRKCA | 0.6470 | 57 | 8 | 1 | 21 | 5 | 0 |
| Depolymerisation of the Nuclear Lamina | 18 | 2967168 | A | T | T | rs145394193 | intronic | 0.0160 | 0.0160 | LPIN2 | 0.7192 | 58 | 8 | 0 | 24 | 2 | 0 |
| Depolymerisation of the Nuclear Lamina | 18 | 3015946 | A | G | G | rs77361064 | upstream | 0.0479 | 0.0479 | LPIN2 | 1.0000 | 57 | 9 | 0 | 22 | 4 | 0 |
| Depolymerisation of the Nuclear Lamina | 18 | 3016562 | T | C | C | rs76840312 | upstream | 0.0473 | 0.0473 | LPIN2 | 1.0000 | 57 | 9 | 0 | 22 | 4 | 0 |
| Depolymerisation of the Nuclear Lamina | 20 | 39968020 | A | T | T | rs117029881 | upstream | 0.0108 | 0.0108 | LPIN3 | 0.7632 | 60 | 5 | 1 | 25 | 1 | 0 |
| Depolymerisation of the Nuclear Lamina | 20 | 39971848 | T | C | C | rs79070826 | intronic | 0.0382 | 0.0382 | LPIN3 | 1.0000 | 61 | 4 | 1 | 25 | 1 | 0 |
| Depolymerisation of the Nuclear Lamina | 20 | 39977650 | C | T | T | rs78339899 | intronic | 0.0345 | 0.0345 | LPIN3 | 1.0000 | 61 | 4 | 1 | 25 | 1 | 0 |
| Depolymerisation of the Nuclear Lamina | 20 | 39982743 | C | T | T | rs7269761 | intronic | 0.0405 | 0.0405 | LPIN3 | 1.0000 | 60 | 4 | 1 | 25 | 1 | 0 |
| Depolymerisation of the Nuclear Lamina | 20 | 39984988 | G | A | A | rs79649817 | intronic | 0.0295 | 0.0295 | LPIN3 | 1.0000 | 61 | 4 | 1 | 25 | 1 | 0 |
| Depolymerisation of the Nuclear Lamina | 20 | 39989181 | C | T | T | rs2235594 | 3_prime_UTR | 0.0462 | 0.0462 | LPIN3 | 1.0000 | 60 | 4 | 1 | 25 | 1 | 0 |
| Disinhibition of SNARE formation | 1 | 109285777 | G | A | A | rs61797290 | upstream | 0.0453 | 0.0453 | STXBP3 | 0.2789 | 60 | 6 | 0 | 22 | 3 | 1 |
| Disinhibition of SNARE formation | 1 | 109290893 | C | T | T | rs1004424 | intronic | 0.0414 | 0.0414 | STXBP3 | 0.2789 | 60 | 6 | 0 | 22 | 3 | 1 |
| Disinhibition of SNARE formation | 1 | 109291952 | G | T | T | rs61797292 | intronic | 0.0458 | 0.0458 | STXBP3 | 0.2789 | 60 | 6 | 0 | 22 | 3 | 1 |
| Disinhibition of SNARE formation | 1 | 109292062 | G | A | A | rs61797293 | intronic | 0.0417 | 0.0417 | STXBP3 | 0.2789 | 60 | 6 | 0 | 22 | 3 | 1 |
| Disinhibition of SNARE formation | 1 | 109293596 | T | C | C | rs61797294 | intronic | 0.0465 | 0.0465 | STXBP3 | 0.2789 | 60 | 6 | 0 | 22 | 3 | 1 |
| Disinhibition of SNARE formation | 1 | 109295260 | C | G | G | rs41299559 | intronic | 0.0458 | 0.0458 | STXBP3 | 0.2699 | 60 | 6 | 0 | 21 | 3 | 1 |
| Disinhibition of SNARE formation | 1 | 109300874 | A | G | G | rs61797297 | intronic | 0.0444 | 0.0444 | STXBP3 | 0.2699 | 60 | 6 | 0 | 21 | 3 | 1 |
| Disinhibition of SNARE formation | 1 | 109308400 | C | G | G | rs61797304 | intronic | 0.0450 | 0.0450 | STXBP3 | 0.2789 | 60 | 6 | 0 | 22 | 3 | 1 |
| Disinhibition of SNARE formation | 1 | 109310759 | A | G | G | rs76989906 | intronic | 0.0458 | 0.0458 | STXBP3 | 0.2699 | 60 | 6 | 0 | 21 | 3 | 1 |
| Disinhibition of SNARE formation | 1 | 109311501 | T | G | G | rs74406644 | intronic | 0.0458 | 0.0458 | STXBP3 | 0.2789 | 60 | 6 | 0 | 22 | 3 | 1 |
| Disinhibition of SNARE formation | 1 | 109316394 | A | C | C | rs61797328 | intronic | 0.0458 | 0.0458 | STXBP3 | 0.2699 | 60 | 6 | 0 | 21 | 3 | 1 |
| Disinhibition of SNARE formation | 1 | 109317182 | C | T | T | rs74727654 | intronic | 0.0459 | 0.0459 | STXBP3 | 0.2699 | 60 | 6 | 0 | 21 | 3 | 1 |
| Disinhibition of SNARE formation | 1 | 109324694 | A | G | G | rs61797330 | intronic | 0.0413 | 0.0413 | STXBP3 | 0.2699 | 60 | 6 | 0 | 21 | 3 | 1 |
| Disinhibition of SNARE formation | 1 | 109334327 | A | C | C | rs61797332 | intronic | 0.0415 | 0.0415 | STXBP3 | 0.2699 | 60 | 6 | 0 | 21 | 3 | 1 |
| Disinhibition of SNARE formation | 1 | 109341648 | T | G | G | rs61797348 | intronic | 0.0458 | 0.0458 | STXBP3 | 0.2699 | 60 | 6 | 0 | 21 | 3 | 1 |
| Disinhibition of SNARE formation | 1 | 109347906 | G | C | C | rs61797351 | intronic | 0.0415 | 0.0415 | STXBP3 | 0.2699 | 60 | 6 | 0 | 21 | 3 | 1 |
| Disinhibition of SNARE formation | 1 | 109351109 | A | G | G | rs61797354 | intronic | 0.0419 | 0.0419 | STXBP3 | 0.2789 | 60 | 6 | 0 | 22 | 3 | 1 |
| Disinhibition of SNARE formation | 16 | 23845860 | G | T | T | rs72777910 | upstream | 0.0300 | 0.0300 | PRKCB | 0.4762 | 60 | 5 | 1 | 22 | 4 | 0 |
| Disinhibition of SNARE formation | 16 | 23849482 | T | C | T | rs2023670 | intronic | 0.9513 | 0.0487 | PRKCB | 0.0099 | 0 | 11 | 55 | 1 | 0 | 25 |
| Disinhibition of SNARE formation | 16 | 23850240 | A | G | A | rs11074581 | intronic | 0.9663 | 0.0337 | PRKCB | 0.0099 | 0 | 11 | 55 | 1 | 0 | 25 |
| Disinhibition of SNARE formation | 16 | 23851956 | T | C | T | rs7189210 | intronic | 0.9663 | 0.0337 | PRKCB | 0.0099 | 0 | 11 | 55 | 1 | 0 | 25 |
| Disinhibition of SNARE formation | 16 | 23852415 | A | T | A | rs2188359 | intronic | 0.9528 | 0.0472 | PRKCB | 0.0099 | 0 | 11 | 55 | 1 | 0 | 25 |
| Disinhibition of SNARE formation | 16 | 23859391 | A | G | G | rs62030647 | intronic | 0.0226 | 0.0226 | PRKCB | 1.0000 | 59 | 7 | 0 | 23 | 3 | 0 |
| Disinhibition of SNARE formation | 16 | 23874933 | A | C | A | rs6497691 | intronic | 0.9663 | 0.0337 | PRKCB | 0.0063 | 0 | 14 | 52 | 1 | 0 | 24 |
| Disinhibition of SNARE formation | 16 | 23876099 | C | T | T | rs79131874 | intronic | 0.0303 | 0.0303 | PRKCB | 1.0000 | 59 | 7 | 0 | 23 | 3 | 0 |
| Disinhibition of SNARE formation | 16 | 23877500 | A | G | A | rs8059885 | intronic | 0.9663 | 0.0337 | PRKCB | 0.0046 | 0 | 14 | 52 | 1 | 0 | 25 |
| Disinhibition of SNARE formation | 16 | 23877606 | A | G | A | rs8060048 | intronic | 0.9644 | 0.0356 | PRKCB | 0.0046 | 0 | 14 | 52 | 1 | 0 | 25 |
| Disinhibition of SNARE formation | 16 | 23877781 | G | A | G | rs8060718 | intronic | 0.9664 | 0.0336 | PRKCB | 0.0046 | 0 | 14 | 52 | 1 | 0 | 25 |
| Disinhibition of SNARE formation | 16 | 23878470 | C | T | C | rs12935004 | intronic | 0.9657 | 0.0343 | PRKCB | 0.0046 | 0 | 14 | 52 | 1 | 0 | 25 |
| Disinhibition of SNARE formation | 16 | 23880851 | C | T | C | rs8061523 | intronic | 0.9664 | 0.0336 | PRKCB | 0.0046 | 0 | 14 | 52 | 1 | 0 | 25 |
| Disinhibition of SNARE formation | 16 | 23881930 | G | A | G | rs8047121 | intronic | 0.9662 | 0.0338 | PRKCB | 0.0046 | 0 | 14 | 52 | 1 | 0 | 25 |
| Disinhibition of SNARE formation | 16 | 23882469 | T | C | T | rs1468129 | intronic | 0.9663 | 0.0337 | PRKCB | 0.0046 | 0 | 14 | 52 | 1 | 0 | 25 |
| Disinhibition of SNARE formation | 16 | 23885608 | A | T | A | rs8044732 | intronic | 0.9664 | 0.0336 | PRKCB | 0.0046 | 0 | 14 | 52 | 1 | 0 | 25 |
| Disinhibition of SNARE formation | 16 | 23885751 | A | G | G | rs62031692 | intronic | 0.0253 | 0.0253 | PRKCB | 1.0000 | 59 | 7 | 0 | 23 | 3 | 0 |
| Disinhibition of SNARE formation | 16 | 23887574 | G | T | T | rs79034087 | intronic | 0.0290 | 0.0290 | PRKCB | 0.3487 | 61 | 4 | 1 | 22 | 4 | 0 |
| Disinhibition of SNARE formation | 16 | 23888354 | C | T | C | rs7404417 | intronic | 0.9664 | 0.0336 | PRKCB | 0.0046 | 0 | 14 | 52 | 1 | 0 | 25 |
| Disinhibition of SNARE formation | 16 | 23889896 | T | C | T | rs8063823 | intronic | 0.9665 | 0.0335 | PRKCB | 0.0046 | 0 | 14 | 52 | 1 | 0 | 25 |
| Disinhibition of SNARE formation | 16 | 23893893 | G | A | G | rs11647359 | intronic | 0.9664 | 0.0336 | PRKCB | 0.0046 | 0 | 14 | 52 | 1 | 0 | 25 |
| Disinhibition of SNARE formation | 16 | 23895034 | A | G | A | rs6497695 | intronic | 0.9665 | 0.0335 | PRKCB | 0.0046 | 0 | 14 | 52 | 1 | 0 | 25 |
| Disinhibition of SNARE formation | 16 | 23895443 | A | G | G | rs62028075 | intronic | 0.0253 | 0.0253 | PRKCB | 1.0000 | 59 | 7 | 0 | 23 | 3 | 0 |
| Disinhibition of SNARE formation | 16 | 23895884 | T | C | T | rs9944348 | intronic | 0.9665 | 0.0335 | PRKCB | 0.0046 | 0 | 14 | 52 | 1 | 0 | 25 |
| Disinhibition of SNARE formation | 16 | 23896089 | T | C | C | rs74572166 | intronic | 0.0245 | 0.0245 | PRKCB | 1.0000 | 59 | 7 | 0 | 23 | 3 | 0 |
| Disinhibition of SNARE formation | 16 | 23896209 | C | A | C | rs9302418 | intronic | 0.9664 | 0.0336 | PRKCB | 0.0046 | 0 | 14 | 52 | 1 | 0 | 25 |
| Disinhibition of SNARE formation | 16 | 23896438 | G | T | T | rs62028076 | intronic | 0.0252 | 0.0252 | PRKCB | 1.0000 | 59 | 7 | 0 | 22 | 3 | 0 |
| Disinhibition of SNARE formation | 16 | 23898605 | A | T | A | rs933290 | intronic | 0.9632 | 0.0368 | PRKCB | 0.0016 | 0 | 17 | 49 | 1 | 0 | 24 |
| Disinhibition of SNARE formation | 16 | 23899211 | A | T | A | rs12926245 | intronic | 0.9632 | 0.0368 | PRKCB | 0.0012 | 0 | 17 | 49 | 1 | 0 | 25 |
| Disinhibition of SNARE formation | 16 | 23899610 | G | A | A | rs17753246 | intronic | 0.0252 | 0.0252 | PRKCB | 1.0000 | 59 | 7 | 0 | 23 | 3 | 0 |
| Disinhibition of SNARE formation | 16 | 23899951 | G | A | A | rs62028077 | intronic | 0.0254 | 0.0254 | PRKCB | 1.0000 | 59 | 7 | 0 | 23 | 3 | 0 |
| Disinhibition of SNARE formation | 16 | 23900716 | T | C | C | rs62028078 | intronic | 0.0252 | 0.0252 | PRKCB | 1.0000 | 59 | 7 | 0 | 23 | 3 | 0 |
| Disinhibition of SNARE formation | 16 | 23901896 | C | T | C | rs6497696 | intronic | 0.9632 | 0.0368 | PRKCB | 0.0015 | 0 | 16 | 50 | 1 | 0 | 25 |
| Disinhibition of SNARE formation | 16 | 23901948 | A | C | A | rs6497697 | intronic | 0.9630 | 0.0370 | PRKCB | 0.0015 | 0 | 16 | 49 | 1 | 0 | 25 |
| Disinhibition of SNARE formation | 16 | 23904058 | A | G | A | rs886115 | intronic | 0.9632 | 0.0368 | PRKCB | 0.0015 | 0 | 16 | 50 | 1 | 0 | 25 |
| Disinhibition of SNARE formation | 16 | 23904781 | G | A | A | rs17753509 | intronic | 0.0253 | 0.0253 | PRKCB | 1.0000 | 59 | 7 | 0 | 23 | 3 | 0 |
| Disinhibition of SNARE formation | 16 | 23905676 | C | T | C | rs7200610 | intronic | 0.9631 | 0.0369 | PRKCB | 0.0015 | 0 | 16 | 50 | 1 | 0 | 25 |
| Disinhibition of SNARE formation | 16 | 23907177 | A | C | C | rs17810011 | intronic | 0.0251 | 0.0251 | PRKCB | 1.0000 | 59 | 7 | 0 | 23 | 3 | 0 |
| Disinhibition of SNARE formation | 16 | 23907765 | C | T | C | rs9925890 | intronic | 0.9632 | 0.0368 | PRKCB | 0.0024 | 0 | 16 | 50 | 1 | 0 | 24 |
| Disinhibition of SNARE formation | 16 | 23912174 | A | G | A | rs12448249 | intronic | 0.9519 | 0.0481 | PRKCB | 0.0015 | 0 | 16 | 50 | 1 | 0 | 25 |
| Disinhibition of SNARE formation | 16 | 23914915 | C | A | C | rs1004186 | intronic | 0.9632 | 0.0368 | PRKCB | 0.0015 | 0 | 16 | 50 | 1 | 0 | 25 |
| Disinhibition of SNARE formation | 16 | 23916258 | G | A | G | rs1004187 | intronic | 0.9632 | 0.0368 | PRKCB | 0.0015 | 0 | 16 | 50 | 1 | 0 | 25 |
| Disinhibition of SNARE formation | 16 | 23916521 | G | C | G | rs1008654 | intronic | 0.9633 | 0.0367 | PRKCB | 0.0015 | 0 | 16 | 50 | 1 | 0 | 25 |
| Disinhibition of SNARE formation | 16 | 23917335 | G | A | G | rs6497699 | intronic | 0.9645 | 0.0355 | PRKCB | 0.0024 | 0 | 16 | 50 | 1 | 0 | 24 |
| Disinhibition of SNARE formation | 16 | 23917465 | C | G | C | rs7186538 | intronic | 0.9645 | 0.0355 | PRKCB | 0.0024 | 0 | 16 | 50 | 1 | 0 | 24 |
| Disinhibition of SNARE formation | 16 | 23917700 | C | A | C | rs7187091 | intronic | 0.9646 | 0.0354 | PRKCB | 0.0024 | 0 | 16 | 50 | 1 | 0 | 24 |
| Disinhibition of SNARE formation | 16 | 23919088 | C | T | T | rs78322646 | intronic | 0.0278 | 0.0278 | PRKCB | 0.7348 | 58 | 8 | 0 | 22 | 4 | 0 |
| Disinhibition of SNARE formation | 16 | 23921083 | C | T | C | rs6497702 | intronic | 0.9647 | 0.0353 | PRKCB | 0.0024 | 0 | 16 | 50 | 1 | 0 | 24 |
| Disinhibition of SNARE formation | 16 | 23925936 | C | G | C | rs11074588 | intronic | 0.9649 | 0.0351 | PRKCB | 0.0015 | 0 | 16 | 50 | 1 | 0 | 25 |
| Disinhibition of SNARE formation | 16 | 23939212 | G | A | G | rs11074590 | intronic | 0.9650 | 0.0350 | PRKCB | 0.0024 | 0 | 16 | 50 | 1 | 0 | 24 |
| Disinhibition of SNARE formation | 16 | 23941628 | C | A | C | rs2005671 | intronic | 0.9647 | 0.0353 | PRKCB | 0.0015 | 0 | 16 | 49 | 1 | 0 | 25 |
| Disinhibition of SNARE formation | 16 | 23943749 | T | C | T | rs9302420 | intronic | 0.9649 | 0.0351 | PRKCB | 0.0024 | 0 | 16 | 50 | 1 | 0 | 24 |
| Disinhibition of SNARE formation | 16 | 23945985 | T | G | T | rs195989 | intronic | 0.9651 | 0.0349 | PRKCB | 0.0015 | 0 | 16 | 50 | 1 | 0 | 25 |
| Disinhibition of SNARE formation | 16 | 23946157 | G | A | A | rs76973283 | intronic | 0.0302 | 0.0302 | PRKCB | 0.5065 | 58 | 8 | 0 | 21 | 5 | 0 |
| Disinhibition of SNARE formation | 16 | 23949175 | G | C | G | rs2560403 | intronic | 0.9657 | 0.0343 | PRKCB | 0.0015 | 0 | 16 | 50 | 1 | 0 | 25 |
| Disinhibition of SNARE formation | 16 | 23949438 | A | G | A | rs195985 | intronic | 0.9658 | 0.0342 | PRKCB | 0.0015 | 0 | 16 | 50 | 1 | 0 | 25 |
| Disinhibition of SNARE formation | 16 | 23953265 | T | C | T | rs2560404 | intronic | 0.9656 | 0.0344 | PRKCB | 0.0024 | 0 | 16 | 50 | 1 | 0 | 24 |
| Disinhibition of SNARE formation | 16 | 23954128 | T | C | C | rs17810486 | intronic | 0.0308 | 0.0308 | PRKCB | 0.3024 | 58 | 7 | 0 | 20 | 5 | 0 |
| Disinhibition of SNARE formation | 16 | 23954253 | G | A | G | rs195994 | intronic | 0.9653 | 0.0347 | PRKCB | 0.0015 | 0 | 16 | 50 | 1 | 0 | 25 |
| Disinhibition of SNARE formation | 16 | 23962258 | G | C | G | rs196000 | intronic | 0.9659 | 0.0341 | PRKCB | 0.0015 | 0 | 16 | 50 | 1 | 0 | 25 |
| Disinhibition of SNARE formation | 16 | 23964858 | T | A | T | rs196003 | intronic | 0.9647 | 0.0353 | PRKCB | 0.0343 | 0 | 15 | 51 | 0 | 1 | 25 |
| Disinhibition of SNARE formation | 16 | 23985814 | C | T | T | rs72779914 | intronic | 0.0487 | 0.0487 | PRKCB | 1.0000 | 53 | 12 | 0 | 21 | 4 | 0 |
| Disinhibition of SNARE formation | 16 | 23987552 | A | G | A | rs169030 | intronic | 0.9709 | 0.0291 | PRKCB | 0.1679 | 0 | 11 | 55 | 0 | 1 | 25 |
| Disinhibition of SNARE formation | 16 | 23988755 | T | C | T | rs196013 | intronic | 0.9681 | 0.0319 | PRKCB | 0.1681 | 0 | 11 | 55 | 0 | 1 | 24 |
| Disinhibition of SNARE formation | 16 | 24009919 | A | G | G | rs75622923 | intronic | 0.0319 | 0.0319 | PRKCB | 0.3348 | 54 | 12 | 0 | 24 | 2 | 0 |
| Disinhibition of SNARE formation | 16 | 24022944 | C | T | T | rs111746132 | intronic | 0.0229 | 0.0229 | PRKCB | 0.7522 | 56 | 10 | 0 | 23 | 3 | 0 |
| Disinhibition of SNARE formation | 16 | 24066378 | G | A | A | rs113426570 | intronic | 0.0216 | 0.0216 | PRKCB | 1.0000 | 57 | 9 | 0 | 23 | 3 | 0 |
| Disinhibition of SNARE formation | 16 | 24100759 | T | A | A | rs11643939 | intronic | 0.0294 | 0.0294 | PRKCB | 0.7522 | 56 | 10 | 0 | 23 | 3 | 0 |
| Disinhibition of SNARE formation | 16 | 24105816 | G | A | A | rs56316329 | intronic | 0.0251 | 0.0251 | PRKCB | 0.6919 | 60 | 4 | 1 | 25 | 0 | 0 |
| Disinhibition of SNARE formation | 16 | 24111853 | T | C | C | rs55959083 | intronic | 0.0431 | 0.0431 | PRKCB | 0.7213 | 58 | 8 | 0 | 23 | 2 | 0 |
| Disinhibition of SNARE formation | 16 | 24112768 | G | A | A | rs117056307 | intronic | 0.0430 | 0.0430 | PRKCB | 0.7192 | 58 | 8 | 0 | 24 | 2 | 0 |
| Disinhibition of SNARE formation | 16 | 24122052 | G | A | A | rs117467859 | intronic | 0.0433 | 0.0433 | PRKCB | 0.7213 | 58 | 8 | 0 | 23 | 2 | 0 |
| Disinhibition of SNARE formation | 16 | 24122492 | C | T | T | rs72779977 | intronic | 0.0457 | 0.0457 | PRKCB | 0.7201 | 57 | 8 | 0 | 23 | 2 | 0 |
| Disinhibition of SNARE formation | 16 | 24123560 | G | A | A | rs60261043 | intronic | 0.0457 | 0.0457 | PRKCB | 0.7201 | 57 | 8 | 0 | 23 | 2 | 0 |
| Disinhibition of SNARE formation | 16 | 24132273 | G | A | A | rs62027458 | intronic | 0.0232 | 0.0232 | PRKCB | 0.1877 | 59 | 7 | 0 | 21 | 4 | 1 |
| Disinhibition of SNARE formation | 16 | 24164042 | G | T | T | rs72779989 | intronic | 0.0487 | 0.0487 | PRKCB | 0.8583 | 57 | 7 | 2 | 22 | 4 | 0 |
| Disinhibition of SNARE formation | 16 | 24197496 | A | T | T | rs79699525 | intronic | 0.0261 | 0.0261 | PRKCB | 0.4631 | 59 | 6 | 0 | 22 | 4 | 0 |
| Disinhibition of SNARE formation | 16 | 24199852 | C | T | T | rs78424166 | intronic | 0.0359 | 0.0359 | PRKCB | 0.4603 | 60 | 6 | 0 | 22 | 4 | 0 |
| Disinhibition of SNARE formation | 17 | 64305051 | A | G | G | rs78357146 | intronic | 0.0186 | 0.0186 | PRKCA | 0.7209 | 57 | 9 | 0 | 23 | 2 | 0 |
| Disinhibition of SNARE formation | 17 | 64315409 | T | C | C | rs80130647 | intronic | 0.0184 | 0.0184 | PRKCA | 0.7043 | 59 | 6 | 0 | 22 | 3 | 0 |
| Disinhibition of SNARE formation | 17 | 64318385 | G | A | A | rs12150623 | intronic | 0.0343 | 0.0343 | PRKCA | 0.7595 | 61 | 4 | 1 | 23 | 2 | 0 |
| Disinhibition of SNARE formation | 17 | 64320040 | C | T | T | rs139317720 | intronic | 0.0229 | 0.0229 | PRKCA | 1.0000 | 62 | 3 | 1 | 24 | 1 | 0 |
| Disinhibition of SNARE formation | 17 | 64326068 | T | G | G | rs12951126 | intronic | 0.0260 | 0.0260 | PRKCA | 1.0000 | 58 | 8 | 0 | 23 | 3 | 0 |
| Disinhibition of SNARE formation | 17 | 64343295 | A | C | C | rs72843901 | intronic | 0.0457 | 0.0457 | PRKCA | 0.4725 | 56 | 9 | 1 | 24 | 1 | 0 |
| Disinhibition of SNARE formation | 17 | 64344650 | C | T | T | rs72846606 | intronic | 0.0454 | 0.0454 | PRKCA | 0.4725 | 56 | 9 | 1 | 24 | 1 | 0 |

| Pathway | Chr | Position | Ref | Alt | Alt2 | rsID | Region | Freq1 | Freq2 | Gene | P-value | N1 | N2 | N3 | N4 | N5 | N6 |
|---|---|---|---|---|---|---|---|---|---|---|---|---|---|---|---|---|---|
| Disinhibition of SNARE formation | 17 | 64344788 | C | T | T | rs72846607 | intronic | 0.0457 | 0.0457 | PRKCA | 0.4725 | 56 | 9 | 1 | 24 | 1 | 0 |
| Disinhibition of SNARE formation | 17 | 64346204 | A | G | G | rs12150367 | intronic | 0.0452 | 0.0452 | PRKCA | 0.4770 | 56 | 9 | 1 | 25 | 1 | 0 |
| Disinhibition of SNARE formation | 17 | 64349104 | C | T | T | rs72846609 | intronic | 0.0497 | 0.0497 | PRKCA | 0.4770 | 56 | 9 | 1 | 25 | 1 | 0 |
| Disinhibition of SNARE formation | 17 | 64351542 | G | C | C | rs28592028 | intronic | 0.0453 | 0.0453 | PRKCA | 0.4770 | 56 | 9 | 1 | 25 | 1 | 0 |
| Disinhibition of SNARE formation | 17 | 64352431 | T | G | G | rs544435459 | intronic | 0.0451 | 0.0451 | PRKCA | 0.4725 | 56 | 9 | 1 | 24 | 1 | 0 |
| Disinhibition of SNARE formation | 17 | 64354597 | T | C | C | rs72846612 | intronic | 0.0451 | 0.0451 | PRKCA | 0.4725 | 56 | 9 | 1 | 24 | 1 | 0 |
| Disinhibition of SNARE formation | 17 | 64358121 | C | G | G | rs72846614 | intronic | 0.0451 | 0.0451 | PRKCA | 0.4770 | 56 | 9 | 1 | 25 | 1 | 0 |
| Disinhibition of SNARE formation | 17 | 64359354 | T | C | C | rs72846615 | intronic | 0.0452 | 0.0452 | PRKCA | 0.4725 | 56 | 9 | 1 | 24 | 1 | 0 |
| Disinhibition of SNARE formation | 17 | 64365593 | C | T | T | rs72846677 | intronic | 0.0452 | 0.0452 | PRKCA | 0.4770 | 56 | 9 | 1 | 25 | 1 | 0 |
| Disinhibition of SNARE formation | 17 | 64367534 | C | T | T | rs72846678 | intronic | 0.0452 | 0.0452 | PRKCA | 0.4725 | 56 | 9 | 1 | 24 | 1 | 0 |
| Disinhibition of SNARE formation | 17 | 64369822 | A | G | G | rs72846681 | intronic | 0.0451 | 0.0451 | PRKCA | 0.4770 | 56 | 9 | 1 | 25 | 1 | 0 |
| Disinhibition of SNARE formation | 17 | 64375194 | T | C | C | rs72846695 | intronic | 0.0452 | 0.0452 | PRKCA | 0.4770 | 56 | 9 | 1 | 25 | 1 | 0 |
| Disinhibition of SNARE formation | 17 | 64377301 | C | A | A | rs113134992 | intronic | 0.0448 | 0.0448 | PRKCA | 0.4770 | 56 | 9 | 1 | 25 | 1 | 0 |
| Disinhibition of SNARE formation | 17 | 64382507 | C | G | G | rs79461368 | intronic | 0.0453 | 0.0453 | PRKCA | 0.4725 | 56 | 9 | 1 | 24 | 1 | 0 |
| Disinhibition of SNARE formation | 17 | 64389524 | C | T | T | rs77682324 | intronic | 0.0403 | 0.0403 | PRKCA | 0.7621 | 59 | 6 | 1 | 25 | 1 | 0 |
| Disinhibition of SNARE formation | 17 | 64395773 | T | C | C | rs78584531 | intronic | 0.0452 | 0.0452 | PRKCA | 0.4770 | 56 | 9 | 1 | 25 | 1 | 0 |
| Disinhibition of SNARE formation | 17 | 64396570 | T | G | G | rs72838208 | intronic | 0.0452 | 0.0452 | PRKCA | 0.4725 | 56 | 9 | 1 | 24 | 1 | 0 |
| Disinhibition of SNARE formation | 17 | 64396695 | A | G | G | rs74329211 | intronic | 0.0432 | 0.0432 | PRKCA | 0.7302 | 58 | 8 | 0 | 21 | 4 | 0 |
| Disinhibition of SNARE formation | 17 | 64396883 | G | A | A | rs72838209 | intronic | 0.0452 | 0.0452 | PRKCA | 0.4770 | 56 | 9 | 1 | 25 | 1 | 0 |
| Disinhibition of SNARE formation | 17 | 64398013 | G | A | A | rs75125285 | intronic | 0.0295 | 0.0295 | PRKCA | 0.7632 | 60 | 5 | 1 | 25 | 1 | 0 |
| Disinhibition of SNARE formation | 17 | 64400648 | C | T | T | rs72838214 | intronic | 0.0452 | 0.0452 | PRKCA | 0.4788 | 55 | 9 | 1 | 25 | 1 | 0 |
| Disinhibition of SNARE formation | 17 | 64402141 | G | T | T | rs10221238 | intronic | 0.0452 | 0.0452 | PRKCA | 0.4770 | 56 | 9 | 1 | 25 | 1 | 0 |
| Disinhibition of SNARE formation | 17 | 64405430 | T | A | A | rs72838216 | intronic | 0.0453 | 0.0453 | PRKCA | 0.4725 | 56 | 9 | 1 | 24 | 1 | 0 |
| Disinhibition of SNARE formation | 17 | 64410564 | C | T | T | rs9972974 | intronic | 0.0390 | 0.0390 | PRKCA | 0.7632 | 60 | 5 | 1 | 25 | 1 | 0 |
| Disinhibition of SNARE formation | 17 | 64412921 | G | A | A | rs62070391 | intronic | 0.0426 | 0.0426 | PRKCA | 0.7621 | 59 | 6 | 1 | 25 | 1 | 0 |
| Disinhibition of SNARE formation | 17 | 64424382 | C | T | T | rs62070395 | intronic | 0.0425 | 0.0425 | PRKCA | 0.7595 | 59 | 6 | 1 | 24 | 1 | 0 |
| Disinhibition of SNARE formation | 17 | 64429928 | G | A | A | rs117729211 | intronic | 0.0272 | 0.0272 | PRKCA | 0.0477 | 60 | 6 | 0 | 19 | 6 | 1 |
| Disinhibition of SNARE formation | 17 | 64431874 | G | A | A | rs72838278 | intronic | 0.0426 | 0.0426 | PRKCA | 0.7642 | 57 | 6 | 1 | 25 | 1 | 0 |
| Disinhibition of SNARE formation | 17 | 64432202 | A | G | G | rs62070397 | intronic | 0.0427 | 0.0427 | PRKCA | 0.7621 | 59 | 6 | 1 | 25 | 1 | 0 |
| Disinhibition of SNARE formation | 17 | 64432881 | A | G | G | rs62070398 | intronic | 0.0426 | 0.0426 | PRKCA | 0.7595 | 59 | 6 | 1 | 24 | 1 | 0 |
| Disinhibition of SNARE formation | 17 | 64438732 | A | T | T | rs113684166 | intronic | 0.0445 | 0.0445 | PRKCA | 0.2744 | 57 | 9 | 0 | 24 | 1 | 0 |
| Disinhibition of SNARE formation | 17 | 64441204 | T | C | C | rs75005068 | intronic | 0.0469 | 0.0469 | PRKCA | 0.2719 | 57 | 9 | 0 | 25 | 1 | 0 |
| Disinhibition of SNARE formation | 17 | 64441759 | C | A | A | rs78330327 | intronic | 0.0465 | 0.0465 | PRKCA | 0.2710 | 56 | 9 | 0 | 25 | 1 | 0 |
| Disinhibition of SNARE formation | 17 | 64445337 | A | C | C | rs118090701 | intronic | 0.0387 | 0.0387 | PRKCA | 1.0000 | 58 | 6 | 1 | 23 | 2 | 0 |
| Disinhibition of SNARE formation | 17 | 64445856 | A | G | G | rs12451388 | intronic | 0.0449 | 0.0449 | PRKCA | 0.2744 | 57 | 9 | 0 | 24 | 1 | 0 |
| Disinhibition of SNARE formation | 17 | 64447721 | G | A | A | rs12452749 | intronic | 0.0462 | 0.0462 | PRKCA | 0.2786 | 57 | 9 | 0 | 23 | 1 | 0 |
| Disinhibition of SNARE formation | 17 | 64450113 | G | C | C | rs80162292 | intronic | 0.0461 | 0.0461 | PRKCA | 0.2710 | 56 | 9 | 0 | 25 | 1 | 0 |
| Disinhibition of SNARE formation | 17 | 64453194 | A | G | G | rs111776777 | intronic | 0.0442 | 0.0442 | PRKCA | 0.8716 | 54 | 10 | 2 | 23 | 3 | 0 |
| Disinhibition of SNARE formation | 17 | 64458036 | C | T | T | rs113153197 | intronic | 0.0339 | 0.0339 | PRKCA | 0.1678 | 54 | 11 | 0 | 25 | 1 | 0 |
| Disinhibition of SNARE formation | 17 | 64462111 | T | G | G | rs79239451 | intronic | 0.0464 | 0.0464 | PRKCA | 0.2744 | 57 | 9 | 0 | 24 | 1 | 0 |
| Disinhibition of SNARE formation | 17 | 64462288 | G | A | A | rs80080003 | intronic | 0.0305 | 0.0305 | PRKCA | 0.1706 | 56 | 10 | 0 | 25 | 1 | 0 |
| Disinhibition of SNARE formation | 17 | 64483589 | G | T | T | rs79070174 | intronic | 0.0391 | 0.0391 | PRKCA | 1.0000 | 59 | 6 | 1 | 23 | 2 | 0 |
| Disinhibition of SNARE formation | 17 | 64487077 | T | G | G | rs78149509 | intronic | 0.0200 | 0.0200 | PRKCA | 0.0381 | 63 | 3 | 0 | 21 | 4 | 1 |
| Disinhibition of SNARE formation | 17 | 64494906 | A | G | G | rs117168126 | intronic | 0.0300 | 0.0300 | PRKCA | 1.0000 | 59 | 6 | 1 | 23 | 2 | 0 |
| Disinhibition of SNARE formation | 17 | 64501943 | C | T | T | rs7217954 | intronic | 0.0422 | 0.0422 | PRKCA | 1.0000 | 59 | 7 | 0 | 23 | 3 | 0 |
| Disinhibition of SNARE formation | 17 | 64519790 | G | A | A | rs11659067 | intronic | 0.0498 | 0.0498 | PRKCA | 0.0125 | 62 | 3 | 0 | 19 | 6 | 0 |
| Disinhibition of SNARE formation | 17 | 64525002 | A | T | T | rs77462363 | intronic | 0.0380 | 0.0380 | PRKCA | 1.0000 | 59 | 6 | 1 | 23 | 2 | 0 |
| Disinhibition of SNARE formation | 17 | 64544723 | C | T | T | rs116879811 | intronic | 0.0274 | 0.0274 | PRKCA | 1.0000 | 60 | 6 | 0 | 24 | 2 | 0 |
| Disinhibition of SNARE formation | 17 | 64559535 | A | G | G | rs227907 | intronic | 0.0478 | 0.0478 | PRKCA | 0.2800 | 55 | 11 | 0 | 21 | 3 | 1 |
| Disinhibition of SNARE formation | 17 | 64561055 | G | A | A | rs62071706 | intronic | 0.0238 | 0.0238 | PRKCA | 0.1546 | 60 | 5 | 0 | 21 | 4 | 1 |
| Disinhibition of SNARE formation | 17 | 64590951 | A | T | T | rs11867591 | intronic | 0.0449 | 0.0449 | PRKCA | 0.2122 | 57 | 8 | 0 | 20 | 6 | 0 |
| Disinhibition of SNARE formation | 17 | 64604776 | C | T | T | rs117539643 | intronic | 0.0479 | 0.0479 | PRKCA | 0.6653 | 51 | 11 | 2 | 23 | 3 | 0 |
| Disinhibition of SNARE formation | 17 | 64608923 | T | C | C | rs72845947 | intronic | 0.0440 | 0.0440 | PRKCA | 0.7355 | 57 | 7 | 2 | 21 | 4 | 0 |
| Disinhibition of SNARE formation | 17 | 64610285 | T | C | C | rs72845948 | intronic | 0.0445 | 0.0445 | PRKCA | 0.8583 | 57 | 7 | 2 | 22 | 4 | 0 |
| Disinhibition of SNARE formation | 17 | 64610480 | A | C | C | rs17759657 | intronic | 0.0444 | 0.0444 | PRKCA | 0.8583 | 57 | 7 | 2 | 22 | 4 | 0 |
| Disinhibition of SNARE formation | 17 | 64612838 | C | T | T | rs16959714 | intronic | 0.0338 | 0.0338 | PRKCA | 0.4353 | 57 | 8 | 0 | 24 | 1 | 0 |
| Disinhibition of SNARE formation | 17 | 64614717 | T | C | C | rs17686540 | intronic | 0.0442 | 0.0442 | PRKCA | 0.7355 | 57 | 7 | 2 | 21 | 4 | 0 |
| Disinhibition of SNARE formation | 17 | 64626385 | G | A | A | rs74352723 | intronic | 0.0460 | 0.0460 | PRKCA | 0.2744 | 57 | 9 | 0 | 24 | 1 | 0 |
| Disinhibition of SNARE formation | 17 | 64628634 | A | G | G | rs117353888 | intronic | 0.0437 | 0.0437 | PRKCA | 0.2521 | 54 | 11 | 0 | 18 | 7 | 0 |
| Disinhibition of SNARE formation | 17 | 64660047 | G | C | C | rs146141011 | intronic | 0.0484 | 0.0484 | PRKCA | 1.0000 | 53 | 9 | 2 | 22 | 4 | 0 |
| Disinhibition of SNARE formation | 17 | 64671047 | C | T | T | rs16959942 | intronic | 0.0496 | 0.0496 | PRKCA | 0.1706 | 56 | 10 | 0 | 25 | 1 | 0 |
| Disinhibition of SNARE formation | 17 | 64712634 | T | C | C | rs78121420 | intronic | 0.0465 | 0.0465 | PRKCA | 1.0000 | 57 | 7 | 2 | 23 | 3 | 0 |
| Disinhibition of SNARE formation | 17 | 64713847 | G | T | T | rs112934229 | intronic | 0.0464 | 0.0464 | PRKCA | 1.0000 | 57 | 7 | 2 | 23 | 3 | 0 |
| Disinhibition of SNARE formation | 17 | 64723880 | G | A | A | rs79547774 | intronic | 0.0264 | 0.0264 | PRKCA | 1.0000 | 57 | 9 | 0 | 23 | 3 | 0 |
| Disinhibition of SNARE formation | 17 | 64738427 | A | C | C | rs117138620 | intronic | 0.0400 | 0.0400 | PRKCA | 0.3055 | 58 | 8 | 0 | 23 | 2 | 1 |
| Disinhibition of SNARE formation | 17 | 64748431 | G | T | T | rs141177250 | intronic | 0.0153 | 0.0153 | PRKCA | 0.7595 | 59 | 6 | 1 | 24 | 1 | 0 |
| Disinhibition of SNARE formation | 17 | 64755018 | T | C | C | rs74831470 | intronic | 0.0226 | 0.0226 | PRKCA | 0.7976 | 56 | 9 | 1 | 23 | 2 | 0 |
| Disinhibition of SNARE formation | 17 | 64760870 | A | G | G | rs77904275 | intronic | 0.0307 | 0.0307 | PRKCA | 0.5233 | 53 | 11 | 1 | 24 | 2 | 0 |
| Disinhibition of SNARE formation | 17 | 64762410 | C | G | G | rs113025478 | intronic | 0.0307 | 0.0307 | PRKCA | 0.6398 | 54 | 11 | 1 | 23 | 2 | 0 |
| Disinhibition of SNARE formation | 17 | 64763235 | T | C | C | rs77635068 | intronic | 0.0306 | 0.0306 | PRKCA | 0.6398 | 54 | 11 | 1 | 23 | 2 | 0 |
| Disinhibition of SNARE formation | 17 | 64771614 | C | T | T | rs80238933 | intronic | 0.0297 | 0.0297 | PRKCA | 0.6398 | 54 | 11 | 1 | 23 | 2 | 0 |
| Disinhibition of SNARE formation | 17 | 64776847 | G | A | A | rs113542727 | intronic | 0.0297 | 0.0297 | PRKCA | 0.5228 | 54 | 11 | 1 | 24 | 2 | 0 |
| Disinhibition of SNARE formation | 17 | 64791836 | G | C | C | rs56884788 | intronic | 0.0390 | 0.0390 | PRKCA | 0.6362 | 55 | 10 | 1 | 23 | 2 | 0 |
| Disinhibition of SNARE formation | 17 | 64792863 | G | C | C | rs72838636 | intronic | 0.0484 | 0.0484 | PRKCA | 0.6470 | 57 | 8 | 1 | 21 | 5 | 0 |
| Disinhibition of SNARE formation | 19 | 54401602 | G | C | C | rs41311973 | intronic | 0.0163 | 0.0163 | PRKCG | 0.7213 | 58 | 8 | 0 | 23 | 2 | 0 |
| Disassembly of the destruction complex and recruitment of AXIN to the membrane | 1 | 212485607 | A | T | A | rs351381 | intronic | 0.9528 | 0.0472 | PPP2R5A | 0.2993 | 0 | 7 | 59 | 0 | 5 | 20 |
| Disassembly of the destruction complex and recruitment of AXIN to the membrane | 1 | 212505662 | A | G | G | rs17665257 | intronic | 0.0348 | 0.0348 | PPP2R5A | 0.6148 | 59 | 6 | 1 | 22 | 4 | 0 |
| Disassembly of the destruction complex and recruitment of AXIN to the membrane | 1 | 212507249 | C | T | T | rs61828730 | intronic | 0.0408 | 0.0408 | PPP2R5A | 0.8011 | 58 | 7 | 1 | 22 | 4 | 0 |
| Disassembly of the destruction complex and recruitment of AXIN to the membrane | 1 | 228204651 | G | A | A | rs12747212 | intronic | 0.0465 | 0.0465 | WNT3A | 0.1984 | 58 | 8 | 0 | 23 | 1 | 1 |
| Disassembly of the destruction complex and recruitment of AXIN to the membrane | 2 | 208631211 | A | T | T | rs74471859 | 3_prime_UTR | 0.0420 | 0.0420 | FZD5 | 0.2352 | 56 | 10 | 0 | 19 | 7 | 0 |
| Disassembly of the destruction complex and recruitment of AXIN to the membrane | 2 | 208632817 | G | A | A | rs35994626 | missense | 0.0418 | 0.0418 | FZD5 | 0.2352 | 56 | 10 | 0 | 19 | 7 | 0 |
| Disassembly of the destruction complex and recruitment of AXIN to the membrane | 3 | 41241993 | A | G | G | rs11564437 | intronic | 0.0255 | 0.0255 | CTNNB1 | 0.2725 | 63 | 2 | 1 | 23 | 3 | 0 |
| Disassembly of the destruction complex and recruitment of AXIN to the membrane | 3 | 119663780 | G | A | A | rs114872182 | intronic | 0.0222 | 0.0222 | GSK3B | 0.8074 | 61 | 4 | 1 | 24 | 1 | 1 |
| Disassembly of the destruction complex and recruitment of AXIN to the membrane | 3 | 119674978 | G | C | C | rs115950158 | intronic | 0.0384 | 0.0384 | GSK3B | 0.2761 | 56 | 10 | 0 | 23 | 1 | 0 |
| Disassembly of the destruction complex and recruitment of AXIN to the membrane | 3 | 119734170 | T | C | C | rs116755725 | intronic | 0.0237 | 0.0237 | GSK3B | 0.0397 | 54 | 11 | 1 | 26 | 0 | 0 |
| Disassembly of the destruction complex and recruitment of AXIN to the membrane | 3 | 119736734 | T | C | C | rs78796384 | intronic | 0.0177 | 0.0177 | GSK3B | 0.1282 | 57 | 8 | 1 | 25 | 0 | 0 |
| Disassembly of the destruction complex and recruitment of AXIN to the membrane | 3 | 119752368 | T | A | A | rs78883820 | intronic | 0.0403 | 0.0403 | GSK3B | 1.0000 | 52 | 13 | 1 | 20 | 5 | 0 |
| Disassembly of the destruction complex and recruitment of AXIN to the membrane | 3 | 119785723 | C | T | T | rs75142045 | intronic | 0.0236 | 0.0236 | GSK3B | 0.0391 | 54 | 11 | 1 | 25 | 0 | 0 |
| Disassembly of the destruction complex and recruitment of AXIN to the membrane | 3 | 119795926 | C | T | T | rs75192446 | intronic | 0.0409 | 0.0409 | GSK3B | 1.0000 | 52 | 13 | 1 | 20 | 5 | 0 |
| Disassembly of the destruction complex and recruitment of AXIN to the membrane | 3 | 119816610 | G | A | A | rs114798987 | upstream | 0.0229 | 0.0229 | GSK3B | 0.0646 | 61 | 5 | 0 | 19 | 6 | 0 |
| Disassembly of the destruction complex and recruitment of AXIN to the membrane | 5 | 112077680 | T | C | C | rs62363988 | intronic | 0.0273 | 0.0273 | APC | 0.7207 | 62 | 3 | 1 | 23 | 2 | 0 |
| Disassembly of the destruction complex and recruitment of AXIN to the membrane | 5 | 112124405 | C | T | T | rs62364020 | intronic | 0.0266 | 0.0266 | APC | 0.7207 | 62 | 3 | 1 | 23 | 2 | 0 |
| Disassembly of the destruction complex and recruitment of AXIN to the membrane | 5 | 112141114 | C | G | G | rs62364022 | intronic | 0.0270 | 0.0270 | APC | 0.7207 | 62 | 3 | 1 | 23 | 2 | 0 |
| Disassembly of the destruction complex and recruitment of AXIN to the membrane | 5 | 112164747 | A | G | G | rs62364023 | intronic | 0.0271 | 0.0271 | APC | 0.7285 | 62 | 3 | 1 | 24 | 2 | 0 |
| Disassembly of the destruction complex and recruitment of AXIN to the membrane | 5 | 133551516 | A | G | G | rs75368011 | intronic | 0.0484 | 0.0484 | PPP2CA | 0.0685 | 61 | 5 | 0 | 20 | 6 | 0 |
| Disassembly of the destruction complex and recruitment of AXIN to the membrane | 5 | 133553213 | T | A | A | rs79758263 | intronic | 0.0483 | 0.0483 | PPP2CA | 0.0646 | 61 | 5 | 0 | 19 | 6 | 0 |
| Disassembly of the destruction complex and recruitment of AXIN to the membrane | 5 | 133554676 | A | G | G | rs71587522 | intronic | 0.0288 | 0.0288 | PPP2CA | 0.7497 | 55 | 11 | 0 | 23 | 3 | 0 |
| Disassembly of the destruction complex and recruitment of AXIN to the membrane | 5 | 137417081 | A | G | G | rs60251892 | upstream | 0.0479 | 0.0479 | WNT8A | 0.4762 | 60 | 5 | 1 | 22 | 4 | 0 |
| Disassembly of the destruction complex and recruitment of AXIN to the membrane | 5 | 137429148 | C | T | T | rs146145625 | downstream | 0.0305 | 0.0305 | WNT8A | 1.0000 | 57 | 8 | 0 | 22 | 3 | 0 |
| Disassembly of the destruction complex and recruitment of AXIN to the membrane | 5 | 148887275 | C | A | A | rs114967510 | intronic | 0.0299 | 0.0299 | CSNK1A1 | 1.0000 | 56 | 10 | 0 | 22 | 4 | 0 |
| Disassembly of the destruction complex and recruitment of AXIN to the membrane | 5 | 148921266 | C | T | T | rs114031822 | intronic | 0.0172 | 0.0172 | CSNK1A1 | 1.0000 | 58 | 7 | 0 | 23 | 3 | 0 |
| Disassembly of the destruction complex and recruitment of AXIN to the membrane | 6 | 42955700 | G | A | A | rs150318932 | intronic | 0.0184 | 0.0184 | PPP2R5D | 0.0188 | 62 | 2 | 1 | 21 | 5 | 0 |
| Disassembly of the destruction complex and recruitment of AXIN to the membrane | 6 | 42956855 | C | G | G | rs139678883 | intronic | 0.0185 | 0.0185 | PPP2R5D | 0.0179 | 63 | 2 | 1 | 21 | 5 | 0 |
| Disassembly of the destruction complex and recruitment of AXIN to the membrane | 6 | 42956915 | G | A | A | rs116911410 | intronic | 0.0185 | 0.0185 | PPP2R5D | 0.0179 | 63 | 2 | 1 | 21 | 5 | 0 |
| Disassembly of the destruction complex and recruitment of AXIN to the membrane | 6 | 42958019 | T | C | C | rs9462861 | intronic | 0.0181 | 0.0181 | PPP2R5D | 0.1111 | 62 | 4 | 0 | 21 | 5 | 0 |
| Disassembly of the destruction complex and recruitment of AXIN to the membrane | 6 | 42958084 | A | G | G | rs114754270 | intronic | 0.0185 | 0.0185 | PPP2R5D | 0.0179 | 63 | 2 | 1 | 21 | 5 | 0 |
| Disassembly of the destruction complex and recruitment of AXIN to the membrane | 6 | 42967569 | G | A | A | rs143336223 | intronic | 0.0182 | 0.0182 | PPP2R5D | 0.0179 | 63 | 2 | 1 | 21 | 5 | 0 |
| Disassembly of the destruction complex and recruitment of AXIN to the membrane | 6 | 42970465 | A | G | G | rs78833196 | intronic | 0.0403 | 0.0403 | PPP2R5D | 0.2977 | 54 | 12 | 0 | 22 | 3 | 1 |
| Disassembly of the destruction complex and recruitment of AXIN to the membrane | 6 | 42978824 | C | T | T | rs41274902 | intronic | 0.0182 | 0.0182 | PPP2R5D | 0.0179 | 63 | 2 | 1 | 21 | 5 | 0 |
| Disassembly of the destruction complex and recruitment of AXIN to the membrane | 7 | 90892588 | C | T | T | rs78521924 | upstream | 0.0215 | 0.0215 | FZD1 | 0.7469 | 57 | 9 | 0 | 21 | 4 | 0 |
| Disassembly of the destruction complex and recruitment of AXIN to the membrane | 7 | 116165233 | C | T | T | rs45498702 | intronic | 0.0462 | 0.0462 | CAV1 | 0.2078 | 58 | 8 | 0 | 20 | 6 | 0 |
| Disassembly of the destruction complex and recruitment of AXIN to the membrane | 7 | 116187330 | G | A | A | rs1474511 | intronic | 0.0342 | 0.0342 | CAV1 | 0.0534 | 59 | 7 | 0 | 18 | 7 | 0 |
| Disassembly of the destruction complex and recruitment of AXIN to the membrane | 7 | 116192330 | C | T | T | rs75022895 | intronic | 0.0478 | 0.0478 | CAV1 | 0.0150 | 58 | 8 | 0 | 16 | 9 | 0 |
| Disassembly of the destruction complex and recruitment of AXIN to the membrane | 7 | 116193508 | A | T | T | rs117052851 | intronic | 0.0304 | 0.0304 | CAV1 | 1.0000 | 56 | 10 | 0 | 21 | 4 | 0 |
| Disassembly of the destruction complex and recruitment of AXIN to the membrane | 7 | 116197908 | C | T | T | rs35902398 | intronic | 0.0348 | 0.0348 | CAV1 | 0.2428 | 58 | 7 | 1 | 26 | 0 | 0 |
| Disassembly of the destruction complex and recruitment of AXIN to the membrane | 10 | 99076980 | G | C | C | rs11189125 | upstream | 0.0428 | 0.0428 | FRAT1 | 1.0000 | 60 | 5 | 1 | 24 | 2 | 0 |
| Disassembly of the destruction complex and recruitment of AXIN to the membrane | 10 | 99081117 | G | A | A | rs11189129 | 3_prime_UTR | 0.0422 | 0.0422 | FRAT1 | 1.0000 | 60 | 5 | 1 | 24 | 2 | 0 |
| Disassembly of the destruction complex and recruitment of AXIN to the membrane | 10 | 99097398 | T | C | C | rs35502594 | upstream | 0.0344 | 0.0344 | FRAT2 | 0.4762 | 60 | 5 | 1 | 22 | 4 | 0 |
| Disassembly of the destruction complex and recruitment of AXIN to the membrane | 11 | 64690070 | G | A | A | rs147029783 | upstream | 0.0444 | 0.0444 | PPP2R5B | 0.3373 | 55 | 11 | 0 | 24 | 2 | 0 |
| Disassembly of the destruction complex and recruitment of AXIN to the membrane | 11 | 64691910 | T | G | G | rs76436719 | upstream | 0.0440 | 0.0440 | PPP2R5B | 0.3373 | 55 | 11 | 0 | 24 | 2 | 0 |
| Disassembly of the destruction complex and recruitment of AXIN to the membrane | 11 | 64692312 | A | G | G | rs60798873 | 5_prime_UTR | 0.0441 | 0.0441 | PPP2R5B | 0.3373 | 55 | 11 | 0 | 24 | 2 | 0 |
| Disassembly of the destruction complex and recruitment of AXIN to the membrane | 11 | 64692356 | C | T | T | rs61024397 | 5_prime_UTR | 0.0441 | 0.0441 | PPP2R5B | 0.3373 | 55 | 11 | 0 | 24 | 2 | 0 |
| Disassembly of the destruction complex and recruitment of AXIN to the membrane | 11 | 64694023 | A | T | T | rs57393537 | intronic | 0.0442 | 0.0442 | PPP2R5B | 0.3373 | 55 | 11 | 0 | 24 | 2 | 0 |
| Disassembly of the destruction complex and recruitment of AXIN to the membrane | 11 | 64695020 | T | G | G | rs885999 | intronic | 0.0443 | 0.0443 | PPP2R5B | 0.3348 | 54 | 11 | 0 | 24 | 2 | 0 |
| Disassembly of the destruction complex and recruitment of AXIN to the membrane | 11 | 64697950 | C | T | T | rs111934356 | splice_region,intronic | 0.0384 | 0.0384 | PPP2R5B | 0.5002 | 55 | 9 | 0 | 24 | 2 | 0 |
| Disassembly of the destruction complex and recruitment of AXIN to the membrane | 11 | 64700976 | T | C | C | rs10488702 | intronic | 0.0458 | 0.0458 | PPP2R5B | 0.7222 | 57 | 9 | 0 | 24 | 2 | 0 |
| Disassembly of the destruction complex and recruitment of AXIN to the membrane | 11 | 68105602 | G | A | A | rs74968229 | intronic | 0.0413 | 0.0413 | LRP5 | 1.0000 | 57 | 9 | 0 | 23 | 3 | 0 |
| Disassembly of the destruction complex and recruitment of AXIN to the membrane | 11 | 68109822 | T | C | C | rs4988311 | intronic | 0.0434 | 0.0434 | LRP5 | 0.8249 | 55 | 10 | 1 | 21 | 5 | 0 |
| Disassembly of the destruction complex and recruitment of AXIN to the membrane | 11 | 68111672 | A | G | G | rs314756 | intronic | 0.0470 | 0.0470 | LRP5 | 0.5316 | 56 | 9 | 0 | 21 | 5 | 0 |
| Disassembly of the destruction complex and recruitment of AXIN to the membrane | 11 | 68176016 | C | T | T | rs76884287 | intronic | 0.0481 | 0.0481 | LRP5 | 0.7302 | 58 | 8 | 0 | 21 | 4 | 0 |
| Disassembly of the destruction complex and recruitment of AXIN to the membrane | 11 | 68192362 | C | G | G | rs75776943 | intronic | 0.0486 | 0.0486 | LRP5 | 0.3091 | 59 | 7 | 0 | 21 | 5 | 0 |
| Disassembly of the destruction complex and recruitment of AXIN to the membrane | 11 | 68205117 | T | C | C | rs137864217 | intronic | 0.0205 | 0.0205 | LRP5 | 0.4984 | 59 | 7 | 0 | 22 | 4 | 0 |
| Disassembly of the destruction complex and recruitment of AXIN to the membrane | 11 | 68214505 | C | G | G | rs78471927 | intronic | 0.0482 | 0.0482 | LRP5 | 0.7227 | 58 | 7 | 0 | 22 | 4 | 0 |
| Disassembly of the destruction complex and recruitment of AXIN to the membrane | 11 | 68214890 | G | A | A | rs79320300 | intronic | 0.0441 | 0.0441 | LRP5 | 0.4984 | 59 | 7 | 0 | 22 | 4 | 0 |
| Disassembly of the destruction complex and recruitment of AXIN to the membrane | 11 | 111624412 | A | G | A | rs4322429 | intronic | 0.9540 | 0.0460 | PPP2R1B | 0.2775 | 2 | 13 | 51 | 0 | 2 | 24 |
| Disassembly of the destruction complex and recruitment of AXIN to the membrane | 12 | 12267520 | G | A | A | rs12826332 | downstream | 0.0468 | 0.0468 | LRP6 | 0.0422 | 61 | 4 | 0 | 20 | 5 | 1 |
| Disassembly of the destruction complex and recruitment of AXIN to the membrane | 12 | 12267823 | T | C | C | rs34372334 | downstream | 0.0471 | 0.0471 | LRP6 | 0.0410 | 62 | 4 | 0 | 20 | 5 | 1 |
| Disassembly of the destruction complex and recruitment of AXIN to the membrane | 12 | 12269960 | G | A | A | rs3741792 | 3_prime_UTR | 0.0471 | 0.0471 | LRP6 | 0.0237 | 62 | 4 | 0 | 19 | 5 | 1 |
| Disassembly of the destruction complex and recruitment of AXIN to the membrane | 12 | 12270560 | A | G | G | rs71457129 | 3_prime_UTR | 0.0470 | 0.0470 | LRP6 | 0.0250 | 61 | 4 | 0 | 19 | 5 | 1 |
| Disassembly of the destruction complex and recruitment of AXIN to the membrane | 12 | 12271574 | C | T | T | rs71457130 | 3_prime_UTR | 0.0469 | 0.0469 | LRP6 | 0.0237 | 62 | 4 | 0 | 19 | 5 | 1 |
| Disassembly of the destruction complex and recruitment of AXIN to the membrane | 12 | 12275297 | A | G | G | rs71457131 | intronic | 0.0317 | 0.0317 | LRP6 | 0.2078 | 58 | 8 | 0 | 20 | 6 | 0 |
| Disassembly of the destruction complex and recruitment of AXIN to the membrane | 12 | 12287013 | T | A | A | rs12819810 | intronic | 0.0495 | 0.0495 | LRP6 | 0.0237 | 62 | 4 | 0 | 19 | 5 | 1 |
| Disassembly of the destruction complex and recruitment of AXIN to the membrane | 12 | 12287589 | C | T | T | rs12819916 | intronic | 0.0496 | 0.0496 | LRP6 | 0.0237 | 62 | 4 | 0 | 19 | 5 | 1 |
| Disassembly of the destruction complex and recruitment of AXIN to the membrane | 12 | 12295552 | T | G | G | rs34694539 | intronic | 0.0496 | 0.0496 | LRP6 | 0.0250 | 61 | 4 | 0 | 19 | 5 | 1 |
| Disassembly of the destruction complex and recruitment of AXIN to the membrane | 12 | 12297091 | A | G | G | rs71457134 | intronic | 0.0496 | 0.0496 | LRP6 | 0.0410 | 62 | 4 | 0 | 20 | 5 | 1 |
| Disassembly of the destruction complex and recruitment of AXIN to the membrane | 12 | 12315857 | T | C | C | rs2300230 | intronic | 0.0496 | 0.0496 | LRP6 | 0.0237 | 62 | 4 | 0 | 19 | 5 | 1 |
| Disassembly of the destruction complex and recruitment of AXIN to the membrane | 12 | 12321526 | G | A | A | rs11054719 | intronic | 0.0210 | 0.0210 | LRP6 | 1.0000 | 58 | 8 | 0 | 23 | 3 | 0 |
| Disassembly of the destruction complex and recruitment of AXIN to the membrane | 12 | 12324058 | G | A | A | rs113794654 | intronic | 0.0315 | 0.0315 | LRP6 | 1.0000 | 60 | 5 | 1 | 24 | 1 | 0 |
| Disassembly of the destruction complex and recruitment of AXIN to the membrane | 12 | 12324841 | A | G | G | rs12833575 | intronic | 0.0327 | 0.0327 | LRP6 | 0.3477 | 57 | 9 | 0 | 20 | 6 | 0 |
| Disassembly of the destruction complex and recruitment of AXIN to the membrane | 12 | 12325932 | C | G | G | rs76038196 | intronic | 0.0382 | 0.0382 | LRP6 | 0.0381 | 63 | 3 | 0 | 21 | 4 | 1 |
| Disassembly of the destruction complex and recruitment of AXIN to the membrane | 12 | 12326027 | C | G | G | rs75049047 | intronic | 0.0381 | 0.0381 | LRP6 | 0.0335 | 63 | 3 | 0 | 20 | 4 | 1 |
| Disassembly of the destruction complex and recruitment of AXIN to the membrane | 12 | 12326128 | C | T | T | rs76299027 | intronic | 0.0380 | 0.0380 | LRP6 | 0.0381 | 63 | 3 | 0 | 21 | 4 | 1 |
| Disassembly of the destruction complex and recruitment of AXIN to the membrane | 12 | 12328826 | T | C | C | rs78024436 | intronic | 0.0382 | 0.0382 | LRP6 | 0.0351 | 62 | 3 | 0 | 20 | 4 | 1 |
| Disassembly of the destruction complex and recruitment of AXIN to the membrane | 12 | 12337956 | C | T | T | rs77595426 | intronic | 0.0383 | 0.0383 | LRP6 | 0.0381 | 63 | 3 | 0 | 21 | 4 | 1 |
| Disassembly of the destruction complex and recruitment of AXIN to the membrane | 12 | 12346016 | G | A | A | rs79359989 | intronic | 0.0383 | 0.0383 | LRP6 | 0.0335 | 63 | 3 | 0 | 20 | 4 | 1 |
| Disassembly of the destruction complex and recruitment of AXIN to the membrane | 12 | 12350029 | A | G | G | rs187847468 | intronic | 0.0383 | 0.0383 | LRP6 | 0.0335 | 63 | 3 | 0 | 20 | 4 | 1 |
| Disassembly of the destruction complex and recruitment of AXIN to the membrane | 12 | 12357470 | T | A | A | rs75431901 | intronic | 0.0382 | 0.0382 | LRP6 | 0.0381 | 63 | 3 | 0 | 21 | 4 | 1 |
| Disassembly of the destruction complex and recruitment of AXIN to the membrane | 12 | 12364411 | T | C | C | rs17374170 | intronic | 0.0382 | 0.0382 | LRP6 | 0.0381 | 63 | 3 | 0 | 21 | 4 | 1 |
| Disassembly of the destruction complex and recruitment of AXIN to the membrane | 12 | 12369072 | C | A | A | rs75798211 | intronic | 0.0468 | 0.0468 | LRP6 | 0.0381 | 63 | 3 | 0 | 21 | 4 | 1 |
| Disassembly of the destruction complex and recruitment of AXIN to the membrane | 12 | 12370932 | C | T | T | rs77970482 | intronic | 0.0381 | 0.0381 | LRP6 | 0.0335 | 63 | 3 | 0 | 20 | 4 | 1 |
| Disassembly of the destruction complex and recruitment of AXIN to the membrane | 12 | 12372047 | T | G | G | rs117001679 | intronic | 0.0383 | 0.0383 | LRP6 | 0.0381 | 63 | 3 | 0 | 21 | 4 | 1 |
| Disassembly of the destruction complex and recruitment of AXIN to the membrane | 12 | 12374101 | A | G | G | rs75253855 | intronic | 0.0383 | 0.0383 | LRP6 | 0.0335 | 63 | 3 | 0 | 20 | 4 | 1 |
| Disassembly of the destruction complex and recruitment of AXIN to the membrane | 12 | 12378128 | C | T | T | rs11054733 | intronic | 0.0209 | 0.0209 | LRP6 | 1.0000 | 57 | 8 | 0 | 23 | 3 | 0 |
| Disassembly of the destruction complex and recruitment of AXIN to the membrane | 12 | 12378158 | C | T | T | rs78815657 | intronic | 0.0383 | 0.0383 | LRP6 | 0.0400 | 62 | 3 | 0 | 21 | 4 | 1 |
| Disassembly of the destruction complex and recruitment of AXIN to the membrane | 12 | 12378767 | T | C | C | rs117299638 | intronic | 0.0379 | 0.0379 | LRP6 | 0.0381 | 63 | 3 | 0 | 21 | 4 | 1 |
| Disassembly of the destruction complex and recruitment of AXIN to the membrane | 12 | 12378920 | C | T | T | rs80214149 | intronic | 0.0383 | 0.0383 | LRP6 | 0.0335 | 63 | 3 | 0 | 20 | 4 | 1 |
| Disassembly of the destruction complex and recruitment of AXIN to the membrane | 12 | 12384668 | G | A | A | rs77157509 | intronic | 0.0376 | 0.0376 | LRP6 | 0.0335 | 63 | 3 | 0 | 20 | 4 | 1 |
| Disassembly of the destruction complex and recruitment of AXIN to the membrane | 12 | 12388356 | A | C | C | rs77588531 | intronic | 0.0383 | 0.0383 | LRP6 | 0.0335 | 63 | 3 | 0 | 20 | 4 | 1 |
| Disassembly of the destruction complex and recruitment of AXIN to the membrane | 12 | 12391576 | T | C | C | rs75155789 | intronic | 0.0380 | 0.0380 | LRP6 | 0.0335 | 63 | 3 | 0 | 20 | 4 | 1 |
| Disassembly of the destruction complex and recruitment of AXIN to the membrane | 12 | 12398849 | T | G | G | rs116946061 | intronic | 0.0380 | 0.0380 | LRP6 | 0.0335 | 63 | 3 | 0 | 20 | 4 | 1 |
| Disassembly of the destruction complex and recruitment of AXIN to the membrane | 12 | 12404270 | C | T | T | rs78533154 | intronic | 0.0384 | 0.0384 | LRP6 | 0.0381 | 63 | 3 | 0 | 21 | 4 | 1 |
| Disassembly of the destruction complex and recruitment of AXIN to the membrane | 12 | 12414109 | G | T | T | rs149872900 | intronic | 0.0383 | 0.0383 | LRP6 | 0.0381 | 63 | 3 | 0 | 21 | 4 | 1 |
| Disassembly of the destruction complex and recruitment of AXIN to the membrane | 12 | 12414920 | T | C | C | rs147798710 | intronic | 0.0383 | 0.0383 | LRP6 | 0.0381 | 63 | 3 | 0 | 21 | 4 | 1 |
| Disassembly of the destruction complex and recruitment of AXIN to the membrane | 12 | 12416595 | A | C | C | rs117775212 | intronic | 0.0383 | 0.0383 | LRP6 | 0.0381 | 63 | 3 | 0 | 21 | 4 | 1 |
| Disassembly of the destruction complex and recruitment of AXIN to the membrane | 14 | 63877780 | T | G | G | rs76445028 | intronic | 0.0431 | 0.0431 | PPP2R5E | 0.3158 | 53 | 13 | 0 | 21 | 4 | 1 |
| Disassembly of the destruction complex and recruitment of AXIN to the membrane | 14 | 63977051 | C | A | A | rs17766076 | intronic | 0.0436 | 0.0436 | PPP2R5E | 0.6433 | 56 | 9 | 1 | 24 | 2 | 0 |
| Disassembly of the destruction complex and recruitment of AXIN to the membrane | 14 | 102253838 | T | C | C | rs151299721 | intronic | 0.0229 | 0.0229 | PPP2R5C | 0.4981 | 56 | 10 | 0 | 24 | 2 | 0 |
| Disassembly of the destruction complex and recruitment of AXIN to the membrane | 14 | 102274571 | A | T | T | rs76294435 | intronic | 0.0481 | 0.0481 | PPP2R5C | 0.2818 | 60 | 6 | 0 | 21 | 5 | 0 |
| Disassembly of the destruction complex and recruitment of AXIN to the membrane | 14 | 102287985 | G | A | A | rs80152381 | intronic | 0.0425 | 0.0425 | PPP2R5C | 1.0000 | 54 | 11 | 0 | 21 | 4 | 0 |
| Disassembly of the destruction complex and recruitment of AXIN to the membrane | 14 | 102326654 | C | T | T | rs80319248 | intronic | 0.0269 | 0.0269 | PPP2R5C | 0.7348 | 58 | 8 | 0 | 22 | 4 | 0 |
| Disassembly of the destruction complex and recruitment of AXIN to the membrane | 14 | 102338112 | A | G | G | rs2476521 | intronic | 0.0458 | 0.0458 | PPP2R5C | 0.0565 | 56 | 10 | 0 | 25 | 0 | 0 |
| Disassembly of the destruction complex and recruitment of AXIN to the membrane | 14 | 102339600 | T | C | C | rs116995481 | intronic | 0.0294 | 0.0294 | PPP2R5C | 0.0569 | 56 | 10 | 0 | 26 | 0 | 0 |
| Disassembly of the destruction complex and recruitment of AXIN to the membrane | 14 | 102344254 | C | T | T | rs8012434 | intronic | 0.0449 | 0.0449 | PPP2R5C | 0.7192 | 58 | 8 | 0 | 24 | 2 | 0 |
| Disassembly of the destruction complex and recruitment of AXIN to the membrane | 14 | 102395021 | T | C | T | rs1741145 | downstream | 0.9636 | 0.0364 | PPP2R5C | 0.7522 | 0 | 10 | 56 | 0 | 3 | 23 |
| Disassembly of the destruction complex and recruitment of AXIN to the membrane | 16 | 337655 | A | G | G | rs45530432 | 3_prime_UTR | 0.0218 | 0.0218 | AXIN1 | 0.5928 | 57 | 7 | 1 | 25 | 1 | 0 |
| Disassembly of the destruction complex and recruitment of AXIN to the membrane | 16 | 350846 | A | G | A | rs214245 | intronic | 0.9604 | 0.0396 | AXIN1 | 0.2078 | 0 | 8 | 58 | 0 | 6 | 20 |
| Disassembly of the destruction complex and recruitment of AXIN to the membrane | 16 | 351517 | A | G | A | rs214242 | intronic | 0.9533 | 0.0467 | AXIN1 | 0.2078 | 0 | 8 | 58 | 0 | 6 | 20 |
| Disassembly of the destruction complex and recruitment of AXIN to the membrane | 16 | 392392 | C | G | G | rs118100148 | intronic | 0.0492 | 0.0492 | AXIN1 | 1.0000 | 51 | 14 | 0 | 21 | 5 | 0 |
| Disassembly of the destruction complex and recruitment of AXIN to the membrane | 17 | 42641345 | G | C | C | rs75575961 | downstream | 0.0180 | 0.0180 | FZD2 | 0.6957 | 60 | 6 | 0 | 21 | 3 | 0 |
| Disassembly of the destruction complex and recruitment of AXIN to the membrane | 19 | 1948218 | C | T | T | rs117285549 | intronic | 0.0293 | 0.0293 | CSNK1G2 | 0.6148 | 59 | 6 | 1 | 22 | 4 | 0 |
| Disassembly of the destruction complex and recruitment of AXIN to the membrane | 19 | 1952808 | G | A | A | rs74415283 | intronic | 0.0236 | 0.0236 | CSNK1G2 | 0.3091 | 59 | 7 | 0 | 21 | 5 | 0 |
| Disassembly of the destruction complex and recruitment of AXIN to the membrane | 19 | 1957031 | G | T | T | rs34956563 | intronic | 0.0245 | 0.0245 | CSNK1G2 | 0.7980 | 57 | 8 | 1 | 24 | 2 | 0 |
| Disassembly of the destruction complex and recruitment of AXIN to the membrane | 19 | 52710886 | A | G | G | rs62109206 | intronic | 0.0217 | 0.0217 | PPP2R1A | 0.8074 | 61 | 4 | 1 | 24 | 1 | 1 |
| Disassembly of the destruction complex and recruitment of AXIN to the membrane | 19 | 52716143 | G | A | A | rs74536039 | intronic | 0.0445 | 0.0445 | PPP2R1A | 0.8249 | 55 | 10 | 1 | 21 | 5 | 0 |
| Disassembly of the destruction complex and recruitment of AXIN to the membrane | 19 | 52720738 | C | T | T | rs77810569 | intronic | 0.0478 | 0.0478 | PPP2R1A | 0.5544 | 55 | 10 | 1 | 20 | 6 | 0 |
| Disassembly of the destruction complex and recruitment of AXIN to the membrane | 19 | 52726288 | G | A | A | rs77891912 | intronic | 0.0456 | 0.0456 | PPP2R1A | 0.5418 | 53 | 13 | 0 | 22 | 3 | 0 |
| Disassembly of the destruction complex and recruitment of AXIN to the membrane | 19 | 52729981 | C | T | T | rs111484780 | 3_prime_UTR | 0.0455 | 0.0455 | PPP2R1A | 0.5425 | 53 | 13 | 0 | 23 | 3 | 0 |
| Disassembly of the destruction complex and recruitment of AXIN to the membrane | 19 | 52734671 | T | C | C | rs76856298 | downstream | 0.0447 | 0.0447 | PPP2R1A | 0.5425 | 53 | 13 | 0 | 23 | 3 | 0 |

## Table S5 (A): SNP-based tests in SYNAPTIC genes (only the top 20 SNPs are shown)

| CHROM | POS | REF | ALT | MINOR ALLELE | ID | CONSEQUENCE | AF | MAF | SYMBOL | P VALUE | Neff=3,704 (Galwey) | Neff=6,074 (Gao) | FWER Neff=6,169 (Li & Ji) | Neff=43,119 (Nyholt/Cheverud) | # RESPONDERS Ref/Ref | Ref/Alt | Alt/Alt | # NON-RESPONDERS Ref/Ref | Ref/Alt | Alt/Alt | % RESPONDERS Ref/Ref | Ref/Alt | Alt/Alt | % NON-RESPONDERS Ref/Ref | Ref/Alt | Alt/Alt | ODDS RATIO Dominant model | Recessive model |
|---|---|---|---|---|---|---|---|---|---|---|---|---|---|---|---|---|---|---|---|---|---|---|---|---|---|---|---|---|
| 4 | 4345379 | T | C | C | rs7695197 | upstream | 0.31 | 0.31 | NSG1 | 3.55E-06 | 0.013 | 0.021 | 0.022 | 0.14 | 34 | 31 | 1 | 7 | 8 | 11 | 50.72 | 46.38 | 2.90 | 27.59 | 31.03 | 41.38 | 2.70 | 23.65 |
| 4 | 4346427 | G | A | A | rs3981 | upstream | 0.32 | 0.32 | NSG1 | 3.84E-06 | 0.014 | 0.023 | 0.023 | 0.15 | 34 | 30 | 1 | 7 | 8 | 11 | 51.47 | 45.59 | 2.94 | 27.59 | 31.03 | 41.38 | 2.78 | 23.29 |
| 4 | 4346465 | T | C | C | rs12641832 | upstream | 0.33 | 0.33 | NSG1 | 3.84E-06 | 0.014 | 0.023 | 0.023 | 0.15 | 34 | 30 | 1 | 7 | 8 | 11 | 51.47 | 45.59 | 2.94 | 27.59 | 31.03 | 41.38 | 2.78 | 23.29 |
| 4 | 4351540 | G | A | A | rs34402391 | intronic | 0.34 | 0.34 | NSG1 | 1.10E-05 | 0.040 | 0.065 | 0.066 | 0.38 | 33 | 30 | 2 | 6 | 8 | 11 | 50.00 | 45.59 | 4.41 | 25.00 | 32.14 | 42.86 | 3.00 | 16.25 |
| 17 | 11348500 | A | G | G | rs9900468 | intronic | 0.21 | 0.21 | SHISA6 | 4.92E-05 | 0.167 | 0.258 | 0.262 | 0.88 | 43 | 14 | 9 | 8 | 18 | 0 | 63.77 | 21.74 | 14.49 | 31.03 | 65.52 | 3.45 | 3.91 | 0.21 |
| 5 | 75425227 | T | G | G | rs10514059 | intronic | 0.22 | 0.22 | SV2C | 1.34E-04 | 0.391 | 0.556 | 0.562 | 1.00 | 41 | 24 | 1 | 5 | 17 | 4 | 60.87 | 36.23 | 2.90 | 20.69 | 62.07 | 17.24 | 5.96 | 6.98 |
| 19 | 15900659 | A | G | G | rs7249737 | upstream | 0.08 | 0.08 | OR10H5 | 2.14E-04 | 0.548 | 0.728 | 0.734 | 1.00 | 53 | 12 | 1 | 10 | 15 | 0 | 78.26 | 18.84 | 2.90 | 39.29 | 57.14 | 3.57 | 5.56 | 1.24 |
| 4 | 4357296 | A | G | A | rs10937867 | intronic | 0.62 | 0.38 | NSG1 | 3.37E-04 | 0.713 | 0.871 | 0.875 | 1.00 | 9 | 33 | 24 | 2 | 3 | 21 | 14.49 | 49.28 | 36.23 | 10.34 | 13.79 | 75.86 | 1.47 | 5.53 |
| 10 | 50815871 | T | C | C | rs75287762 | upstream | 0.08 | 0.08 | SLC18A3 | 3.74E-04 | 0.750 | 0.897 | 0.901 | 1.00 | 59 | 7 | 0 | 14 | 12 | 0 | 86.96 | 11.59 | 1.45 | 51.72 | 44.83 | 3.45 | 6.22 | 2.43 |
| 19 | 15906476 | G | C | C | rs62106069 | downstream | 0.08 | 0.08 | OR10H5 | 3.90E-04 | 0.764 | 0.906 | 0.910 | 1.00 | 52 | 11 | 1 | 11 | 15 | 0 | 79.10 | 17.91 | 2.99 | 41.38 | 55.17 | 3.45 | 5.36 | 1.16 |
| 13 | 47440800 | T | C | C | rs9567743 | intronic | 0.18 | 0.18 | HTR2A | 3.92E-04 | 0.766 | 0.908 | 0.911 | 1.00 | 28 | 31 | 7 | 21 | 2 | 1 | 42.03 | 46.38 | 11.59 | 81.48 | 11.11 | 7.41 | 0.16 | 0.61 |
| 11 | 84340842 | T | G | G | rs17807944 | intronic | 0.22 | 0.22 | DLG2 | 4.15E-04 | 0.785 | 0.920 | 0.923 | 1.00 | 30 | 32 | 4 | 21 | 2 | 3 | 44.93 | 47.83 | 7.25 | 75.86 | 10.34 | 13.79 | 0.26 | 2.05 |
| 7 | 157825498 | C | T | C | rs6953551 | intronic | 0.70 | 0.30 | PTPRN2 | 4.37E-04 | 0.802 | 0.930 | 0.932 | 1.00 | 1 | 26 | 38 | 2 | 1 | 22 | 2.94 | 39.71 | 57.35 | 10.71 | 7.14 | 82.14 | 0.25 | 3.42 |
| 19 | 15855434 | C | A | A | rs56108924 | downstream | 0.10 | 0.10 | OR10H3 | 5.33E-04 | 0.861 | 0.961 | 0.963 | 1.00 | 52 | 13 | 1 | 10 | 15 | 1 | 76.81 | 20.29 | 2.90 | 37.93 | 55.17 | 6.90 | 5.42 | 2.48 |
| 6 | 34036642 | G | T | T | rs1906954 | intronic | 0.06 | 0.06 | GRM4 | 5.86E-04 | 0.886 | 0.972 | 0.973 | 1.00 | 62 | 2 | 0 | 18 | 8 | 0 | 94.03 | 4.48 | 1.49 | 65.52 | 31.03 | 3.45 | 8.29 | 2.36 |
| 19 | 15910817 | T | C | C | rs55908984 | downstream | 0.07 | 0.07 | OR10H5 | 6.49E-04 | 0.910 | 0.981 | 0.982 | 1.00 | 53 | 12 | 1 | 11 | 15 | 0 | 78.26 | 18.84 | 2.90 | 41.38 | 55.17 | 3.45 | 5.10 | 1.20 |
| 19 | 15907908 | C | T | T | rs62106071 | downstream | 0.08 | 0.08 | OR10H5 | 6.54E-04 | 0.911 | 0.981 | 0.982 | 1.00 | 52 | 12 | 1 | 11 | 15 | 0 | 77.94 | 19.12 | 2.94 | 41.38 | 55.17 | 3.45 | 5.01 | 1.18 |
| 19 | 15908978 | C | T | T | rs72995365 | downstream | 0.07 | 0.07 | OR10H5 | 6.54E-04 | 0.911 | 0.981 | 0.982 | 1.00 | 52 | 12 | 1 | 11 | 15 | 0 | 77.94 | 19.12 | 2.94 | 41.38 | 55.17 | 3.45 | 5.01 | 1.18 |
| 6 | 44192158 | T | C | C | rs1886884 | intronic | 0.47 | 0.47 | SLC29A1 | 6.72E-04 | 0.917 | 0.983 | 0.984 | 1.00 | 9 | 33 | 23 | 14 | 8 | 4 | 14.71 | 50.00 | 35.29 | 51.72 | 31.03 | 17.24 | 0.16 | 0.38 |
| 2 | 50882657 | C | T | C | rs9752732 | intronic | 0.82 | 0.18 | NRXN1 | 7.31E-04 | 0.933 | 0.988 | 0.989 | 1.00 | 2 | 11 | 53 | 2 | 13 | 10 | 4.35 | 17.39 | 78.26 | 10.71 | 50.00 | 39.29 | 0.38 | 0.18 |

Table S5 (B): Gene-based tests in SYNAPTIC genes

| SYMBOL | # MUTATIONS | P VALUE | FWER |
|---|---|---|---|
| DLG2 | 208 | 6.25E-07 | 0.000137 |
| SLC6A4 | 2 | 1.86E-03 | 0.336693 |
| DAGLA | 5 | 8.14E-03 | 0.834428 |
| SHISA9 | 15 | 3.74E-02 | 0.999771 |
| SCRIB | 2 | 4.25E-02 | 0.999928 |
| SV2B | 17 | 6.11E-02 | 0.999999 |
| PPFIA4 | 3 | 6.33E-02 | 0.999999 |
| PRKCZ | 5 | 6.61E-02 | 1.000000 |
| NAAA | 1 | 7.03E-02 | 1.000000 |
| SYT17 | 4 | 7.47E-02 | 1.000000 |
| USP46 | 3 | 7.60E-02 | 1.000000 |
| GRIA1 | 38 | 8.74E-02 | 1.000000 |
| FNTA | 2 | 9.10E-02 | 1.000000 |
| HRH2 | 7 | 9.10E-02 | 1.000000 |
| GRM1 | 17 | 9.16E-02 | 1.000000 |
| AP3D1 | 3 | 9.35E-02 | 1.000000 |
| LYPD6B | 7 | 9.43E-02 | 1.000000 |
| DTNBP1 | 16 | 9.86E-02 | 1.000000 |
| RAP1A | 16 | 9.99E-02 | 1.000000 |
| SLC6A7 | 1 | 1.04E-01 | 1.000000 |
| SRC | 1 | 1.20E-01 | 1.000000 |
| APBA1 | 3 | 1.22E-01 | 1.000000 |
| DNM3 | 36 | 1.29E-01 | 1.000000 |
| STX3 | 5 | 1.29E-01 | 1.000000 |
| KIF17 | 4 | 1.49E-01 | 1.000000 |
| OR56A4 | 2 | 1.62E-01 | 1.000000 |
| ATP1A2 | 2 | 1.62E-01 | 1.000000 |
| DOC2B | 8 | 1.63E-01 | 1.000000 |
| SV2C | 12 | 1.71E-01 | 1.000000 |
| DAG1 | 4 | 1.72E-01 | 1.000000 |
| SNCG | 1 | 1.75E-01 | 1.000000 |
| FRRS1L | 4 | 1.76E-01 | 1.000000 |
| BRSK1 | 2 | 1.78E-01 | 1.000000 |
| GLUL | 2 | 1.80E-01 | 1.000000 |
| RAB11A | 2 | 1.92E-01 | 1.000000 |
| WNT7A | 3 | 1.93E-01 | 1.000000 |
| HTR1D | 2 | 2.01E-01 | 1.000000 |
| GABBR1 | 30 | 2.02E-01 | 1.000000 |
| STXBP1 | 10 | 2.02E-01 | 1.000000 |
| SLC6A15 | 4 | 2.10E-01 | 1.000000 |
| SHISA6 | 15 | 2.18E-01 | 1.000000 |
| DNM2 | 12 | 2.25E-01 | 1.000000 |
| DDC | 6 | 2.28E-01 | 1.000000 |
| HTR1E | 29 | 2.32E-01 | 1.000000 |
| FLOT1 | 2 | 2.37E-01 | 1.000000 |
| RPH3A | 13 | 2.58E-01 | 1.000000 |
| KCNC4 | 3 | 2.62E-01 | 1.000000 |

| Gene | Count | P-value | Adj |
|---|---|---|---|
| EFNB2 | 2 | 2.67E-01 | 1.000000 |
| COLQ | 7 | 2.83E-01 | 1.000000 |
| SLC1A7 | 8 | 2.87E-01 | 1.000000 |
| CTNND1 | 4 | 2.88E-01 | 1.000000 |
| SYN3 | 46 | 2.88E-01 | 1.000000 |
| NUMB | 10 | 2.90E-01 | 1.000000 |
| GRM4 | 2 | 2.92E-01 | 1.000000 |
| SLC6A9 | 3 | 3.00E-01 | 1.000000 |
| KIF5C | 5 | 3.01E-01 | 1.000000 |
| UNC13A | 7 | 3.15E-01 | 1.000000 |
| SNAPIN | 1 | 3.22E-01 | 1.000000 |
| SLC18A1 | 3 | 3.27E-01 | 1.000000 |
| SYT1 | 41 | 3.29E-01 | 1.000000 |
| ALDH5A1 | 6 | 3.29E-01 | 1.000000 |
| HTR4 | 12 | 3.93E-01 | 1.000000 |
| PDE1B | 2 | 3.97E-01 | 1.000000 |
| HTR5A | 3 | 4.00E-01 | 1.000000 |
| HSPA8 | 3 | 4.05E-01 | 1.000000 |
| AP2B1 | 5 | 4.14E-01 | 1.000000 |
| HRAS | 1 | 4.16E-01 | 1.000000 |
| CNIH2 | 2 | 4.16E-01 | 1.000000 |
| ICA1 | 32 | 4.20E-01 | 1.000000 |
| CPLX2 | 8 | 4.25E-01 | 1.000000 |
| CACNA1B | 5 | 4.45E-01 | 1.000000 |
| CLSTN1 | 5 | 4.64E-01 | 1.000000 |
| NPTXR | 2 | 4.68E-01 | 1.000000 |
| CHRM5 | 5 | 4.74E-01 | 1.000000 |
| AKAP9 | 1 | 4.86E-01 | 1.000000 |
| BAIAP3 | 1 | 4.86E-01 | 1.000000 |
| CACNG2 | 14 | 4.89E-01 | 1.000000 |
| SLC6A13 | 3 | 4.93E-01 | 1.000000 |
| CHRM3 | 37 | 4.97E-01 | 1.000000 |
| ADRB1 | 1 | 4.98E-01 | 1.000000 |
| CHRM1 | 1 | 4.98E-01 | 1.000000 |
| CHAT | 1 | 5.03E-01 | 1.000000 |
| LRTOMT | 1 | 5.06E-01 | 1.000000 |
| HRH1 | 6 | 5.09E-01 | 1.000000 |
| SYT2 | 3 | 5.09E-01 | 1.000000 |
| SLC6A18 | 2 | 5.11E-01 | 1.000000 |
| NF1 | 6 | 5.13E-01 | 1.000000 |
| GAD1 | 6 | 5.25E-01 | 1.000000 |
| SLC30A1 | 2 | 5.50E-01 | 1.000000 |
| SLC17A6 | 3 | 5.69E-01 | 1.000000 |
| SLC6A20 | 10 | 5.75E-01 | 1.000000 |
| SNCA | 3 | 5.80E-01 | 1.000000 |
| STX1A | 3 | 5.83E-01 | 1.000000 |
| SLC6A12 | 3 | 5.83E-01 | 1.000000 |
| CACNG7 | 2 | 5.83E-01 | 1.000000 |

| Gene | Count | p-value | q-value |
|---|---|---|---|
| NRG1 | 103 | 5.83E-01 | 1.000000 |
| UNC13B | 5 | 5.85E-01 | 1.000000 |
| NPTX1 | 2 | 5.90E-01 | 1.000000 |
| SLC1A2 | 12 | 6.06E-01 | 1.000000 |
| LYNX1 | 4 | 6.33E-01 | 1.000000 |
| DOC2A | 1 | 6.47E-01 | 1.000000 |
| EPS8 | 7 | 6.54E-01 | 1.000000 |
| ATP2A2 | 2 | 6.58E-01 | 1.000000 |
| ARHGAP44 | 47 | 6.59E-01 | 1.000000 |
| SLC22A1 | 2 | 6.72E-01 | 1.000000 |
| SLC38A1 | 7 | 6.81E-01 | 1.000000 |
| RHOT1 | 8 | 6.95E-01 | 1.000000 |
| NRXN1 | 77 | 6.98E-01 | 1.000000 |
| SLC32A1 | 1 | 7.19E-01 | 1.000000 |
| SLC5A7 | 2 | 7.23E-01 | 1.000000 |
| OR5T3 | 1 | 7.30E-01 | 1.000000 |
| SYT12 | 3 | 7.31E-01 | 1.000000 |
| SYT7 | 4 | 7.36E-01 | 1.000000 |
| PRKCI | 2 | 7.44E-01 | 1.000000 |
| SLC6A17 | 7 | 7.51E-01 | 1.000000 |
| CALY | 1 | 7.52E-01 | 1.000000 |
| HTR6 | 2 | 7.52E-01 | 1.000000 |
| PPFIA1 | 2 | 7.55E-01 | 1.000000 |
| GABRA2 | 18 | 7.59E-01 | 1.000000 |
| PPP1R9A | 15 | 7.60E-01 | 1.000000 |
| ALDH9A1 | 3 | 7.65E-01 | 1.000000 |
| DRD3 | 1 | 7.68E-01 | 1.000000 |
| SLC18A2 | 2 | 7.74E-01 | 1.000000 |
| NRXN2 | 5 | 7.79E-01 | 1.000000 |
| SLC1A3 | 14 | 7.88E-01 | 1.000000 |
| HOMER1 | 5 | 7.88E-01 | 1.000000 |
| CHRM2 | 12 | 7.88E-01 | 1.000000 |
| SLC6A2 | 6 | 7.92E-01 | 1.000000 |
| SLC1A6 | 3 | 7.94E-01 | 1.000000 |
| DRD4 | 1 | 7.96E-01 | 1.000000 |
| AP2M1 | 1 | 7.98E-01 | 1.000000 |
| HIP1 | 17 | 7.99E-01 | 1.000000 |
| PATE1 | 4 | 8.06E-01 | 1.000000 |
| CPLX1 | 5 | 8.09E-01 | 1.000000 |
| CACNG4 | 3 | 8.09E-01 | 1.000000 |
| LYPD6 | 4 | 8.11E-01 | 1.000000 |
| CHRNB4 | 2 | 8.17E-01 | 1.000000 |
| PPFIA3 | 3 | 8.17E-01 | 1.000000 |
| GPER1 | 4 | 8.18E-01 | 1.000000 |
| GSG1L | 19 | 8.23E-01 | 1.000000 |
| NPTX2 | 2 | 8.24E-01 | 1.000000 |
| DAGLB | 1 | 8.25E-01 | 1.000000 |
| ADAM10 | 12 | 8.30E-01 | 1.000000 |

| Gene | Count | P-value | Adj |
|---|---|---|---|
| KIF5A | 3 | 8.38E-01 | 1.000000 |
| SLC6A16 | 2 | 8.45E-01 | 1.000000 |
| MCTP1 | 45 | 8.46E-01 | 1.000000 |
| PRIMA1 | 7 | 8.48E-01 | 1.000000 |
| DRD2 | 7 | 8.50E-01 | 1.000000 |
| NIPSNAP1 | 2 | 8.58E-01 | 1.000000 |
| SLC6A6 | 7 | 8.61E-01 | 1.000000 |
| RAB8A | 2 | 8.65E-01 | 1.000000 |
| GAD2 | 2 | 8.78E-01 | 1.000000 |
| LIN7A | 12 | 8.80E-01 | 1.000000 |
| GPHN | 62 | 8.92E-01 | 1.000000 |
| GRIP1 | 24 | 8.98E-01 | 1.000000 |
| YWHAE | 23 | 9.12E-01 | 1.000000 |
| SYT11 | 3 | 9.16E-01 | 1.000000 |
| PATE4 | 7 | 9.24E-01 | 1.000000 |
| RIMS1 | 17 | 9.24E-01 | 1.000000 |
| CACNA1A | 19 | 9.26E-01 | 1.000000 |
| SYNJ1 | 3 | 9.27E-01 | 1.000000 |
| GRIN1 | 5 | 9.29E-01 | 1.000000 |
| SLC6A3 | 4 | 9.32E-01 | 1.000000 |
| SNAP25 | 8 | 9.33E-01 | 1.000000 |
| CAMK2A | 5 | 9.38E-01 | 1.000000 |
| SLC6A5 | 6 | 9.38E-01 | 1.000000 |
| RIMS2 | 39 | 9.39E-01 | 1.000000 |
| SLC17A8 | 6 | 9.41E-01 | 1.000000 |
| SNCAIP | 5 | 9.43E-01 | 1.000000 |
| CACNG3 | 7 | 9.48E-01 | 1.000000 |
| SLC22A2 | 3 | 9.50E-01 | 1.000000 |
| HTR7 | 5 | 9.51E-01 | 1.000000 |
| SLC6A19 | 2 | 9.53E-01 | 1.000000 |
| ITGB3 | 9 | 9.55E-01 | 1.000000 |
| NSG1 | 6 | 9.61E-01 | 1.000000 |
| RIMS4 | 6 | 9.62E-01 | 1.000000 |
| MAPK10 | 19 | 9.70E-01 | 1.000000 |
| SYN2 | 56 | 9.79E-01 | 1.000000 |
| SYT5 | 3 | 9.82E-01 | 1.000000 |
| EPS15 | 8 | 9.84E-01 | 1.000000 |
| PAH | 4 | 9.85E-01 | 1.000000 |
| NRXN3 | 187 | 9.90E-01 | 1.000000 |
| DGKI | 40 | 9.90E-01 | 1.000000 |
| AGTPBP1 | 6 | 9.93E-01 | 1.000000 |
| HTR2A | 26 | 9.94E-01 | 1.000000 |
| LRRC7 | 34 | 9.95E-01 | 1.000000 |
| GPC6 | 98 | 9.97E-01 | 1.000000 |
| ADORA1 | 12 | 9.98E-01 | 1.000000 |
| COMT | 5 | 9.98E-01 | 1.000000 |
| NAALAD2 | 23 | 9.98E-01 | 1.000000 |
| PPFIA2 | 47 | 9.98E-01 | 1.000000 |

| Gene | Count | p-value | q-value |
|---|---|---|---|
| MEF2C | 39 | 9.99E-01 | 1.000000 |
| SLC6A11 | 21 | 9.99E-01 | 1.000000 |
| PTPRN2 | 90 | 1.00E+00 | 1.000000 |
| SNPH | 4 | 1.00E+00 | 1.000000 |
| NOS1 | 11 | 1.00E+00 | 1.000000 |
| GRM5 | 13 | 1.00E+00 | 1.000000 |
| DLG1 | 72 | 1.00E+00 | 1.000000 |
| CACNG5 | 12 | 1.00E+00 | 1.000000 |
| ABAT | 14 | 1.00E+00 | 1.000000 |
| CPT1C | 1 | 1.00E+00 | 1.000000 |
| DLG4 | 1 | 1.00E+00 | 1.000000 |
| DNM1 | 1 | 1.00E+00 | 1.000000 |
| GRASP | 1 | 1.00E+00 | 1.000000 |
| GRIP2 | 4 | 1.00E+00 | 1.000000 |
| HRH3 | 1 | 1.00E+00 | 1.000000 |
| HRH4 | 1 | 1.00E+00 | 1.000000 |
| HTR1F | 1 | 1.00E+00 | 1.000000 |
| KCNMB4 | 3 | 1.00E+00 | 1.000000 |
| LGI1 | 1 | 1.00E+00 | 1.000000 |
| OR11H4 | 1 | 1.00E+00 | 1.000000 |
| PPT1 | 1 | 1.00E+00 | 1.000000 |
| PSCA | 1 | 1.00E+00 | 1.000000 |
| RAB3A | 1 | 1.00E+00 | 1.000000 |
| RIMS3 | 2 | 1.00E+00 | 1.000000 |
| SHISA7 | 1 | 1.00E+00 | 1.000000 |
| SLC29A1 | 1 | 1.00E+00 | 1.000000 |
| SLC6A1 | 4 | 1.00E+00 | 1.000000 |
| SNAP47 | 2 | 1.00E+00 | 1.000000 |
| SYT4 | 1 | 1.00E+00 | 1.000000 |

Table S5 (C): SNP-based tests in EPILEPSY genes (only the top 20 SNPs are shown)

| | | | | | | | | | | | | | FWER | | | # RESPONDERS | | | # NON-RESPONDERS | | | % RESPONDERS | | | % NON-RESPONDERS | | | ODDS RATIO | |
|---|---|---|---|---|---|---|---|---|---|---|---|---|---|---|---|---|---|---|---|---|---|---|---|---|---|---|---|---|---|
| CHROM | POS | REF | ALT | MINOR ALLELE | ID | CONSEQUENCE | AF | MAF | SYMBOL | P VALUE | Neff=4,205 (Galwey) | Neff=6,797 (Gao) | Neff=7,141 (Li & Ji) | Neff=55,488 (Nyholt/Cheverud) | Ref/Ref | Ref/Alt | Alt/Alt | Ref/Ref | Ref/Alt | Alt/Alt | Ref/Ref | Ref/Alt | Alt/Alt | Ref/Ref | Ref/Alt | Alt/Alt | Dominant model | Recessive model |
| 7 | 75690988 | A | C | C | rs10216079 | intronic | 0.20 | 0.20 | MDH2 | 2.65E-05 | 0.105 | 0.165 | 0.172 | 0.770 | 37 | 26 | 2 | 26 | 0 | 0 | 55.88 | 39.71 | 4.41 | 93.10 | 3.45 | 3.45 | 0.09 | 0.77 |
| 7 | 75677578 | G | T | T | rs2286828 | intronic | 0.21 | 0.21 | MDH2 | 2.66E-05 | 0.106 | 0.165 | 0.173 | 0.771 | 38 | 26 | 2 | 26 | 0 | 0 | 56.52 | 39.13 | 4.35 | 93.10 | 3.45 | 3.45 | 0.10 | 0.79 |
| 7 | 75677739 | G | A | A | rs2286829 | intronic | 0.21 | 0.21 | MDH2 | 2.66E-05 | 0.106 | 0.165 | 0.173 | 0.771 | 38 | 26 | 2 | 26 | 0 | 0 | 56.52 | 39.13 | 4.35 | 93.10 | 3.45 | 3.45 | 0.10 | 0.79 |
| 7 | 75686007 | G | A | A | rs867973 | intronic | 0.18 | 0.18 | MDH2 | 2.66E-05 | 0.106 | 0.165 | 0.173 | 0.771 | 38 | 26 | 2 | 26 | 0 | 0 | 56.52 | 39.13 | 4.35 | 93.10 | 3.45 | 3.45 | 0.10 | 0.79 |
| 7 | 75688672 | G | A | A | rs78587454 | intronic | 0.18 | 0.18 | MDH2 | 2.66E-05 | 0.106 | 0.165 | 0.173 | 0.771 | 38 | 26 | 2 | 26 | 0 | 0 | 56.52 | 39.13 | 4.35 | 93.10 | 3.45 | 3.45 | 0.10 | 0.79 |
| 7 | 75695081 | G | A | A | rs3779419 | intronic | 0.23 | 0.23 | MDH2 | 2.66E-05 | 0.106 | 0.165 | 0.173 | 0.771 | 38 | 26 | 2 | 26 | 0 | 0 | 56.52 | 39.13 | 4.35 | 93.10 | 3.45 | 3.45 | 0.10 | 0.79 |
| 16 | 90008296 | T | C | C | rs11644157 | downstream | 0.06 | 0.06 | TUBB3 | 1.30E-04 | 0.420 | 0.586 | 0.604 | 0.999 | 62 | 3 | 0 | 16 | 10 | 0 | 92.65 | 5.88 | 1.47 | 58.62 | 37.93 | 3.45 | 8.89 | 2.39 |
| 9 | 87388848 | G | A | G | rs4877288 | intronic | 0.63 | 0.37 | NTRK2 | 2.36E-04 | 0.629 | 0.799 | 0.814 | 1.000 | 16 | 36 | 14 | 1 | 7 | 16 | 24.64 | 53.62 | 21.74 | 7.41 | 29.63 | 62.96 | 4.09 | 6.12 |
| 16 | 90003359 | G | A | A | rs4534840 | downstream | 0.11 | 0.11 | TUBB3 | 2.41E-04 | 0.637 | 0.805 | 0.821 | 1.000 | 62 | 4 | 0 | 15 | 10 | 0 | 91.30 | 7.25 | 1.45 | 57.14 | 39.29 | 3.57 | 7.88 | 2.52 |
| 16 | 78331490 | A | G | G | rs73574528 | intronic | 0.14 | 0.14 | WWOX | 2.80E-04 | 0.692 | 0.851 | 0.865 | 1.000 | 31 | 29 | 6 | 23 | 2 | 0 | 46.38 | 43.48 | 10.14 | 85.71 | 10.71 | 3.57 | 0.14 | 0.33 |
| 16 | 89991963 | C | T | T | rs62052213 | intronic | 0.08 | 0.08 | TUBB3 | 3.05E-04 | 0.723 | 0.874 | 0.887 | 1.000 | 59 | 5 | 0 | 15 | 11 | 0 | 89.55 | 8.96 | 1.49 | 55.17 | 41.38 | 3.45 | 6.96 | 2.36 |
| 16 | 89998157 | G | A | A | rs62052214 | intronic | 0.06 | 0.06 | TUBB3 | 3.34E-04 | 0.754 | 0.896 | 0.908 | 1.000 | 62 | 4 | 0 | 16 | 10 | 0 | 91.30 | 7.25 | 1.45 | 58.62 | 37.93 | 3.45 | 7.41 | 2.43 |
| 16 | 90007025 | A | G | G | rs11640500 | downstream | 0.06 | 0.06 | TUBB3 | 3.34E-04 | 0.754 | 0.896 | 0.908 | 1.000 | 62 | 4 | 0 | 16 | 10 | 0 | 91.30 | 7.25 | 1.45 | 58.62 | 37.93 | 3.45 | 7.41 | 2.43 |
| 16 | 90008896 | G | A | A | rs74251586 | downstream | 0.06 | 0.06 | TUBB3 | 3.34E-04 | 0.754 | 0.896 | 0.908 | 1.000 | 62 | 4 | 0 | 16 | 10 | 0 | 91.30 | 7.25 | 1.45 | 58.62 | 37.93 | 3.45 | 7.41 | 2.43 |
| 16 | 90008928 | C | A | A | rs79886664 | downstream | 0.06 | 0.06 | TUBB3 | 3.34E-04 | 0.754 | 0.896 | 0.908 | 1.000 | 62 | 4 | 0 | 16 | 10 | 0 | 91.30 | 7.25 | 1.45 | 58.62 | 37.93 | 3.45 | 7.41 | 2.43 |
| 9 | 87382724 | T | C | T | rs2378671 | intronic | 0.53 | 0.47 | NTRK2 | 4.61E-04 | 0.856 | 0.956 | 0.963 | 1.000 | 16 | 36 | 14 | 1 | 8 | 16 | 24.64 | 53.62 | 21.74 | 7.14 | 32.14 | 60.71 | 4.25 | 5.56 |
| 9 | 87384534 | G | A | G | rs10746750 | intronic | 0.59 | 0.41 | NTRK2 | 4.61E-04 | 0.856 | 0.956 | 0.963 | 1.000 | 16 | 36 | 14 | 1 | 8 | 16 | 24.64 | 53.62 | 21.74 | 7.14 | 32.14 | 60.71 | 4.25 | 5.56 |
| 9 | 87392582 | C | T | C | rs10465180 | intronic | 0.63 | 0.37 | NTRK2 | 4.61E-04 | 0.856 | 0.956 | 0.963 | 1.000 | 16 | 36 | 14 | 1 | 8 | 16 | 24.64 | 53.62 | 21.74 | 7.14 | 32.14 | 60.71 | 4.25 | 5.56 |
| 9 | 87394720 | G | C | G | rs1778971 | intronic | 0.59 | 0.41 | NTRK2 | 4.61E-04 | 0.856 | 0.956 | 0.963 | 1.000 | 16 | 36 | 14 | 1 | 8 | 16 | 24.64 | 53.62 | 21.74 | 7.14 | 32.14 | 60.71 | 4.25 | 5.56 |
| 9 | 87395810 | A | G | A | rs1778970 | intronic | 0.59 | 0.41 | NTRK2 | 4.61E-04 | 0.856 | 0.956 | 0.963 | 1.000 | 16 | 36 | 14 | 1 | 8 | 16 | 24.64 | 53.62 | 21.74 | 7.14 | 32.14 | 60.71 | 4.25 | 5.56 |

# Table S5 (D): Gene-based tests in EPILEPSY genes

| SYMBOL | # MUTATIONS | P VALUE | FWER |
|---|---|---|---|
| RTN4IP1 | 19 | 0.0056 | 0.7923 |
| DYRK1A | 30 | 0.0069 | 0.8579 |
| PEX6 | 4 | 0.0150 | 0.9856 |
| KCNB1 | 35 | 0.0210 | 0.9974 |
| TBC1D24 | 2 | 0.0212 | 0.9976 |
| PRRT2 | 1 | 0.0303 | 0.9998 |
| SLC13A5 | 3 | 0.0343 | 0.9999 |
| PRICKLE1 | 8 | 0.0420 | 1.0000 |
| GALC | 8 | 0.0475 | 1.0000 |
| ALG6 | 4 | 0.0541 | 1.0000 |
| SETD5 | 5 | 0.0563 | 1.0000 |
| MDH2 | 2 | 0.0597 | 1.0000 |
| SURF1 | 2 | 0.0683 | 1.0000 |
| CHRNB2 | 1 | 0.0685 | 1.0000 |
| PIGO | 1 | 0.0697 | 1.0000 |
| PSAP | 1 | 0.0701 | 1.0000 |
| SYNGAP1 | 3 | 0.0712 | 1.0000 |
| NHLRC1 | 6 | 0.0723 | 1.0000 |
| PAFAH1B1 | 4 | 0.0857 | 1.0000 |
| GRIN2B | 55 | 0.0901 | 1.0000 |
| PCDH12 | 1 | 0.0941 | 1.0000 |
| IDH2 | 2 | 0.0960 | 1.0000 |
| RNASEH2C | 1 | 0.1006 | 1.0000 |
| EPM2A | 11 | 0.1089 | 1.0000 |
| KCNJ10 | 2 | 0.1098 | 1.0000 |
| ALG8 | 2 | 0.1136 | 1.0000 |
| CLN8 | 1 | 0.1162 | 1.0000 |
| COQ9 | 2 | 0.1162 | 1.0000 |
| GABRB2 | 95 | 0.1235 | 1.0000 |
| ARFGEF2 | 2 | 0.1289 | 1.0000 |
| NDUFV1 | 3 | 0.1289 | 1.0000 |
| NTRK2 | 111 | 0.1497 | 1.0000 |
| PEX7 | 2 | 0.1528 | 1.0000 |
| DYNC1H1 | 7 | 0.1535 | 1.0000 |
| DHX30 | 1 | 0.1630 | 1.0000 |
| SCN2A | 8 | 0.1671 | 1.0000 |
| FRRS1L | 4 | 0.1757 | 1.0000 |
| GLUL | 2 | 0.1796 | 1.0000 |
| TBCK | 10 | 0.1802 | 1.0000 |
| FAR1 | 1 | 0.1809 | 1.0000 |
| LYST | 3 | 0.1836 | 1.0000 |
| SLC12A5 | 2 | 0.1890 | 1.0000 |
| COL4A2 | 21 | 0.1898 | 1.0000 |
| STXBP1 | 10 | 0.2022 | 1.0000 |
| AKT1 | 2 | 0.2033 | 1.0000 |
| CIC | 2 | 0.2096 | 1.0000 |
| OCLN | 4 | 0.2120 | 1.0000 |

| Gene | Count | Value | P |
|---|---|---|---|
| MFF | 2 | 0.2171 | 1.0000 |
| KCNK4 | 1 | 0.2176 | 1.0000 |
| HLCS | 26 | 0.2329 | 1.0000 |
| TRPM6 | 6 | 0.2398 | 1.0000 |
| PCCB | 5 | 0.2403 | 1.0000 |
| TRAK1 | 4 | 0.2448 | 1.0000 |
| GOSR2 | 2 | 0.2558 | 1.0000 |
| AP3B2 | 5 | 0.2578 | 1.0000 |
| HACE1 | 6 | 0.2670 | 1.0000 |
| GTPBP2 | 1 | 0.2672 | 1.0000 |
| RMND1 | 2 | 0.2780 | 1.0000 |
| MOCS2 | 1 | 0.2781 | 1.0000 |
| SMARCC2 | 1 | 0.2818 | 1.0000 |
| LIAS | 1 | 0.2834 | 1.0000 |
| SETBP1 | 35 | 0.2839 | 1.0000 |
| AIMP1 | 2 | 0.2845 | 1.0000 |
| NAGA | 4 | 0.2919 | 1.0000 |
| DENND5A | 2 | 0.2945 | 1.0000 |
| ASPA | 2 | 0.2947 | 1.0000 |
| EIF3F | 4 | 0.2955 | 1.0000 |
| KIF5C | 5 | 0.3007 | 1.0000 |
| SCN3A | 2 | 0.3009 | 1.0000 |
| AKT3 | 36 | 0.3086 | 1.0000 |
| EEF1A2 | 3 | 0.3161 | 1.0000 |
| TUBB3 | 1 | 0.3163 | 1.0000 |
| ALDH5A1 | 6 | 0.3294 | 1.0000 |
| EPG5 | 9 | 0.3296 | 1.0000 |
| ARID1B | 52 | 0.3372 | 1.0000 |
| CNNM2 | 6 | 0.3418 | 1.0000 |
| GNB1 | 4 | 0.3438 | 1.0000 |
| DNM1L | 6 | 0.3441 | 1.0000 |
| SLC25A12 | 2 | 0.3477 | 1.0000 |
| NSD1 | 6 | 0.3554 | 1.0000 |
| STAG1 | 22 | 0.3685 | 1.0000 |
| COQ4 | 1 | 0.3739 | 1.0000 |
| TANGO2 | 1 | 0.3842 | 1.0000 |
| EIF2B5 | 15 | 0.3876 | 1.0000 |
| HSD17B4 | 7 | 0.3885 | 1.0000 |
| SLC25A22 | 2 | 0.4034 | 1.0000 |
| COL4A1 | 36 | 0.4116 | 1.0000 |
| SCN8A | 10 | 0.4148 | 1.0000 |
| AMT | 1 | 0.4157 | 1.0000 |
| HRAS | 1 | 0.4157 | 1.0000 |
| NDE1 | 2 | 0.4209 | 1.0000 |
| NPRL3 | 3 | 0.4213 | 1.0000 |
| PIK3R2 | 10 | 0.4223 | 1.0000 |
| RORB | 6 | 0.4236 | 1.0000 |
| HCN2 | 6 | 0.4346 | 1.0000 |

| Gene | Count | Value | P |
|---|---|---|---|
| TSEN2 | 3 | 0.4435 | 1.0000 |
| IFIH1 | 6 | 0.4476 | 1.0000 |
| POMT1 | 1 | 0.4554 | 1.0000 |
| CACNA1E | 41 | 0.4576 | 1.0000 |
| ADAR | 2 | 0.4603 | 1.0000 |
| DEAF1 | 3 | 0.4644 | 1.0000 |
| KARS | 2 | 0.4671 | 1.0000 |
| FBXO11 | 3 | 0.4680 | 1.0000 |
| RHOBTB2 | 1 | 0.4762 | 1.0000 |
| PDHX | 1 | 0.4770 | 1.0000 |
| PHACTR1 | 13 | 0.4879 | 1.0000 |
| ATP1A1 | 4 | 0.4892 | 1.0000 |
| D2HGDH | 2 | 0.4905 | 1.0000 |
| PEX19 | 4 | 0.4909 | 1.0000 |
| ALG11 | 1 | 0.4981 | 1.0000 |
| WDR62 | 1 | 0.4984 | 1.0000 |
| ZBTB18 | 1 | 0.4999 | 1.0000 |
| CAD | 1 | 0.5027 | 1.0000 |
| GABRA1 | 6 | 0.5037 | 1.0000 |
| PTS | 1 | 0.5065 | 1.0000 |
| ROGDI | 2 | 0.5065 | 1.0000 |
| TSC2 | 2 | 0.5134 | 1.0000 |
| BOLA3 | 2 | 0.5275 | 1.0000 |
| MTHFR | 1 | 0.5316 | 1.0000 |
| PRMT7 | 1 | 0.5321 | 1.0000 |
| TREX1 | 1 | 0.5357 | 1.0000 |
| RTTN | 6 | 0.5359 | 1.0000 |
| RNASET2 | 4 | 0.5373 | 1.0000 |
| KCTD3 | 2 | 0.5374 | 1.0000 |
| TSC1 | 1 | 0.5425 | 1.0000 |
| SCARB2 | 5 | 0.5434 | 1.0000 |
| MFSD8 | 2 | 0.5449 | 1.0000 |
| TSEN54 | 2 | 0.5494 | 1.0000 |
| PACS1 | 10 | 0.5497 | 1.0000 |
| CPA6 | 14 | 0.5516 | 1.0000 |
| NDUFAF2 | 17 | 0.5585 | 1.0000 |
| STRADA | 3 | 0.5586 | 1.0000 |
| MTOR | 4 | 0.5639 | 1.0000 |
| FARS2 | 41 | 0.5675 | 1.0000 |
| SZT2 | 6 | 0.5676 | 1.0000 |
| CHRNA2 | 3 | 0.5718 | 1.0000 |
| SEPSECS | 2 | 0.5833 | 1.0000 |
| ALDH7A1 | 6 | 0.5897 | 1.0000 |
| KCNJ11 | 2 | 0.5934 | 1.0000 |
| NACC1 | 4 | 0.5977 | 1.0000 |
| GABRG2 | 4 | 0.6056 | 1.0000 |
| SLC1A2 | 12 | 0.6065 | 1.0000 |
| PACS2 | 7 | 0.6074 | 1.0000 |

| Gene | Count | Value | P |
|---|---|---|---|
| RORA | 44 | 0.6079 | 1.0000 |
| STAMBP | 5 | 0.6188 | 1.0000 |
| BTD | 8 | 0.6306 | 1.0000 |
| SUCLA2 | 5 | 0.6335 | 1.0000 |
| GM2A | 3 | 0.6363 | 1.0000 |
| SCN9A | 10 | 0.6665 | 1.0000 |
| NARS2 | 8 | 0.6707 | 1.0000 |
| TBCD | 13 | 0.6804 | 1.0000 |
| PLAA | 18 | 0.6829 | 1.0000 |
| NRXN1 | 77 | 0.6977 | 1.0000 |
| EHMT1 | 13 | 0.7075 | 1.0000 |
| CHRNA4 | 14 | 0.7080 | 1.0000 |
| PHGDH | 4 | 0.7089 | 1.0000 |
| CTNNA2 | 104 | 0.7096 | 1.0000 |
| KCNQ2 | 9 | 0.7130 | 1.0000 |
| PPP3CA | 26 | 0.7133 | 1.0000 |
| PEX10 | 1 | 0.7184 | 1.0000 |
| MMADHC | 1 | 0.7209 | 1.0000 |
| FGF12 | 40 | 0.7212 | 1.0000 |
| PEX2 | 7 | 0.7237 | 1.0000 |
| ARG1 | 1 | 0.7302 | 1.0000 |
| ARV1 | 1 | 0.7302 | 1.0000 |
| GRIN2D | 1 | 0.7302 | 1.0000 |
| MOGS | 1 | 0.7348 | 1.0000 |
| GABBR2 | 52 | 0.7352 | 1.0000 |
| ALG9 | 3 | 0.7375 | 1.0000 |
| MOCS1 | 3 | 0.7376 | 1.0000 |
| WDR45B | 2 | 0.7491 | 1.0000 |
| COG7 | 1 | 0.7495 | 1.0000 |
| BCKDHB | 6 | 0.7555 | 1.0000 |
| NGLY1 | 7 | 0.7558 | 1.0000 |
| PNPO | 1 | 0.7632 | 1.0000 |
| KCNQ5 | 73 | 0.7653 | 1.0000 |
| UBA5 | 3 | 0.7681 | 1.0000 |
| PEX13 | 1 | 0.7700 | 1.0000 |
| ST3GAL5 | 21 | 0.7868 | 1.0000 |
| RELN | 35 | 0.7913 | 1.0000 |
| MAP2K2 | 1 | 0.7929 | 1.0000 |
| UBE3A | 13 | 0.7931 | 1.0000 |
| BRAT1 | 8 | 0.7952 | 1.0000 |
| CC2D2A | 1 | 0.8053 | 1.0000 |
| PEX1 | 3 | 0.8154 | 1.0000 |
| FUT8 | 10 | 0.8211 | 1.0000 |
| ALPL | 6 | 0.8235 | 1.0000 |
| ATP6V1A | 1 | 0.8279 | 1.0000 |
| BRAF | 8 | 0.8333 | 1.0000 |
| PCCA | 10 | 0.8353 | 1.0000 |
| EML1 | 17 | 0.8395 | 1.0000 |

| Gene | Count | Value1 | Value2 |
|---|---|---|---|
| PMM2 | 2 | 0.8427 | 1.0000 |
| CACNA1D | 18 | 0.8588 | 1.0000 |
| MACF1 | 10 | 0.8590 | 1.0000 |
| KCNQ3 | 42 | 0.8659 | 1.0000 |
| GNAO1 | 2 | 0.8733 | 1.0000 |
| PIGN | 6 | 0.8744 | 1.0000 |
| EMX2 | 2 | 0.8790 | 1.0000 |
| TBL1XR1 | 11 | 0.8817 | 1.0000 |
| RARS2 | 3 | 0.8826 | 1.0000 |
| MAP2K1 | 4 | 0.8827 | 1.0000 |
| KCNC1 | 2 | 0.8856 | 1.0000 |
| NDUFS4 | 3 | 0.8906 | 1.0000 |
| GPHN | 62 | 0.8916 | 1.0000 |
| MLC1 | 4 | 0.8939 | 1.0000 |
| RALA | 9 | 0.8992 | 1.0000 |
| ZEB2 | 2 | 0.9072 | 1.0000 |
| ACOX1 | 27 | 0.9095 | 1.0000 |
| PIGG | 5 | 0.9100 | 1.0000 |
| MTR | 9 | 0.9148 | 1.0000 |
| NBEA | 114 | 0.9155 | 1.0000 |
| DOCK7 | 10 | 0.9209 | 1.0000 |
| GRIN2A | 15 | 0.9219 | 1.0000 |
| CACNA1A | 19 | 0.9256 | 1.0000 |
| SYNJ1 | 3 | 0.9266 | 1.0000 |
| GRIN1 | 5 | 0.9294 | 1.0000 |
| SPTAN1 | 3 | 0.9369 | 1.0000 |
| KCNT1 | 2 | 0.9374 | 1.0000 |
| UNC80 | 6 | 0.9424 | 1.0000 |
| CREBBP | 8 | 0.9467 | 1.0000 |
| PIK3CA | 21 | 0.9479 | 1.0000 |
| GLDC | 6 | 0.9496 | 1.0000 |
| SLC6A19 | 2 | 0.9526 | 1.0000 |
| KCNA2 | 15 | 0.9546 | 1.0000 |
| SLC2A1 | 3 | 0.9552 | 1.0000 |
| CACNA1G | 2 | 0.9572 | 1.0000 |
| NDUFAF5 | 18 | 0.9579 | 1.0000 |
| KRAS | 7 | 0.9583 | 1.0000 |
| FBXL4 | 13 | 0.9583 | 1.0000 |
| RNASEH2B | 4 | 0.9592 | 1.0000 |
| SMARCA2 | 15 | 0.9601 | 1.0000 |
| QDPR | 8 | 0.9604 | 1.0000 |
| WASF1 | 6 | 0.9627 | 1.0000 |
| FKTN | 3 | 0.9627 | 1.0000 |
| WWOX | 224 | 0.9631 | 1.0000 |
| CHD2 | 7 | 0.9687 | 1.0000 |
| GCH1 | 4 | 0.9728 | 1.0000 |
| TCF4 | 27 | 0.9782 | 1.0000 |
| DIAPH1 | 10 | 0.9782 | 1.0000 |

| Gene | Count | Value1 | Value2 |
|---|---|---|---|
| DEPDC5 | 103 | 0.9810 | 1.0000 |
| DPYD | 46 | 0.9836 | 1.0000 |
| CYFIP2 | 8 | 0.9836 | 1.0000 |
| PAH | 4 | 0.9849 | 1.0000 |
| NDUFA10 | 4 | 0.9886 | 1.0000 |
| KIF1A | 13 | 0.9888 | 1.0000 |
| MBD5 | 56 | 0.9889 | 1.0000 |
| GLB1 | 8 | 0.9929 | 1.0000 |
| PTEN | 6 | 0.9957 | 1.0000 |
| HECW2 | 55 | 0.9965 | 1.0000 |
| PLCB1 | 55 | 0.9971 | 1.0000 |
| GRIA4 | 14 | 0.9984 | 1.0000 |
| HCN1 | 14 | 0.9985 | 1.0000 |
| MEF2C | 39 | 0.9991 | 1.0000 |
| KIF2A | 7 | 0.9994 | 1.0000 |
| GNAQ | 16 | 0.9996 | 1.0000 |
| MAGI2 | 175 | 0.9998 | 1.0000 |
| GABRB3 | 20 | 0.9999 | 1.0000 |
| GNB5 | 33 | 0.9999 | 1.0000 |
| CNTNAP2 | 372 | 1.0000 | 1.0000 |
| ABAT | 14 | 1.0000 | 1.0000 |
| AMPD2 | 1 | 1.0000 | 1.0000 |
| CLTC | 1 | 1.0000 | 1.0000 |
| DNM1 | 1 | 1.0000 | 1.0000 |
| EFTUD2 | 1 | 1.0000 | 1.0000 |
| FGFR3 | 1 | 1.0000 | 1.0000 |
| GSS | 1 | 1.0000 | 1.0000 |
| HEPACAM | 3 | 1.0000 | 1.0000 |
| HEXA | 2 | 1.0000 | 1.0000 |
| ITPA | 1 | 1.0000 | 1.0000 |
| LGI1 | 1 | 1.0000 | 1.0000 |
| MBOAT7 | 1 | 1.0000 | 1.0000 |
| OTUD6B | 1 | 1.0000 | 1.0000 |
| POMGNT1 | 1 | 1.0000 | 1.0000 |
| PPT1 | 1 | 1.0000 | 1.0000 |
| RAB18 | 1 | 1.0000 | 1.0000 |
| RFT1 | 3 | 1.0000 | 1.0000 |
| SLC35A1 | 2 | 1.0000 | 1.0000 |
| SLC6A1 | 4 | 1.0000 | 1.0000 |
| TUBB2A | 1 | 1.0000 | 1.0000 |
| VARS | 1 | 1.0000 | 1.0000 |
| WDR73 | 1 | 1.0000 | 1.0000 |

# Table S5 (E): Lists of SYNAPTIC and EPILEPSY genes

| SYNAPTIC | EPILEPSY |
|---|---|
| AC136616.1 | AARS |
| RAP1A | ADSL |
| SYT12 | ALDH7A1 |
| HTR7 | ALG11 |
| PPFIA2 | ALG13 |
| DNM3 | ARHGEF9 |
| SLC6A15 | ARX |
| SLC17A8 | ATP1A3 |
| KCNMB4 | ATP6V0A2 |
| DOC2A | ATRX |
| SLC30A1 | BRAT1 |
| CHRM3 | BSCL2 |
| SYN1 | CACNA1D |
| PRIMA1 | CDKL5 |
| SLC10A4 | CHD2 |
| ATP1A2 | CHRNA2 |
| CPT1C | CHRNA4 |
| IQSEC2 | CHRNB2 |
| PNKD | CIC |
| SLC6A13 | CLN8 |
| CACNG3 | CLTC |
| NF1 | CNKSR2 |
| HTR6 | CNTNAP2 |
| SLC6A11 | CPA6 |
| EFNB2 | CYFIP2 |
| SV2B | DEPDC5 |
| SYNJ1 | DIAPH1 |
| HRH4 | DNM1 |
| ARHGAP44 | DOCK7 |
| FLOT1 | DPYD |
| DTNBP1 | DYRK1A |
| SYT8 | EHMT1 |
| CASK | EML1 |
| P2RY11 | EPG5 |
| CPLX3 | EPM2A |
| CHAT | FOXG1 |
| SLC6A6 | GABBR2 |
| GRIP2 | GABRA1 |
| PPT1 | GABRB3 |
| SLC18A3 | GABRG2 |
| SLC6A20 | GLYCTK |
| FRRS1L | GNAO1 |
| CTNND1 | GPAA1 |
| PRKCI | GRIN1 |
| SLC6A1 | GRIN2A |
| RIMS1 | GRIN2B |
| NRXN3 | GSS |
| GPC6 | HCN1 |
| SLC18A1 | HECW2 |
| LRTOMT | HMGCL |
| DDC | HNRNPH2 |
| SV2C | HNRNPU |
| HTR1E | HTRA2 |
| BRSK1 | IDH2 |
| RIMS2 | IER3IP1 |
| HRH1 | IQSEC2 |
| DAGLB | ITPA |
| EPS15 | KCNA2 |
| COMT | KCNB1 |
| KIF17 | KCNC1 |
| NUMB | KCNJ10 |
| SLC6A4 | KCNQ2 |
| GFAP | KCNQ3 |
| CAMK2A | KCNT1 |
| SLC6A7 | KCTD7 |
| SHISA9 | KIF1BP |
| FNTA | LGI1 |
| LGI1 | MBD5 |
| SNCG | MBOAT7 |
| STX1A | MDH2 |
| DLG4 | MECP2 |
| MAOA | MEF2C |
| ALDH5A1 | MFF |
| MAOB | MOGS |
| SRC | MTOR |
| SLC1A2 | NACC1 |
| SLC6A19 | NEXMIF |
| AP2B1 | PCDH19 |
| HCRT | PIGA |
| GPER1 | PIGN |
| CHRM5 | PIGT |
| SLC6A18 | PLCB1 |
| DRD3 | PLPBP |
| BAIAP3 | PNKP |
| SHISA6 | PNPO |
| NAALAD2 | POLG |
| CPLX4 | PRODH |
| OR11H4 | PRRT2 |
| NRXN2 | PURA |
| SLC1A3 | QARS |
| CACNG5 | RANBP2 |
| CACNG4 | SCARB2 |
| SHISA7 | SCN1A |
| LYNX1 | SCN1B |
| SLURP2 | SCN2A |
| STX3 | SCN8A |
| SYN3 | SCN9A |
| PRKCZ | SETD5 |
| UNC13B | SIK1 |
| NSG1 | SLC12A5 |
| SYT17 | SLC13A5 |
| UNC13A | SLC16A2 |
| GCHFR | SLC1A2 |
| NPTXR | SLC25A1 |
| DRD1 | SLC25A22 |
| PPFIA1 | SLC2A1 |
| CHRNB4 | SLC35A2 |
| GRM1 | SLC6A1 |
| SLC6A3 | SLC6A19 |
| APBA1 | SLC9A6 |
| HRH2 | SPTAN1 |
| CPLX2 | STRADA |
| YWHAE | STX1B |
| PDZD11 | STXBP1 |
| OR56A5 | SUOX |
| CPLX1 | SYNGAP1 |
| OR56A4 | SYNJ1 |
| OR56A1 | SZT2 |
| SLC6A9 | TBC1D24 |
| RAB3A | TCF4 |
| ADAM10 | TPP1 |
| DRD2 | TRAK1 |
| SLC1A6 | TRPM6 |
| VAMP2 | UBE2A |
| SLC6A8 | UBE3A |
| RHOT1 | WDR45 |
| DNM1 | WDR45B |
| ATP2A2 | WWOX |
| DLG3 | ZEB2 |
| SV2A | ABAT |
| CACNG2 | ACOX1 |
| SLC6A16 | ADAR |
| STX1B | ADGRG1 |
| NPTX2 | ADPRHL2 |
| GRIA1 | AIMP1 |
| HTR1D | AKT1 |
| NPTX1 | AKT3 |
| STX4 | ALDH5A1 |
| OR6T1 | ALG1 |
| ITGB3 | ALG3 |
| SLC38A1 | ALG6 |
| SNAP25 | ALG8 |
| SYT7 | ALG9 |
| ARC | ALPL |
| PSCA | AMPD2 |
| SCRIB | AMT |
| RAB3GAP1 | AP3B2 |
| GABBR1 | ARFGEF2 |
| PPP1R9A | ARG1 |
| DNAJC5 | ARID1B |
| GRIPAP1 | ARV1 |
| HTR2C | ASPA |
| CACNG7 | ATP1A1 |
| RIMS3 | ATP6V1A |
| HRAS | ATP7A |
| CACNG8 | BCKDHA |
| CHRM2 | BCKDHB |
| PATE1 | BCS1L |
| PATE4 | BOLA3 |
| DAGLA | BRAF |
| OR10H2 | BTD |
| OR10H3 | C12orf57 |
| GRM4 | CACNA1A |
| OR10H5 | CACNA1E |
| SLC6A14 | CACNA1G |
| GABRQ | CAD |
| AC114267.1 | CASK |
| EPS8 | CC2D2A |
| SYN2 | CLCN4 |
| CACNA1B | CLN3 |
| OR13F1 | CNNM2 |
| DGKI | COG7 |
| HRH3 | COL18A1 |
| WNT7A | COL4A1 |
| LIN7B | COL4A2 |
| NRG1 | COQ2 |
| GRASP | COQ4 |
| GAD2 | COQ9 |
| HPCA | CREBBP |
| HOMER1 | CSTB |
| PORCN | CTNNA2 |
| HTR4 | CTSD |
| OR10H4 | D2HGDH |
| KIF5C | DCX |
| SNAPIN | DDX3X |
| SLC17A7 | DEAF1 |
| LYPD6B | DENND5A |
| ACHE | DHCR7 |
| PPFIA3 | DHDDS |
| KIF5A | DHX30 |
| AP3D1 | DNM1L |
| GPHN | DPAGT1 |
| HTR1F | DPM1 |
| LYPD6 | DYNC1H1 |
| FLOT2 | EARS2 |
| SNCA | EEF1A2 |
| RIMS4 | EFTUD2 |
| LRRC7 | EIF2B2 |
| ICA1 | EIF2B4 |
| TH | EIF2B5 |
| CNIH2 | EIF2S3 |
| DRD4 | EIF3F |
| RPH3A | EMX2 |
| SLC29A1 | ETHE1 |
| DVL1 | EXOSC3 |
| GHSR | FAR1 |
| NAAA | FARS2 |
| COLQ | FBXL4 |
| CLSTN1 | FBXO11 |
| SLC22A2 | FGF12 |
| SYT5 | FGFR3 |
| HIP1 | FH |
| DNM2 | FKTN |
| SLC1A7 | FLNA |
| CHRM1 | FOLR1 |
| SLC18A2 | FRRS1L |
| ASIC1 | FUCA1 |
| DLG2 | FUT8 |
| ABAT | GABRB2 |
| AKAP9 | GALC |
| SYT4 | GAMT |
| SLC29A2 | GBA |
| RAB8A | GCH1 |
| HTR1B | GFAP |
| DOC2B | GFM1 |
| NRXN1 | GLB1 |
| OPHN1 | GLDC |
| DMD | GLUD1 |
| OR5T2 | GLUL |
| OR5T3 | GM2A |
| OR5T1 | GNAQ |
| SLC17A6 | GNB1 |
| GABRA2 | GNB5 |
| DRD5 | GOSR2 |
| SLC6A2 | GPHN |
| SNCAIP | GRIA4 |
| MEF2C | GRIN2D |
| AGTPBP1 | GTPBP2 |
| STXBP1 | HACE1 |
| SLC22A1 | HAX1 |
| CALY | HCFC1 |
| VPS35 | HCN2 |
| PAH | HEPACAM |
| SYT2 | HEXA |
| SLC6A5 | HLCS |
| ZNF219 | HRAS |
| GRIP1 | HSD17B4 |
| HSPA8 | IFIH1 |
| GRIN1 | IKBKG |
| SLC32A1 | IRF2BPL |
| ALDH9A1 | KARS |
| SLC6A17 | KCNA1 |
| SNAP47 | KCNJ11 |
| KCNC4 | KCNK4 |
| PPFIA4 | KCNQ5 |
| RAB11A | KCTD3 |
| HTR5A | KIF1A |
| NOS1 | KIF2A |
| ADORA1 | KIF5C |
| HTR2B | KRAS |
| CHRM4 | LIAS |
| ADRB1 | LYST |
| ERBIN | MACF1 |
| HTR2A | MAF |
| CLN3 | MAGI2 |
| MCTP1 | MAP2K1 |
| HTR1A | MAP2K2 |
| CACNA1A | MED12 |
| SLC6A12 | MFSD8 |
| DAG1 | MLC1 |
| GRM5 | MMACHC |
| GRIN3B | MMADHC |
| MAPK10 | MOCS1 |
| NIPSNAP1 | MOCS2 |
| SNPH | MPDU1 |
| SLC5A7 | MTHFR |
| USP46 | MTR |
| PDE1B | NAGA |
| LIN7A | NARS2 |
| SYT11 | NBEA |
| GLUL | NDE1 |
| FMR1 | NDUFA1 |
| PTPRN2 | NDUFA10 |
| PRKN | NDUFAF2 |
| SYT1 | NDUFAF5 |
| GAD1 | NDUFS4 |
| PEBP1 | NDUFS8 |
| LIN7C | NDUFV1 |
| DLG1 | NGLY1 |
| OR10J5 | NHLRC1 |
| GSG1L | NPRL3 |
| GPC4 | NRXN1 |
| TSPOAP1 | NSD1 |
| AP2M1 | NSDHL |
| | NTRK2 |
| | OCLN |
| | OPHN1 |
| | OTUD6B |
| | PACS1 |
| | PACS2 |
| | PAFAH1B1 |
| | PAH |
| | PCCA |
| | PCCB |
| | PCDH12 |
| | PDHA1 |
| | PDHX |
| | PET100 |
| | PEX1 |
| | PEX10 |
| | PEX12 |
| | PEX13 |
| | PEX19 |
| | PEX2 |
| | PEX3 |
| | PEX6 |
| | PEX7 |
| | PHACTR1 |
| | PHGDH |
| | PIGG |
| | PIGO |
| | PIGW |
| | PIK3CA |
| | PIK3R2 |
| | PLAA |
| | PMM2 |
| | POMGNT1 |
| | POMT1 |
| | PPP3CA |
| | PPT1 |
| | PRICKLE1 |
| | PRMT7 |
| | PSAP |
| | PTEN |

PTPN23
PTS
QDPR
RAB11B
RAB18
RALA
RARS2
RELN
RFT1
RHOBTB2
RMND1
RNASEH2A
RNASEH2B
RNASEH2C
RNASET2
ROGDI
RORA
RORB
RTN4IP1
RTTN
SAMHD1
SCN3A
SCO1
SCO2
SEPSECS
SETBP1
SLC1A4
SLC25A12
SLC35A1
SLC6A8
SMARCA2
SMARCC2
SMS
SNORD118
ST3GAL5
STAG1
STAMBP
SUCLA2
SURF1
SYN1
TANGO2
TBCD
TBCK
TBL1XR1
TMEM70
TREX1
TRIM8
TSC1

TSC2
TSEN2
TSEN54
TUBA1A
TUBB2A
TUBB2B
TUBB3
TUBB4A
TUBG1
UBA5
UFM1
UNC80
VARS
WASF1
WDR62
WDR73
YWHAG
ZBTB18
ATN1
CSTB

# Table S6

| CHROM | POS | REF | ALT | MINOR ALLELE | ID | CONSEQUENCE | AF | MAF | SYMBOL | P VALUE | FWER | | | | rPPA (flat prior) | | | rPPA (empirical prior) | | |
|---|---|---|---|---|---|---|---|---|---|---|---|---|---|---|---|---|---|---|---|---|
| | | | | | | | | | | | Neff=160,965 (Galwey) | Neff=258,715 (Gao) | Neff=357,717 (Li & Ji) | Neff=3,577,371 (Nyholt/Cheverud) | $\pi$=1e-4 | $\pi$=1e-5 | $\pi$=1e-6 | $\pi$=1e-4 | $\pi$=1e-5 | $\pi$=1e-6 |
| 17 | 4350407 | G | C | G | rs2047231 | intronic | 0.82 | 0.18 | SPNS3 | 1.60E-07 | 0.026 | 0.041 | 0.056 | 0.437 | 0.80 | 0.29 | 0.04 | 0.97 | 0.78 | 0.26 |
| 15 | 50553000 | G | A | A | rs7182203 | intronic | 0.26 | 0.26 | HDC | 2.43E-07 | 0.038 | 0.061 | 0.083 | 0.581 | 0.96 | 0.73 | 0.21 | 0.98 | 0.84 | 0.35 |
| 4 | 8291102 | C | T | T | rs7683388 | intronic | 0.39 | 0.39 | HTRA3 | 1.52E-06 | 0.217 | 0.326 | 0.420 | 0.996 | 0.74 | 0.22 | 0.03 | 0.95 | 0.65 | 0.16 |
| 17 | 4350182 | T | C | T | rs2047233 | intronic | 0.88 | 0.12 | SPNS3 | 1.58E-06 | 0.224 | 0.335 | 0.431 | 0.996 | 0.25 | 0.03 | 0.00 | 0.77 | 0.25 | 0.03 |
| 5 | 179619074 | T | C | C | rs34570575 | intronic | 0.10 | 0.10 | RASGEF1C | 2.87E-06 | 0.370 | 0.524 | 0.642 | 1.000 | 0.22 | 0.03 | 0.00 | 0.64 | 0.15 | 0.02 |
| 3 | 37604012 | T | G | G | rs9825420 | intronic | 0.18 | 0.18 | ITGA9 | 3.32E-06 | 0.414 | 0.577 | 0.695 | 1.000 | 0.23 | 0.03 | 0.00 | 0.68 | 0.17 | 0.02 |
| 4 | 4345379 | T | C | C | rs7695197 | upstream | 0.31 | 0.31 | NSG1 | 3.55E-06 | 0.436 | 0.601 | 0.720 | 1.000 | 0.63 | 0.15 | 0.02 | 0.60 | 0.13 | 0.01 |
| 4 | 4346427 | G | A | A | rs3981 | upstream | 0.32 | 0.32 | NSG1 | 3.84E-06 | 0.461 | 0.630 | 0.747 | 1.000 | 0.61 | 0.14 | 0.02 | 0.57 | 0.12 | 0.01 |
| 4 | 4346465 | T | C | C | rs12641832 | upstream | 0.33 | 0.33 | NSG1 | 3.84E-06 | 0.461 | 0.630 | 0.747 | 1.000 | 0.61 | 0.14 | 0.02 | 0.57 | 0.12 | 0.01 |
| 4 | 113961039 | G | T | G | rs10015551 | intronic, non-coding transcript | 0.57 | 0.43 | RP11-650J17.1 | 4.72E-06 | 0.532 | 0.705 | 0.815 | 1.000 | 0.46 | 0.08 | 0.01 | 0.69 | 0.19 | 0.02 |
| 15 | 99100580 | A | G | A | rs2311753 | intergenic | 0.79 | 0.21 | | 5.49E-06 | 0.587 | 0.758 | 0.860 | 1.000 | 0.47 | 0.08 | 0.01 | 0.84 | 0.34 | 0.05 |
| 15 | 99101462 | A | G | A | rs2871859 | intergenic | 0.79 | 0.21 | | 5.49E-06 | 0.587 | 0.758 | 0.860 | 1.000 | 0.47 | 0.08 | 0.01 | 0.84 | 0.34 | 0.05 |
| 17 | 4350367 | A | G | A | rs2047232 | intronic | 0.88 | 0.12 | SPNS3 | 5.64E-06 | 0.597 | 0.768 | 0.867 | 1.000 | 0.09 | 0.01 | 0.00 | 0.45 | 0.08 | 0.01 |
| 2 | 156765630 | T | C | C | rs16839725 | intergenic | 0.29 | 0.29 | | 5.90E-06 | 0.613 | 0.783 | 0.879 | 1.000 | 0.48 | 0.08 | 0.01 | 0.29 | 0.04 | 0.00 |
| 14 | 47620548 | C | T | C | rs1952220 | intronic | 0.87 | 0.13 | MDGA2 | 6.25E-06 | 0.634 | 0.801 | 0.893 | 1.000 | 0.19 | 0.02 | 0.00 | 0.09 | 0.01 | 0.00 |
| 15 | 99102128 | A | G | A | rs11853887 | intergenic | 0.76 | 0.24 | | 6.52E-06 | 0.650 | 0.815 | 0.903 | 1.000 | 0.45 | 0.08 | 0.01 | 0.83 | 0.32 | 0.05 |
| 4 | 113961811 | C | T | C | rs10433900 | intronic, non-coding transcript | 0.52 | 0.48 | RP11-650J17.1 | 6.81E-06 | 0.666 | 0.828 | 0.912 | 1.000 | 0.44 | 0.07 | 0.01 | 0.66 | 0.16 | 0.02 |
| 3 | 37603952 | T | A | A | rs9825270 | intronic | 0.19 | 0.19 | ITGA9 | 7.07E-06 | 0.679 | 0.839 | 0.920 | 1.000 | 0.08 | 0.01 | 0.00 | 0.39 | 0.06 | 0.01 |
| 4 | 16428073 | G | A | A | rs73224601 | intronic, non-coding transcript | 0.10 | 0.10 | RP11-446J8.1 | 7.46E-06 | 0.699 | 0.855 | 0.931 | 1.000 | 0.07 | 0.01 | 0.00 | 0.55 | 0.11 | 0.01 |
| 3 | 72015276 | C | A | A | rs2036594 | intergenic | 0.17 | 0.17 | | 8.48E-06 | 0.745 | 0.889 | 0.952 | 1.000 | 0.11 | 0.01 | 0.00 | 0.57 | 0.12 | 0.01 |
| 19 | 46569619 | A | G | A | rs713409 | intronic | 0.53 | 0.47 | IGFL4 | 8.91E-06 | 0.762 | 0.900 | 0.959 | 1.000 | 0.37 | 0.06 | 0.01 | 0.70 | 0.19 | 0.02 |
| 19 | 46569630 | A | G | A | rs713411 | intronic | 0.53 | 0.47 | IGFL4 | 8.91E-06 | 0.762 | 0.900 | 0.959 | 1.000 | 0.37 | 0.06 | 0.01 | 0.70 | 0.19 | 0.02 |
| 1 | 156486009 | G | A | G | rs6685228 | downstream | 0.76 | 0.24 | RP11-284F21.8 | 9.82E-06 | 0.794 | 0.921 | 0.970 | 1.000 | 0.06 | 0.01 | 0.00 | 0.31 | 0.04 | 0.00 |